\documentclass[11pt,axodraw]{cernrepNum}
\usepackage{graphicx}
\usepackage{here}
\usepackage{color}
\usepackage{amsmath}
\usepackage{longtable}
\usepackage{epsfig}
\usepackage{axodraw}
\newcommand{\raw}{\rightarrow}
\newcommand{\nn}{\nonumber}

\newcommand{\gev}{ {\rm GeV} }

\newcommand{\km}{ {\rm km} }

\newcommand{\meter}{ {\rm m} }

\newcommand{\cm}{ {\rm cm} }

\newcommand{\bea}{\begin{eqnarray}}
\newcommand{\eea}{\end{eqnarray}}
\newcommand{\be}{\begin{equation}}
\newcommand{\ee}{\end{equation}}
\newcommand{\tetaot}{\mbox{$\theta_{13}$}}

\newcommand{\tsun}{\mbox{$\theta_{12}$}}

\newcommand{\delot}{\mbox{$\Delta_{13}$}}

\newcommand{\datm}{\mbox{$\Delta_{atm}^2$}}

\newcommand{\numu}{\mbox{$\nu_{\mu}$}}
\newcommand{\nutau}{\mbox{$\nu_{\tau}$}}
\newcommand{\numubar}{\mbox{$\overline{\nu}_{\mu}$}}
\newcommand{\nue}{\mbox{$\nu_{e}$}}
\newcommand{\nuebar}{\mbox{$\overline{\nu}_{e}$}}
\newcommand{\muminus}{\mbox{$\mu^{-}$}}
\newcommand{\muplus}{\mbox{$\mu^{+}$}}

\newcommand{\num}{\mbox{${\nu}_{\mu}$}}

\newcommand{\gsim}{\mbox{\raisebox{-1.ex}
{$\stackrel{\textstyle >}{\textstyle \sim}$}}}

\def\etal{{\it et al.}}

\def\tetaot{\mbox{$\theta_{13}$}}

\def\delot{\mbox{$\Delta_{23}$}}

\def\datm{\mbox{$\Delta_{atm}^2$}}

\begin{document}
\title{OSCILLATION PHYSICS WITH A NEUTRINO FACTORY}
\author{
M.~Apollonio$^1$, A.~Blondel$^2$, A.~Broncano$^3$,M.~Bonesini$^4$, 
J.~Bouchez$^5$, A.~Bueno$^6$,
J.~Burguet-Castell$^7$, M.~Campanelli$^{2,*}$, D.~Casper$^8$,
G.~Catanesi$^9$,  A.~Cervera$^{10}$,
S.~Cooper$^{11}$,M.~Donega$^2$,A.~Donini$^{12}$,
A.~de~Gouv\^ea$^{13}$, A.~de~Min$^{14}$,
R.~Edgecock$^{15,16}$, J.~Ellis$^{17}$,
M.~Fechner$^{18}$, E.~Fernandez$^{19}$,
F.~Ferri$^4$,
B.~Gavela$^3$, G.~Giannini$^1$, D.~Gibin$^{14}$, S.~Gilardoni$^{2,16}$,
J.~J.~G\'omez-Cadenas$^{2,7}$, P.~Gruber$^{16}$, A.~Guglielmi$^{14}$,
P.~Hern\'andez$^{17}$,
P.~Huber$^{20}$, M.~Laveder$^{14}$, 
M.~Lindner$^{20}$, S.~Lola$^{17,\dagger}$, D.~Meloni$^{12}$, O.~Mena$^3$,
H.~Menghetti$^{21}$, M.~Mezzetto$^{14}$,
P.~Migliozzi$^{22}$,
S.~Navas-Concha$^6$,
V.~Palladino$^{23}$, I.Papadopoulos$^{24}$, K.Peach$^{15}$, 
E.~Radicioni$^9$, S.~Ragazzi$^4$, S.~Rigolin$^{17}$, A.~Romanino$^{25}$, 
J.~Rico$^6$, A.~Rubbia$^6$, G.~Santin$^1$, G.~Sartorelli$^{21}$, 
M.~Selvi$^{21}$, M.Spiro$^5$, T.~Tabarelli$^4$,
A.~Tonazzo$^4$,
M.~Velasco$^{26}$, G.~Volkov$^{27}$, W.~Winter$^{20}$,
P.~Zucchelli$^{10,28}$
}
\vspace{5cm}
\institute{\begin{center}
~\\
$^1$ University of Trieste and INFN Trieste, Italy\\
$^2$ DPNC, University of Geneva, Switzerland\\
$^3$ Dep. de F\'{\i}sica Te\'orica, Univ. Aut\'onoma de Madrid, Madrid, 
Spain\\
$^4$ University of Milano 2 Bicocca and INFN Milano, Italy\\
$^5$ DAPNIA, CEA Saclay, France\\
$^6$ ETH Zurich, Switzerland\\
$^7$ University of Valencia and IFIC Valencia, Spain\\
$^8$ University of California, Irvine, California, USA\\ 
$^9$ INFN Bari, Italy\\
$^{10}$ CERN EP, Geneva, Switzerland\\
$^{11}$ Oxford University, Oxford, United Kingdom\\
$^{12}$ University of Roma I and INFN Roma I, Italy\\
$^{13}$ Fermilab, Batavia, Illinois, USA\\
$^{14}$ University of Padova and INFN Padova, Italy\\
$^{15}$ Rutherford Appleton Laboratory, United Kingdom\\
$^{16}$ CERN PS, Geneva, Switzerland\\
$^{17}$ CERN TH, Geneva, Switzerland\\
$^{18}$ D\'epartement de Physique de l'Ecole Normale Sup\'erieure,
Paris, France\\
$^{19}$ Universidad Autonoma de Barcelona, Spain\\
$^{20}$ Tech. U. M\"unchen, Munich, Germany\\
$^{21}$ University of Bologna and INFN Bologna, Italy\\
$^{22}$ INFN Napoli, Italy\\
$^{23}$ University of Napoli, Italy\\
$^{24}$ CERN IT, Geneva, Switzerland\\
$^{25}$ Scuola Normale Superiore, Pisa, Italy\\
$^{26}$ Northwestern University, Evanston, Illinois, USA\\
$^{27}$ IFVE, Protvino, Russia\\
$^{28}$ INFN Ferrara, Italy\\
$^\dagger$ now at HR division, CERN, Geneva, Switzerland\\
~\\
$^*$ Editor: {\tt Mario.Campanelli@cern.ch}
\end{center}}
\maketitle
\vspace{-.5cm}
\begin{flushright}
{CERN-TH/2002-208, hep-ph/0210192}
\end{flushright}
\vspace{-.5cm}

\newpage

\begin{abstract}

A generation of neutrino experiments have established that neutrinos mix
and probably have mass. The mixing phenomenon points to processes beyond
those of the Standard Model, possibly at the Grand Unification energy
scale. A extensive sequence of of experiments will be required to measure
precisely all the parameters of the neutrino mixing matrix, culminating
with the discovery and study of leptonic CP violation. As a first step,
extensions of conventional pion/kaon decay beams, such as off-axis beams
or low-energy super-beams, have been considered. These could yield first
observations of $\nu_\mu \to \nu_e$ transitions at the atmospheric
frequency, which have not yet been observed, and a first measurement of
$\theta_{13}$.  Experiments with much better flux control can be envisaged
if the neutrinos are obtained from the decays of stored particles. One
such possibility is the concept of beta beams provided by the decays of
radioactive nuclei, that has been developed within the context of these
studies. These would provide a pure (anti-)electron-neutrino beam of a few
hundred MeV, and beautiful complementarity with a high-intensity,
low-energy conventional beam, enabling experimental probes of T violation
as well as CP violation. Ultimately, a definitive and complete set of
measurements would offered by a Neutrino Factory based on a muon storage
ring. This powerful machine offers the largest reach for CP violation,
even for very small values of $\theta_{13}$.

\end{abstract}
 
\newpage
\tableofcontents
\newpage
\section{INTRODUCTION}

Neutrino experiments over 30 years~\cite{Nobel}, culminating with the 
Super-Kamiokande 
atmospheric neutrino data~\cite{SK_atm}, have provided,
for the first time, unambiguous evidence for the existence of physics
beyond the Standard Model. This comes from the fact that the $\nu_{\mu}$
to $\nu_e$ flux ratio is far from theoretical expectations, in combination
with the nontrivial angular dependence of the atmospheric $\nu_{\mu}$
flux. This `atmospheric neutrino puzzle' cannot be explained by standard
means, such as changing the cosmic ray spectrum, or improving the
atmospheric neutrino flux computations. Furthermore, the Super-Kamiokande
result is in good agreement with other, less precise, measurements of the
atmospheric neutrino flux \cite{osc_data_atm}.

On a different front, solar neutrino experiments
\cite{osc_data_solar,SNO_res} have consistently been measuring solar
$\nu_e$ fluxes which are significantly smaller than those predicted by
theory \cite{SSM}.  It is equally hard to explain this `solar
neutrino puzzle' by traditional means (dramatically modifying the
currently accepted solar models, questioning the estimation of systematic
effects by some of the experiments, etc.).  The recent measurement of the
solar neutrino flux via neutrino -- deuteron scattering performed by the
SNO Collaboration in both charged-current and neutral-current processes
provides unambiguous evidence (at the five-sigma level) that there are
active neutrinos other than $\nu_e$ coming from the Sun
\cite{SNO_res,SNO_nc}.

Neutrino oscillations provide the simplest and most elegant solution to
{\sl both} the atmospheric and solar neutrino puzzles. Neutrino
oscillations take place if neutrinos have non-degenerate masses and,
similar to what happens in the quark sector, the neutrino mass eigenstates
differ from the neutrino weak, or flavour, eigenstates. Although less
standard solutions to the atmospheric and the solar neutrino puzzles, such
as exotic neutrino decays, or flavour-violating
interactions~\cite{nonstandard} may still be advocated, no satisfactory
single solution to both anomalies other than neutrino oscillations is
known.

The implications of the neutrino data are extremely interesting, since
they point towards non-zero neutrino masses, which are {\it prima facie}
evidence for physics beyond the Standard Model. In the absence of
right-handed neutrinos, $\nu_R$, no Dirac neutrino mass can be generated,
while the transformation properties of the left-handed neutrinos, $\nu_L$,
under ${ SU(2) \times U(1)}$ also forbid a renormalisable Majorana mass
term. On the other hand, non-zero neutrino masses arise naturally in many
extensions of the Standard Model, which generically contain an extended
lepton and/or Higgs sector, and possibly new lepton number-violating
interactions.

A large number of analyses of the solar, atmospheric and reactor neutrino
data can be found in the literature \cite{guid}, including two-flavour and
three-flavour analyses of the solar data \cite{osc_anal_solar},
two-flavour analyses of the atmospheric data \cite{osc_anal_atm},
three-flavour analyses of the combined atmospheric and reactor data
\cite{osc_anal_atm_reac} and combined analyses of all neutrino data
\cite{osc_anal_comb}.  It turns out that both the solar and the
atmospheric neutrino deficits can be accommodated in minimal schemes with
three light neutrinos, which may have either of the following hierarchical 
patterns of masses:

(a) The normal hierarchy, in which the masses are fixed by the 
mass differences required for
the atmospheric and solar deficits. The atmospheric neutrino data require
$m_{3} \approx 10^{-1} \; {\rm to} \; 10^{-1.5}$ eV, while $m_{2}$ is
determined by the solar neutrino squared-mass difference.

(b) Inverted hierarchy solutions, in which $|m_1| \sim |m_2| \gg |m_3|$,
where $m^2_{1,2} \sim \Delta m^2_{atm}$ and $\Delta m_{12} = \Delta
m^{2}_{sun}$. 

Since oscillation experiments are only sensitive to mass differences of
two neutrino species and not to the absolute values of the masses, normal
and inverted solutions with near-degenerate masses are also allowed.
 
Information about the absolute neutrino mass scale can be derived from 
direct searches, and experiments looking for
neutrinoless double-beta decay if neutrinos are
Majorana particles, are sensitive to neutrino masses ${\cal O} 
({\rm eV})$ or below. If the masses are close to the upper boundary of
this limit, neutrinos may still provide a significant component of hot
dark matter;
next generation experiments looking for direct neutrino mass will probe
this scenario, which is however disfavoured by the most recent
cosmological observations.

Neutrino oscillations (and other types of new physics in the neutrino
sector) can potentially be observed in terrestrial neutrino experiments,
by studying, for example, the flux of $\bar{\nu}_e$ coming from nuclear
reactors \cite{osc_data_reactor} or studying the $\nu_{\mu}$ flux from
pion or muon decays \cite{osc_data_baseline}. The K2K experiment has
reported first results that are consistent with the atmospheric neutrino
data~\cite{K2K}. So far, no other terrestrial evidence for oscillations 
has been
confirmed, although the current results significantly constrain the
neutrino oscillation parameter space. However, the LSND Collaboration has
reported an anomalous flux of $\bar{\nu}_{e}$ from $\mu^+$ decays, which
may be interpreted as evidence for
$\bar{\nu}_{\mu}\leftrightarrow\bar{\nu}_{e}$ oscillations~\cite{LSND}.
This experimental evidence has not yet been independently confirmed, but
will be put to the test in the near future~\cite{miniboone}.

The minimal schemes with only three neutrino masses allow only two
independent mass differences, and thus the oscillation interpretation
of the LSND result cannot be
considered, unless a light sterile neutrino is introduced
\cite{osc_anal_sterile}. In this case, one has to take into account the
constraints from cosmological Big Bang Nucleosynthesis: a sterile neutrino
that mixes with an active one, thus being in equilibrium at the time of
nucleosynthesis, can change the abundance of primordially produced
elements, such as $^{4}$He and deuterium. The larger the mixing and the
mass differences between the sterile and active neutrinos, the bigger the
deviations from the observed light element abundances. This implies that
models where the sterile component contributes to solar rather than
atmospheric neutrino oscillations are accommodated more easily within the
standard nucleosynthesis scenarios. However, recent data on solar neutrino
oscillations disfavor the sterile solution also in this case. In our
discussion, we focus on the minimal schemes with only three light
neutrinos. The physics programme of the Neutrino Factory would be even 
richer if there were more light neutrinos.

We can hope that future neutrino data will
provide important information on the possible models.
Indeed, the main goals of the next generation of neutrino experiments 
include the determination of neutrino mass-squared 
differences and leptonic mixing angles. If neutrino
oscillations are indeed the solution to the neutrino puzzles, the measurement of
these fundamental parameters is of utmost importance.
Moreover, in general we may expect that
$P(\bar\nu_\alpha\to\bar\nu_\beta) \neq P(\nu_\alpha\to\nu_\beta)$
and that we may look for CP violation in a neutrino factory~\cite{CPviol}
by measuring observables of the type
\begin{equation}
A_{CP}\equiv\frac{P(\nu_e\rightarrow \nu_\mu)-
P(\bar\nu_e\rightarrow \bar\nu_\mu)}{P(\nu_e\rightarrow
 \nu_\mu)+
P(\bar\nu_e\rightarrow \bar\nu_\mu)}.
\end{equation}

In the near future, it is essential to:

$\bullet$~Confirm the atmospheric neutrino puzzle in a terrestrial
experiment and determine the `atmospheric' mixing parameters. This is
one of the driving forces behind the accelerator-based long-baseline
neutrino experiments, namely the K2K experiment~\cite{K2K}, which started
taking data in 1999 and already has presented some results, the Fermilab
NUMI/MINOS~\cite{MINOS} project, which is under construction and is
supposed to start data-taking in 2005, and the CNGS (CERN neutrinos to
Gran Sasso) project~\cite{CNGS}, which should start running in 2006. The
CNGS effort also aims to determine whether
$\nu_{\mu}\leftrightarrow\nu_{\tau}$ is the dominant oscillation channel
responsible for the atmospheric neutrino anomaly, by searching for the
appearance of $\tau$ leptons from an initially pure $\nu_{\mu}$ beam.

$\bullet$~Establish that the solar neutrino puzzle is indeed due to
neutrino oscillations and determine what are the `solar' mixing
parameters. This is one of the goals of the SNO experiment \cite{SNO}.
Other experiments will also contribute significantly to these goals, such
as KamLAND~\cite{KamLAND}, which may provide terrestrial confirmation for
the solar neutrino puzzle as long as the solar oscillation parameters lie
in the LMA region, and GNO, which may observe anomalous seasonal
variations of the solar neutrino flux~\cite{GNO}. New experiments, such as
Borexino~\cite{borexino}, which is already under construction and should
start taking data in 2003, and a possible upgrade of the
KamLAND~\cite{KamLAND} experiment, may also observe anomalous seasonal
variations~\cite{seasonal} or a day-night variation~\cite{day-night} of
the $^7$Be solar neutrino flux. It should also be mentioned that a
`background-free' version of Borexino, or an experiment to measure the
$pp$ solar neutrino flux, similar to the HELLAZ~\cite{HELLAZ} proposal,
should also be able to detect the presence of neutrinos other than $\nu_e$
coming from the Sun \cite{recoil_spectrum}.

It is very likely that, after the present and the next rounds of neutrino
experiments, both the atmospheric and solar neutrino puzzles will be
unambiguously established as signals for new physics. Moreover, neutrino
flavour conversions would also be confirmed.  Neutrino oscillations, and
therefore unambiguous evidence for neutrino masses and mixing, will be
more difficult to establish explicitly by the observation of an
oscillating pattern. However, the K2K data offer a hint of an oscillation
effect in their energy spectrum, KamLAND~\cite{KamLAND} has an opportunity
when studying reactor neutrinos, and this is also an objective of the
MINOS experiment. For atmospheric neutrinos, a possible option for better
observations in the future is to build a new generation of larger
atmospheric neutrino detectors with higher performance. Current detector
studies include target materials as different as iron~\cite{MONOLITH},
liquid argon~\cite{CNGS} and water~\cite{UNO}.

We set out in this review a multi-step programme for exploring neutrino 
oscillations and related physics. This includes:

$\bullet$~An intense hadron source, such as could be provided by the 
Superconducting Proton Linac (SPL) project at CERN, which could yield a 
low-energy neutrino super-beam.

$\bullet$~The beta-beam concept, which envisages the production of pure 
$\nu_e$ and ${\bar\nu_e}$ beams via the decays of radioactive nuclei 
stored in a ring.

$\bullet$~The Neutrino Factory itself, in which pure $\nu_e$ and 
${\bar\nu_\mu}$ beams, or pure $\nu_\mu$ and ${\bar\nu_e}$ beams, are 
produced via the decays of muons stored in a ring.

As we discuss in this report, this multi-step programme represents a
systematic scheme for exploring neutrino physics. There are significant
synergies between successive steps in the programme. It also offers unique
prospects for short-baseline neutrino physics and studies of rare decays
of slow and stopped muons, that are discussed
elsewhere~\cite{GGetal,MMetal}. Moreover, in the longer term, the Neutrino
Factory is an essential stepping-stone towards possible muon colliders, as
also discussed elsewhere~\cite{muoncollcoll}.

\newpage
\def\baselinestretch{1.15}

\section{CURRENT STATUS OF NEUTRINO MASSES AND OSCILLATIONS}

\subsection{\bf Neutrino Masses}

\subsubsection{Laboratory limits}

Direct laboratory limits on neutrino masses are obtained from
kinematical studies.
The most stringent current upper limit is that on the $\bar{\nu}_e$ mass, 
coming from studies of the end-point of the electron energy spectrum in
Tritium beta decay~\cite{me}
\[
m_{\bar {\nu_{e}}} \leq 2.5  ~{\rm eV}
\]
For some time, these experiments tended to prefer a negative mass 
squared, but this problem has now disappeared, as has the previous
report of a spectral feature near the end-point. The proposed
KATRIN experiment aims to improve the sensitivity to 
$m_{\bar {\nu_{e}}} \sim 0.3$~eV \cite{KATRIN}.
Constraints on the mass of $\nu_{\mu}$ are derived
from the decay $\pi^{+} \rightarrow    \mu^{+} + \nu_{\mu}$,
which leads to the bound \cite{mmu}
\[
m_{\nu_{\mu}} \leq  170 \, {\rm keV}
\]
This upper limit could be improved by careful studies using the 
high fluxes provided at the front end of a neutrino factory.
Finally, the mass of $\nu_\tau$ is
constrained by $\tau$ decays into multihadron final states:
$\tau^{-} \rightarrow    2 \pi^{-}  \pi^{+}  \nu_{\tau}$
and 
$\tau^{-} \rightarrow    3 \pi^{-} 2 \pi^{+} \pi^{0} \nu_{\tau}$.
The current limit is~\cite{mtau}
\[
m_{\nu_{\tau}}
\leq 15.5 \, {\rm MeV}
\]
 and experiments at B factories may be sensitive to 
$m_{\nu_{\tau}} < 10$~MeV.
We note that the distinction between neutrino flavour and
mass eigenstates is unimportant for the above direct upper limits, as long 
as the mass
differences indicated by oscillation experiments are much smaller.

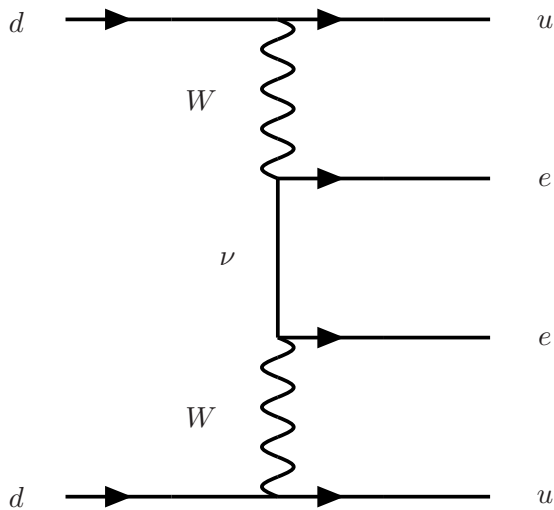
\begin{figure}[ht]  
\hspace*{4.5 cm}
{
\unitlength=2.0 pt
\SetScale{2.0}
\SetWidth{0.7}      
\begin{picture}(100,100)(0,0)
\ArrowLine(0,0)(20,0)
\Line(20,0)(40,0)
\ArrowLine(40,0)(60,0)
\Line(60,0)(80,0)
\ArrowLine(0,90)(20,90)
\Line(20,90)(40,90)
\ArrowLine(40,90)(60,90)
\Line(60,90)(80,90)
\ArrowLine(40,30)(60,30)
\Line(60,30)(80,30)
\ArrowLine(40,60)(60,60)
\Line(60,60)(80,60)
\Photon(40,0)(40,30){3}{4}
\Photon(40,60)(40,90){3}{4}
\Line(40,30)(40,60)
\Text(-10,0)[]{ $d$}
\Text(-10,90)[]{ $d$}
\Text(90,0)[]{ $u$}
\Text(90,90)[]{ $u$}
\Text(90,60)[]{ $e$}
\Text(90,30)[]{ $e$}
\Text(30,45)[]{ $\nu$}
\Text(25,15)[]{ $W$}
\Text(25,75)[]{ $W$}
\end{picture}
}
\caption{\it Diagrammatic representation of the possible role of a
Majorana neutrino mass in generating neutrinoless double-$\beta$
decay.}
\label{fig:dbeta}
\end{figure}

An important constraint on Majorana neutrino masses 
arises from neutrinoless double-$\beta$
decay~\cite{ndb}, in which an $(A,Z)$ nucleus decays to $(A,Z+2)
+ 2 \ e^-$, without any neutrino emission. This could be
generated by the following quark-level interaction:
\[
d + d \rightarrow    u + u + e^- + e^-
\]
which violates lepton number by two units $(\Delta L =2)$.
Such a transition could be generated by an exotic,
beyond the Standard Model interaction, and any such
interaction would necessarily generate a non-zero
Majorana neutrino mass~\cite{Majorana}. Assuming that
this dominates the neutrinoless double-$\beta$ decay matrix 
element as illustrated in Fig.~\ref{fig:dbeta}, 
it can be used to constrain the combination
\begin{equation}
<m_{ee}> \equiv |\sum U^{*2}_{ei} m_i|,
\label{dbeta}
\end{equation}
which involves a coherent sum over all the different
Majorana neutrino masses $m_i$, weighted by their
mixings with the electron flavour eigenstate, which may 
include CP-violating phases, as discussed below. This 
observable is
therefore distinct from the quantity observed in Tritium 
$\beta$ decay.

The interpretation of neutrinoless
double-$\beta$ decay data depends on calculations of the  
nuclear matrix elements entering in this process.
The strictest limit that had been reported until recently
came from a study of the
$^{76}{\rm Ge}$ isotope by the Heidelberg-Moscow 
Collaboration~\cite{Baudis}:
\[
<m_{ee}> ~\leq ~{\rm 0.2 ~eV}
\]
Subsequently, the data of this experiment have been reanalysed in~\cite{KK},
where evidence for neutrinoless double-$\beta$ decay at a rate
corresponding to a mass
\[
<m> = (0.11 - 0.56) ~{\rm eV}, 
\]
has been reported, with a preferred value of 0.39~eV. However, this
interpretation is not yet generally accepted. We note that there are proposals
capable of improving the sensitivity of neutrinoless double-$\beta$ decay
experiments by an order of magnitude.

\subsubsection{Astrophysical and cosmological constraints on neutrino 
masses}

For a review of cosmological and astrophysical
limits on neutrino masses, see~\cite{Dolgov}.
Neutrinos much lighter than 1~MeV have relic number densities that are
essentially independent of their masses, yielding a relic energy
density that is linear in the sum of their masses \cite{ger-zel}:
\begin{equation}
\Omega_\nu h^2 \; = \; \left( { \sum_i m_{\nu_i} \over 94~{\rm eV}} \right),
\label{Omega}
\end{equation}
where $\Omega_\nu \equiv \rho_{\nu} / \rho_c$, $\rho_\nu$ is the neutrino
energy density and $\rho_c$ the critical
density, $h$ parametrises the uncertainty in the 
Hubble parameter and is probably in the range $0.6 \leq h \leq 0.9$, and
the sum in (\ref{Omega}) is over all conventional electroweak doublet 
neutrinos, which are assumed to be metastable~\footnote{There are also constraints 
on the lifetimes of unstable neutrinos obtained by requiring that the energy 
density of the relativistic decay products be below the critical density, and 
late decays should also satisfy constraints from the cosmic microwave background
radiation and light-element abundances.}. Cosmic microwave 
background data and large-redshift supernovae indicate a total 
matter density $\Omega_m < 0.5$, corresponding to 
$\sum_i m_{\nu_i} \lappeq 30$~eV. Theories of large-scale structure
formation suggest that $\Omega_\nu \ll \Omega_m$, with the remainder of
$\Omega_m$ dominated by cold dark matter \cite{Fukug}, and a recent global analysis of
data on the cosmic microwave background radiation, large-scale structure,
big-bang nucleosynthesis and large-redshift supernovae yields
\[
\sum_i m_{\nu_i} \leq 2.5 ~{\rm eV}.
\]
The small differences in neutrino masses-squared indicated by the
oscillation data discuss later then imply that each neutrino species must
weigh $\lappeq 0.9$~eV.

Cosmological nucleosynthesis additionally imposes constraints on possible
oscillations between conventional active neutrinos and any additional
light sterile neutrinos \cite{subir}. If these are sufficiently strong that the
relativistic density of additional neutrinos is comparable to that of
the active neutrinos, the rate of expansion of the universe is affected
during nucleosynthesis, and hence the produced abundance of
$^{4}$He in particular. The success of conventional calculations of
primordial nucleosynthesis suggests that
\[
\delta m^2 ~\sin^2\theta < 10^{-6} ~{\rm eV}^2
\]
for the mass-squared difference $\delta m^2$ and mixing angle $\theta$
between any active and sterile species.

Additional limits on neutrino masses and mixing are obtained
from astrophysical processes, such as supernova physics~\cite{SNova-g,SNova}.
The arrival times of neutrinos from SN 1987a have been used to derive
upper limits of about 20~eV on neutrino masses. Oscillations inside a
supernova must be considered in connection with the $r$-process, and
also for the interpretation of any future neutrino signal from a galactic
supernova.
As a final point in this short discussion, we note that one of the
favoured scenarios for generating the baryon asymmetry in the universe is
leptogenesis~\cite{FY}. In this scenario, CP violation in the decays of
massive singlet neutrinos create a lepton asymmetry, which is subsequently
recycled by non-perturbative electroweak interactions into a baryon
asymmetry. We discuss later the possibility that the parameters of such a
leptogenesis scenario may be related to neutrino and/or charged-lepton
parameters that may be measurable in experiments.

\subsubsection{General principles for neutrino masses}

There is no fundamental theoretical reason why neutrinos should not have
masses and mix with one another. It is generally thought that particles
are massless if and only if they are associated with an exact gauge
symmetry. Examples include the photon and gluon, which are thought to be
massless because of $U(1)$ and $SU(3)$ gauge symmetries, respectively.  
There is no exact gauge symmetry associated with lepton number $L$, so it
is expected that lepton number should be violated at some level. If $L$ is
violated by two units: $\Delta L = 2$, then neutrino masses may arise.

On the other hand, the particle content of the minimal Standard Model, 
in conjunction with gauge invariance and renormalisability, allows
neither a Dirac nor a  Majorana neutrino mass term. A Dirac mass 
of the form $m (\bar{\nu_{L}}\nu_{R} + \bar{\nu_{R}}\nu_{L})$
cannot arise in the absence of a singlet $\nu_{R}$ field, whilst a 
Majorana term $m \nu_{L}^{T} \sigma_{2} \nu_{L}$ has weak isospin
$I = 1$, and hence would violate $SU(2)$ gauge invariance. However, it
is possible to introduce Majorana neutrino masses into the Standard Model,
even without postulating any new particles, at the price of postulating a 
higher-order non-renormalizable 
interaction linking two left-handed doublet neutrino fields and two Higgs 
doublets:
\begin{equation}
{(H.L) (H.L) \over M},
\label{eq:mnuSM}
\end{equation}
where $H$ denotes a Higgs doublet field, $L$ denotes a generic lepton
doublet field, and $M$ is some (large) mass parameter that is required
for dimensional reasons. An interaction of the form (\ref{eq:mnuSM})
would yield neutrino masses of order
\begin{equation}
m_\nu \sim {< 0 | H | 0 >^2 \over M},
\label{eq:H2overM}
\end{equation}
which would be $\ll m_{q,\ell}$ if $M \gg < 0 | H | 0 >$.
However, an interaction of the type (\ref{eq:mnuSM}) 
is not satisfactory from a 
theoretical point of view, because it is non-renormalizable.
Therefore, we need to understand the presence of such a term
in the framework of well-motivated extensions of the Standard Model.

\begin{figure}[ht]
\centerline{\epsfig{figure=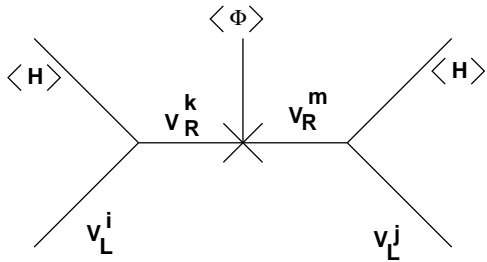,width=0.4\textwidth,clip=}}
\caption{\it Diagrammatic representation of the seesaw mechanism
for generating small neutrino masses.}
\label{fig:N}
\end{figure}

The minimal such possibility is to add three heavy singlet-neutrino
fields ${N^c}_i$ to the Standard Model, often called right-handed 
neutrinos, without necessarily expanding
the gauge group. Then the following neutrino Dirac and Majorana 
mass terms are allowed in the Lagrangian:
\begin{eqnarray}
\label{w}
{\cal L} = N^{c}_i (M_{\nu_D})_{ij} L_j 
  + \frac{1}{2}{N^c}_i (M_{\nu_R})_{ij} N^c_j + h.c. 
\label{MseesawM}
\end{eqnarray}
where the indices $i,j$ run over three generations,
$M_{\nu_D} = Y_\nu <0|H|0>$ is the Dirac mass matrix, and
$M_{\nu_R}$ is the Majorana mass matrix for the right-handed 
isosinglet neutrino sector. The most general neutrino mass matrix is then
\begin{eqnarray}
{\cal M}  = \left(
\begin{array}{cc}
0 & M_{\nu_D}\\
M_{\nu_D}^{T} & M_{\nu_R}
\end{array}
\right).
\label{eq1ne}
\end{eqnarray}
The entries in ${M_{\nu_D}}$ require electroweak symmetry
breaking, and so must be $\lappeq m_W$, whereas the entries
in ${M_{\nu_R}}$ may be arbitrarily large.
Assuming that ${M_{\nu_R}} \gg {M_{\nu_D}}$,
the light eigenvalues of $M$ 
are given by
\begin{eqnarray}
m^{light}_{\nu} \simeq \frac{M_{\nu_D}^{2}}{M_{\nu_R}}  \nonumber
\end{eqnarray}
and therefore are extremely suppressed, as
is also obvious from the associated diagram shown in 
Fig.~\ref{fig:N}. This is the well-known seesaw mechanism~\cite{seesaw},
which explains naturally why the neutrinos
are so much lighter than the other known fermions. 

Neutrino mixing arises from the mismatch between the mass
eigenstates and the current eigenstates that couple via the weak
interactions to charged leptons of definite flavour.
The neutrino mixing matrix is therefore given by 
\begin{equation}
V = V_{\nu}^{\dagger}V_{\ell},
\label{VnuVell}
\end{equation} 
where $V_{\ell}$ diagonalizes the charged-lepton mass matrix, and
$V_{\nu}$ diagonalizes the light neutrino mass matrix 
$m^{light}_{\nu}$~\cite{mns}. As a unitary $3 \times 3$ matrix,
$V$ would seem to have 9 parameters {\it a priori}, but
3 of these can be absorbed by phase transformations of the
charged-lepton fields, yielding a net total of 6 parameters,
of which 3 are light-neutrino mixing angles $\theta_{ij}: 1 \le i \ne
j \le 3$ and 3 are CP-violating light-neutrino mixing phases,
including the oscillation phase $\delta$ and two Majorana phases
$\phi_{1,2}$ that appear in the neutrinoless double-$\beta$
decay observable. Thus, the total number of neutrino parameters that are 
in principle observable at low energies is 9: 3 light-neutrino 
masses $m^{light}_{\nu}$, 3 real mixing 
angles and 3 CP-violating phases.

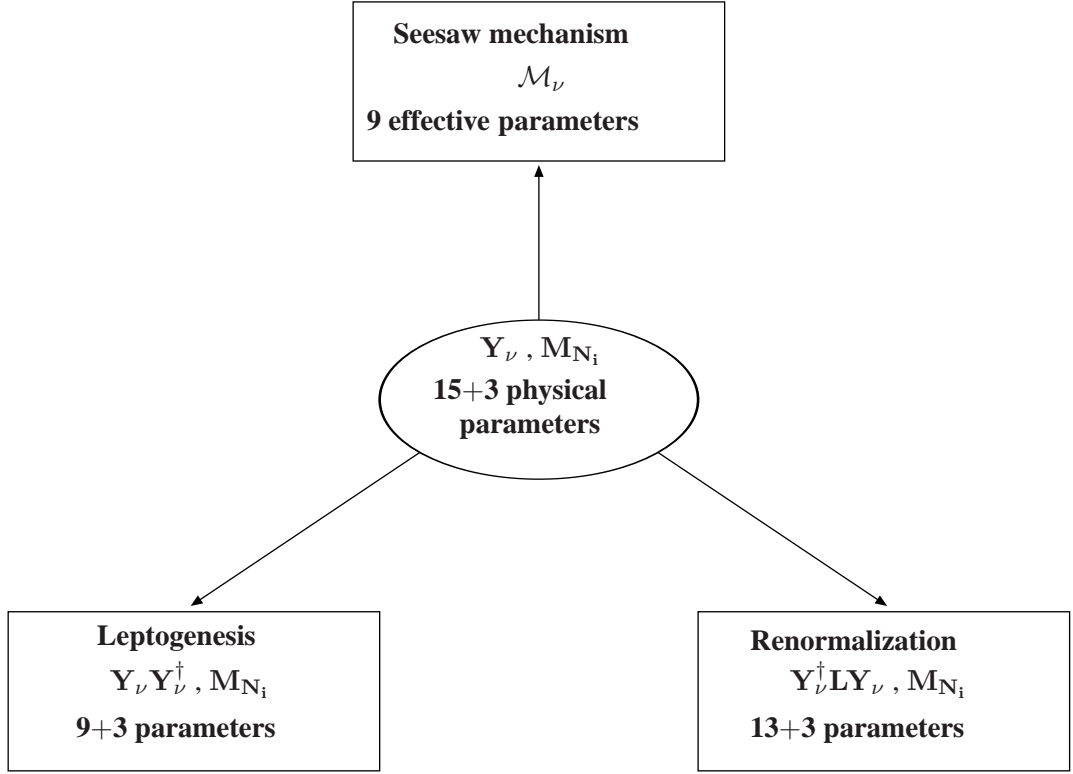
\begin{figure}[t]
\begin{center}
\begin{picture}(400,300)(-200,-150)
\Oval(0,0)(30,60)(0)
\Text(-25,13)[lb]{ ${\bf Y_\nu}$  ,  ${\bf M_{N_i}}$}
\Text(-40,-2)[lb]{{\bf 15$+$3 physical}}
\Text(-30,-15)[lb]{{\bf parameters}}
\EBox(-70,90)(70,150)
\Text(-55,135)[lb]{{\bf Seesaw mechanism}}
\Text(-8,117)[lb]{${\bf {\cal M}_\nu}$}
\Text(-65,100)[lb]{{\bf 9 effective parameters}}
\EBox(-200,-140)(-60,-80)
\Text(-167,-95)[lb]{{\bf Leptogenesis}}
\Text(-165,-113)[lb]{ ${\bf Y_\nu Y_\nu^\dagger}$ , ${\bf M_{N_i}}$}
\Text(-175,-130)[lb]{{\bf 9$+$3 parameters}}
\EBox(60,-140)(200,-80)
\Text(80,-95)[lb]{{\bf Renormalization}}
\Text(95,-113)[lb]{${\bf Y_\nu^\dagger L Y_\nu}$ , ${\bf M_{N_i}}$}
\Text(80,-130)[lb]{{\bf 13$+$3 parameters}}
\LongArrow(0,30)(0,87)
\LongArrow(-45,-20)(-130,-77)
\LongArrow(45,-20)(130,-77)
\end{picture}
\end{center}
\caption{\it
Roadmap for the physical observables derived from $Y_\nu$ and $N_i$.
}
\label{fig:map}
\end{figure}

However, the minimal seesaw model contains a total of 18 parameters, even 
after
taking into account the possible field redefinitions. In addition to the 9
parameters mentioned above, there are additionally 3 heavy Majorana
neutrino mass eigenvalues ${M}_{\nu_R}$, 3 more real mixing angles and 3
more phases associated with the heavy-neutrino sector~\cite{parameters}.
As illustrated in Fig.~\ref{fig:map}, 12 of these parameters play a role
in generating the baryon number of the universe via leptogenesis, but
these do not include $\theta_{ij}$, $\delta$ and $\phi_{1,2}$. Since the
lepton number density involves a unitary sum over the light neutrino and
lepton species, it is insensitive to the values of these light-neutrino
mixing angles and phases. However, as also shown in Fig.~\ref{fig:map}, 16
of the 18 neutrino parameters contribute to renormalization effects that
are, in principle, measurable in a supersymmetric version of the minimal
seesaw model.

\subsubsection{Aspects of models for neutrino masses and mixing}

Our experimental knowledge of the 18 neutrino parameters
introduced above is limited so far to 4: 2 neutrino mass-squared
differences, and 2 mixing angles. As we discuss later in more detail, the
data indicate that these 2 measured neutrino mixing angles are probably
both quite large, possibly even maximal, whilst the third angle is
relatively small. Since the seesaw mechanism suggests that the origins of
neutrino masses and mixings are different from, and more complicated than
those of quarks, we should, in retrospect, perhaps not have been so
surprised that some neutrino mixing angles are larger than those in the
quark sector. The spate of experimental information on neutrino masses and
mixing, and their unexpected nature, has spawned many theoretical models,
some of whose general features we review below. The neutrino factory is
uniquely well placed to provide crucial input for distinguishing between
such models, in particular via neutrino-oscillation experiments. Models
for neutrino masses and mixing typically incorporate ideas for family (or
generation) symmetries and/or grand unification schemes linking quarks and
leptons, and we now give examples of each.

{\it $U(1)$ Flavour Symmetries}

The left- and right-handed components of quarks and leptons may be 
assigned
various $U(1)$ charges, as are the Higgs fields, such that only one or a
limited number of entries in the mass matrices can be generated by
renormalizable terms in the effective Lagrangian \cite{FN}. 
Additional entries
become permitted if the symmetry is broken, for example by one or more
vacuum expectation values ${\cal V}$ for fields that appear in 
non-renormalizable
terms in the mass matrices, scaled by inverse powers of larger masses $M$.
This type of scheme provides a perturbative expansion in powers of some
small parameter $\epsilon \equiv {\cal V}/ M$. In a popular class of
realizations, only the (3,3)  element of the associated mass matrix is
non-zero at leading order, whilst other terms arise with various powers of
$\epsilon$, that are fixed by the $U(1)$ charge assignments:

\begin{equation}
{\cal M}\sim \left( 
\begin{array}{ccc}
\epsilon^m & \epsilon^n & \epsilon^p \\ 
\epsilon^q & \epsilon^r & \epsilon^s \\ 
\epsilon^t & \epsilon^u & 1
\end{array}
\right) 
\label{U1model}
\end{equation}
In such models, each term in the mass matrix has a numerical 
coefficient that may be calculated only with a more complete theory. 

In its absence, there are numerical ambiguities in the predictions of 
such a model, but mixing angles are generically powers of $\epsilon$. 
There are two possible ways to obtain large mixing angles 
in such a perturbative $U(1)$ framework. One of the mixing 
matrices may contain more than one entry appearing with the same power 
of $\epsilon$, for example perhaps the (3,~3) and (3,~2) entries in 
(\ref{U1model}) might both be ${\cal O}(1)$. Alternatively, the numerical 
coefficients might be such as to compensate for the `small' expansion 
parameter $\epsilon$.

{\it Grand Unified Theories}

In a generic scheme, there is a lot of freedom in assigning the various
$U(1)$ flavour charges. However in Grand Unified Theories, quark and lepton
fields that belong in the same GUT multiplets have the same flavour
charges. This introduces additional constraints, thus increasing
predictivity \cite{GUTS}.  For instance, in $SO(10)$ models, all quarks and leptons
are accommodated in a single ${\mathbf {16}}$ representation of the group,
implying left-right symmetric mass matrices with similar structures for
all fermions. As a consequence, such GUT models predict $ V_{\mu {\tau
}}\approx V_{cb} $, which is inconsistent with the data. Hence, in order
to construct a viable $SO(10)$ solution, one must consider the effects of
the additional Higgs multiplets that break $SO(10)$ down to $SU(3) \times
SU(2) \times U(1)$. However, this introduces additional parameters, thus
decreasing predictivity.

The situation is different in  $SU(5)$ unification, where the
$(q,u^{c},e^{c})_{i}$ fields (for left-handed quarks, right-handed up 
quarks and right-handed charged leptons, respectively)
belong to ${\mathbf {10}}$ representations, the $(\ell,d^{c})_{i}$
fields (for left-handed leptons and right-handed down quarks,
respectively) belong to ${\mathbf {\overline 5}}$ representations,
and the $N_{i}$ (singlet neutrinos) to singlet 
representations of the group. In this case,
the up-quark mass matrix is symmetric, there is a lot of freedom
in chosing the neutrino mass matrices, and
the charged-lepton mass matrix is the transpose of the 
down-quark one. Hence the mixing of the left-handed leptons is related
to that of the right-handed down quarks, and not with the small CKM 
mixing of the left-handed quarks. Thus, $SU(5)$ can in principle
accommodate large MNS neutrino mixing at the same time
as small CKM quark mixing, without any tuning of parameters.

In left-right symmetric models, one has identical $U(1)$ flavour charges 
for the left- and  right-handed fields, as in $SO(10)$ unification, 
but quarks and leptons need not be correlated. This leads to  symmetric
quark, lepton and neutrino mass matrices, but allows the
lepton mixing to be independent of that in the quark sector.

{\it Non-Abelian Flavour Symmetries}

Specific $U(1)$ flavour models may yield one relatively large
neutrino mixing angle, but tend to favour small values of the
other mixing angles and hierarchical neutrino masses. Thus, such models 
are comfortable with the
small value of $\theta_{13}$, but could be embarrassed if the
current preference for the LMA solar solution is confirmed.
This feature of models with Abelian flavour symmetries may be
traced to the lack of charge quantization and arbitrary coefficients,
making it difficult to obtain accurate cancellations between
the various entries of the mass matrices, unless extra assumptions
are made, for example in the heavy singlet-neutrino sector. \\

The situation is reversed in non-Abelian models. To illustrate this,
consider a simple case in which the lepton fields are $SO(3)$ triplets. In
this case, degenerate lepton textures are to be expected. Subsequently,
one may break $SO(3)$ so that there are large mass splittings for charged
leptons, but not for the light neutrinos~\cite{nonab}. This means that, in
schemes with non-Abelian flavour symmetries, solutions with degenerate
neutrinos and bimaximal mixing can be generated in a natural way. On the
other hand, the understanding of any small mixing angles and phases might
then become an issue.

\subsubsection{Testing models of neutrino masses}

The neutrino oscillation measurements that can be made at a neutrino
factory, notably the magnitudes of $\theta_{13}$ and $\delta$, will
provide important constraints on models of neutrino masses. For example,
as just discussed,
the expected size of $\delta$ and other CP-violating phases may be
rather different in Abelian and non-Abelian flavour models.

\begin{figure}[tbp]
\vspace*{-4.4 cm}
\centerline{\epsfig{figure=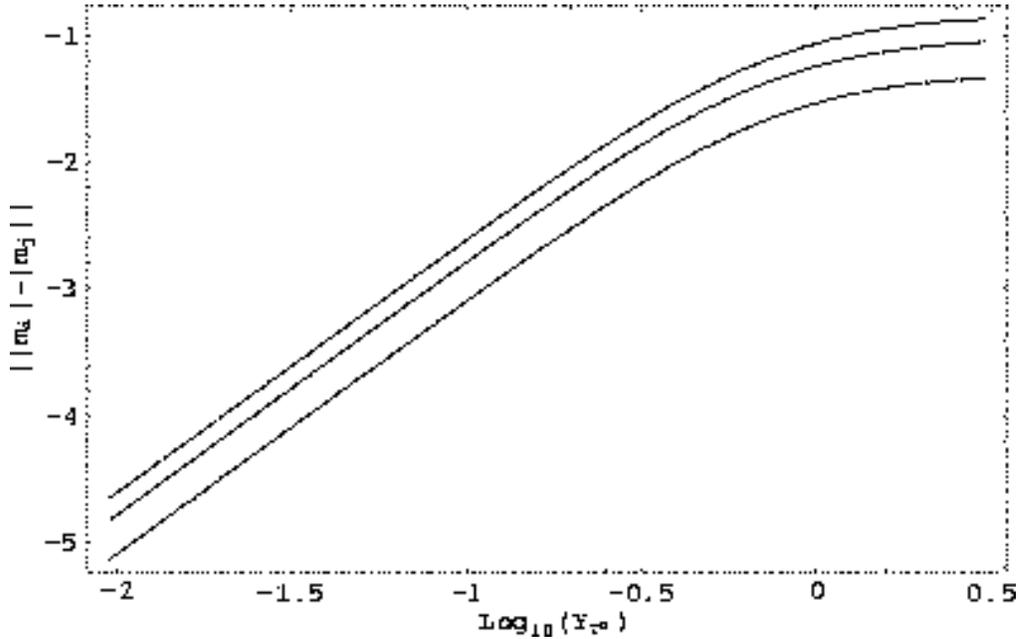,width=1.2\textwidth,clip=,bb=50 -80 600 650}}
\vspace*{-12.0 cm}
\caption{{\it
Renormalization of $m_{eff}$ eigenvalues 
in a model which would seem 
to lead to bimaximal mixing. The range of 
initial values of the $\tau$ Yukawa coupling 
chosen correspond to values of tan$\beta$ in the range 1 to 58,
assuming $M_N = 10^{13}$~GeV.}}
\label{runs0}
\end{figure}

When comparing model predictions with data, care must be taken to
include the effects of quantum corrections in the neutrino 
parameters, which cause them to vary as functions of energy
\cite{neut-RGEs}. To leading order,
entries in the neutrino mass matrix are renormalized by 
multiplicative factors  \cite{EL}:
\begin{equation}
{{m^{light}_\nu}(Q)}^{ij} =
{{m^{light}_\nu}(M_N)}^{ij} \times I_i \times I_j \,,
\label{RGis} 
\end{equation}
where the ${{m^{light}_\nu}(M_N)}^{ij}$, $i = e,\mu,\tau$ denote the initial mass
entries at the high-energy energy scale where the neutrino mass matrix is
generated, and the
\begin{equation}
I_{i} \equiv \exp[\frac 1{16\pi^2}\int_{t_0}^t Y_{i}^2 dt]:
 ~t \equiv \ln\mu
\end{equation}
are integrals determined by the running of the charged-lepton
Yukawa couplings $Y_i$ as functions of the renormalisation
scale $\mu$. This effect implies that the neutrino mixing can be 
amplified or even destroyed, as we go from high to low
energies, and the neutrino mass eigenvalues may also be
altered significantly. Although
these renormalization effects are not significant for
schemes with hierarchical neutrino masses, they
are potentially large in models with degenerate neutrinos 
and may be important in models with bimaximal mixing.
For instance, for the neutrino mass texture
\[
{m^{light}_{nu}} \propto 
\left (
\begin{array}{ccc}
0 &  {1\over\sqrt2} &  {1\over\sqrt2} \\
{1\over\sqrt2} &  {1\over2} &  -{1\over2} \\
{1\over\sqrt2} &  -{1\over2} &  {1\over2} \\
\end{array}
\right ) 
\]
which would seem to lead to bimaximal mixing,
Fig. \ref{runs0} indicates a large change of the eigenvalues
when running from high to low energies \cite{EL}, and analogous large 
effects may also occur in the mixing angles.

In supersymmetric seesaw models, quantum corrections may provide other
ways of measuring neutrino parameters and testing models via
lepton-flavour-violating processes. In the minimal seesaw model without
supersymmetry, the amplitudes for processes such as $\mu\to e\gamma$,
$\mu$-$e$ conversion in nuclei, $\mu\to eee$ and $\tau \to \ell \gamma, 3
\ell$~\cite{KunOk,GGetal,LFVref} where $\ell = e, \mu$ denotes a generic light
charged lepton, are proportional to the neutrino mass-squared difference,
and so are many orders of magnitude smaller than the existing experimental
bounds. This is no longer the case in supersymmetric theories, due to the
existence of loop diagrams with internal sparticles, that may violate
charged-lepton numbers. This may happen either because sfermion masses are
not diagonal in the lepton flavour basis already at the input scale
$M_{GUT}$, in which case very large lepton-flavour violating rates are
generically predicted \cite{KunOk}, or because flavour-violating effects
are induced by quantum corrections \cite{bm}.

These are generic in the Minimal Supersymmetric Standard Model
with heavy singlet neutrinos, because the Dirac Yukawa couplings to
neutrinos and charged leptons cannot, in general, be diagonalized 
simultaneously. Since both these sets of lepton Yukawa
couplings appear in the renormalization-group equations, 
the slepton mass matrices receive off-diagonal
contributions.  In the leading-logarithmic approximation,
in the basis where the charged-lepton Yukawa couplings are diagonal, 
these are given by
\begin{eqnarray}
\left( \delta{m}_{\tilde{\ell}}^2 \right)^i_j \propto \frac 1{16\pi^2} (3 + a^2)
\left( Y_\nu^{\dagger}\right)^i_k L_k \left( 
Y_\nu \right)^k_j m_{0}^2,
\label{offdiagonal}
\end{eqnarray}
where $L_k \equiv \ln\frac{M_{GUT}}{M_{N_k}}$, the singlet-neutrino mass 
$M_{N_k}$ is the scale above which
the Dirac Yukawa coupling $Y_\nu$ appears in the renormalization-group
equations, $a$ is  related to the trilinear mass parameter, and
$m_{0}$ is the common value of the scalar masses at the GUT
scale.

In this class of models, the rates predicted for processes violating
charged-lepton flavour may be close to the current experimental bounds,
and are in principle sensitive to up to 16 of the 18 parameters in the
minimal seesaw model, as indicated in
Fig.~\ref{fig:map}~\cite{parameters}.  Therefore, different models predict
in general different rates for the different charged-lepton
flavour-violating processes: the larger the lepton mixing and the larger
the neutrino mass scales, the larger the rates. Consequently, schemes with
degenerate eV neutrinos and bimaximal mixing generally yield significantly
larger effects than schemes with hierarchical neutrinos and small mixing
angles.

The prospects for measuring some of these processes at the front end of a
neutrino factory are discussed elsewhere~\cite{GGetal}. Here we just note
that the present experimental upper limits on the most interesting of
these processes are $BR(\mu \to e \gamma) < 1.2 \times
10^{-11}$~\cite{Brooks} and $R(\mu^- Ti \to e^- Ti) < 6.1 \times
10^{-13}$~\cite{Wintz}, and that projects are underway to improve these
upper limits significantly. An experiment with a sensitivity ${BR}(\mu\to
e\gamma)\sim 10^{-14}$ is being prepared at PSI \cite{PSI} and the MECO
experiment with a sensitivity to $R(\mu^- N \to e^- N) \sim 10^{-17}$ has
been proposed for BNL~\cite{MECO}, whilst the PRISM project~\cite{PRISM}
and a neutrino factory~\cite{GGetal} may reach sensitivities to ${BR}(\mu\to
e\gamma) \sim 10^{-15}$ and ${Br}(\mu\to eee) \sim 10^{-16}$. The latter
sensitivity may open the way to measuring the T-odd, CP-violating
asymmetry $A_T(\mu\to eee)$. Other measurements that may be sensitive to
CP violation in the lepton sector include those of the electric dipole
moments of the electron and muon, and a neutrino factory would also have
unique sensitivity to the latter~\cite{GGetal}.

In this way, the front end of the neutrino factory could contribute to
determining as many as 16 of the 18 parameters in the neutrino sector, via
their renormalization of soft supersymmetry-breaking parameters (see
Fig.~\ref{fig:map}). Any such information would therefore provide useful
constraints on neutrino models. Moreover, the combination of such
front-end data with oscillation data from the neutrino factory may enable
the baryon number of the universe to be calculated, if it is due to
leptogenesis in the minimal supersymmetric version of the seesaw model.

\subsection{\bf Oscillation Physics}\label{sec:oscphys}

Neutrino oscillations in vacuum would arise if neutrinos were massive and
mixed \cite{pontecorvo} similar to what happen in the quark sector. If
neutrinos have masses, the weak eigenstates, $\nu_\alpha$
($\alpha=e,\mu,\tau, ...$), produced in a weak interaction are, in
general, linear combinations of the mass eigenstates $\nu_i$ ($i=1,2,3,
...$).  We now review the basic physics of neutrino oscillations in vacuum
and in matter.

\subsubsection{Relativistic approach to the two-family formula}
In the simpler case of two-family mixing, one has: 
\begin{equation}
\left( \begin{array}{c}
 \nu_\alpha \\ 
 \nu_\beta \\ 
\end{array} \right) = 
\left(\begin{array}{cc} 
 \cos\theta & \sin\theta \\  
 -\sin\theta & \cos\theta \\ 
\end{array} \right)
\left(\begin{array}{c} 
 \nu_1 \\ 
 \nu_2 \\ 
\end{array} \right) 
\end{equation}
We use the standard approximation that $|\nu\rangle$ is a plane wave and 
consider its propagation in a one-dimensional space. A mass eigenstate produced 
at t,x=0, will evolve in space and in time as:
\begin{eqnarray}
|\nu_{i}(t,x) \rangle &=& e^{i(p_i x- E_i t)} |\nu_{i}\rangle \qquad 
{\rm for}~i=(1,2).
\end{eqnarray}
Starting from a flavour eigenstate $|\nu_\alpha \rangle$, the probability 
for detecting a state $\langle \nu_\beta|$ at a distance $L$ and time $t$ 
is given by:
\begin{eqnarray}
P(\nu_\alpha\to\nu_\beta) &\equiv & |\langle \nu_\beta | \nu_\alpha(t,L) \rangle |^2 
 = \sin^2 2\theta \ \sin^2 \left[\frac{(p_1 - p_2) L - (E_1 - E_2) t}{2} \right]~. 
\label{prob2a}
\end{eqnarray}
In the `same-energy prescription', which is consistent with the 
wave-packet
treatment~\footnote{For a detailed discussion, see for 
example~\cite{lipkin} and~\cite{takeuchi}.},
one assumes that the two neutrino mass eigentstates have the same energy, 
$E_{1}=E_{2}=E$, but different momentum:
\begin{eqnarray}
p_{1} = \sqrt{E^{2} -m_{1}^{2}}~, \qquad  
p_{2} = \sqrt{E^{2} -m_{1}^{2} + \Delta m^2_{12 }} \qquad ({\rm with} \qquad 
\Delta m^2_{12} \equiv m_1^2 - m_2^2)
\end{eqnarray}
which leads to: 
\begin{eqnarray}
\label{prob2b}
P(\nu_\alpha\to\nu_\beta) 
 = \sin^2 2\theta\sin^2\frac{(p_1- p_2) L}{2}
 \simeq \sin^2 2\theta\sin^2 \left(\kappa \frac{\Delta m^2_{12} L}{E}\right),
\end{eqnarray}
In (\ref{prob2b}) $\kappa$ is 1/4 in natural units ($\hbar=c=1$) or 
$1.27$ if we consider practical units where the energy is expressed 
in GeV, the distance in km and the mass difference squared in eV$^2$.
 
\subsubsection{Three families in vacuum}

\begin{figure}[tb]
  \begin{center}
    \epsfig{file=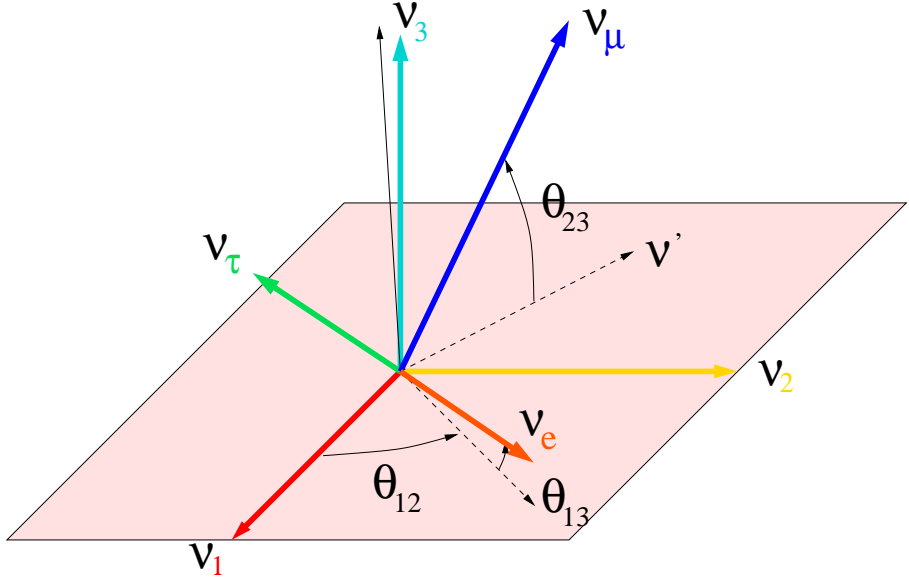,width=12cm}
    \caption{\it Representaion of the 3-dimensional rotation between the 
flavor and mass neutrino eigenstates.}
    \label{fig:rotation}
  \end{center}
\end{figure}

In the three-family scenario, the general relation between the flavour 
eigenstates $\nu_\alpha$  and the mass eigenstates
$\nu_i$ is given by the 3x3 mixing matrix $V$: 
\begin{equation}
V = U A, 
\end{equation}
where the matrix $A$ contains the Majorana phases
\begin{eqnarray}
A=\left(\begin{array}{ccc}
e^{i\alpha}&0&0\\
0&e^{i\beta}&0\\
0&0&1\\ \end{array} \right)
\end{eqnarray}
that are not observable in oscillation experiments, and $U$ is the MNSP 
matrix~\cite{pontecorvo,maki}, which is usually parameterized 
by~\cite{pdg}~\footnote{Notice, 
though, that our convention for the sign of $\delta$ is opposite from that
used in this reference. The MNSP matrix is the leptonic analogue to the 
CKM~\cite{CKM} matrix of the quark sector.}:
\begin{eqnarray}
U = \left( \begin{array}{ccc}
U_{e1}&U_{e2}&U_{e3}\\
U_{\mu 1}&U_{\mu 2}&U_{\mu 3}\\
U_{\tau 1}&U_{\tau 2}&U_{\tau 3}\\
\end{array}\right)=
\left( \begin{array}{ccc}
c_{12}c_{13}&s_{12}c_{13}&s_{13}e^{-i\delta}\\
-s_{12}c_{23}-c_{12}s_{13}s_{23}e^{i\delta}&
c_{12}c_{23}-s_{12}s_{13}s_{23}e^{i\delta}&c_{13}s_{23}\\
s_{12}s_{23}-c_{12}s_{13}c_{23}e^{i\delta}&
-c_{12}s_{23}-s_{12}s_{13}c_{23}e^{i\delta}&c_{13}c_{23}\\
\end{array}
\right)
\end{eqnarray}
where, for the sake of brevity, we write $ s_{ij}\equiv\sin \theta_{ij}$, 
$ c_{ij}\equiv\cos\theta_{ij}$. The relevant angles can be derived from 
the matrix elements via the following relations:
\begin{eqnarray}
\tan^2\theta_{23}&\equiv&\frac{|U_{\mu3}|^2}{|U_{\tau3}|^2}, \\
\tan^2\theta_{12}&\equiv&\frac{|U_{e2}|^2}{|U_{e1}|^2}, \\
\sin^2\theta_{13}&\equiv&|U_{e3}|^2, \\
\sin\delta&\equiv&
\frac{8\ Im({U_{e2}^*U_{e3}U_{\mu2}U_{\mu3}^*})}
{\sin 2\theta_{12}\sin 
2\theta_{23}\sin2\theta_{13}\cos\theta_{13}}.
\end{eqnarray}
The relations between mass and flavor eigenstates can be visualized
as rotations in a three-dimensional space, with the angles defined
as in Fig.~\ref{fig:rotation}.\par

With a derivation analogous to the two-family case, the oscillation
probability for neutrinos reads
\begin{eqnarray}
\label{probnu}
P(\nu_\alpha\to\nu_\beta)=\sum_{jk} J_{\alpha\beta j k}
e^{-i\Delta m^2_{jk}L/2E}
\end{eqnarray}
where $J_{\alpha\beta jk}=U_{\beta j}U^*_{\beta k}U^*_{\alpha j}
U_{\alpha k}$. For antineutrinos, the probability is obtained with the 
substitution $J_{\alpha\beta jk} \to J_{\alpha\beta jk}^*$. As 
$J_{\alpha\beta jk}$ is not real in general, due to the phase 
$\delta$, neutrino and antineutrino oscillation probabilities are different, 
and therefore CP is violated in the neutrino mixing sector. The CP-even 
and CP-odd contributions in the oscillation probability can easily be
distinguished by separating the real and imaginary parts of 
$J_{\alpha\beta jk}$:
\begin{eqnarray}
P(\nu_\alpha\to\nu_\beta)=\delta_{\alpha\beta}-4\sum_{j>k}
Re(J_{\alpha\beta jk})\sin^2\frac{\Delta m^2_{jk}L}{4E}+
2\sum_{j>k}Im(J_{\alpha\beta jk})\sin\frac{\Delta m^2_{jk}L}{2E}.
\end{eqnarray}
The Jarlskog determinant $J$~\cite{Jarl} is defined by:
\begin{eqnarray}
Im(J_{\alpha\beta jk}) &=& J \sum_{\gamma , l} \epsilon_{\alpha \beta \gamma} 
\ \epsilon_{j k l} 
\end{eqnarray}
and can be expressed in terms of the mixing parameters, as: 
\begin{eqnarray}
J = c_{13}\sin 2\theta_{12}\sin 2\theta_{13}\sin 2\theta_{23} \sin\delta.
\end{eqnarray}
We define a complex quantity $\tilde{J}$ whose real part is $J$:
\begin{eqnarray}
\tilde{J} = c_{13}\sin 2\theta_{12}\sin 2\theta_{13}\sin 2\theta_{23} 
e^{i\delta}.
\end{eqnarray}
Oscillations in a three-generation scenario are consequently described by 
six 
independent parameters: two mass differences ($\Delta m^{2}_{12}$ and 
$\Delta m^{2}_{23}$), three Euler angles ($\theta_{12}$, $\theta_{23}$ and 
$\theta_{13}$) and one CP-violating phase $\delta$.
As an example, we give the full oscillation probability for the oscillation
$\nu_e\to\nu_\mu$:
\begin{eqnarray}
P(\nu_e\to\nu_\mu)=P(\bar{\nu}_\mu\to\bar{\nu}_e) &=& \nonumber \\
4 c^2_{13}[\sin^2\Delta_{23}s^2_{12}s^2_{13}s^2_{23} &+& c^2_{12}
(\sin^2\Delta_{13}s^2_{13}s^2_{23}+\sin^2\Delta_{12}s^2_{12}
(1-(1+s^2_{13})s^2_{23}))] \nonumber \\
&-& \frac{1}{4} |\tilde{J}| \cos\delta[\cos2\Delta_{13}-\cos 2\Delta_{23}-2
\cos 2\theta_{12}\sin^2\Delta_{12}] \nonumber \\
&+& \frac{1}{4} |\tilde{J}| \sin\delta[\sin 2\Delta_{12}-\sin 2\Delta_{13}
+\sin 2\Delta_{23}],
\label{eq:nuenumuvacuum}
\end{eqnarray}
where we have used the contracted notation $\Delta_{jk}\equiv\Delta 
m^2_{jk}L/4E$~\cite{cptoptim}. The second and third lines are the CP-violating
terms, and are proportional to the imaginary and real parts of 
$\tilde{J}$, respectively.

The present experimental knowledge on neutrino oscillation parameters 
indicates $\Delta m^2_{12}$ $\ll$ $\Delta m^2_{23}$ and small values 
for $\tetaot$. In this situation, a good and simple approximation 
for the $\nu_e \raw \nu_\mu$ transition probability is obtained by expanding to 
second order in the small parameters, $\tetaot, \Delta_{12} / \Delta_{13}$ and 
$\Delta_{12}$ \cite{golden}:
\begin{eqnarray}
P_{\nu_ e\nu_\mu} & = & 
s_{23}^2 \, \sin^2 2 \tetaot \, \sin^2 \delot + 
c_{23}^2 \, \sin^2 2 \theta_{12} \, \sin^2 \Delta_{12} 
+ |\tilde{J}| \, \cos \left (\delta - \delot \right ) \; \Delta_{12} \sin \delot~. 
\label{vacexpand} 
\end{eqnarray} 
We refer to the three terms
in (\ref{vacexpand}) as the atmospheric $P^{atm}_{\nu ( \bar \nu 
) }$, solar $P^{sol}$, 
and interference $P^{inter}_{\nu ( \bar \nu) }$ terms, 
respectively~\cite{jordi}. It is easy to show that
\bea
 |P^{inter}_{\nu(\bar \nu)}| \leq P^{atm}_{\nu ( \bar \nu ) } + P^{sol},
\eea
implying two very different regimes. When $\theta_{13}$ is relatively 
large or $\Delta m^2_{12}$ small, the probability is dominated by the 
atmospheric term, since $P^{atm}_{\nu (\bar \nu )}\gg P^{sol}$. We will 
refer to this situation as the atmospheric regime. Conversely, when 
$\theta_{13}$ is very small or $\Delta m^2_{12}$ large, the solar 
term dominates: $P^{sol} \gg P^{atm}_{\nu (\bar \nu )}$. This is the solar
regime. The interference term is relevant only if either $\Delta m^2_{12}$ 
or $\theta_{13}$ are non-negligible. The possibility of observing CP 
violation in neutrino oscillations is connected to the possibility of
separating the interference term from the solar and atmospheric 
contributions 
in (\ref{vacexpand}).

If we neglect completely $\Delta m^2_{12}$, setting it to zero\footnote{This 
is a good approximation if LOW or VO solutions happen to be the one chosen by nature
for the solar oscillation.}, oscillation probabilities take the simplified form
\begin{eqnarray}
  P(\nu_e\to\nu_e)=& 1-\sin^2 2\theta_{13} \sin^2 \Delta_{23} \\
  P(\nu_e\to\nu_\mu)=& \sin^2 2\theta_{13}\sin^2 \theta_{23} \sin^2 \Delta_{23} 
     \label{eq:nuenumusimp} \\
  P(\nu_e\to\nu_\tau)=& \sin^2 2\theta_{13}\cos^2\theta_{23}\sin^2\Delta_{23} \\
  P(\nu_\mu\to\nu_\mu)=&1-4\cos^2\theta_{13}\sin^2\theta_{23}(1-
\cos^2\theta_{13}\sin^2\theta_{23})\sin^2\Delta_{23}\\
  P(\nu_\mu\to\nu_\tau)=&\cos^4\theta_{13}\sin^2 2\theta_{23}\sin^2\Delta_{23}.
\end{eqnarray}
As expected CP-violating effects are absent since they require the interference of 
oscillations from two mass differences.

\subsubsection{Oscillations in matter}
\label{sect:oscmatter}

\begin{figure}[tb]
\begin{minipage}[h]{0.43\textwidth}
\begin{center}
\epsfig{file=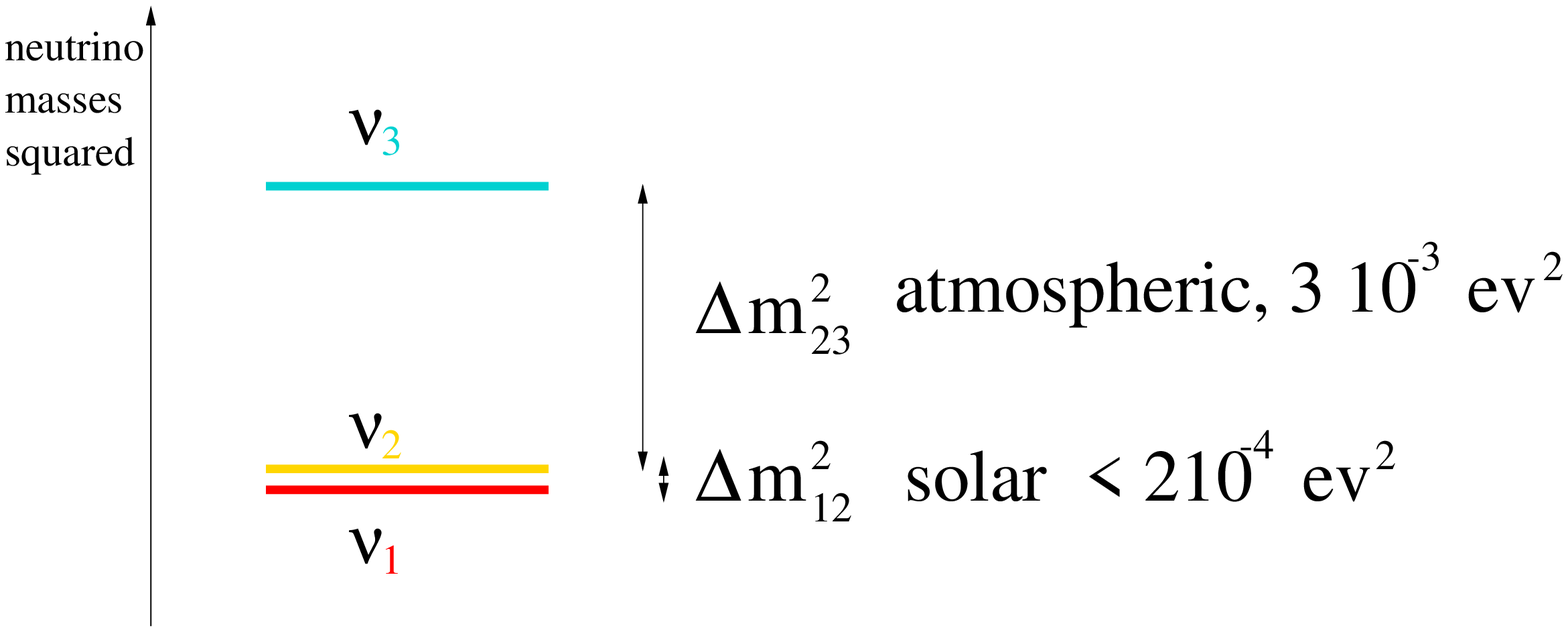,width=8cm}
\end{center}
\end{minipage}\hspace{.5cm}
\begin{minipage}[h]{0.43\textwidth}
\begin{center}
\epsfig{file=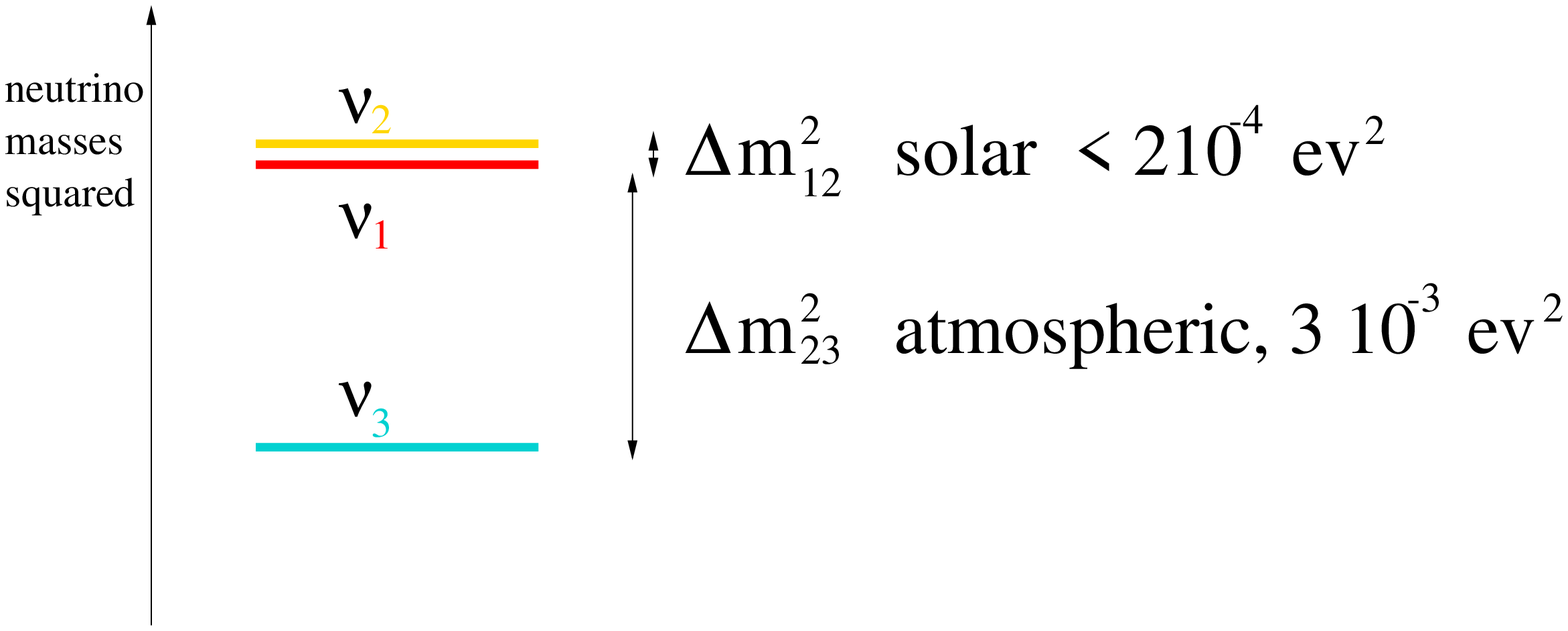,width=8cm}
\end{center}
\end{minipage}
\caption{\it Two possible configurations for the neutrino mass texture: 
direct (left) or inverted (right).}
\label{fig:signd}
\end{figure}

So far, we have analyzed neutrino oscillations in vacuum. 
When neutrinos travel through matter (e.g. in the Sun, Earth, or a supernova),
their coherent forward scattering from particles they encounter along the 
way can significantly modify their propagation. As a result, the probability 
for changing flavor can be rather different from its vacuum value. This 
is known as the Mikheyev-Smirnov-Wolfenstein (MSW) effect~\cite{MSW}.  

The reason is simple: matter contains electrons but no muons or taus.
Neutrinos of the three flavors interact with the electrons, the protons 
and the neutrons of the matter via neutral currents (NC). Electron neutrinos 
in addition interact with electrons via the charged current (CC). Since
the $W$ mass is much larger than typical center-of-mass energies for
neutrino scattering, the relevant parts of the Lagrangian for electron 
neutrinos can be written in the Fermi approximation
\begin{eqnarray}
{\cal L}=\bar{\nu}_e(i \partial\!\!\!/ 
-m)\nu_e-2\sqrt{2}G_F(\bar{\nu}_{eL}
\gamma_\mu\nu_{eL})(\bar{e}_L\gamma^\mu e_L).
\end{eqnarray}
The latter term modifies the effective mass of the $\nue$ when it goes 
through matter and, consequently, the transition probability changes. 
In the electron reference frame only the $\gamma_0$ part of the electron 
current contributes, being just the number operator:
\begin{eqnarray}
<\bar{e}_L\gamma_0 e_L> = n_e/2
\end{eqnarray}
with $n_e$ the electron density given by: 
\begin{eqnarray}
n_{e} = N_{A}\times Y_{e}\times \rho(r)~, 
\end{eqnarray}
and $Y_{e}= Z/A$ and $\rho$ are the electron fraction and the density
of the medium. Both of them depend on the physical propierties of the
medium traversed by the neutrinos. For the Earth's mantle, $Y_{e}= 0.494$, 
and for the Earth's core $Y_{e}= 0.466$~\cite{ms}. 
The contribution to the $\nue$ effective Hamiltonian due to the CC 
interaction with 
matter is consequently given by:
\begin{eqnarray}
\label{Heff}
A =\sqrt{2}\,G_F\,n_e~. 
\end{eqnarray}
For antineutrinos $A$ has to be replaced by $-A$, so the behaviour in 
matter of neutrinos and antineutrinos is different (matter is not CP 
invariant). 
In the treatment of oscillations, the extra potential only appears for
electron neutrinos, so the oscillatory terms will no longer be proportional 
to the mass-squared differences between the three families, but to 
effective 
masses-squared that result from the diagonalization of the 
Hamiltonian:
\begin{eqnarray}
U\left( \begin{array}{ccc}
m_1^2&0&0\\
0&m_2^2&0\\
0&0&m_3^2\\
\end{array}\right)U^\dagger +
\left( \begin{array}{ccc}
D&0&0\\
0&0&0\\
0&0&0\\
\end{array}
\right),
\end{eqnarray}
with $D = \pm 2 A E_{\nu}$ (depending if neutrinos or antineutrinos are 
considered).\par
The diagonalization of the 3-family Hamiltonian in the presence
of the extra matter term modifies the effective mixing angles 
appearing in oscillation formulae.
It has to be noticed that the ordering of the masses in the Hamiltonian
is relevant, and the effect of matter is swapped
between neutrinos and antineutrinos in the cases of direct and reversed 
mass hierarchy, as seen in Fig.~\ref{fig:signd}. Thus, the 
effect of matter in a neutrino oscillation
experiment allows the determination of the sign of $\Delta m^2_{23}$.
\par
\begin{figure}[tbh]
  \begin{center}
    \epsfig{file=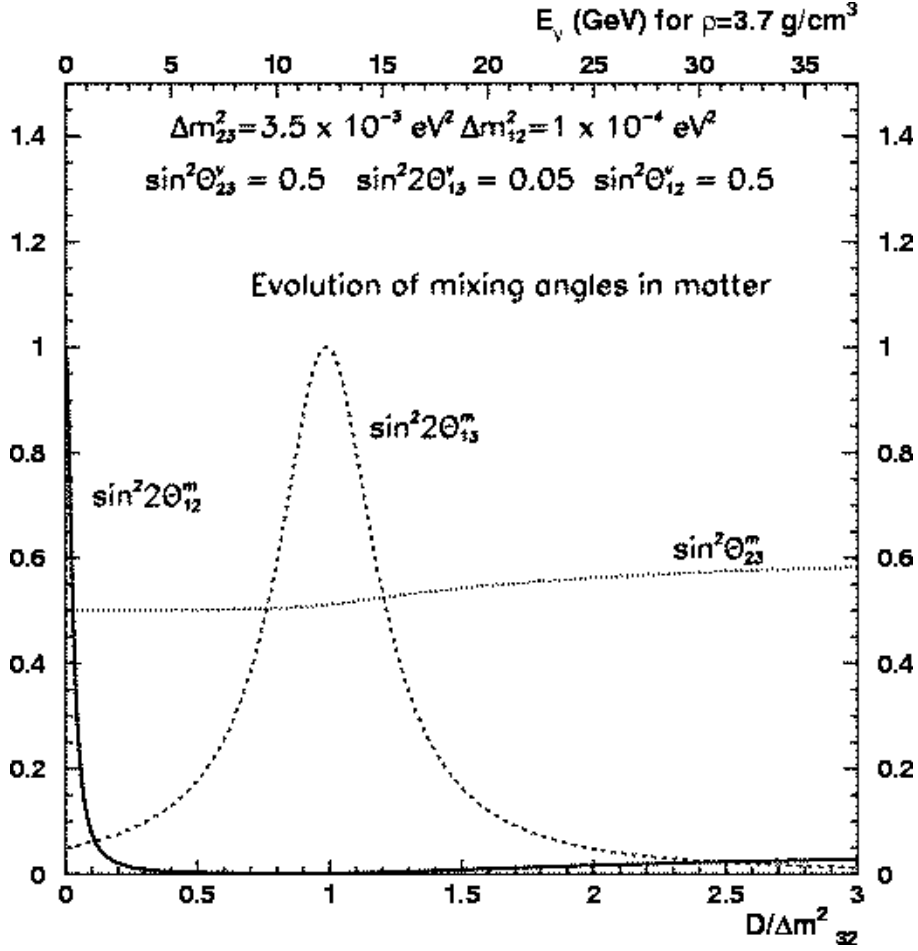,width=12cm}
    \caption{\it Evolution of the effective angles in matter as a function 
of $D$.}
    \label{fig:anglesmatter}
  \end{center}
\end{figure}

The full formulae for the effective quantities (mass differences and
angles) in a three-family scenario are quite complicated,
and can be found in~\cite{zaglauer} and \cite{physpot}. 
However, it can be shown that,
in the case $|\Delta m^2_{12}|\ll|\Delta m^2_{13}|\approx
|\Delta m^2_{23}|$, the three-family mixing decouples into two
independent two-family mixing scenarios, even in the presence
of matter effects~\cite{CPviol}. For energies not too small with respect to
that of the oscillation maximum
(where $|\Delta m^2_{23} L/E|\approx 1$), the effective oscillation
parameters will become (see Fig.~\ref{fig:anglesmatter}~\cite{physpot}):

\begin{eqnarray}
 \sin^2\theta^m_{23}(D)&\simeq&\sin^2\theta_{23}\\
 \sin^2 2\theta^m_{12}(D)&\simeq& \frac{\Delta m^2_{12}}{D} \sin2\theta_{12}\\
 \sin^2 2\theta^m_{13}(D)&\simeq& \frac{\sin^2 2\theta_{13}}{F} \\
 \delta^m &\simeq& \delta \\
 \Delta M^2_{12} &\simeq& D\\
 \Delta M^2_{23} &\simeq&\Delta m^2_{23}\sqrt{F}\\
\label{eqs:matterparam}
\end{eqnarray}
where we have defined
\begin{eqnarray}
  F \equiv \sin^2 2\theta_{13}+(\frac{D} {\Delta m^2_{23}}-\cos 
  2\theta_{13})^2.
\end{eqnarray}
As expected, we see explicitly that all the angles return to their 
original values in the limit of small matter density $D\to 0$ (i.e. 
$F \to 1$) in the limit $\Delta m_{12}\ll \Delta m_{13}$ 
and $\Delta m_{12}\ll D$.

At the Mikheyev-Smirnov-Wolfenstein (MSW) resonance energy
\begin{eqnarray}
  E^{res}_\nu(GeV)=\frac{\cos 2\theta_{13}\Delta m^2_{23}}
 {2\sqrt{2} G_F n_e}\simeq\frac{1.32\times 10^4\cos 2\theta_{13}
 \Delta m^2_{23}(eV^2)}{\rho(g/cm^3)}
\end{eqnarray}
the value of $\sin^2 2\theta^m_{13}$ becomes equal to 1,
independently of the actual value of the angle in vacuum. Also
the effective value of $\Delta m^2_{32}$ is significantly modified.
If we take $\Delta m^2_{23}=2.5\times 10^{-3}$ eV$^2$, $\theta_{13}\ll 1$ 
and $\rho=2.8$ g/cm$^3$, the energy of this MSW resonance is 11.8 GeV.\par
To see how this affects the oscillation probability, we can 
take the simplified formula for the $\nu_e\to\nu_\mu$ oscillation in 
vacuum, as appears in (\ref{vacexpand}), and neglect the mass difference 
$\Delta_{12}$. Replacing the vacuum parameters with their effective 
values, one obtains the approximate $\nu_e\to\nu_\mu$ oscillation 
probability in matter:
\begin{eqnarray}
  P(\nu_e\to\nu_\mu)=\sin^2 2\theta_{13}^m \sin^2\theta_{23}^m \sin^2
\Delta^m_{23}=
\frac{\sin^2 2\theta_{13}}{F} \sin^2\theta_{23}\sin^2
\left(\Delta m^2_{23}\sqrt{F}\frac{L}{4E}\right),
\label{eq:nuenumumatter}
\end{eqnarray}
and the oscillation probability at the MSW resonance energy becomes
\begin{eqnarray}
  P(\nu_e\to\nu_\mu;E^{res})= \sin^2\theta_{23}\sin^2
\left(\Delta m^2_{23}\sin 2\theta_{13}\frac{L}{4E^{res}}\right).
\end{eqnarray}
If we assume that $\sin 2\theta_{13}$ is small, and if the
distance $L$ is not too far away from that of the first maximum,
the probability reads:
\begin{eqnarray}
  P(\nu_e\to\nu_\mu;E^{res})\approx s_{23}^2\sin^2 2\theta_{13}
\times \left(\Delta m^2_{23}\frac{L}{4E^{res}}\right).
\label{eq:mattosc}
\end{eqnarray}
In the limit $\Delta m^2_{23} L/4E\ll 1$, this coincides with the vacuum 
expression of (\ref{eq:nuenumusimp}). 
This is why it is not possible
to measure the sign of $\Delta m^2_{23}$ if the baseline is too
short. Matter effects become visible at distances where the 
argument of the $\sin$ function approaches $\pi/4$. The 
oscillation maximum for the value $E^{res}=11.8$ GeV and $\Delta m^2_{23}=
2.5\times 10^{-3}$ eV$^2$ is $L = 2300$~km, so matter effects start
to be visible as an enhancement of the oscillation probability
around the MSW resonance energy for baselines longer than about 
1000 km. For antineutrinos the effective mixing angle is smaller
than the real one. For small baselines, a similar compensation as
in (\ref{eq:mattosc}) takes place, but at distances where matter
effects start to play a role the oscillation probability in matter
is smaller than that in vacuum.\par
The above considerations are valid near the resonance energy, in
other words for $D\approx \Delta m^2_{23}$. If $D>\Delta m^2_{23}$
i.e., $E_\nu>2E^{res}$, the effective mixing angle is always
smaller than the vacuum one, and the effect of matter is again a
suppression of the oscillation probability.

\def\beq{ \begin{equation}}
\def\eeq{\end{equation} }
\def\bea{\begin{eqnarray}}
\def\eea{\end{eqnarray}}
\def\r{\gamma}
\def\eps{\epsilon}
\def\be{\bar{\epsilon}}
\def\d{\delta}
\def\a{\alpha}
\def\b{\beta}
\def\n{\nu}
\def\la{\lambda}
\def\ra{\rightarrow}
\def\k{m_{eff}}
\def\e{\epsilon}
\def\p{\pi}
\def\th{\theta}
\hyphenation{author another created financial paper re-commend-ed Post-Script}


\subsubsection{Current status of neutrino mixing parameters}

We assume that active neutrino oscillations are indeed the solution 
to the neutrino puzzles, and that there are only
three light neutrino species. As discussed above, in this case the 
Standard Model is augmented
by at least 9 new parameters, of which 3 are neutrino
masses, 3 are real mixing angles, there is one oscillation phase
 and two additional Majorana phases, which exist only if the neutrinos are
Majorana particles. Neutrino oscillation experiments, however, are not
sensitive to the Majorana phases, and we set them to zero henceforth.

The angles $\theta_{23}$ and $\theta_{12}$ are thought to be mainly
responsible for solving, respectively, the atmospheric and solar neutrino
puzzles, as long as $|\Delta m^2_{23}|>|\Delta m^2_{12}|$ (normal neutrino
mass hierarchy).  If the opposite happens to be true ($|\Delta
m^2_{23}|<|\Delta m^2_{12}|$, an inverted mass hierarchy), the definitions
above can still be related to the appropriate neutrino puzzles, as long as
the mass eigenstates are relabelled $3\rightarrow 2\rightarrow
1\rightarrow 3$. We make this assumption henceforth, and refer to
$\theta_{23}\equiv\theta_{atm}$, $\theta_{12}\equiv\theta_{sun}$.  
Thus, the third mass eigenstate is defined as the one whose mass
squared is `further away' from the other two, which are arranged in
ascending order of masses-squared, without any loss of generality. Given
this choice, the inverted hierarchy differs from the normal hierarchy by
the sign of $\Delta m^2_{23}$. The entire oscillation parameter space is
spanned by varying the three mixing angles from 0 to $\pi/2$, keeping
$\Delta m^2_{12}>0$~\footnote{In two-flavour solar data analyses, it is
important also to keep the mixing angle from 0 to $\pi/2$ if $\Delta
m^2>0$ is fixed, in order to cover the entire parameter space
\cite{day-night,dark_side}.}, $\Delta m^2_{23}$ (including the sign), and
varying $\delta$ from $-\pi$ to $\pi$.

The current knowledge of the oscillation parameters can be summarized as
follows. There is good evidence that $|\Delta
m^2_{23}|\gg\Delta m^2_{12}$, and therefore $\Delta m^2_{23}\simeq\Delta
m^2_{13}$. This being the case, the current atmospheric data are sensitive
to $|\Delta m^2_{23}|$, $\tan^2\theta_{\rm atm}$, and $|U_{e3}|^2$, while
the current solar data are sensitive to $\Delta m^2_{12}$,
$\tan^2\theta_{sun}$, and $|U_{e3}|^2$. The Chooz and Palo Verde reactor 
experiments, when
considered in combination with the solar data, constrain $|U_{e3}|^2$ as a
function of $|\Delta m^2_{23}|$. The experimentally allowed range
of the parameters depends on a number of assumptions: which
experimental data are considered, how many neutrino species participate in
the oscillation, what was the statistical recipe used to define
allowed regions, etc. We comment briefly later on some of these points.
In this Section, we simply quote the current `standard' results, 
with some appropriate references.

\begin{figure}[htb]
\parbox{0.65\textwidth}{\includegraphics[height=7.0cm]{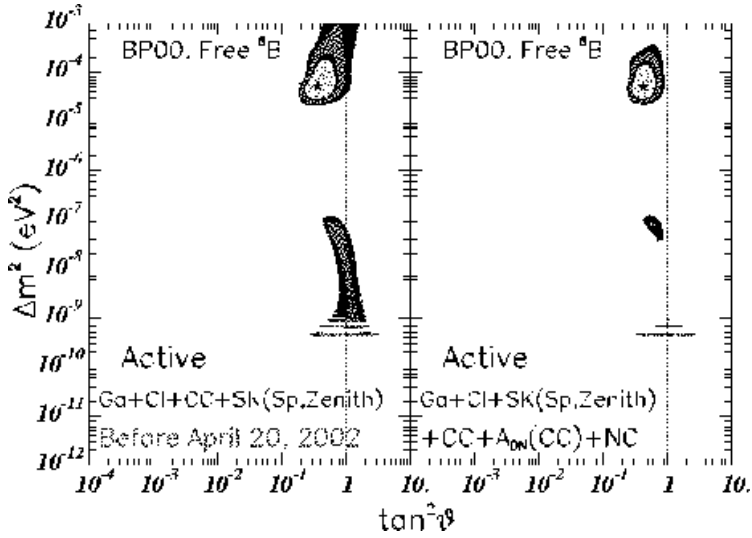}}
 \parbox{0.35\textwidth}{\caption{\it Global fits to current solar 
neutrino data~\cite{bahcalletal}, leaving free the $^8$B neutrino flux. 
The best-fit point is marked by the star, while the allowed regions are
shown at 90\%, 95\%, 99\%, and 99.73\% CL.  
Here, $\Delta m^2\equiv\Delta m^2_{12}$ and 
$\tan^2\theta\equiv\tan^2\theta_{sun}$. 
This analysis was performed assuming a two-flavour oscillation scenario. 
In the left panel,
only data announced before April 20, 2002 is analyzed, while in the right 
panel all the current
neutrino data is considered, including the recent neutral current 
measurements from SNO~\cite{SNO_nc}.}
\label{solar_fig}}
\end{figure}

For the atmospheric and reactor parameters, one 
obtains~\cite{osc_anal_comb} at the 99\% confidence level (CL),
\begin{eqnarray}
&|U_{e3}|^2<0.06, \\
&0.4<\tan^2\theta_{\rm atm}<2.5, \\
&1.2\times 10^{-3}~{\rm eV}^2<|\Delta m^2_{23}|<6.3\times 10^{-3}~{\rm eV}^2.
\end{eqnarray}
The situation of the solar parameters is less certain, and is best
described graphically. Fig.~\ref{solar_fig} depicts the result of the most
recent analysis of all solar neutrino data~\cite{bahcalletal}, which uses
the current version of the standard solar model~\cite{SSM}, but allows the
$^8$B neutrino flux to float in the fit. This particular analysis is
performed assuming two-flavour oscillations, a scenario which is realised
in the three-flavour case for $|U_{e3}|^2=0$.  The changes to the allowed
regions are not qualitatively significant for values of $|U_{e3}|^2$ up to
the upper bound quoted above: in fact, the allowed regions grow, even for
$|U_{e3}|^2=0$, simply because the confidence level contours are defined
for three instead of two degrees of freedom. The right 
panel of Fig.~\ref{solar_fig} exhibits
the impact of the recent neutral-current measurement from SNO on the
preferred solar neutrino oscillation parameters.

There are still several disjoint regions of the parameter space which
satisfy the current solar neutrino data at some CL. These include the
large mixing angle (LMA) solution, the low $\Delta m^2_{12}$ MSW solution
(LOW), various vacuum (VAC) solutions at very small values of $\Delta
m^2_{12}$, and (at a lower CL) the small mixing angle (SMA) MSW~\cite{MSW}
solution. Of the four regions, two (LMA, and LOW) are very robust, and
appear in different types of data analysis.  The VAC solutions are rather
unstable, and can disappear if the data are analysed in a different
fashion.  Note that the LOW solution is no longer connected to the VAC
solutions, as used to be the case~\cite{dark_side} before the recent SNO
neutral-current data. The SMA solution is currently excluded at more than
the three-sigma level, but cannot be completely discarded yet. A word of
caution is in order: it has recently been pointed out by several different
groups~\cite{stat_critic} that the standard statistical treatment of the
solar neutrino data - which is usually analysed via a $\chi^2$ method, and
includes the definition of confidence level contours - is not appropriate.
However, a $\chi^2$ method should be perfectly adequate when more data
available become available in the near future.

In summary, while some of the oscillation parameters are known with some
precision (the atmospheric mass-squared difference is known within a
factor of roughly six), the information regarding other parameters is very
uncertain. In particular, $\tan^2\theta_{sun}$ can be either very small
($\sim 10^{-4}-10^{-3}$), or close to unity, while $\Delta m^2_{12}$ can
take many different values, from around $10^{-10}$~eV$^2$ to more than
$10^{-4}$~eV$^2$. Finally, there is absolutely no information on the
CP-violating phase $\delta$, nor on the sign of $\Delta m^2_{23}$, while
for $|U_{e3}|^2$ only a moderate upper bound has been
established~\footnote{It is interesting to note that $|U_{e3}|$ can still 
be as large as the sine of the Cabibbo angle.}.
 
\subsubsection{Motivations for new physics}

So far, we have not discussed the potential implications of the LND data.
If the three experimental results from solar, atmospheric and LSND
neutrinos are all correct, there must be three different mass-squared
differences:  $\Delta m^2_{sun} << \Delta m^2_{atm} << \Delta m^2_{lsnd}$,
which cannot be accommodated with just three neutrinos.  The simplest
option for incorporating all the data would be to introduce a fourth light
neutrino, which must be {\it sterile}, i.e., having interactions with
Standard Model particles much weaker than the conventional weak
interaction, in order not to affect the invisible $Z$ decay width that was
measured very precisely at LEP.

In this case, one has to take into account the constraints from
cosmological Big Bang Nucleosynthesis: a sterile neutrino that mixes with
an active one, and is thus in equilibrium at the time of nucleosynthesis,
can change the abundance of primordially produced elements, such as
$^{4}$He and deuterium. The larger the mixing and the mass differences
between the sterile and active neutrinos, the bigger the deviations from
the successful standard calculations of the light element abundances. This
implies that models where the sterile component contributes to solar
rather than atmospheric neutrino oscillations would be accommodated more
easily. However, recent data from the two solar neutrino experiments - SNO
on charged-current and neutral-current scattering - and Super-Kamiokande -
on electron scattering - provide very strong evidence that there are
additional active neutrinos coming from the Sun. A large mixing between
the sterile and active neutrinos is excluded if the SSM calculation of the
$^8B$ flux is correct \cite{Bahcall:2002hv}.

Altogether, we conclude that current neutrino data and standard big bang
nucleosynthesis disfavour the four-neutrino mixing scheme.

An alternative approach accounts for the LSND results within the
three-neutrino framework, by invoking CPT
violation~\cite{Barenboim:2002hx}. However, this hypothesis involves a
dramatic change in the Standard Model that requires more motivation from
the data. Another possibility is that small non-standard weak interactions
of leptons may instead provide a simultaneous solution to the three
neutrino anomalies without introducing a sterile neutrino or invoking CPT
violation~\cite{Babu:2002ic}. According to this hypothesis, $\Delta L = 2$
lepton-number-violating muon decays are invoked to account for the LSND
events. These anomalous decays could easily be tested at a Neutrino
Factory using a detector capable of charge
discrimination~\cite{Bueno:2000jy}, as discussed later in this report.

\subsubsection{Prospects for the near future}

More information will be needed for the detailed planning of neutrino
factory experiments, much of which may be provided by near-future
experiments, as we now discuss. The precision with which some neutrino
oscillation parameters are known will improve significantly in the near
future, and it is almost certain that the ambiguity in determining the
solution to the solar neutrino puzzle will disappear. In this Section, we
review briefly how open questions in the current analysis may be resolved
by future experiments, discussing subsequently more details of the 
experiments.

The values of $|\Delta m^2_{23}|$ and $\tan^2\theta_{\rm atm}$ should be
better determined by long-baseline neutrino
experiments~\cite{K2K,MINOS,CNGS}.  In particular, the MINOS
experiment~\cite{MINOS} aims at 10\% uncertainties, while the CNGS
programme~\cite{CNGS} may achieve slightly better precision. The
sensitivity to $|U_{e3}|^2$, on the other hand, is expected to be limited
to at most a few \%: perhaps $|U_{e3}|^2=0.01$ could be obtained at
the first-generation long-baseline beams, with JHF doing an order of
magnitude better.

In the solar sector, different solutions will be explored by different
experiments. We concentrate first on strong `smoking gun' analyses, and
comment later on the prospects for other experiments.

{\it The LMA solution} to the solar neutrino puzzle will soon be either
established or excluded by the KamLAND reactor experiment~\cite{KamLAND}.
Furthermore, if LMA happens to be the correct solution, KamLAND should be
able to measure the oscillation parameters $\tan^2\theta_{sun}$ and
$\Delta m^2_{12}$ with good precision by analysing the $\bar{\nu}_e$
energy spectrum, as has been recently investigated by different
groups~\cite{kamland_results,kamland_plus_solar,prospects_summary}.  
Three years of KamLAND running should allow one to determine, at the
three-sigma level, $\Delta m^2_{12}$ within 5\% and $\sin^2 2\theta_{sun}$
within 0.1. A combination of KamLAND reactor data and solar data should
start to address the issue whether $\theta_{sun}$ is smaller or greater
than $\pi/4$ \cite{kamland_plus_solar}.

{\it The LOW solution} will be either excluded or unambiguously
established by the Borexino experiment~\cite{day-night} (and possibly by
an upgrade of the KamLAND experiment such that it can be used to see
$^7$Be solar neutrinos), using the zenith-angle distribution of the $^7$Be
solar neutrino flux. This should allow one to measure, at the three-sigma
level, $\Delta m^2_{12}$ within a factor of three (say, in the range 1 to
$3\times 10^{-7}$~eV$^2$) and $\tan^2\theta_{sun}$ within
0.2~\cite{day-night,prospects_summary}.  These estimates are very
conservative and do not depend, for example, on the solar model prediction
for the $^7$Be neutrino flux~\cite{day-night}.
 
{\it VAC solutions} with $\Delta m^2_{12}$ less than a few $\times
10^{-9}$~eV$^2$ and greater than a few $\times 10^{-11}$~eV$^2$, and
$\tan^2\theta_{sun}$ between roughly 0.01 and 100, will also be either
excluded or established by experiments capable of measuring the $^7$Be
solar neutrino flux. In this region of parameter space, the flux of $^7$Be
solar neutrinos depends very strongly on the Earth--Sun distance, and
anomalous seasonal variations should be readily observed, for example, at
GNO or Borexino. Estimates of what will be inferred from the data of these 
experiments and KamLAND 
indicate~\cite{seasonal}
that, even if the $^7$Be solar neutrino flux is conservatively assumed to
be unknown, $\Delta m^2_{12}$ can be measured at better than the percent
level (see also~\cite{prospects_summary}).

The situation with {\it the SMA solution} is less clear. If SMA is indeed
correct, something must be wrong with part of the current solar neutrino
data, or we have been extremely `unlucky'. Nonetheless, the SMA solution
would be favoured if none of the above `smoking gun' signatures were
observed. There are also `smoking gun' signatures for the SMA solution,
but their non-observation would not necessarily exclude the SMA region.  
One characteristic feature of the SMA solution is a spectrum distortion
for $^8$B neutrinos, which can be measured at Super-Kamiokande and SNO. 
The
current Super-Kamiokande spectrum data are consistent with a constant
suppression of the $^8$B neutrino flux, and an analysis of the
Super-Kamiokande data alone excludes the SMA solution at more than 95\%
CL~\cite{superk_only}.  Combined analyses of all the solar data render the
SMA solution even less likely. The SMA solution also predicts that the
$^7$Be solar neutrino flux should be very suppressed with respect to
standard solar model results. Therefore, the measurement of a very small
$^7$Be neutrino flux could be interpreted as a `smoking gun' signature for
the SMA solution.  However, background-suppression methods would be needed
to measure a very suppressed flux, and some independent checks of the
reliability of such techniques are inconclusive~\cite{seasonal}.
Furthermore, in order to relate a small measured flux to neutrino
oscillations, one must rely on predictions from solar physics, which we
should prefer to avoid.

\begin{figure}[htb]
\parbox{0.45\textwidth}{\includegraphics[height=11.0cm]{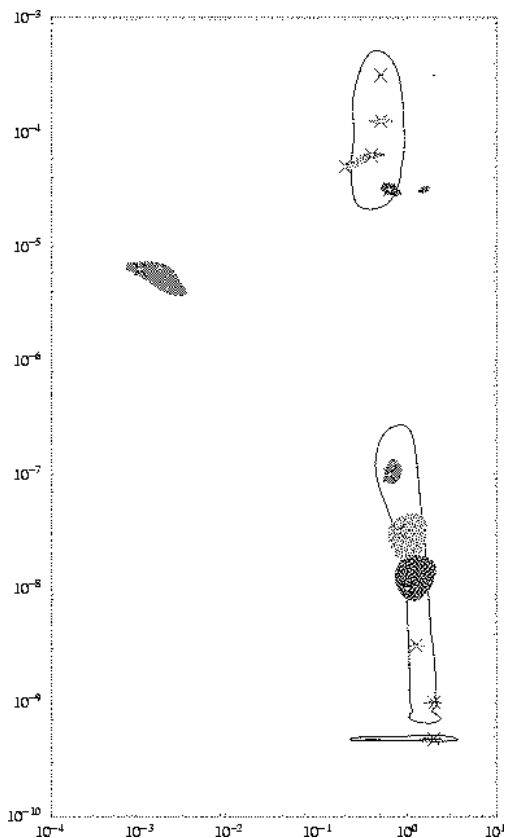}}
 \parbox{0.55\textwidth}{\caption{{\it The 99\% confidence level contours
obtained after fitting simulated future solar data (including KamLAND and Borexino), 
for a few points (marked with an $\times$) 
in the $(\Delta m^2_{12}/$eV$^2, \tan^2\theta_{sun})$ plane. The 
continuous lines
indicate the 99\% CL contours obtained after analysing the solar
data available earlier in 2002. Note that not only is the true solution 
(SMA, LMA, etc) identified, but 
the values of $\Delta m^2_{12}/$eV$^2$ and $\tan^2\theta_{sun}$ are also 
measured with good accuracy. See~\cite{prospects_summary} for details.} 
\label{various_measurements}}}
\end {figure}

Fig.~\ref{various_measurements} depicts the expected 99\% CL contours in
the $(\Delta m^2_{12}/$eV$^2, \tan^2\theta_{sun})$ plane for different
candidate solutions to the solar neutrino puzzle, after the advent of
KamLAND and Borexino. We see that the large degeneracies in the parameter
space will be lifted (with a few exceptions in the LMA region, as pointed
out in~\cite{kamland_plus_solar}), and that reasonably precise
measurements of the oscillation parameters are to be expected.

Supplementary possibilities are also available. The SNO experiment may 
also
provide sufficient further information to resolve the ambiguities in the
solar neutrino sector~\cite{SNO_spectrum,SNO_generic}. Further information
may also be obtained if neutrinos from a nearby supernova are
detected~\cite{supernova}. Finally, it is important to mention that
non-oscillation experiments can also contribute to the understanding of
neutrino masses and leptonic mixing angles. In particular, future searches
for neutrinoless double beta decay \cite{doublebeta_exp} are not only
capable of measuring a particular combination of the Majorana neutrino
phases, but can also help piece together the solar neutrino
puzzle~\cite{doublebeta_theo}.

We conclude that near-future experiments will provide much of the key 
information needed for planning neutrino-factory experiments, in 
particular whether the LMA solar solution is correct.   

\newpage
\section{CONVENTIONAL NEUTRINO BEAMS}

\subsection{\bf First-Generation Long-Baseline Neutrino Projects}

The small values of $\Delta m^2$ measured in oscillations of atmospheric
and solar neutrinos have focused the interest of the community of neutrino
physicists on long-baseline projects using neutrinos of artificial
origin. As an example, for a value of $\Delta m^2_{23}=2.5\times
10^{-3}eV^2$, the maximum of the oscillation probability for 1~GeV
neutrinos, a typical energy produced in accelerators, occurs at a distance
of about 500~km. The $\Delta m^2$ value for solar neutrinos is about two
orders of magnitude smaller, so oscillations of reactor neutrinos, whose
energy is of the order of 10 MeV, can be observed at the same distance. We
describe in this chapter the different approved long-baseline projects.

\subsubsection{KamLAND}

We have seen that solar neutrino experiments suggest oscillations with
parameters in one of four possible regions, usually referred to as the
large mixing angle (LMA), small mixing angle (SMA), LOW and vacuum (VAC)
solutions. Only the large mixing angle solution would allow the practical
possibility of discovering CP violation in neutrino oscillations. This
region will be probed over the next few years by the KamLAND reactor
neutrino experiment~\cite{KamLAND}, which started taking data in January
2002. After a few years of data taking, KamLAND will be capable of either
excluding the entire LMA region or, in the case of a positive signal for
oscillations, not only of establishing $\nu_{e}\leftrightarrow \nu_{\rm
other}$ oscillations in the solar parameter region, but also of measuring
the oscillation parameters $(\tan^2\theta_{12},\Delta m^2_{12})$ with
unprecedented precision~\cite{kamland_results,Kam_KAM,Carlos-DeGouvea}.

KamLAND is located in the old Kamiokande site in the Kamioka
mine in Japan. Its neutrino source consists of 16 nuclear power plants at 
distances of a few hundred km. The
apparatus consists of approximately 1~kt of liquid scintillator
that detects reactor neutrinos through the reaction
\begin{equation}
\label{eqn:nureac} p + \overline{\nu}_{e} \rightarrow n + e^{+}.
\end{equation}
The positron is then detected via scintillation light and its
annihilation with an electron.  This annihilation, in delayed
coincidence with the $\gamma$-ray from neutron capture, represents
a very clean signature.

To estimate the precision to which KamLAND could measure the oscillation
parameters, the authors of~\cite{murayama} considered a 3 kt-y exposure
during which the reactors operated at 78\% of their maximum capacity.
Backgrounds were neglected and perfect detector efficiency was assumed.
Systematic effects, in particular those connected with chemical
composition and flux uncertainties, have also been studied.

\begin{figure}[bthp]
\begin{center}
\epsfig{file=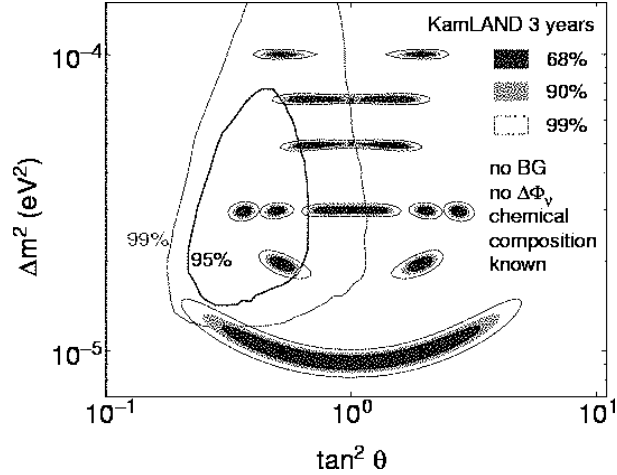,width=8cm}
\end{center}
\caption{\it The expected measurement of $\Delta m^2$ and $\tan^2
\theta$ in the KamLAND experiment. Contours for 68\%, 90\% and 99\% CL are 
shown, and the LMA
solution to the solar neutrino problem is overlaid~\cite{murayama}.} 
\label{fig:contours}
\end{figure}

The measurements that KamLAND could be expected to perform with
these assumptions are shown in Fig.~\ref{fig:contours}, taken 
from~\cite{murayama}. Since the oscillations depend on the mixing
angle only through $\sin^2 2\tsun$, there is a two-fold degeneracy
in the measurement, hence the reflection symmetry about $\tan^2
\tsun=1$. The LMA solution, which is overlaid, does not
possess such a symmetry, so it is necessary to plot against $\tan^2
\tsun$ and not $\sin^2 2\tsun$~\cite{darkside}, in order to have a visual
combination of the two measurements.
%

It has been shown in~\cite{Carlos-DeGouvea} that, combining KamLAND with
the solar data, one can solve the two-fold degeneracy, i.e., find out
whether $\theta_{12}<\pi/4$ or $\theta_{12}>\pi/4$ at the 95\% confidence
level (CL) for most of the LMA region. In this reference, a more
conservative analysis than in~\cite{murayama} was performed, with
comparable results.

\subsubsection{Long-baseline accelerator-based experiments}

Whilst the KamLAND experiment will help understand oscillations in
the solar region, the study of the atmospheric region is best performed
with artificial neutrino beams from particle accelerators.
Three such beams exist or are planned in three different
continents: 
\begin{itemize}
\item The K2K beam from KEK to Super-Kamiokande,
\item The NuMI beam from Fermilab to the Soudan mine,
\item The CNGS beam from CERN to the Gran Sasso laboratory.
\end{itemize}
These three projects are quite different in scope and capabilities,
due to the different characteristics of the accelerators that produce
the beams, and of the proposed detectors. The K2K beam
has lower energy and intensity, and its main aim is to
confirm the $\nu_\mu$ disappearance observed in Super-Kamiokande.
The NuMI beam will have more intensity, and the focusing system
chosen will produce neutrinos peaked at an energy of about 2 GeV,
to achieve a better precision on the measurement of the oscillation
parameters. The CNGS beam will have smaller intensity but
higher energy than NuMI, and will be used for the $\tau$
search, to confirm that the atmospheric deficit is due to
$\nu_\mu\to\nu_\tau$ oscillations. Characteristics of the
different beams are shown in Table~\ref{tab:lblbeams}.
\begin{table}[htbp]
  \begin{center}
    \begin{tabular}{|c|c|c|c|c|}\hline
      Beam&$E_{protons}$(GeV)&Power (MW)&$<E_\nu>$&L (km)\\ \hline
      K2K&12&0.005&1.3&250\\
      NuMI (LE)&120&0.41&3.5&734\\
      CNGS&400&0.3&18&732\\
      JHF&50&0.75&0.7&295\\ \hline
    \end{tabular}
    \caption{\it Characteristics of long-baseline beams}
    \label{tab:lblbeams}
  \end{center}
\end{table}

A large community is gathering around the possibility of building a
beamline for neutrinos at the future Japanese Hadron Facility (JHF), and
sending an off-axis beam to Super-Kamiokande, and to a planned future
large \v{C}erenkov detector (Hyper-Kamiokande). Details of this
second-generation beamline, that will mainly be used to look for a
non-zero value of the electron neutrino mixing angle $\theta_{13}$, are
given below. A more detailed analysis is present
in~\cite{gomezharris}

\subsubsection{K2K}

The only long-baseline experiment to have taken data so far is
the K2K experiment, which was in the middle of its run when an
unfortunate accident in the Super-Kamiokande detector~\footnote{See 
{\tt http://www-sk.icrr.u-tokyo.ac.jp/doc/sk/index.html}
for details.} stopped their data taking until January 2003. The goal of
K2K is to confirm the evidence of oscillations presented by
Super-Kamiokande by looking for the disappearance of $\nu_\mu$'s. 
The neutrino beam has a mean energy of about 1\gev and points toward the
Super-Kamiokande detector, which is located some 250 \km from the
neutrino source. Thus, the $\Delta^2 m$ to which they are
sensitive is:
\[
\Delta m^2 \sim \frac{1}{250} \sim 4 \times 10^{-3} eV^2
\]
near the maximum of the atmospheric oscillation.
Before the accident in Super-Kamiokande, K2K had about $5 \times 10^{19}$ 
of their planned final $10^{20}$ protons on target. They had observed $56$ 
events, compared with the $80 \pm 6$ expected in the case of no oscillations. 
These numbers and their measured energy spectrum are consistent 
with the Super-Kamiokande data. However, the K2K data 
alone could not yet be regarded as an independent confirmation of the 
atmospheric oscillation effect.

\subsubsection{NuMI}

\begin{figure}[bthp]
\begin{center}
\epsfig{file=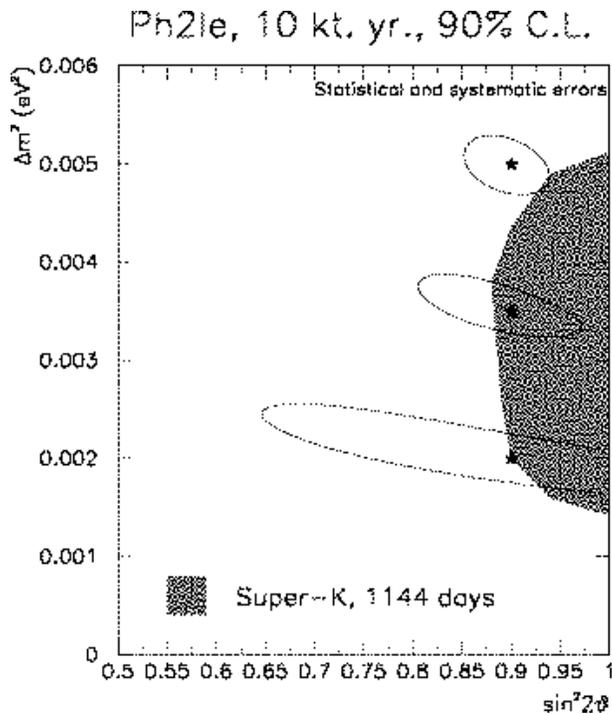,width=8cm}
\end{center}
\caption{\it Expected 90\% C.L. determination of the atmospheric 
oscillation 
parameters in a 10 kt-year exposure of MINOS to the low-enegy NuMI beam.} 
\label{fig:minos-sensi}
\end{figure}

The MINOS detector~\cite{MINOS} for the NUMI~\cite{NuMI} beam between
Fermilab and the Soudan mine in Minnesota is currently under construction.
Its goal is to make a precision measurement of the parameters governing
the atmospheric neutrino oscillation. The experiment has a baseline of 735
\km, and can be modified to produce beams of mean energies up to 14\gev.
The low-energy beam, with an average neutrino energy of 3.5 \gev, yields
about the same $L/E$ than the K2K experiment, and due to its higher flux
is expected to provide a better precision on the atmospheric oscillation
angle $\Delta m^2_{23}$. The far detector is a 5~kt sandwich of magnetized
steel and scintillator planes. The near detector will be located at 1040
\meter from the proton target, and will be a scaled-down 1~kt
version of MINOS. Systematic errors from the difference in the neutrino
spectrum between near and far location are expected to be at the level of
few \%. The run is expected to start in early 2005, and after a two-year
run MINOS should be able to perform a 10\% measurement of the atmospheric
parameters, unless $\datm$ happens to be too low, as shown in
Fig.~\ref{fig:minos-sensi}.

\subsubsection{CERN-Gran Sasso}

Due to the higher energy of the SPS accelerator, the beam
from CERN to Gran Sasso is the ideal place to look for $\tau$
appearance, and this has been the main objective in designing the 
detectors
at Gran Sasso.\par
The first approved project aimed at $\tau$ identification is
the OPERA experiment~\cite{OPERA}. The requirement for a large 
mass and a space resolution of the
order of 1~$\mu$m has led to the choice of hybrid emulsion as the
basic detector component.
Lead-emulsion sandwiches are arranged in bricks, walls, modules and
supermodules, for a total mass of 1.8~kt. Tracking devices are
placed behind each brick to localize where interactions take place,
enabling one to remove and develop in real time the interesting 
brick.\par
Another detector that will observe neutrinos from the CNGS beam
is ICARUS~\cite{ICARUS}. So far, only a 600~t module is approved,
which has already been built and successfully tested in summer 2001 on 
the surface in
Pavia, and a proposal for a five-fold mass increase is pending
approval. The detector consists of a large liquid Argon TPC, 
that allows a three-dimensional reconstruction of neutrino 
interactions with excellent imaging and energy reconstruction.\par
The $\tau$ identification strategy of the two detectors is 
complementary. The OPERA approach is topological, the $\tau$
being identified by the direct observation of the decay kink, while
ICARUS will exploit the different kinematics of $\tau$ decays
(especially in the golden $\tau\to e$ channel) compared to the
backgrounds from charged currents and $\nu_e$ in the beam.
The $\tau$ identification capabilities of the two detectors 
(assuming the 3~kt option for ICARUS) are quite similar.
Each experiment expects to be able to observe about 10 $\tau$ events
after 5 years of running, with an expected background of $0.7$ events,
assuming $\Delta m_{23}^2$ is $2.8\times 10^{-3}eV^2$.

\subsubsection{JHF-Super-Kamiokande}

We have seen that the three projects K2K, NuMI and CNGS will give us a
better understanding of oscillations in the atmospheric region, measuring
precisely the oscillation parameters and establishing the
$\nu_\mu\to\nu_\tau$ nature of such oscillations. Some progress will be
also made in the search for a non-zero participation of the electron
neutrino in the third flavour eigenstate, namely the element $|U_{e3}|$ of 
the leptonic mixing matrix.  The search for a
nonzero value of $\sin\theta_{13}$ is a fundamental step towards
defining the future of neutrino physics. A further step in this direction
will require high-intensity neutrino beams (super-beams).

The first neutrino super-beam will probably be built from the future
Japanese Hadron Facility (JHF) in the JAERI lab at Tokai-mura (60 km NE of
KEK) to the Kamioka site~\cite{jhf}.  The 50~GeV JHF machine, scheduled to
start operation in 2006, is designed to deliver $3.3 \times 10^{14}$
protons every $3.4$ seconds with a high beam power of 0.77~MW, that could
later be upgraded to 4~MW~\cite{mori}.  The beam axis will be tilted by 2
degrees with respect to the position of the far location. This off-axis
beam will be almost monochromatic around an energy of 800 MeV, and will
have a smaller $\nu_e$ contamination than the direct beam. The baseline
between JHF and Super-Kamiokande is about 295~km.

\begin{figure}[bthp]
\centerline{\epsfig{file=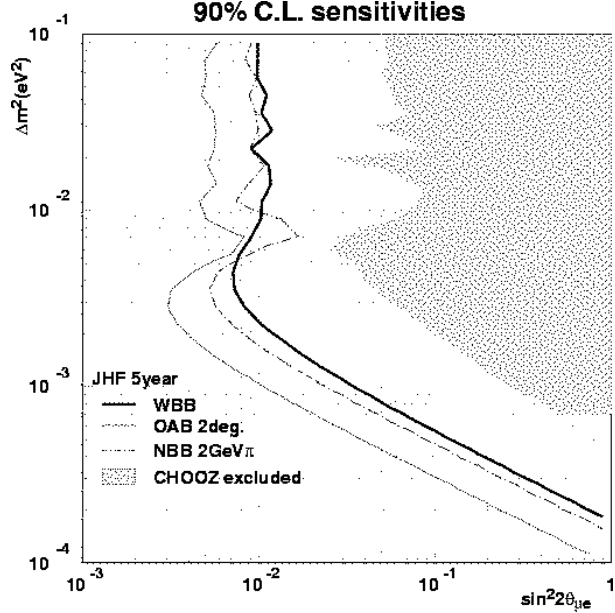,width=8cm} } 
\caption{
\it The expected 90\% CL sensitivity contours for 5-year exposures in 
various
beam configurations for the JHF-Super-Kamiokande project. The 90\%
CL excluded region of the Chooz experiment is plotted for comparison. 
Note
that the plotted quantity is $\sin^22\theta_{\mu e} \equiv 0.5
\sin^2 2\theta_{13}$.} 
\label{nueapp:contours}
\end{figure}

The first phase of the project could start as early as 2007. Given
the great intensity of the beam and the large mass of
Super-Kamiokande, in a 5-year run this project would provide the
best knowledge of the neutrino mixing matrix parameters before the 
Neutrino Factory. The authors of~\cite{jhf} claim:

\begin{itemize}
\item A precision measurement of the atmospheric oscillation
pattern, yielding $\sin^22\theta_{23}$ with $1 \%$ precision and $\Delta
m^2_{atm}$ with a precision better than $10^{-4}$~$\rm eV^2$,
\item Sensitivity to $\theta_{13}$ as small as
$3^\circ (\sin^22\theta_{13} \sim 0.01)$, by searching for $\nu_\mu
\rightarrow \nu_e$ appearance (see \cite{jhf} for details
on background reduction, taking advantage of the narrow beam and
the good electron discrimination capabilities of
Super-Kamiokande). This is illustrated in 
Fig.~\ref{nueapp:contours} (notice that the plotted quantity is
$\sin^22\theta_{\mu e} \equiv 0.5 \sin^2 2\theta_{13}$). The
expected sensitivity is an order of magnitude better than the
current limit set by the Chooz experiment, namely $\sin^22\theta_{13} < 
0.1$.
\end{itemize}


\subsubsection{Possible scenario}

After operation of the first-generation neutrino beams
and further progress in the JHF project, one could expect the 
experimental situation to be as follows:

\begin{itemize}
\item KamLAND will have established or excluded the LMA solution
for the solar oscillations,
\item MINOS will have measured the
atmospheric oscillation parameters with a few \% precision,
\item The reality of oscillations will have been proven beyond any doubt 
by the observation of $\tau$ appearance at Gran Sasso,
\item The angle $\theta_{13}$ will be measured if it is larger than about
$2^\circ-3^\circ$.
\end{itemize}

\subsection{\bf Second-Generation Long-Baseline Neutrino Beams}

Extremely intense neutrino beams could have very large discovery
potential, in the search for a non-zero value of $\theta_{13}$
and possibly CP violation, for a favourable choice of parameters.
Several studies have been performed to assess the discovery reach
of such machines, also in comparison with Neutrino Factories, 
since normally the proton driver used to produce such a super-beam
would be the first step in the construction of a Neutrino Factory.\par
We have already seen that the JHF-Super-Kamiokande project can be
considered as the first neutrino super-beam. However, it is
foreseen to increase the intensity of that beam by almost a factor 5, and
to use as a target a next-generation water \v{C}erenkov detector 
(Hyper-Kamiokande), 
with an order of magnitude more mass than the present 
Super-Kamiokande. Such a large water detector would not only be
used as a neutrino target, but, due to its large mass, would be able
to push the search for nucleon decay into an unexplored region, and
also be a unique laboratory for supernova neutrino astronomy. Another 
possible location for a large water detector is the Fr\'ejus tunnel
between France and Italy. This site is located at 130~km from CERN,
from where it could be illuminated by a low-energy neutrino beam.
Another option has been studied in the United States, in the context of
a high-power proton driver for the Fermilab complex. In that case
the energy is larger than the previous proposals, and different 
possible detectors have been studied.\par
A different way of building a neutrino beam, conceptually similar
to that of the Neutrino Factory, is to produce neutrinos from the
decays of heavy ions circulating in a storage ring. Compared to
storing muons, the quality factor, defined as the ratio of the 
relativistic $\gamma$ factor and the free energy $E_0$ of the reaction
is much higher, thus producing narrower beams, and large fluxes
even for relatively small beam currents. The resulting $\beta$ beam would
consist of pure $\nu_e$ (or $\bar{\nu}_e$, depending on the isotope
used for neutrino production), of relatively low energy. The physics
capabilities of such a beam are similar to those of the super-beam,
but complementary since electron neutrinos are involved instead of
muon neutrinos. Very interesting opportunities may instead rise from
the possibility of running simultaneously, but with different timing,
a super-beam and a radioactive decay beam (beta beam). The possibility
of having $\nu_e$ and $\nu_\mu$ in the same beamline, but separated 
by timing, is extremely interesting since, unlike the Neutrino Factory,
oscillated events can be separated from those of the original beam
by mean of flavour tagging without the need of lepton charge 
identification. The most appealing possibility of this option
is the direct search 
for T violation, comparing the $\nu_e\to\nu_\mu$ and $\nu_\mu\to\nu_e$
oscillation probabilities.

\subsubsection{JHF-Hyper-Kamiokande}

The JHF neutrino project already foresees the possibility of an 
upgrade to higher neutrino flux (4~MW with respect to the initial 
0.77~MW) and the beam will be pointed in a direction intermediate 
between the present
Super-Kamiokande location and the main candidate site for a future 
Hyper-Kamiokande experiment, in such a way that both locations will be
illuminated by a narrow-band beam, off-axis by 2 degrees.
The Hyper-Kamiokande far detector would be composed of
several almost cubic modules of size $45\times 45\times 46$, 
for a total fiducial volume of 70 kt per module; 16 compartments
would make a total fiducial mass of above 1 Mt, read by about
200,000 photomultipliers, if the spacing between them is 1 meter.
\par
\begin{figure}[tbhp]
\begin{center}
\epsfig{file=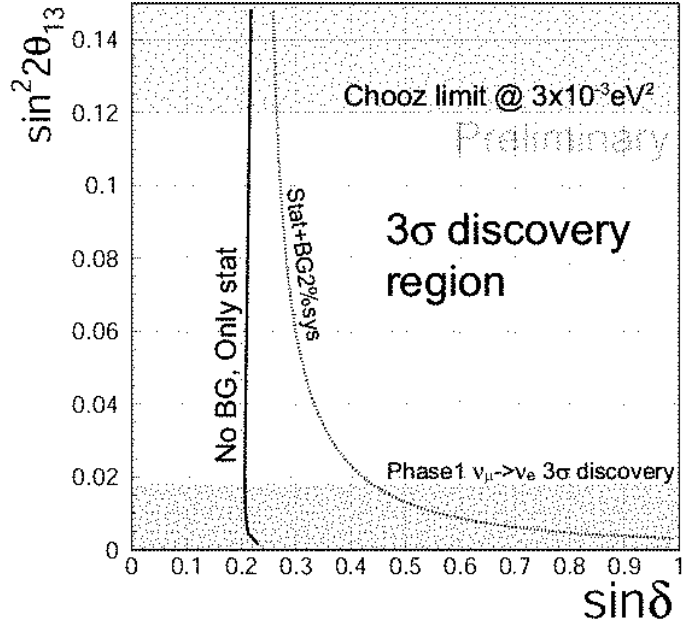,width=9cm}
\end{center}
\caption{\it The JHF-Hyper-Kamiokande sensitivity for $\sin^2 
2\theta_{13}$ 
= 0.01, 0.04 and 0.10.} 
\label{fig:jhfcp}
\end{figure}

A combination of Hyper-Kamiokande and a 4 MW beam would
provide the possibility of
exploring CP violation in the leptonic sector. For this purpose,
it is necessary to compare oscillation probabilities for neutrinos
and antineutrinos. The latter can be produced by focusing negative
instead of positive pions in the horn system. However, since they 
are smaller in number, and since antineutrinos have a smaller cross
section than neutrinos, the antineutrino run would have to be
considerably longer than that with neutrinos, in order to 
have comparable statistical accuracy. For instance, 
for an experimental setup with a 4 MW beam, 1 Mt far
detector, 2 years of $\nu_\mu$ and 6.8 years of $\bar{\nu}_\mu$ 
running, the 3$\sigma$ sensitivity extends to $\sin\delta>0.25$
(14 degrees) at large $\sin^2 2\theta_{13}$ and $\sin\delta>0.55$
for $\sin^2 2\theta_{13}=0.01$, as seen in Fig.~\ref{fig:jhfcp}.\par

\subsubsection{Possible off-axis experiments in the CNGS beam}
Proposals have been made for an off-axis experiment exposed to the
CNGS beam, also with the aim of improving the sensitivity to
$\theta_{e3}$.  One suggestion is to install a water \v{C}erenkov
experiment in the Gulf of Taranto, whose distance from CERN could be
varied to optimize the sensitivity to the oscillation
pattern~\cite{Dydak}. Another suggestion is a surface experiment close to
the Gran Sasso laboratory, using the atmosphere as a
target~\cite{Vannucci}.

\subsubsection{Studies for a high-energy super-beam at FNAL}
\newcommand{\pr}[1]{Phys. Rev. #1}
\newcommand{\prl}[1]{Phys. Rev. Lett. #1}

\def\beq{\begin{equation}}
\def\eeq{\end{equation}}
\def\bea{\begin{eqnarray}}
\def\eea{\end{eqnarray}}
\def\bq{\begin{quote}}
\def\eq{\end{quote}}
\def\ben{\begin{enumerate}}
\def\een{\end{enumerate}}
\def\nn{\nonumber}
\def\fr{\frac}
\def\hl{\hline}
\def\ra{\rightarrow}

\def\dm{\Delta m}
\def\dtm{\Delta \tilde{m}}

\def\etal{{\it et al.}}
\def\ie{{\it i.e.}}
\def\eg{{\it e.g.}}
\def\lesssim{\mathrel{\mathpalette\vereq<}}
\def\gtrsim{\mathrel{\mathpalette\vereq>}}
\makeatletter
\def\vereq#1#2{
\lower3pt\vbox{\baselineskip1.5pt \lineskip1.5pt
\ialign{$\m@th#1\hfill##\hfil$\crcr#2\crcr\sim\crcr}}}
\makeatother
 
\renewcommand{\thefootnote}{\fnsymbol{footnote}}
\def\beq{\begin{equation}}
\def\eeq{\end{equation}}
\def\bea{\begin{eqnarray}}
\def\eea{\end{eqnarray}}
\def\bq{\begin{quote}}
\def\eq{\end{quote}}
\def\ben{\begin{enumerate}}
\def\een{\end{enumerate}}
\def\nn{\nonumber}
\def\fr{\frac}
\def\hl{\hline}
\def\ra{\rightarrow}


An efficient way of producing a super-beam at FNAL would be to take
advantage of the NuMI~\cite{MINOS} investment and make an upgrade of the
proton driver~\cite{booster,pdriver,foster}. One could then envisage a new
generation of experiments to include shooting the intense conventional
muon-type neutrino beam, produced with NuMI and the upgraded proton
driver, towards a detector located suitably off the main beam direction.  
These off-axis beams have several advantages with respect to their on-axis
counterparts, as discussed earlier. They are more intense, narrower and
lower energy beams without a large high-energy tail, and they provide
clean access to $\nu_e$ appearance, given the existence of a suitable
detector with good electron identification capabilities. In order to fully
take advantage of such tools, however, two different beams are essential,
one predominantly $\nu_{\mu}$ and the other predominantly
$\bar{\nu}_{\mu}$. The antineutrino beam, however, provides an extra
experimental challenge: as discussed in connection with JHF, the pion
production and antineutrino scattering cross sections are suppressed with
respect to those for neutrinos. In order to obtain a statistically
comparable antineutrino and neutrino data set, the running time ratio
between antineutrinos and neutrinos should be 2 or 3 to 1.

Studies show how well experiments in the NuMI beamline with a proton
driver upgrade plus an off-axis detector can address these
issues~\cite{our_paper}. In order to estimate properly the experimental
response, e.g., signal efficiency as well as beam and detector-induced
backgrounds, a realistic simulation of the NuMI beam was performed,
including the response of a 20~kt highly segmented iron detector, followed
by a detailed `data' analysis. A 5~kt Liquid Argon TPC in the same NUMI
location would have comparable sensitivity to $|U_{e3}|^2$.

\begin{figure}
\centerline{\epsfxsize 14.2cm \epsffile{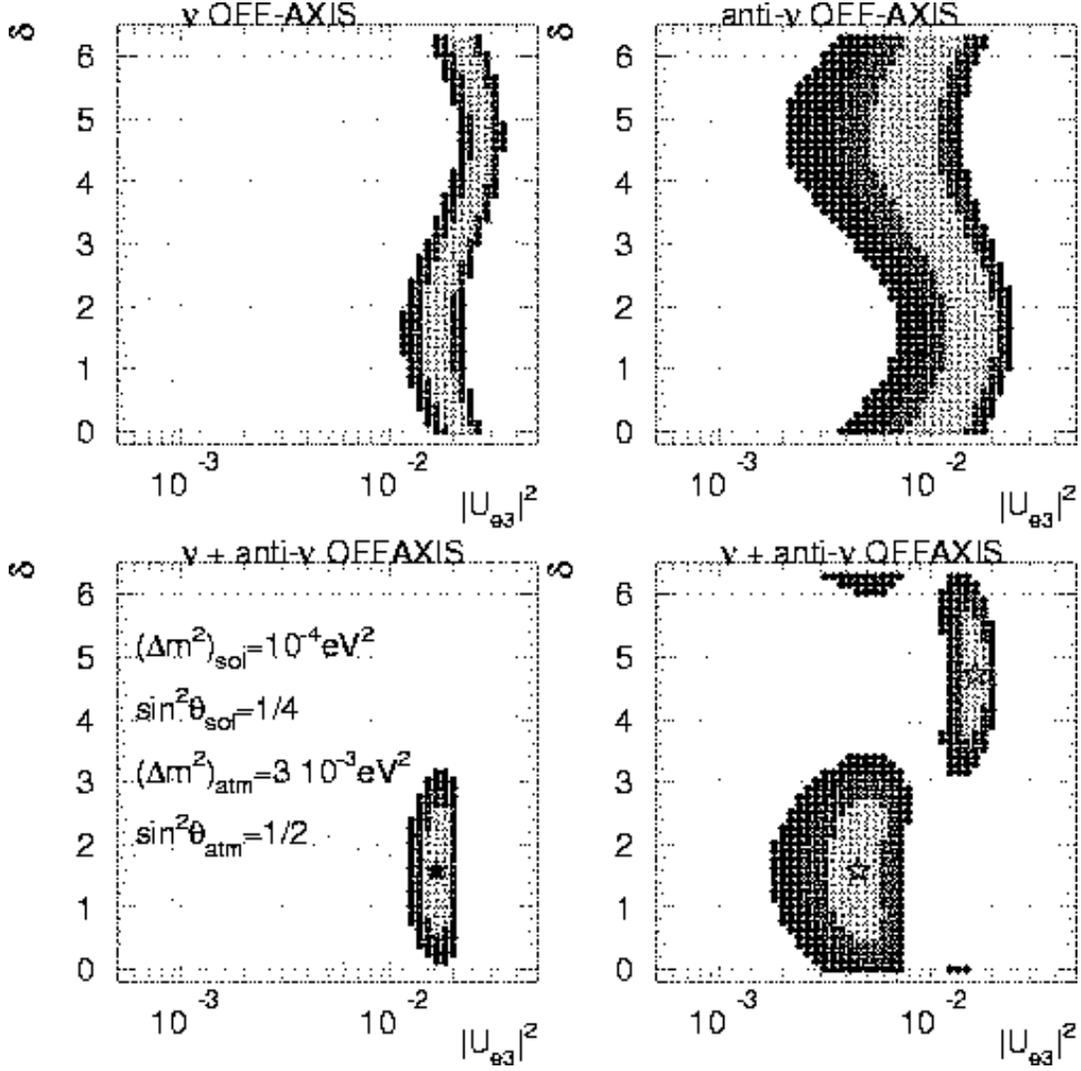}}
\caption{\it Top panel: one-, two-, and three-sigma measurement contours 
in 
the
$(|U_{e3}|^2, \delta)$ plane, after a five-year programme (120~kt-y
of neutrino-beam running (left) and  300~kt-y of antineutrino-beam 
running (right)). The simulated data is consistent with $|U_{e3}|^2=0.017$ 
and $\delta=\pi/2$. Bottom left panel: same as above, after the two data 
sets are 
combined. The solid star indicates the simulated input. Bottom right 
panel:
same as before, for different simulated data points (indicated by the 
stars): $\Delta m^2_{13}=+3\times10^{-3}$~eV$^2$,
$\sin^2\theta_{\rm atm}=1/2$, $\Delta m^2_{12}=1\times 10^{-4}$~eV$^2$,
$\sin^2\theta_{sun}=1/4$.}
\label{measure_lma}
\end{figure}

To assess properly the capabilities of such a set up, it is crucial to
explore all the different physics scenarios, which we hope will be
distinguished by the current KamLAND reactor experiment~\cite{KamLAND}.
For different values of the solar mass-squared difference, we obtain
different results for a five-year programme with an upgraded NuMI beam and
a 900~km long baseline off-axis experiment. The mean neutrino energy at
this location is 2~GeV, while the $\nu_e$ and $\bar{\nu_e}$ fraction in
the beam is less then 0.5\% in the signal region:

\begin{enumerate}
\item If KamLAND does not observe a suppression of the reactor neutrino 
flux,
$|U_{e3}|^2$ can be measured with very good precision and the neutrino mass 
pattern can be established, as long as $|U_{e3}|^2\gtrsim \rm few \times 
10^{-3}$.

\item If KamLAND sees a distortion of the reactor neutrino spectrum,
one should be capable of measuring $|U_{e3}|^2$ with good precision
and obtaining a rather strong hint for CP violation,
as long as $|U_{e3}|^2\gtrsim \rm few \times 10^{-3}$ and
$\delta$ is close to either $\pi/2$ or $3\pi/2$, as seen in
Fig.~\ref{measure_lma}, assuming that the neutrino mass hierarchy is 
already known. 

\item If KamLAND sees an oscillation signal but is not able to measure 
$\Delta m^2_{sun}$, one should be capable of measuring $|U_{e3}|^2$ 
with some precision and obtaining a strong hint for CP violation
as long as $|U_{e3}|^2\gtrsim 10^{-2}$ and $\delta$ is close to either 
$\pi/2$ or 
$3\pi/2$, even if $\Delta m^2_{sun}$ is poorly known. However, $\Delta 
m^2_{sun}$ 
cannot be measured with any reasonable precision, even assuming  that the 
neutrino mass hierarchy is already known.
\end{enumerate}

Other physics goals of the NuMI off-axis project are to measure with an
order of magnitude better precision, compared to MINOS and CNGS estimates,
the atmospheric parameters: $\delta(\Delta m_{\rm atm}^2)\lesssim
10^{-4}$~eV$^2$ and $\delta(\sin^2 2\theta_{\rm atm})\lesssim 0.01$,
confirm $\nu_\mu \leftrightarrow \nu_\tau$ oscillations, or discover
sterile neutrinos by measuring the neutral-current event rate, and to
improve by a factor of 20 the sensitivity to $\nu_\mu \to \nu_e$
appearance. After a five-year programme using NuMI with a 20~kt off-axis
detector, one could establish a two-sigma confidence level on the $\nu_\mu
\to \nu_e$ transition of $|U_{e3}|^2>0.00085~(0.0015)$ with~(without) an
upgrade to the proton driver, assuming $\Delta m^2_{sun}\ll
10^{-4}$~eV$^2$ and a normal neutrino mass hierarchy.

\subsubsection{The SPL super-beam}
\newcommand{\pizero}{\ensuremath{\pi^\circ}}
\newcommand{\numunue}{\ensuremath{\nu_\mu \rightarrow \nu_e}}
\newcommand{\nubarmu}{\ensuremath{\overline{\nu}_\mu}}
\newcommand{\pnuenue}{\ensuremath{P(\nue \rightarrow \nue)}}
\newcommand{\pnumunue}{\ensuremath{P(\nu_\mu \rightarrow \nu_e)}}
\newcommand{\pnumunumu}{\ensuremath{P(\nu_\mu \rightarrow \nu_\mu)}}
\newcommand{\dmsun}{\ensuremath{\delta m^2_{sun}\ }}
\newcommand{\dmatm}{\ensuremath{\delta m^2_{atm}}}
\newcommand{\senl}{\ensuremath{{\rm sin}^2(1.27 \cdot \dmsun  \cdot L/E)}}
\newcommand{\senh}{\ensuremath{{\rm sin}^2(1.27 \cdot \dmatm  \cdot L/E)}}
\newcommand{\send}{\ensuremath{{\rm sin}^2(1.27 \cdot (\dmatm + \dmsun)
\cdot L/E)\,}}
\newcommand{\thetaot}{\ensuremath{\theta_{13}}}
\newcommand{\thetatt}{\ensuremath{\theta_{23}}}
\newcommand{\sthetaot}{\ensuremath{{\rm sin}^2(2\theta_{13})}}
\newcommand{\sthetatt}{\ensuremath{{\rm sin}^2(2\theta_{23})}}
\newcommand{\nubare}{\ensuremath{\overline{\nu}_{e}\  }}
\newcommand{\numunutau}{\ensuremath{\nu_\mu \rightarrow \nu_\tau}}
\newcommand{\nueovernumu}{\ensuremath{\nue/\numu}}
\renewcommand{\thefootnote}{\alph{footnote}}

The planned Super Proton Linac (SPL)
would be a 2.2 GeV proton beam of 4 MW power~\cite{spl},
which could be used as a first stage of the Neutrino Factory
complex at CERN~\cite{CERN:nufact}. It would work with a repetition rate
of 75~Hz, delivering $1.5\cdot10^{14}$ protons per pulse, 
corresponding to $10^{23}$ protons
on target (pot) in a conventional year of $10^{7}$~s. These 
characteristics are also suitable for the production of a super-beam.

Pions would be produced by the interactions of the
2.2 GeV proton beam with a liquid mercury target and focused 
with a magnetic horn. The target and horn assumed in the following
are just those studied for the Neutrino Factory,
and no optimization has been attempted for this super-beam.
The pions next traverse a cylindrical decay tunnel
of 1 meter radius and 20 meters length.
These dimensions have been optimized in order to keep the
$\nu_e$~ contribution low and the \numu~ flux as high as possible.

\begin{figure}[tbh]
  \vspace*{-0.1cm}
  \epsfig{file=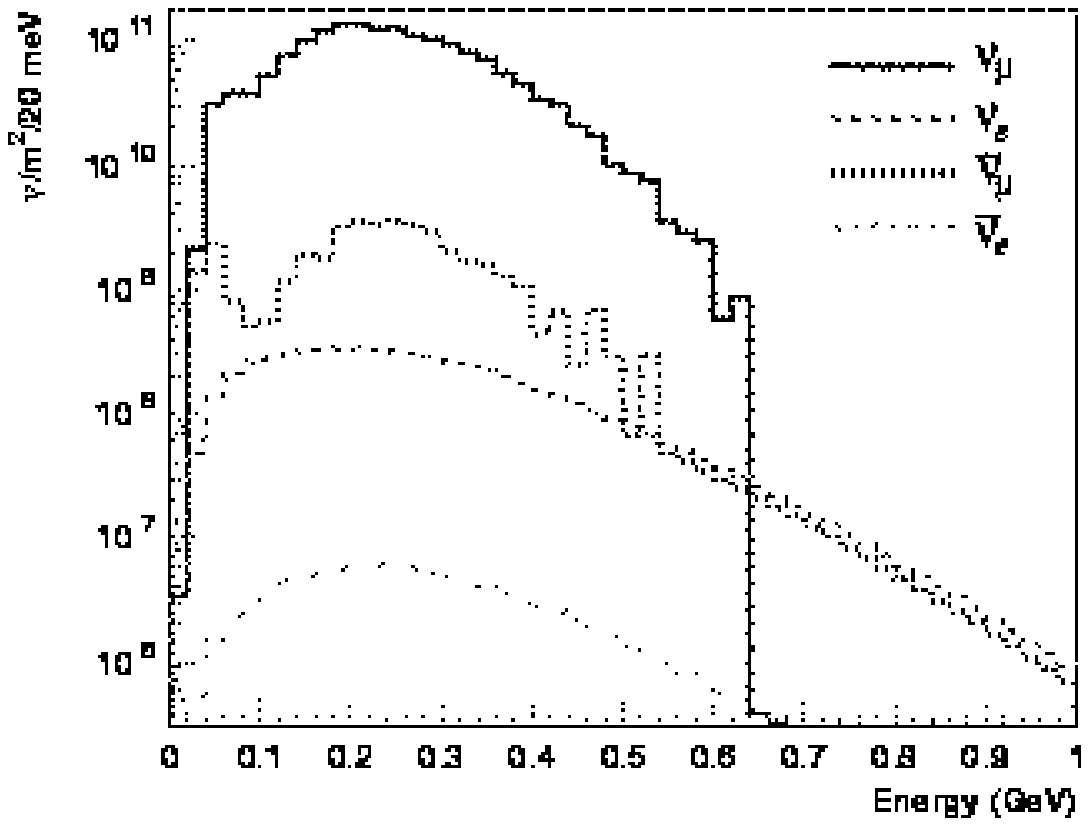,width=0.48\textwidth}
  \epsfig{file=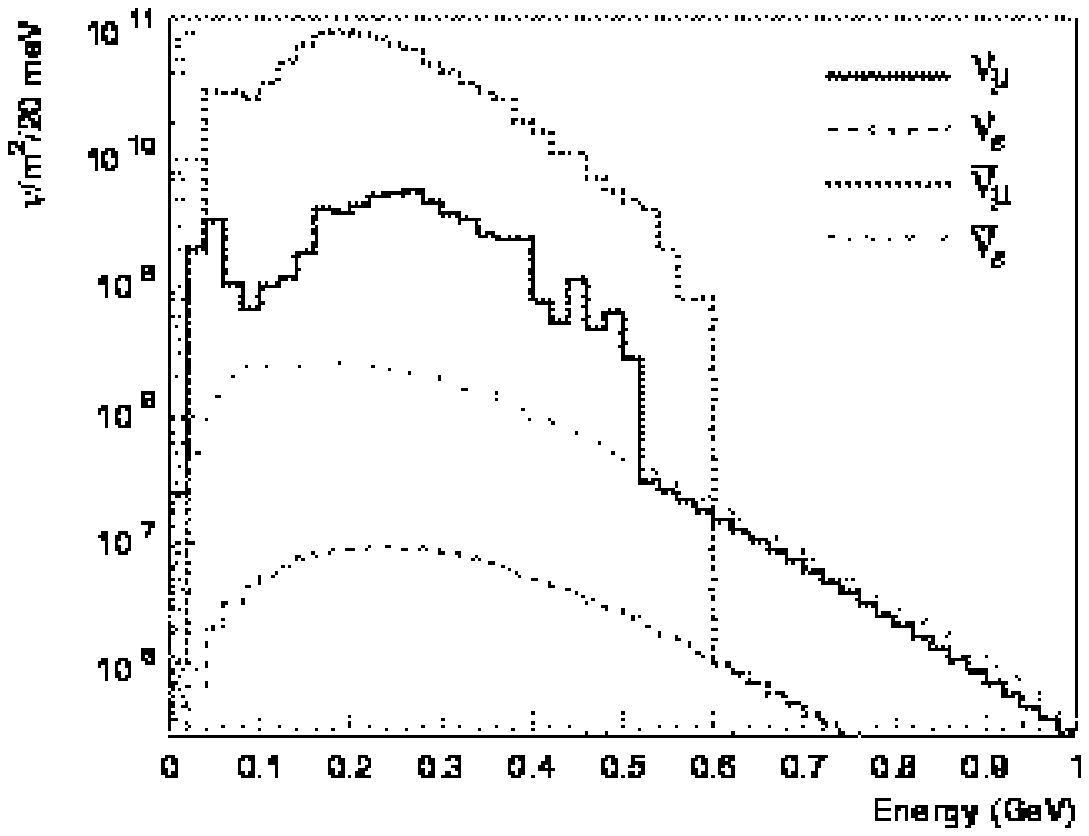,width=0.48\textwidth}
  \vspace*{-0.2cm}
   \caption{\it The SPL neutrino spectra, for $\pi^+$ (left) and 
   $\pi^-$ (right) focused in the horn,
   computed at 50~km from the target.}
   \label{fig:spl-spectra}
\end{figure}

The {\tt MARS}
program\cite{mars} has been used to generate and track pions. Polarization
effects on $\mu$ decays were added, and
analytical calculations used to follow the $\pi$ and
$\mu$ decays and particle trajectories~\cite{donega}.
The work presented in the following was previously published in
somewhat more detail in~\cite{venice}.
The resulting neutrino spectra are shown in Fig.~\ref{fig:spl-spectra}
and Table~\ref{tab:fluxes}.
\begin{table}[!t]
   \caption{\it The SPL neutrino fluxes, for $\pi^+$ (left) and 
   $\pi^-$ (right) focused in the horn,
   computed at 50~km from the target.}
\begin{tabular}{|c|ccc||c|ccc|}
\hline
\multicolumn{4}{|c||}{$\pi^+$ focused beam} &
      \multicolumn{4}{|c|}{$\pi^-$ focused beam} \\
\hline
Flavor & Absolute Flux & Relative & $\langle E_\nu \rangle$ &
Flavor & Absolute Flux & Relative & $\langle E_\nu \rangle$ \\
        & ($\nu/10^{23}{\rm pot}/{\rm m^2}$) & (\%) & (GeV) &
        & ($\nu/10^{23}{\rm pot}/{\rm m^2}$) &  (\%)& (GeV)  \\
\hline
$\nu_\mu$ & $1.7 \cdot 10^{12}$ & 100 & 0.26 &
             \nubarmu & $1.1 \cdot 10^{12}$  & 100 & 0.23 \\
\nubarmu & $4.1 \cdot 10^{10}$  & 2.4 & 0.24 &
             $\nu_\mu$ & $6.3 \cdot 10^{10}$ & 5.7 & 0.25 \\
$\nu_e$ & $6.1 \cdot 10^{9}$ & 0.36 & 0.24 &
             \nubare & $4.3 \cdot 10^{9}$ & 0.39 & 0.25 \\
\nubare & $1.0 \cdot 10^{8}$ & 0.006  & 0.29 &
             $\nu_e$ & $1.6 \cdot 10^{8}$ & 0.15 & 0.29 \\
\hline
\end{tabular} 
\label{tab:fluxes}
\end{table}
The average energy of the neutrinos is around 250 MeV, and the 
\nue~ contamination $\sim 0.4\%$.

The \nue~ in the beam can be produced only by $\mu$ decays, since 
the center-of-mass 
energy is below the kaon production threshold.
The \numu~ and \nue~ come from the same decay chain:
$\pi^+ \rightarrow \numu \mu^+ \rightarrow e^+ \nue \nubarmu$,
so the \nue~ flux can be predicted by a direct measure
of the \numu~ themselves.
It has been estimated that, combining the measurement of the \numu~ 
interaction rate at
the far detector with the \numu~ and \nue~ rates in a nearby detector of
0.5~kt at 1~km from the target, the \nue~ contamination can
be established with 2\% systematic and statistical errors.
\subsubsection{Detector scenarios}
\label{sec:det}
\begin{figure}[!tb]
  \vspace*{-0.1cm}
  \centerline{\epsfig{file=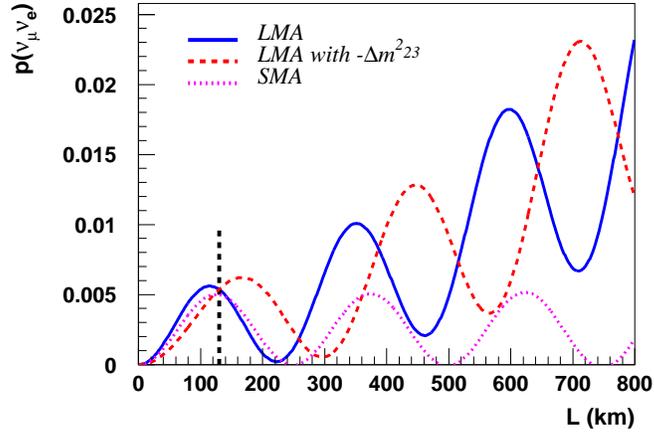,width=0.55\textwidth}}
  \vspace*{-0.15cm}
  \caption{\it The oscillation probability $P(\nu_\mu\rightarrow \nu_e)$ 
computed for the SPL neutrino beam, with
  $\langle E_{\nu_\mu} \rangle =0.26$ GeV,
  $\Delta m^2_{atm}=2.5\cdot 10^{-3}eV^2$, $\Delta m^2_{sun}=5\cdot
  10^{-5}$, $6\cdot10^{-6}$ eV$^2$,
   $\pm \Delta m^2_{23}$, $\sin^2{\thetaot}=0.01$,
  $\delta=0$.} 
  \label{fig:osc}
\end{figure}

Fig.~\ref{fig:osc} shows the oscillation probability
\pnumunue~ as a function of the distance.
Notice that the
first maximum of the oscillation at $L\sim 130$~km
matches the distance between CERN
and the Modane laboratory in the Fr\'ejus tunnel, where one could
locate a large neutrino detector~\cite{mosca}.
The detection of
low-energy neutrinos at $\cal{O}$(100) km from the source requires a
massive target with high efficiency. Moreover, a search for
$\nu_e$ appearance demands excellent rejection of physics
backgrounds, namely $\mu$ misidentification and neutral current
$\pi^0$ production, which should be controlled to a lower level
than the irreducible beam-induced background.
Backgrounds from atmospheric neutrinos remain negligible if the
accumulator foreseen in the Neutrino Factory complex is used, which 
provides a duty cycle of 4000.

Here two detector technologies are considered, which have
demonstrated excellent performance in the low-energy regime, while
being able to provide massive targets. These are water \v{C}erenkov
detectors such as Super-Kamiokande\cite{Fukuda:1998tw}, and diluted liquid
scintillator detectors such as used by the
LSND~\cite{LSND:dif} and 
MiniBOONE experiments~\cite{miniboone}, where both \v{C}erenkov
and scintillation light are measured.\par
We have considered an apparatus of 40~kt fiducial mass and
sensitivity identical to the Super-Kamiokande experiment. The
response of the detector to the neutrino beam
was studied using the NUANCE~\cite{casper} 
neutrino
physics generator and detector simulation and reconstruction
algorithms developed for the Super-Kamiokande atmospheric neutrino
analysis. These algorithms, and their agreement with real neutrino
data, have been described
elsewhere~\cite{Fukuda:1998tw,Messier:1999kj,Shiozawa:1999sd}.

The Super-Kamiokande standard algorithms for $\mu/e$ separation
are very effective in the SPL-super-beam energy regime~\footnote{We have 
only tightened the cut on the particle
identification criterion, which is based on a maximum likelihood
fit of both the $\mu$-like and e-like hypotheses ($P_{\mu}$ and $P_e$,
respectively.).
We use $P_e > P_{\mu}+1$ instead of the default $P_e > P_{\mu}$.}.
Events with a $\pi^0$ are easily rejected 
in events where the two rings are found by a standard
$\pi^0$ search algorithm, \`a la Super-Kamiokande.
To reject events where only one $\gamma$ has been initially
identified, an algorithm~\cite{Barszczak:thesis} forces the
identification of a candidate for a
second ring, which, if the primary ring is truly a single
electron, is typically either very low energy, or extremely
forward.  
By requiring that the invariant mass formed by the primary and
the secondary rings is less than $45 \,
\hbox{MeV/$c^2$}$, almost all remaining $\pi^0$ contamination 
is removed.
Since the SPL super-beam would be cleaner and at lower energies than the
JHF beam~\cite{jhf}, we have not attempted  to introduce more
aggressive \pizero rejection algorithms.
The data reduction is summarized in Table~\ref{tbl:Events}, 
where we see that contamination by the intrinsic beam $\nu_e$ 
is dominant.

\begin{table}[tbh]
\caption{\it Summary of simulated data samples for $\pi^+$ and
$\pi^-$  focused neutrino beams.
The numbers in the rightmost column
(after all cuts) represent the sample used to estimate the
oscillation sensitivity for a 200~kt-y exposure.}
\begin{center}
\begin{tabular}{|c|c|c|c|c|c|c|}
\hline
 &&  & Fit in fid. vol. & Tight & &  \\
 & Initial& Visible & Single-ring & particle & No & $m_{\gamma\gamma}$ \\
Channel & sample & events & $100 - 450 \, {\rm MeV/c^2}$ & ID &
$\mu \rightarrow e$ & $\le 45 \, {\rm MeV/c^2}$ \\
\hline
\hline
\multicolumn{7}{|c|}{$\pi^+$  focused beam} \\
\hline
$\nu_{\mu}$ & 3250& 887 & 578.4 & 5.5 & 2.5 & 1.5 \\
$\nu_e$& 18&  12. & 8.2 & 8.0 & 8.0 & 7.8 \\
NC & 2887& 36.9 & 8.7 & 7.7 & 7.7 & 1.7 \\ \hline $\nu_{\mu}
\rightarrow \nu_{e}$ & 100\% & 82.4\% & 77.2\% & 76.5\% & 70.7\% & 70.5\%\\
\hline
\hline
\multicolumn{7}{|c|}{$\pi^-$  focused beam} \\
\hline
$\stackrel{-}{\nu}_{\mu}$ & 539& 186 & 123 & 2.3 & 0.7 & 0.7 \\
$\stackrel{-}{\nu}_e$ & 4& 3.3 & 3 & 2.7 & 2.7 & 2.7 \\
NC & 687& 11.7 & 3.3 & 3 & 3 & 0.3 \\ \hline
 $\nubarmu
\rightarrow \nubare$ & 100\% & 79.3\% & 74.1\% & 74.0\%
& 67.1\% & 67.0\% \\ \hline
\end{tabular}
\end{center}
 \label{tbl:Events}
\end{table}

The energy range (50 MeV-1 GeV) and the rejections against background
needed by MiniBOONE nicely match the requirements of our
study. The neutrino-$^{12}$C cross sections are taken 
from~\cite{RPA}~\footnote{They come from an upgraded version of the
continuous random phase approximation method,
and on average they are lower by about
$\sim 15\%$ from those quoted by the MiniBOONE experiment.}.
Table~\ref{tab:events} shows the data reduction
assuming no \numu-\nue~ oscillation, for
a 200~kt-y exposure.
As before, intrinsic beam \nue~ contamination turns out to
be the dominant background.

\begin{table}[!tb]
\caption{\it Summary of data samples 
in a 40 kton  liquid scintillator
detector at $L=130$ km for a 200~kt-y exposure.} 
\begin{center}
\begin{tabular}{|c|c|c||c|c|c|}
\hline \multicolumn{3}{|c||}{$\pi^+$ focused beam} &
\multicolumn{3}{|c|}{$\pi^-$ focused beam}\\
\hline
Channel & Initial sample & Final sample & Channel & Initial sample & Final sample \\
\hline
$\numu^{CC}$ & 2538 & 2.5 & $\nubarmu^{CC}$ & 451 & 0.5 \\
$\nue^{CC}$ & 12 & 6 & $\nubare^{CC}$ & 2.3 & 1.0\\
NC (visible) & 48 & 0.5 & NC & 10 & $<0.1$ \\
\hline
\numunue & 100\% & 50\% & $\nubarmu \rightarrow \nubare$ & 100\% & 50 \%  \\
\hline
\end{tabular}
\end{center}
\label{tab:events}
\end{table}
 
As one can see comparing Tables~\ref{tbl:Events} and \ref{tab:events},
the performances of both devices are quite similar.
The conclusion is that one would prefer for this experiment
a water detector, where truly gargantuan sizes can be afforded more 
easily.

\subsubsection{Physics pote ntial}\label{sec:spl}

\begin{figure}[!tb]
  \vspace*{-0.5cm}
  \epsfig{figure=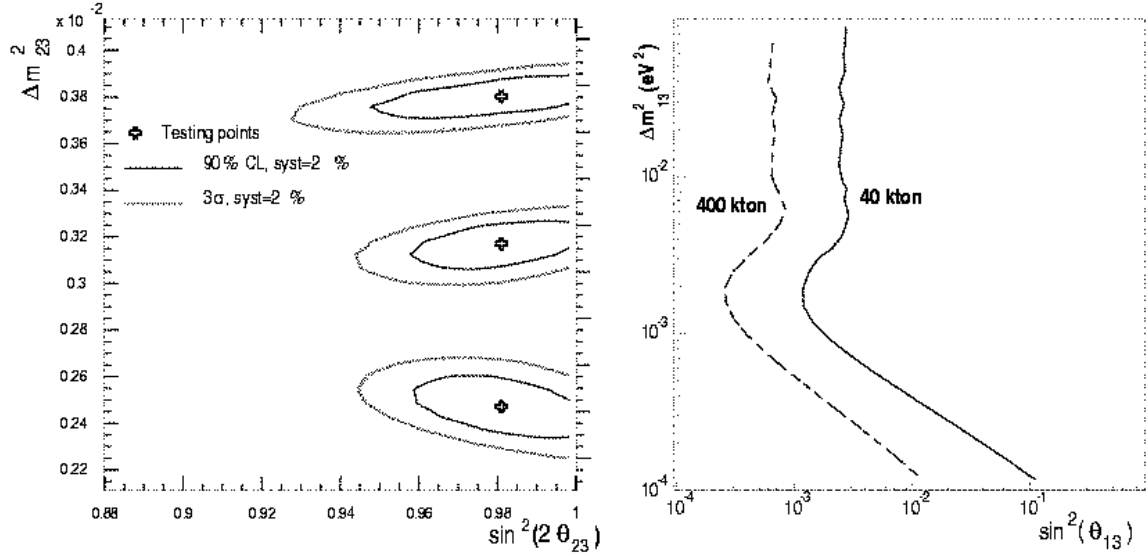,width=\textwidth}
  \caption{\it Left panel: Fits of \dmatm (eV$^2$),\sthetatt~ after
  a 200~kt-y exposure at $L=130$km, assuming
  2\% systematic errors. 
   The crosses mark the initial points
  $(0.98, 3.8\cdot10^{-3})$, $(0.98, 3.2\cdot10^{-3})$,
  $(0.98, 2.5\cdot10^{-3})$ in the \dmatm, \sthetatt~ coordinates.
  Right panel: The \numunue~ oscillation sensitivity for $\pi^+$
  focused neutrino beams with a 40 (400) kt detector.}
  \label{fig:result130}
\end{figure}

To illustrate the attainable precision in measuring \dmatm~ and \thetatt~,
Fig.~\ref{fig:result130} shows the
result of a 200~kt-y exposure \numu~ disappearance experiment using
a liquid scintillator detector. The computation is
performed defining 4 energy bins in the 0.1-0.7 GeV energy range
and including Fermi motion, which is by far the most important limiting
factor for energy reconstruction at these energies: see~\cite{mauro}
for more details. We find that $\Delta m^2_{23}$ can be measured
with a standard deviation of  $1\cdot10^{-4}\quad {\rm eV^2}$,
while $\sin^2 2\thetatt$ is measured at the $\sim 1\%$ precision level.

In the case of a water \v{C}erenkov detector, the first energy bin would
be lost, since the $\mu$ would be below threshold. In this case the
resolution on the atmospheric parameters is slightly worsened,
especially in the case of low $\Delta m^2$.
The different solar solutions do not
affect \pnumunue~ very much for $L=130$~km, as can be seen 
in Fig.~\ref{fig:osc}. Here we use SMA parameters, a hierarchical mass 
model, and set $\delta=0$.
Given the low statistics and the poor energy resolution, we
consider in this case a counting experiment, not exploiting spectral
distortions.
The right panel of Fig.~\ref{fig:result130} shows the expected
sensitivity for a 5-year run with a 40~kt (400~kt)  water
target at a distance of 130 km.

The sensitivity of this search turns out to be about 15 (60) times better
than the present experimental limit of $\sin^2{\thetaot}$ coming
from the Chooz experiment\cite{osc_data_reactor}.
In the rest of this subsection we assume
$\delta m_{12}^2 =
10^{-4}$~eV$^2$ (in the upper part of the LMA region) and
$\sin^2{2\theta_{12}}=1$.
Since the $\nubarmu + \,^{16}$O cross-section is
approximately six times less than that for $\nu + {}^{16}$O at
$E_\nu \simeq 0.25$ GeV,
we have considered a 10~y run with focused $\pi^-$ and a 2~y run 
with focused $\pi^+$.

We follow the approach in~\cite{golden,jordi}, and fit
simultaneously the number of detected electron-like events to
the CP phase $\delta$ and \thetaot. Notice that,
although we apply a full three-family treatment to our
calculations, including matter effects, these are not important at
the short distances and low energies considered. 

\begin{figure}[!tb]
  \epsfig{figure=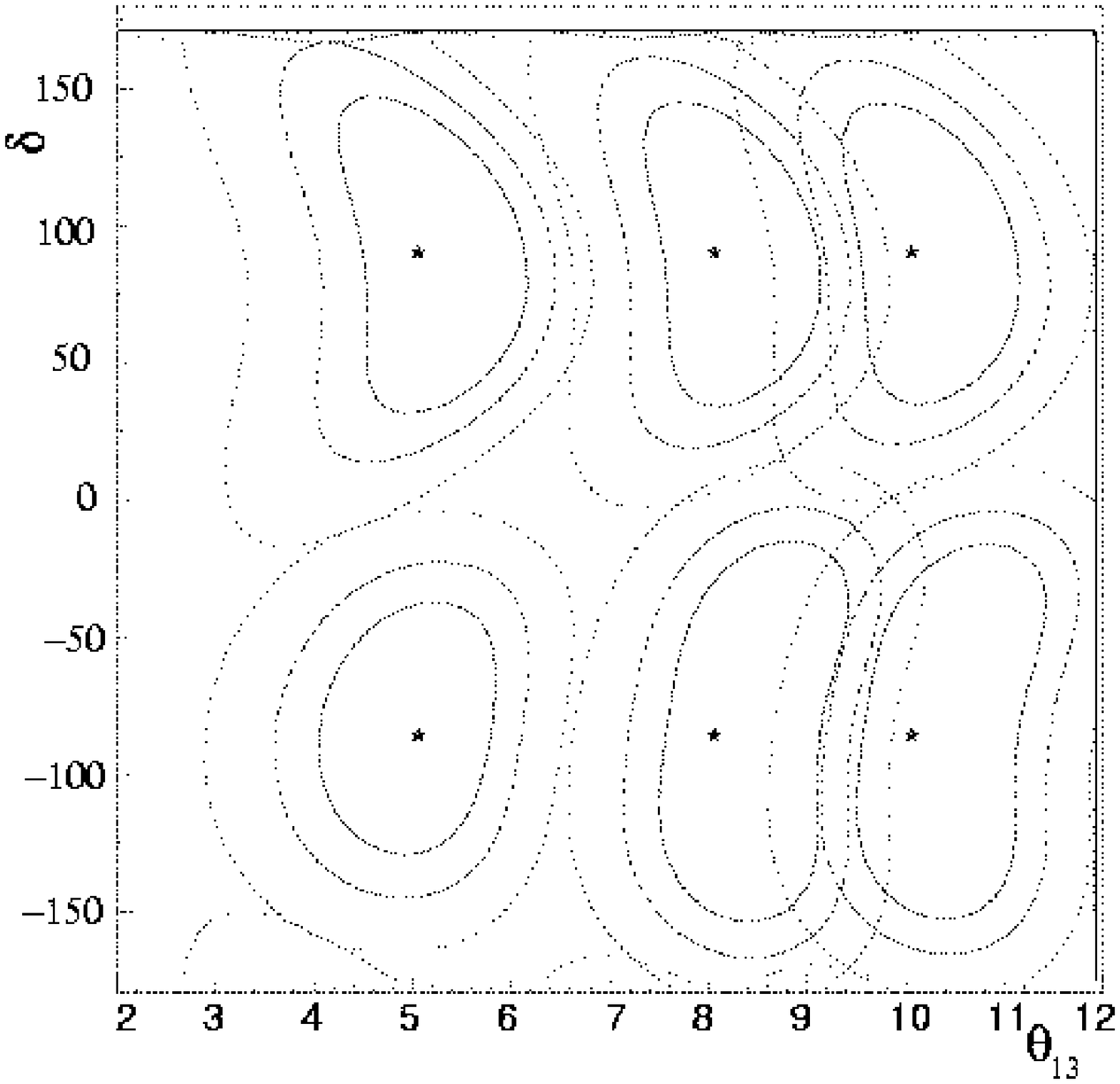,width=0.48\textwidth}
  \epsfig{figure=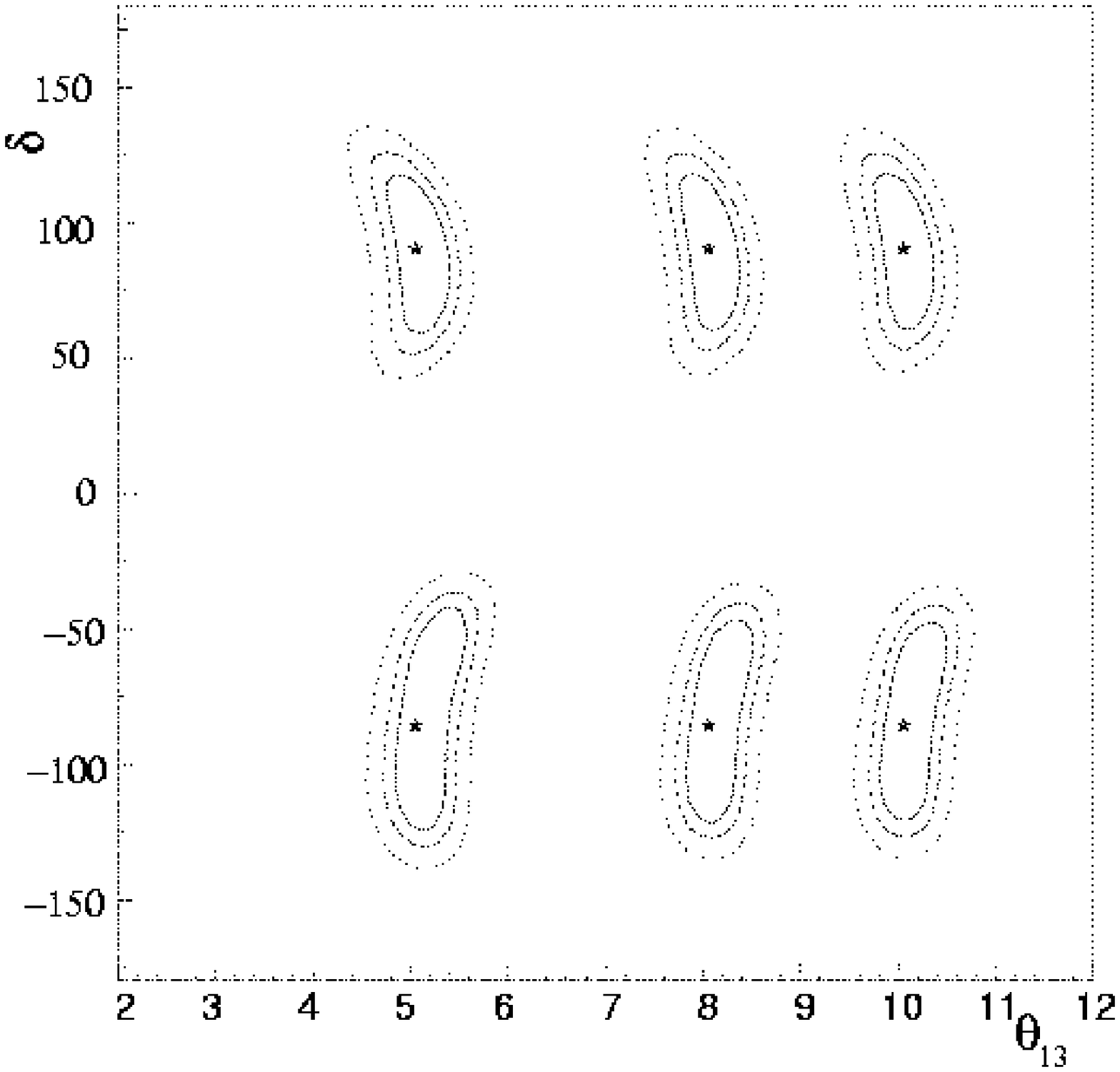,width=0.48\textwidth}
  \caption{\it The $1 \,\sigma$, 90 \%CL and 99 \%CL
   intervals resulting from a
   simultaneous fit to the \thetaot~ and $\delta$ parameters. The
   generated values were $\thetaot=5^\circ, 8^\circ, 10^\circ$,
   $\delta=\pm 90^\circ$. These results were
   computed for a 10~y anti-neutrino run
   and a 2-y neutrino run, using a 40~kt detector (left)
   or a 400~kt detector (right) at 130~km. }
  \label{fitcp}
\end{figure}
Fig.~\ref{fitcp}~(left) shows the confidence level contours 
including statistical and background subtraction errors.
A maximal CP-violating phase: $\delta =
\pm 90^\circ$ would just be distinguishable from a zero CP-violating phase 
($\delta = 0^\circ$).
Fig.~\ref{fitcp}~(right) shows the result of the same fit, now
assuming a very large water detector, such as the proposed
UNO~\cite{UNO}, with a fiducial mass of
400~kt. Clearly, the prospects to observe CP violation are much
improved.

{\it Summarizing}, we have examined the physics potential of a low-energy 
super-beam
which could be produced by the CERN Super Proton Linac. Water
\v{C}erenkov and liquid oil scintillator detectors have been
considered. 
\begin{itemize}
\item{
The low energy of the beam studied has several advantages. The beam
systematics are kept low, because 2.2 GeV protons are below the
kaon production threshold. Furthermore, $e/\mu$ and $e/\pi^0$
separation in a water (liquid oil) detector is near optimal at
these low energies. The drawback is the small anti neutrino cross
section, which is more than a factor five smaller than the neutrino
cross section. These cross sections are in addition quite uncertain
and would certainly need to be measured.}
\item{The peak of the oscillation is at a distance of about 130~km, where 
an ideal location exists, the Modane laboratory
in the Fr\'ejus tunnel.}
\item{A `moderate-size' detector (`only' twice as big as
Super-Kamiokande) at this baseline could, in a five-year run,
improve the precision in the determination 
of the atmospheric parameters by about one
order of magnitude with respect to the expected precision of 
next-generation neutrino experiments such as MINOS. It could also
improve by more than one order of magnitude the sensitivity to
$\sin^2{\thetaot}$, compared to the present experimental limits.}
\item{Such a detector could also, if the solution to the solar neutrino
problem lies in the upper part of the LMA region, detect 
a maximal CP-violating phase. 
For CP-violation studies, a very large detector \`a la UNO (with 400~kt
fiducial mass) is mandatory.}

\end{itemize}

\subsubsection{Effect of the inclusion of neutrino spectral information}

In the previous subsection, the parameters of neutrino oscillations were
extracted either by counting the number of events or by fitting smeared
generated neutrino energies. Here we describe a reconstruction of the
neutrino energy starting from the information available in a water
\v{C}erenkov detector, i.e., the lepton energy $E_l$, its momentum $P_l$
and angle, also considering detector smearing, Fermi motion and Pauli
blocking.  

In the ideal case, it is possible to reconstruct the neutrino
energy from the lepton kinematics in a water \v{C}erenkov detector using

\begin{equation}
E_{\nu}=\frac{m_n E_l + \frac{m_p^2 -m_n^2 -m_{l}^2}{2}}{m_n - E_l 
+P_l \cos(\theta)},
\label{eq:qereconst}
\end{equation}
where $\theta$ is the angle between the outgoing lepton and the neutrino
direction. This angle can be measured via the knowledge of the beam
direction, and the reconstructed lepton direction.
The simulation used here is the fast Monte Carlo described in \cite{max}. 
The technique
presented in this paragraph aims at unfolding detector and nuclear physics
effects, using a Monte Carlo reweighting method.  We have used the 
spectrum calculated for the CERN SPL super-beam.  The fast simulation only 
deals with CC quasi-elastic
interactions.  To describe nuclear effects, neutrons are generated
isotropically in a sphere with Fermi momentum $k_F = 225$ MeV/c.  The
nuclear potential well is taken to be 50~MeV deep, and the double
differential cross section from~\cite{Gaisser} was used.  Pauli blocking
was implemented by requiring that the outgoing proton lies outside the
Fermi sphere of radius $k_F$. Based on Super-Kamiokande 
data~\cite{Casper,subgev}, the momentum resolution for
electrons is set to 2.5\%/$\sqrt{E\mathrm{(GeV)}}$+0.5\%; for muons it is
constant and equal to 3\% in this energy range; the angular resolution is
estimated to be 3 degrees, constant for electrons and muons.

The
analysis method is based on the construction, for each data event, of a
distribution of all the Monte Carlo events with a visible energy 
close
to that of the data event. It is based on the method described 
in~\cite{reweight}. For each data event, a box is defined with a small
volume around its reconstructed energy. All MC events whose {
reconstructed} energy lie close to that of the data event are considered
to be good approximation to the data event, and their value of the {
generated} energy is used in the likelihood function. The total 
likelihood can
be found in~\cite{max}, and maximizes information from the total number of
events and unfolded spectral information. The weight of the
each MC event in the box is the ratio between the
number of events for the value of the {candidate} oscillation
parameter set and the number of events for the value of the oscillation
parameters {at which the MC sample was produced}.

\begin{figure}[tbh]
\centering
\begin{minipage}[h]{0.45\textwidth}
\begin{center}
\includegraphics[scale=0.4]{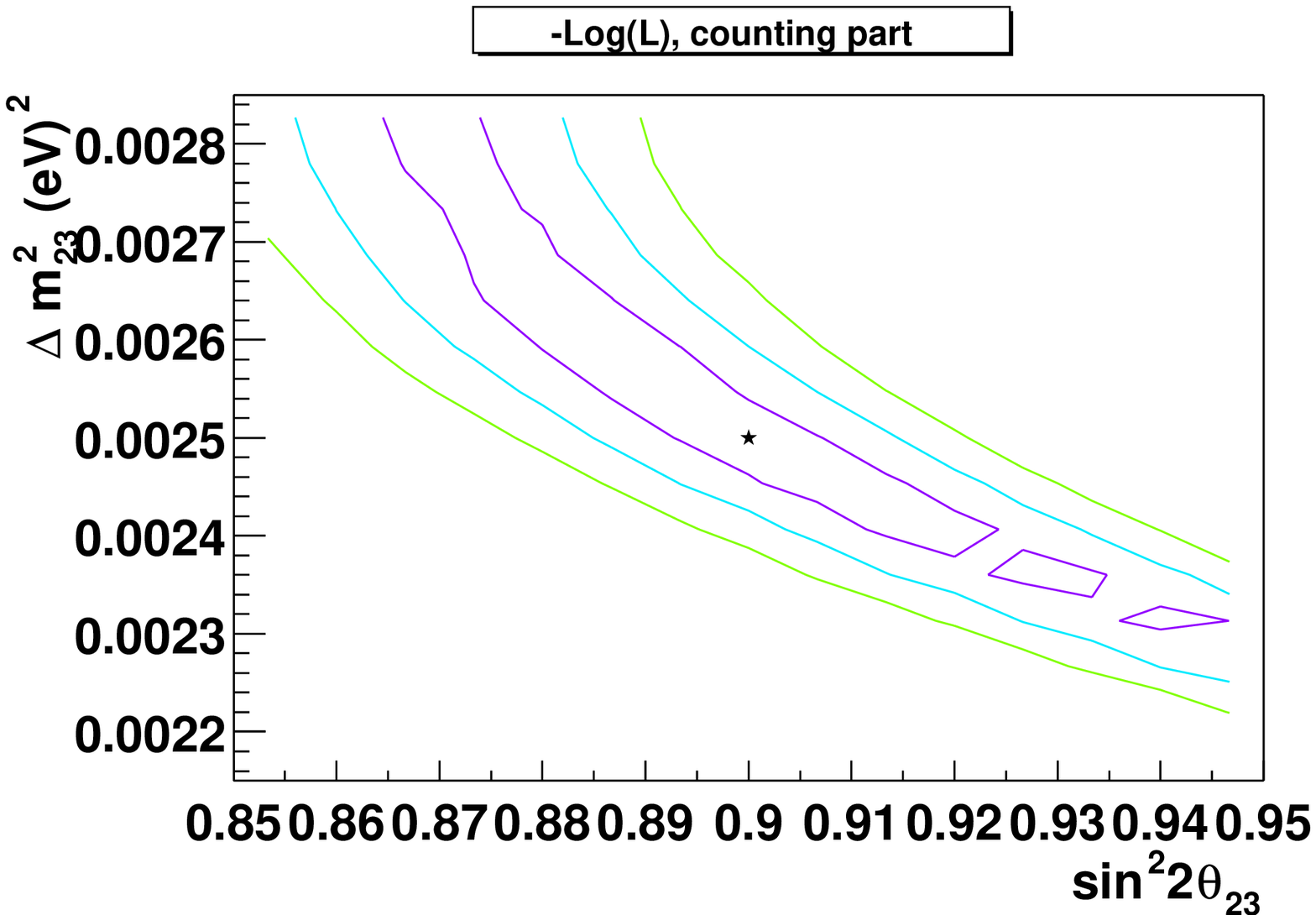}
\end{center}
\end{minipage}
\begin{minipage}[h]{0.45\textwidth}
\begin{center}
\includegraphics[scale=0.4]{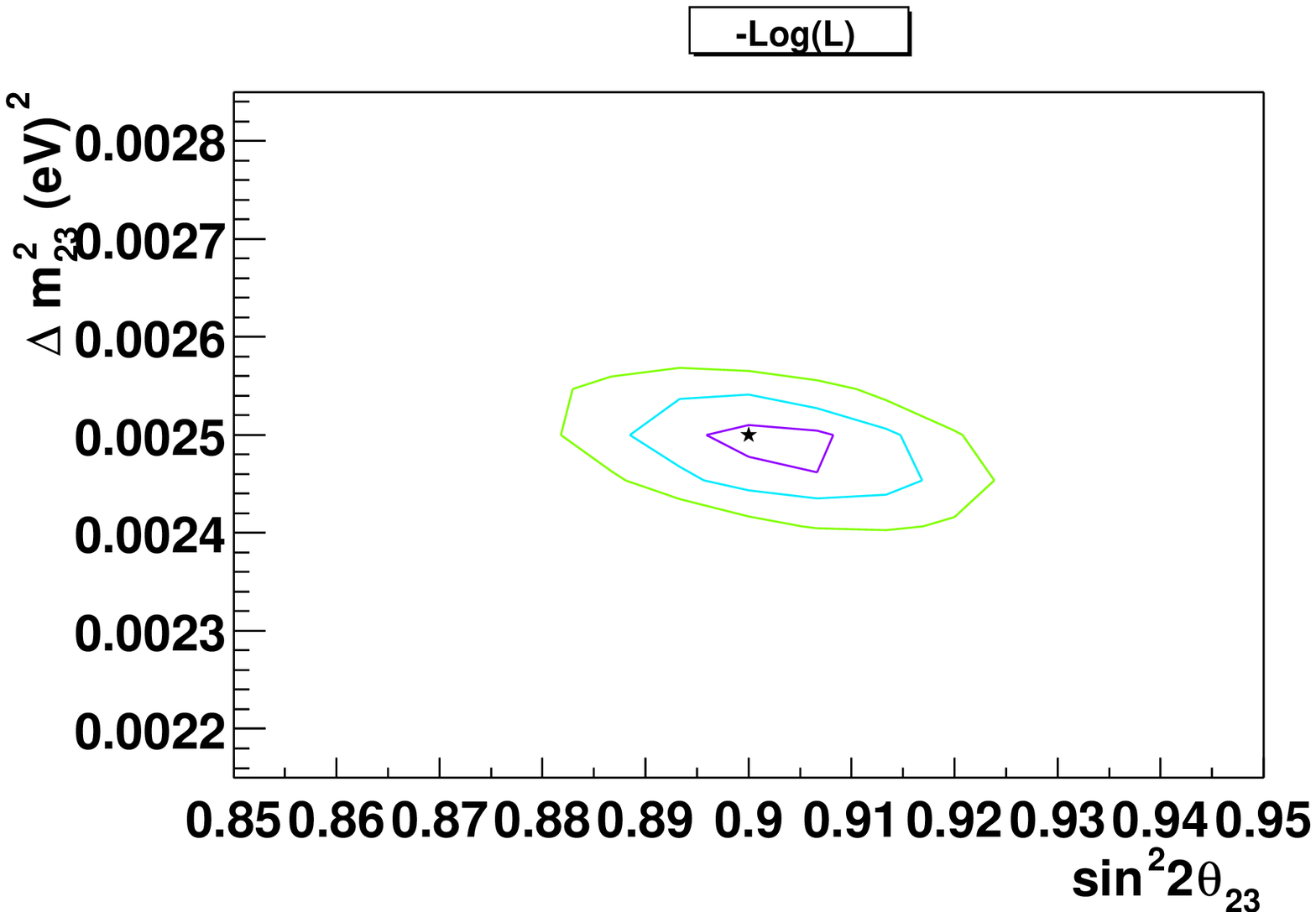}
\end{center}
\end{minipage}
\caption{\it Contour plots of the different terms and the total of the log 
likelihood function. The star indicates the true value of the experimental 
parameters. The contours are the 1$\sigma$, 2$\sigma$ and $3\sigma$ confidence 
levels.Both plots are made
with the ($2.5\times10^{-3}\ \mathrm{eV}^{2}$, 0.9) sample.  In each case,
the star indicates the parameters at which the `experimental'
sample was produced. Left panel: contours of the likelihood using 
counting information only. Right panel: contours of the total likelihood.}
\label{fig:maximtheta23}
\end{figure}

The results of a two-dimensional fit of the likelihood for the atmospheric
parameters are displayed in Fig.~\ref{fig:maximtheta23}. It is clear 
that the
atmospheric parameters are correlated, therefore counting the number of
events is not enough to obtain a measurement. The use of the reweighting
method dramatically improves the precision, showing the efficiency of the
unfolding procedure: $\Delta m^{2}_{23} $ can be measured with a standard
deviation of $\sim 0.7\times\,10^{-4}\ \mathrm{eV}^{2} $, and
$\sin^{2}2\theta_{23}$ is measured with a precision of $\sim 1.5$\%.

\begin{figure}[tbh]
\label{fig:CPantibox}
\centering
\begin{minipage}[h]{0.45\textwidth}
\begin{center}
\includegraphics[scale=0.4]{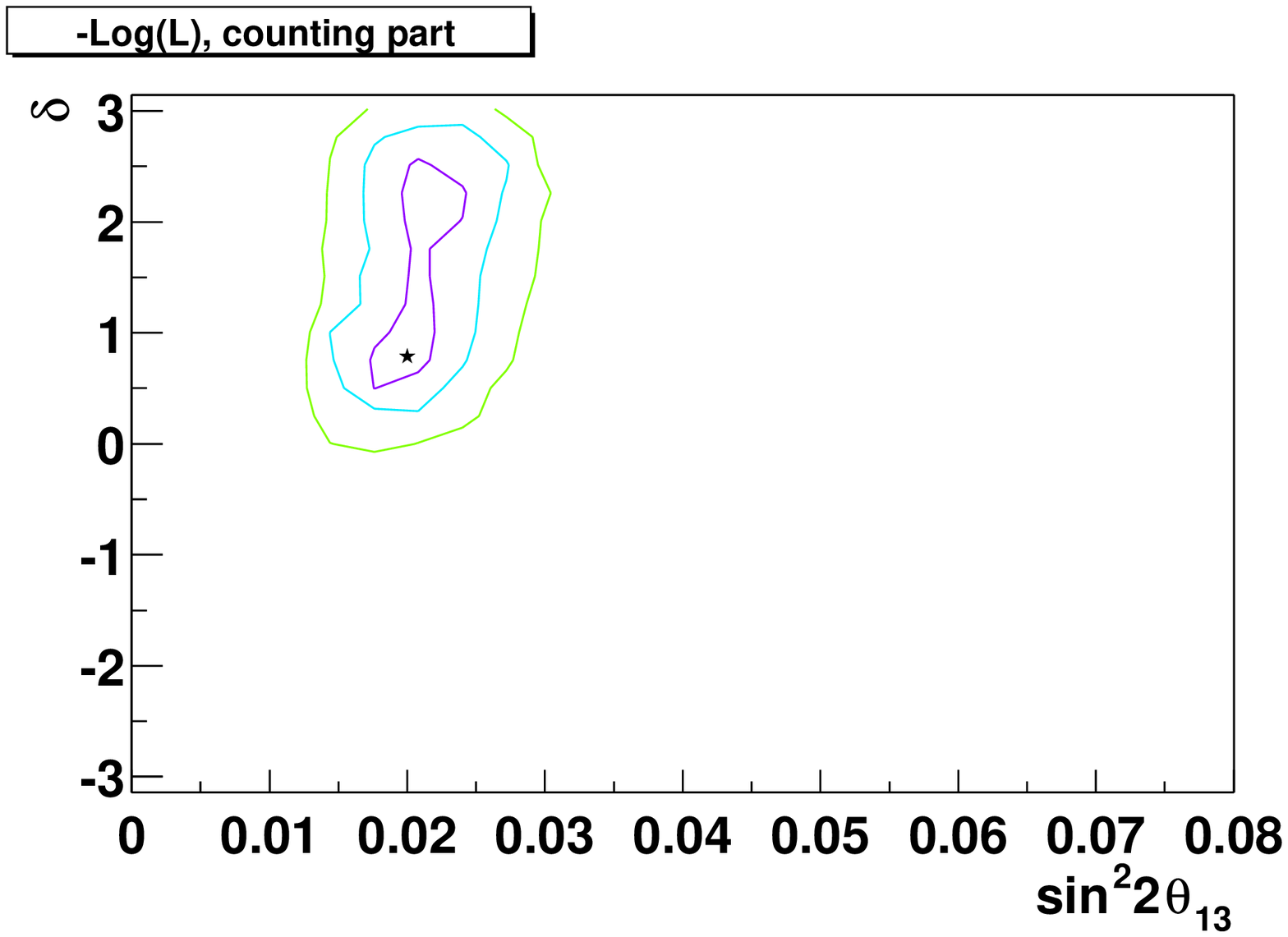}
\end{center}
\end{minipage}
\begin{minipage}[h]{0.45\textwidth}
\begin{center}
\includegraphics[scale=0.4]{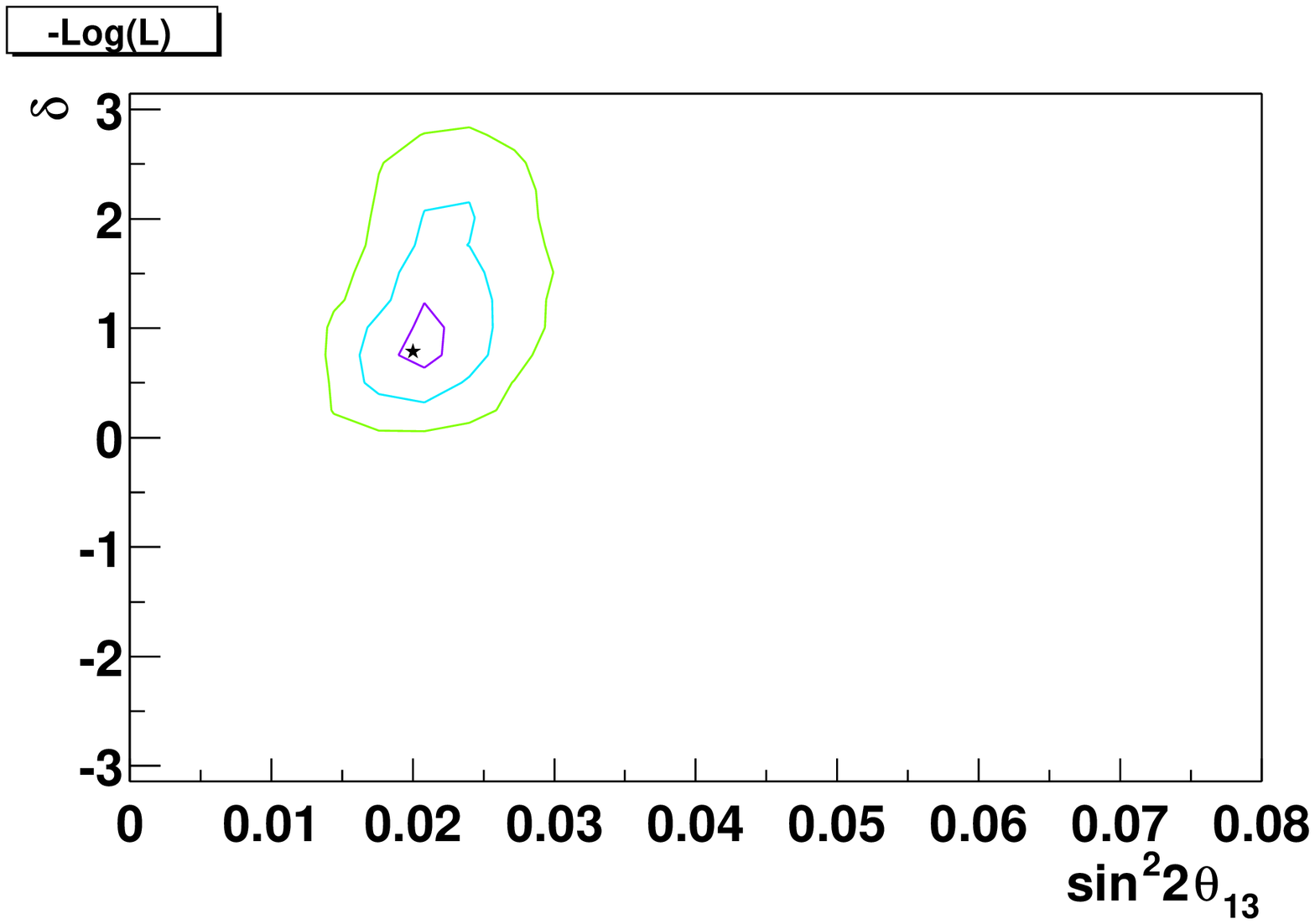}
\end{center}
\end{minipage}
\caption{\it Contours of the likelihood, assuming 200 kt-y of electron 
neutrino data and 1000 kt-y of electron antineutrino data.
The star indicates the position at which the `experimental' sample was 
produced, and $\delta$ is measured in radians.
Left panel: contours of the likelihood using counting information only. 
Right panel: contours of the total likelihood.}
\end{figure}

The results are also very promising for CP violation in the neutrino
sector: using neutrinos with an exposure of 200~kt-y and antineutrinos
with 1000 kt-y to compensate for the lower cross section, we have obtained
the results plotted in Fig.~\ref{fig:CPantibox}. In this case, the number
of events (left-hand side plot) contains most of the information, but
reconstructing the spectral information reduces the 1$\sigma$ error on
$\delta$ by roughly a factor three, from 120$^{\circ}$ to 35$^{\circ}$. We
find that $\sin^{2}2\theta_{13}$ can be measured with precision $\sim
20$\%.

\newpage
\def\dm{\ensuremath{\Delta m}}
\def\dm2{\ensuremath{\Delta m^{2}\ }}
\def\sen2th{\ensuremath{ \sin^{2}(2\theta)\ }}
\def\(({\left(}
\def\)){\right)}

\def\nubar{$\overline{\nu}\ $}
\def\nue{\ensuremath{\nu_{e}\ }}
\def\nubare{\ensuremath{\overline{\nu}_{e}\ }}
\def\nubarecc{$\overline{\nu}_{e}^{CC}\ $}
\def\numu{\ensuremath{\nu_{\mu}\ }}
\def\nubarmu{\ensuremath{\overline{\nu}_{\mu}\ }}
\def\nubarmucc{$\overline{\nu}_{\mu}^{CC}\ $}
\def\nutau{\ensuremath{\nu_{\tau}\ }}
\def\nubartau{\ensuremath{\overline{\nu_{\tau}}\ }}
\def\nulep{$\nu^{l}\ $}
\def\nubarlep{$\overline{\nu^{l}}\ $}

\def\nuef{\nu_{e}}
\def\nubarf{\overline{\nu}}
\def\nubaref{\overline{\nu}_{e}}
\def\numuf{\nu_{\mu}}
\def\nubarmuf{\overline{\nu_{\mu}}}
\def\nutauf{\nu_{\tau}}
\def\nubartauf{\overline{\nu_{\tau}}}
\def\nulepf{\nu^{l}}
\def\nubarlepf{\overline{\nu^{l}}}
\def\pzero{\ensuremath{\pi^0\ }}

\newcommand{\mphad}{\ensuremath{p_{had}}}
\newcommand{\plep}{\ensuremath{\vec{p}_{lep}}}
\newcommand{\pele}{\ensuremath{\vec{p}_{ele}}}
\newcommand{\mpele}{\ensuremath{p_{lep}}}
\newcommand{\evis}{\ensuremath{E_{vis}\,}}
\newcommand{\pte}{\ensuremath{p^T_e}}
\newcommand{\ptlep}{\ensuremath{p^T_{lepton}}}
\newcommand{\nuecc}{\ensuremath{\nu_e^{CC}\,}}
\newcommand{\numucc}{\ensuremath{\nu_\mu^{CC}\,}}
\newcommand{\numunc}{\ensuremath{\nu_\mu^{NC}}}
\newcommand{\antinuecc}{\ensuremath{\overline{\nu_e}^{CC}\,}}
\newcommand{\antinumucc}{\ensuremath{\overline{\nu_\mu}^{CC}}}

\newcommand{\chisq}{\ensuremath{\chi^{2}\ }}
\newcommand{\REDBLA}[1]{\red {#1} \black}
\def\mc2{\multicolumn{2}{c|}}
\newcommand{\pnuenumu}{\ensuremath{p(\nue \rightarrow \numu)\,}}
\newcommand{\nuenumu}{\ensuremath{\nue \rightarrow \numu\,}}
\newcommand{\nubarenubarmu}{\ensuremath{\overline{\nu}_e \rightarrow \overline{\nu}_\mu\,}}
\newcommand{\dmot}{\ensuremath{\delta m^2_{12}\,}}
\newcommand{\dmtt}{\ensuremath{\delta m^2_{23} \,}}

\def\He{^6{\mathrm{He}}}
\def\Li{^6{\mathrm{Li}}}
\def\Ne{^{18}{\mathrm{Ne}}}
\def\anue{\overline{{\mathrm\nu}}_{\mathrm e}}
\def\anumu{\overline{{\mathrm\nu}}_{\mathrm \mu}}
\section{BETA BEAMS}
\subsection{\bf Introduction}

The demand for better neutrino beams is correlated with the considerable
improvement in neutrino detectors, and to the recent exciting claims of
evidence for neutrino oscillations by various experiments. The largest
existing detectors are based on the water \v{C}erenkov detection
technique, that seems to be scalable to the $\approx$~Mton order of
magnitude~\cite{uno}. Such a detector is motivated independently by the
exploration of extra-galactic neutrinos, solar neutrinos, atmospheric
neutrinos and proton decay. The additional possibility of a conventional
high-intensity beam, called a super-beam would allow one to extend the
search for the mixing angle $\theta_{13}$, limited mainly by the beam
purity that can be obtained.

A high-intensity neutrino source of a single flavour, with improved
backgrounds and known energy spectrum and intensity, could extend further
the measurement of the missing lepton mixing parameter, $\theta_{13}$, and
allow one to explore the CP-violation phase $\delta$. The beta beam
described here could provide such a source.

The beta beam could be exploited well by non-magnetic water \v{C}erenkov
detectors, and is very complementary to a super-beam. Combining a
super-beam together with a beta beam, the CP-violation phase $\delta$
could be measured in complementary ways: by a CP-violation measurement and
a T-violation measurement, as is explained below in detail.

\subsection{\bf Machine Issues}

It is proposed to produce a collimated $\anue$ beam or $\nue$ beam, by
accelerating to high energy radioactive ions that will decay through a
beta process~\cite{pieroz}.

Radioactive ion production and acceleration to low energies of several MeV
have already been performed for nuclear studies, and various techniques
have been developed, e.g., at the CERN ISOLDE facility~\cite{isol}.
Moroever, the acceleration of positively-charged ions to about 150
GeV/nucleon has already been done in the CERN PS/SPS accelerators for the
experimental programme on heavy-ion collisions. Storage of radioactive-ion
bunches in a storage ring would be very similar, in principle, to what is
being studied for the muon-based neutrino factory scheme~\cite{garoby}.
The combination of these capabilities would provide a unique physics
opportunity for CERN.

The neutrino beam resulting from radioactive-ion decays has three 
distinctive and novel features:

\begin{itemize}

\item[$\bullet$] A single neutrino flavour, which is essentially 
background-free;

\item[$\bullet$] A well-known energy spectrum and intensity;

\item[$\bullet$] Low energy in the ion centre of mass which, when given a
large Lorentz boost, results in strong collimation. This feature is
particularly important for long-baseline neutrino studies.

\end{itemize}

\subsubsection{Nuclear beta decays}

\begin{figure}[t]
\begin{center}
      \includegraphics[width=10cm]{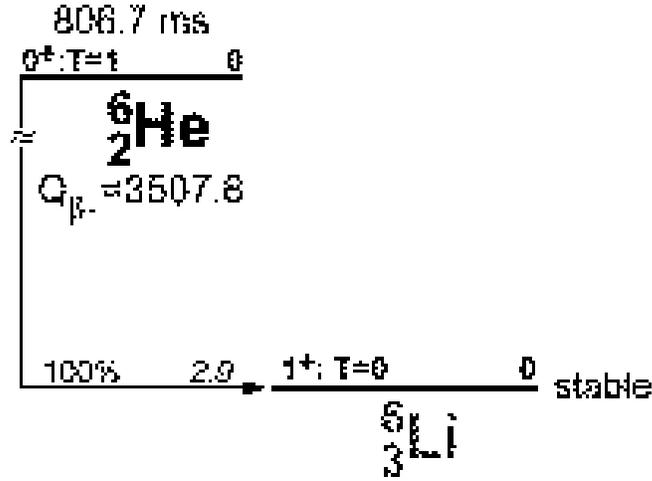}
      \caption{\it Allowed transitions of $\He$~\cite{nucltabl}.}
        \label{hedecay}
\end{center}
\end{figure}

As a guideline, we consider a textbook atomic $\beta^-$ decay which has
well-known characteristics and good features for neutrino production:

$$ ^6_2\mathrm{He}^{++} \rightarrow \;\;\;  ^6_3\mathrm{Li}^{+++} \; 
\mathrm{e}^- \anue \;\;\;.$$
The half-life $T_{1/2}$ of $^6$He is 0.8067~s and the $Q$ value of the
decay is 3.5078~MeV~\cite{nucltabl}. The energy spectrum of the electron
produced in $\He$ beta decay has been measured extensively, and is well
described theoretically (without corrections due to the Coulomb attraction
between the nucleus and the electron) by the simple analytic formula:

$$ N(E) \mathrm{d}E \approx E^2(E-Q)^2 \mathrm{d}E\;\;\;,$$
where $E$ is the electron kinetic energy. The neutrino spectrum may be
completely determined by laboratory measurements of the associated
electron, without involving a neutrino measurement, since $E_\mathrm{e}
+E_\nu \approx Q$ because of the relatively large mass of the nucleus. The
average energy of the neutrino from $\He$ decay is 1.937 MeV, and the
neutrino is emitted isotropically, since the parent ion is spinless. The
beta decay of $\He$ is an extremely clean process, and in fact is the only
allowed nuclear transition, as seen in Fig.~\ref{hedecay}.

Increasing the atomic number, various potential candidates emerge with
different features: lifetime, neutrino energy, and $Z/A$. A list of
possible neutrino and antineutrino emitter candidates is reported in
Table~\ref{candidates}, as well as the possible neutrino energy if are
accelerated to the maximum energy achievable by a 450~GeV proton machine
such as the CERN SPS.

\begin{table*}[th]
\begin{center} 
\caption{\it Possible characteristics of some beta-beam emitter 
candidates. The laboratory neutrino energies
are computed for emitters accelerated to the maximum energy achievable 
by a 450~GeV proton accelerator such as the CERN SPS. All energies are in 
MeV units.
}
\label{candidates}
 \begin{tabular}{|l|l|l|l|l|l|l|l|}
\hline
\hline
Nucleus & Z/A & $T_{\frac{1}{2}}$ (s)& $Q_\beta$& $ Q_{\beta}^{eff}$& $E_{\beta}$& $E_{\nu}$& $E_{LAB}$ \\
\hline
$\anue$ decay &&&&&&& \\
\hline 
$^6_2$He&3.0&0.807&3.5&3.5&1.57&1.94&582 \\
$^8_2$He&4.0&0.119&10.7&9.1&4.35&4.80&1079 \\
$^8_3$Li&2.7&0.838&16.0&13.0&6.24&6.72&2268 \\
$^9_3$Li&3.0&0.178&13.6&11.9&5.73&6.20&1860 \\
$^{16}_6$C&2.7&0.747&8.0&4.5&2.05&2.46&830 \\
$^{18}_7$N&2.6&0.624&13.9&8.0&5.33&2.67&933 \\
$^{25}_{10}$Ne&2.5&0.602&7.3&6.9&3.18&3.73&1344 \\
$^{26}_{11}$Na&2.4&1.072&9.3&7.2&3.34&3.81&1450 \\
\hline
$\nue$ decay &&&&&&& \\
\hline
$^8_5$B &1.6 &0.77 &17.0 &13.9 &6.55 &7.37 &4145 \\
$^{10}_6$C &1.7 &19.3 &2.6 &1.9 &0.81 &1.08 &585 \\
$^{18}_{10}$Ne &1.8 &1.67 &3.4 &3.4 &1.50 &1.86 &930 \\
$^{33}_{18}$Ar &1.8 &0.173 &10.6 &8.2 &3.97 &4.19 &2058 \\
$^{34}_{18}$Ar &1.9 &0.845 &5.0 &5.0 &2.29 &2.67 &1270 \\
$^{35}_{18}$Ar &1.9 &1.775 &4.9 &4.9 &2.27 &2.65 &1227 \\
$^{37}_{19}$K &1.9 &1.226 &5.1 &5.1 &2.35 &2.72 &1259 \\
$^{80}_{37}$Rb &2.2 &34 &4.7 &4.5 &2.04 &2.48 &1031 \\
\hline
\hline
  \end{tabular}
\end{center}
\end{table*} 

\begin{figure}[t]
\begin{center}
      \includegraphics[width=8cm]{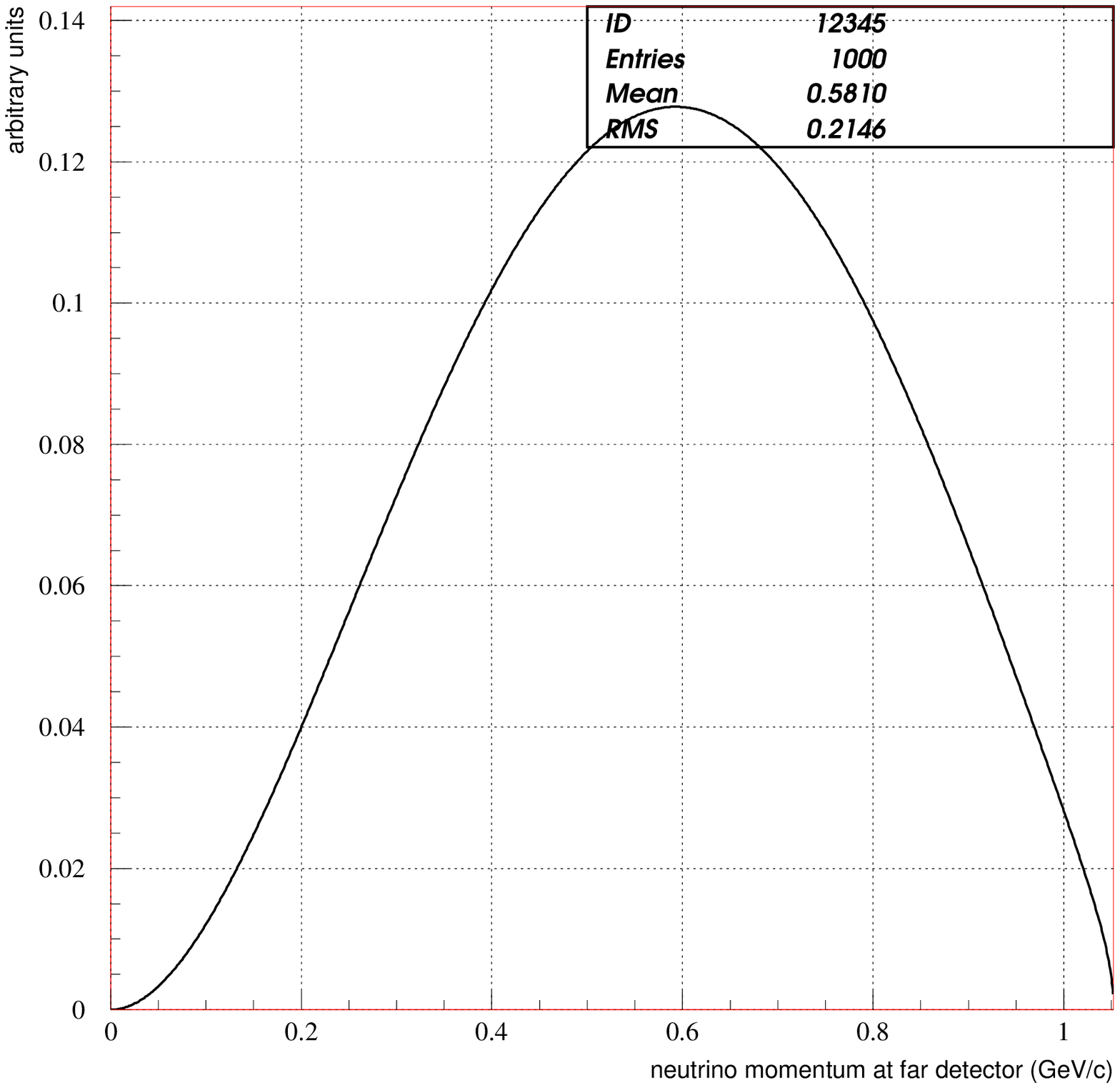}
      \caption{\it The boosted spectrum of $\He$ neutrinos at the far 
detector.}
        \label{betaboost}
\end{center}
\end{figure}

\subsubsection{The relativistic effect}

We suppose that the $\He$ ion is accelerated to $\gamma = 150$,
corresponding to the typical energy-per-nucleon currently obtained in the
heavy-ion runs of the CERN SPS. In the laboratory frame, the neutrino
momentum transverse to the beam axis is identical to that observed when
the atom is at rest: 1.937 MeV on average. In contrast, the average
longitudinal momentum is multiplied by a factor corresponding to $\gamma$,
and therefore neutrinos have typical decay angles of $1/\gamma$, in our
case 7~mrad. In the forward direction, the centre-of-mass neutrino energy
is multiplied by the factor $2\gamma$, so that the average neutrino energy
at a far detector is 582 MeV, as seen in Fig.~\ref{betaboost}.

As the lateral dimensions of the far detector are typically much smaller
than $1/\gamma$ multiplied by the distance $L$, the neutrino spectrum has
essentially no radial dependence. The neutrino flux per parent decay and
unit area is obtained by a Lorentz transformation of the centre-of-mass
distribution of neutrino emission into the laboratory system. For the
value $\gamma=150$, the relative neutrino flux computed at distances above
$\approx$ 1~km varies according to the $1/L^2$ scaling law, and at 100~km
distance, corresponding to $\left<E\right>/L = 5.9 \times 10^{-3}$~GeV/km,
its value is $\Phi=7.2\times10^{-7}\; \mathrm{m}^{-2}$. It is important to
compare the focusing properties of a beta beam with those of a muon-based
neutrino-factory beam. The comparison should be made for identical values
of $\left<E\right>/L$: if we choose arbitrarily the value
$\left<E\right>/L=5.9\times10^{-3}$ GeV/km previously mentioned, the
neutrino-factory detector would be located 5750~km from a 50~GeV muon
storage ring~\cite{garoby}. The relative flux of neutrinos reaching the
far detector of the muon-based neutrino factory is
$5.7\times10^{-9}/\mathrm{m}^2$, 128 times less than in a beta beam. It
should also be said that the relative neutrino flux comparison is
essentially independent of the $\gamma$ factor, if the comparison is made
under identical $\left<E\right>/L$ conditions, since both fluxes are
proportional to $1/L^2$. This is strictly correct if the polarization
effects due to the muon spin are neglected. The polarization of the muon
affects the energy spectrum and the relative flux of both neutrino
flavours produced in the muon decay~\cite{lion}.

Another significant neutrino beam parameter is the number of neutrino
interactions when $E/L \approx \dm2$. This is the parameter that 
determines
the overall statistics collected by an oscillation disappearance
experiment, and is also indicative of the appearance signal intensity,
since 
$$I \propto \mathrm{sin}^2(1.27 \frac{ \dm2 [\mathrm{eV}^2]
L[\mathrm{km}]}{E[\mathrm{GeV}]}).$$ 
Assuming the neutrino cross sections
to be proportional to the neutrino energy~\footnote{This approximation is
good for electron and muon neutrinos at the energies under
discussion.}, and taking into account the fact that the focusing 
properties of the neutrino beam depend solely on the $\gamma$ factor, 
the neutrino interaction rate $N_{\mathrm{int}}$ in the far detector is:

$$N_{\mathrm{int}} \propto (\dm2)^2 \times \frac{\gamma}{E_{\mathrm{cms}}} $$
where $E_{\mathrm{cms}}$ is the neutrino energy
in the frame where its parent is at rest.

The quality factor $ \gamma / E_{\mathrm{cms}} $ characterizes the
neutrino beam and now includes the interaction probability in the
approximation previously described. It is straightforward to see that --
despite the larger probability that the high-energy neutrinos from muon
decay have to interact -- a $\He$ beta beam accelerated to $\gamma=150$ is
more than five times as efficient as a neutrino beam from muons
accelerated to $\gamma=500$.

\subsubsection{Baseline, energy and intensity considerations} 

Order-of-magnitude performances of the acceleration scheme can be
estimated from the current efficiencies of existing machines.  The
acceleration scheme and the target production yields which are assumed are
described elsewhere in this report. Table~\ref{beta-beam} shows the
possible neutrino flux of a beta beam, which can be easily scaled to
different detector distances as already discussed. This table is estimated
for the highest energy possible with $\He$ acceleration in a facility like
the CERN SPS. One may also accelerate other ions, and in particular
$\beta^+$ emitters that produce a $\nue$ in the final state. A good
candidate is $\Ne$, and features of its acceleration scheme are summarized
in Table~\ref{betabeam1}.

\begin{table*}[th]
\begin{center} 
\caption{\it Possible characteristics of a beta beam optimized for 
the $\anue$ interaction rate.}
\label{beta-beam}
 \begin{tabular}{|l|l|}
\hline
\hline
$\He$ ion production  & $5\times10^{13}$/s every 8 s \\
$\He$ collection efficiency& 20\%  \\
$\He$ accelerator efficiency& 65\%  \\
$\He$ maximum final energy & 150 GeV/nucleon  \\
$\anue$ average energy & 582 MeV  \\
\hline
Storage ring total intensity & $1\times 10^{14}$ $\He$ ions \\
Straight section relative length & 36\%  \\
\hline
Running time/year & $10^7$  s  \\
Detector distance & 100 km  \\
$\left<E\right>/L$ & $5.9\times10^{-3}$ GeV/km  \\
\hline
$\anue$ interaction rate on $\mathrm{H}_2\mathrm{O}$ & 69/kton/year  \\
 \hline
  \end{tabular}
\end{center}
\end{table*}

\begin{table*}[th]
\begin{center} 
\caption{\it Possible characteristics of a beta beam optimized for the 
$\nue$ interaction rate.}
\label{betabeam1}
 \begin{tabular}{|l|l|}
\hline
\hline
$\Ne$ ion production  & $1\times10^{12}$/s every 4 s \\
$\Ne$ collection efficiency& 50\%  \\
$\Ne$ accelerator efficiency& 82\%  \\
$\Ne$ maximum final energy & 75 GeV/nucleon  \\
$\nue$ average energy & 279 MeV  \\
\hline
Storage ring total intensity & $1.3\times 10^{13}$ $\Ne$ ions \\
Straight section relative length & 36\%  \\
\hline
Running time/year & $10^7$  s  \\
Detector distance & 130 km  \\
$\left<E\right>/L$ & $2.1\times10^{-3}$ GeV/km  \\
\hline
$\nue$ interaction rate on $\mathrm{H}_2\mathrm{O}$ & 3.1/kton/year  \\
 \hline
  \end{tabular}
\end{center}
\end{table*}

\subsection{\bf Physics Reach of the Beta Beam}

\subsubsection{Signal}

The signal in a beta beam looking for \nuenumu oscillations would be the
appearance of \numu charged-current events, mainly via quasi-elastic
interactions. To select such events the following criteria have been used:

\begin{itemize}
        \item 
        A single-ring event,
                \item 
        The track identified as a muon track using the standard 
Super-Kamiokande identification algorithms,
                        \item 
        The detection of the muon decay into an electron.
\end{itemize}

The signal efficiency as function of the true neutino energy, in the case
of \nubarmu interactions, is shown in Fig.~\ref{fig:beta:eff}(left), with
and without the request of the detection of the decay electron.
The backgrounds and signal efficiency have been studied in a
full simulation, using the NUANCE code~\cite{casper}, reconstructing
events in a Super-Kamiokande-like detector.

\subsubsection{Backgrounds}

The beta beam is intrinsically free from contamination by any different
flavour of neutrino. However, background can be generated by imperfections
in particle identification or external sources, such as single-pion
production in neutral-current \nue(\nubare) interactions, electrons
(positrons) mis-identified as muons, and atmospheric neutrino
interactions.

\begin{figure}[ht]
  \epsfig{file=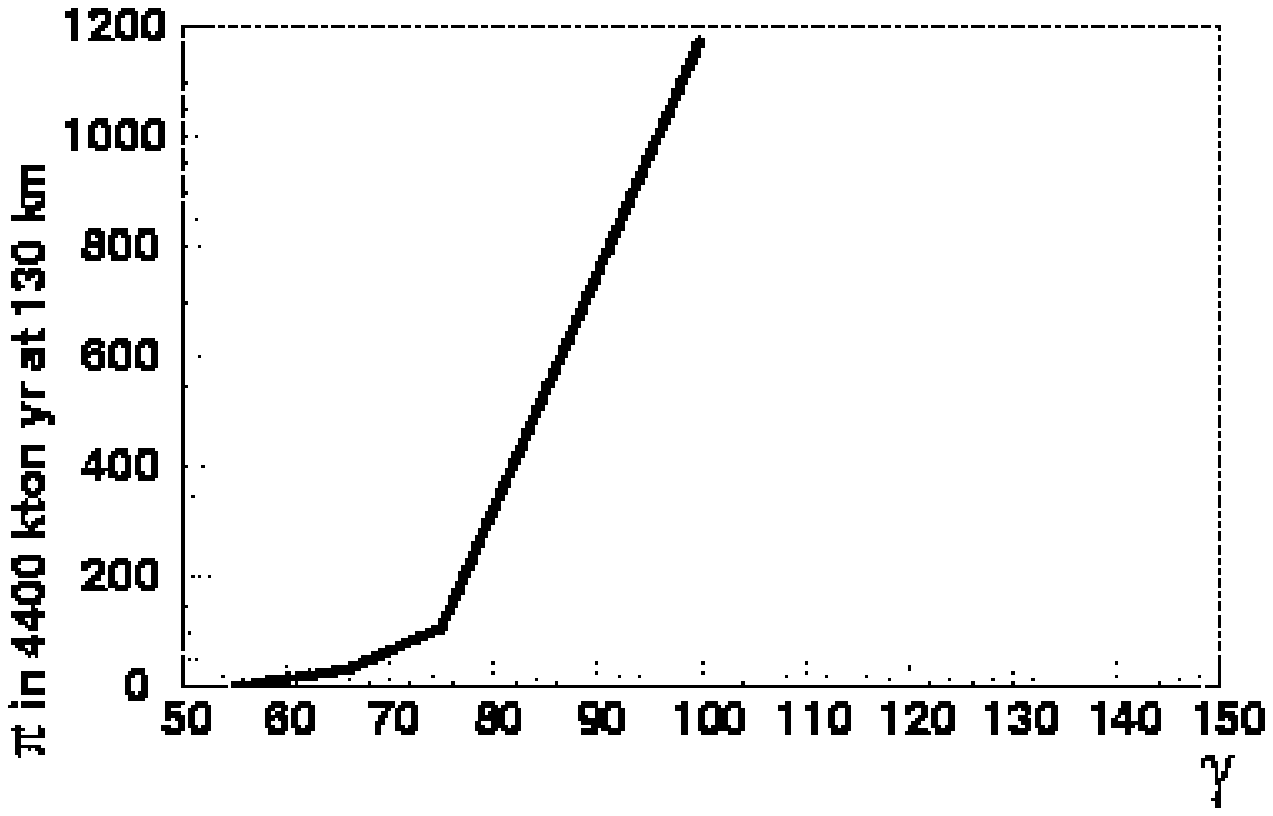,width=0.47\textwidth}
  \epsfig{file=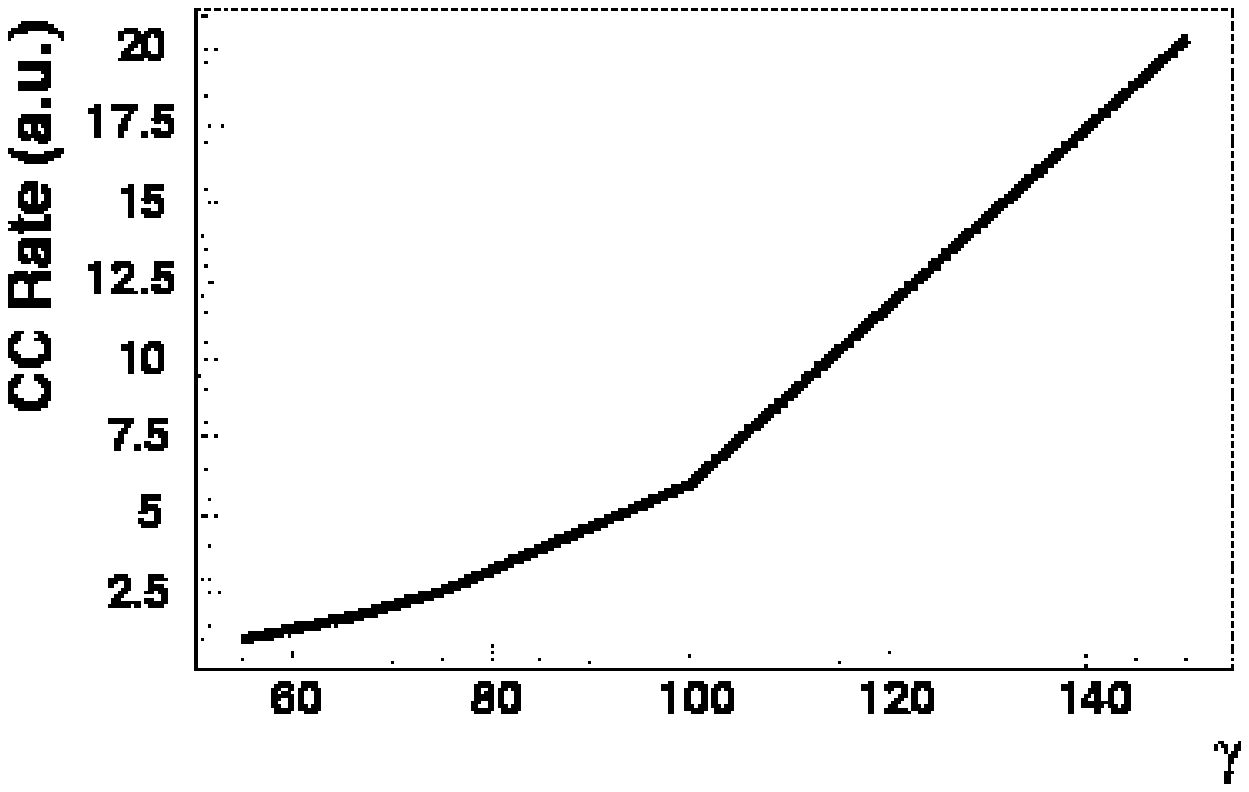,width=0.47\textwidth}
\caption{\it Left panel: Backgrounds from single-pion production for a 
4400~kt-y
exposure, as function of $\gamma$. Right panel: The charged-current \nue
interaction rate as function of $\gamma$ in arbitrary units.}
\label{fig:beta:rates}
\end{figure}

Neutral-current (NC) backgrounds come from the resonant processes
\begin{equation}
\nu_e(\nubare) + p \rightarrow \nu_e(\nubare)   \pi^+ n
\label{eq:bck1}
\end{equation}
and
\begin{equation}
\nu_e(\nubare)   + n \rightarrow \nu_e(\nubare)   \pi^- p
\label{eq:bck2}
\end{equation}
In a water \v{C}erenkov detector, pions below 1~GeV can be distinguished
from muons only by requiring the decay electron signature to be associated
to an identified muon. However, at low neutrino energy this background is
intrinsically highly suppressed, since the threshold for these processes
is about~400 MeV, and furthermore the outgoing pion must be above the
\v{C}erenkov threshold. As a net result, at $\gamma$ values below 60, NC
pion production is suppressed. At higher $\gamma$ values, single-pion
production rises quite fast, as shown in Fig.~\ref{fig:beta:rates}(left).
Electrons (positrons) produced by \nue (\nubare) can be mis-identified
as muons, therefore giving a fake signal.
Standard algorithms for particle identification in water \v{C}erenkov
detectors are quite effective to suppress such backgrounds. Furthermore,
the signal of the decay electron in muon tracks can be used to reinforce
the muon identification.
%

Atmospheric neutrino interactions are extimated to be $\sim$~50/kton/yr in
the energy range of interest for the experiment, a number of interactions
that far exceeds the oscillation signal. The atmospheric neutrino
backround has to be reduced mainly by timing of the parent ion bunches.
The time structure of the interactions in the detector is identical to the
time structure of the parent ions in the decay ring. The location of the
beta decay does not matter, since the parents have essentially the same
speed as the neutrinos. The far-detector duty cycle is therefore given by
the ratio of the bunch length to the ring length. For a decay ring of
6.9~km and a bunch length of 10~ns, which seems feasible, this ratio
provides a rejection factor of $2\cdot 10^4$. The directionality of the
(anti)neutrinos can be used to suppress further the atmospheric neutrino
background by a factor $\sim 4$. With these rejection factors, the
atmospheric neutrino background can be reduced to the order of 1 event/440
kton/yr. Moreover, out-of-spill neutrino interactions can be used to
normalize this background to the 1\% accuracy level.

\subsubsection{Systematic errors}

The cross-section estimates for signal and background production at
energies below 1 GeV are quite uncertain, the systematic errors being of
the order of 20 and 30 \% respectively. Systematic errors of this size
would seriously reduce the sensitivity of the experiment. On the other
hand, a beta beam is ideal for measuring these cross sections, provided
that a close detector at the distance of about 1~km from the decay tunnel
and of proper size (1~kt at least) can be built:

\begin{itemize}
        \item
The energy and the flux of the neutrino beam is completely defined
by the acceleration complex, and can be known with a precision better
than 1\%, the limiting factor being knowledge of the number of accelerated
ions in the machine.
        \item
The near/far residual error is extremely reduced, because in a beta beam
the divergence
of the beam is completely defined by the decay properties of the parent
ion.
        \item
The $\gamma$ factor of the accelerated ions can be varied, in particular
a scan can be initiated below the production threshold of the backgrounds,
allowing a precise measure of the cross sections for resonant processes.
\end{itemize}
It is extimated that a residual systematic error of 2\% will be
the final precision with which both the signal and the backgrounds
can be evaluated.

\subsubsection{Beam optimization}

The Lorentz boost factor $\gamma$ of the ion accelerator can be tuned to
optimize the sensitivity of the experiment to CP violation in the neutrino
sector. We perform the optimization assuming the value $\delta m^2_{atm} =
2.5\cdot10^{-3}$~eV$^2$ for the atmospheric neutrinos, a baseline of
130~km, the CERN-Fr\'ejus distance, and a $^6$He beam. In principle,
baselines in the range 100-250~km are possible, given the characteristics
of the acceleration scheme.

There are competitive processes that must be balanced to find the
optimal sensitivity:

\begin{enumerate}
        \item
The number of quasi-elastic events in the far detector scales roughly
as $\gamma^{3}$, see Fig.~\ref{fig:beta:rates}(right).
This factor comes from the
beam focussing, which is $\propto \gamma^2$, and the cross section, 
which is $\propto \gamma$.
The number of quasi-elastic events at a given L/E is then proportional
to $\gamma$. 
        \item
The number of background events from the processes 
(\ref{eq:bck1}, \ref{eq:bck2}) increases
with $\gamma$, as shown in Fig.~\ref{fig:beta:rates}(left).
An excessive background severely limits the
sensitivity of the experiment, so background contamination
prefers the lowest possible $\gamma$.
        \item
The signal efficiency as function of energy, 
see in Fig.~\ref{fig:beta:eff}(left),
severely disfavours neutrino energies below 300~MeV. We note that
the signal efficiency decreases slightly above 0.7~MeV. This is due
to the request for single-ring events: at higher values of the $\nu$ 
energy, the fraction of quasi-elastic events decreases.
\begin{figure}[ht]
  \epsfig{file=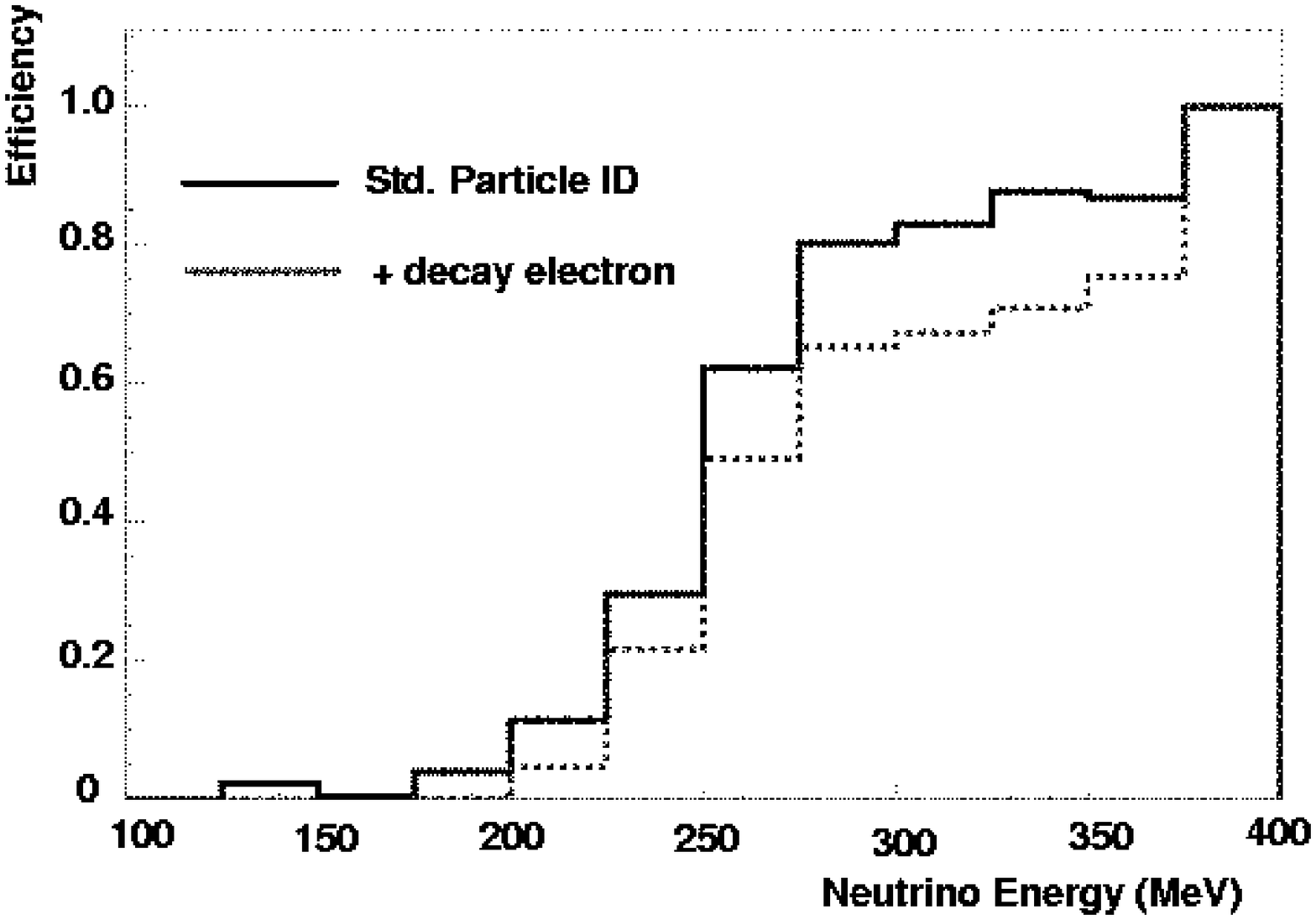,width=0.47\textwidth}
  \epsfig{file=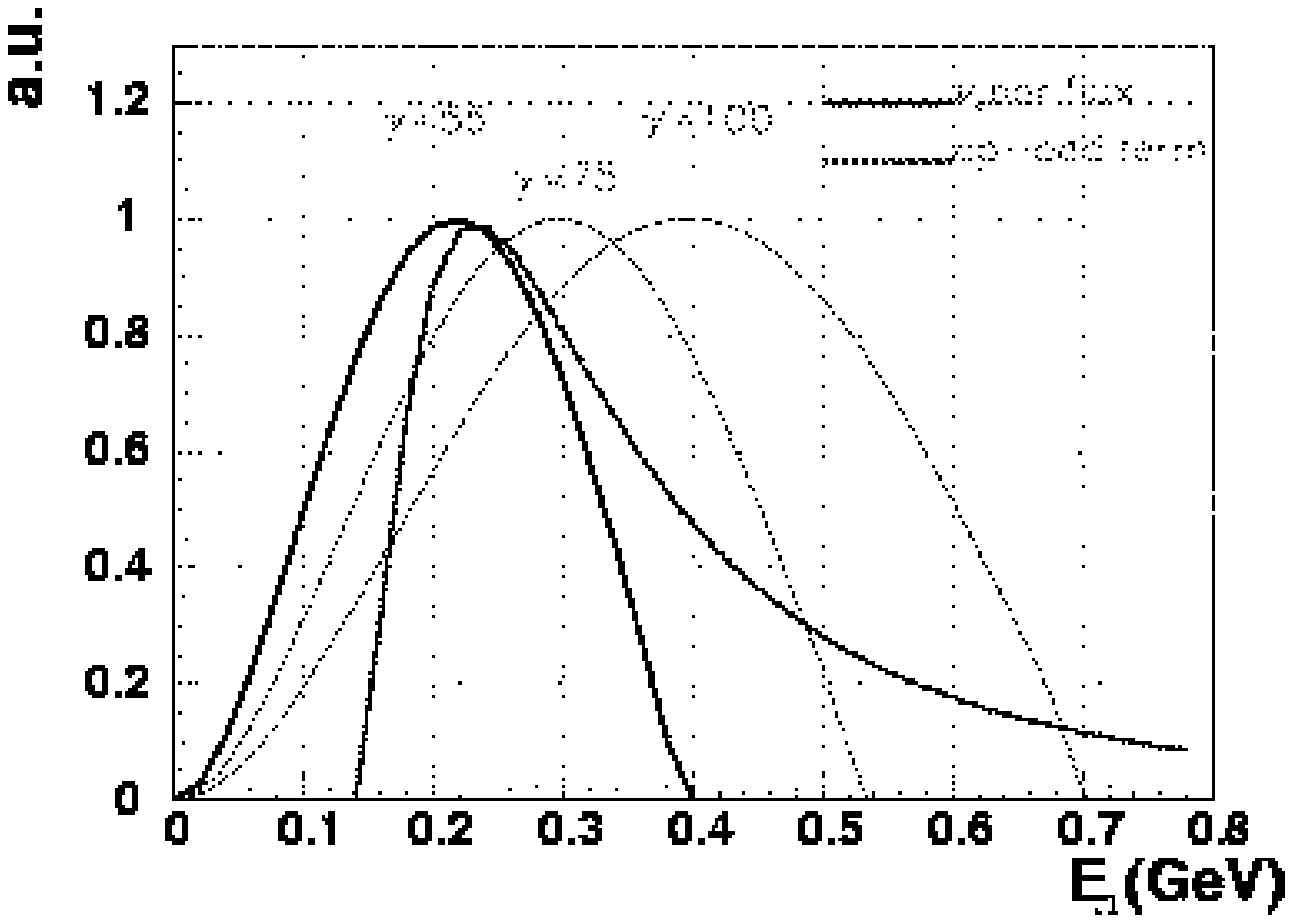,width=0.47\textwidth}
\caption{\it The signal efficiency as function of the neutrino energy 
(left). Energy spectra for different values of $\gamma$ compared with 
the probability function for CP-odd oscillations (right).}
\label{fig:beta:eff}
\end{figure}
        \item
The true signal of the experiment consists of the observation of those 
\numu produced by the CP-odd term in \pnuenumu.
In Fig.~\ref{fig:beta:eff}(right), this term is compared with the neutrino 
energy spectrum
produced at three different $\gamma$ values: $\gamma=50, 75, 100$, all
at the baseline of 130~km. It is evident that 
the neutrino energy is de-tuned with respect to the oscillating term
when $\gamma \geq 100$.
\end{enumerate}
Based on these considerations, a $\gamma$ value of 75 seems to approach
the optimal value for CP sensitivity.
We report In Table~\ref{tab:beta:rates} signal and background rates for a 
4400~kt-y exposure to  $^6$He and $^{18}$Ne beams.

\begin{table}[ht]
\begin{center}
\begin{tabular}{|l|rr|}
\hline
10 years        & $^6He$  &   $^{18}Ne$    \\
                &  $\gamma=75$    &   $\gamma=75$ \\
\hline
CC events (no osc, no cut) & 40783   &  18583\\
Total oscillated           &  34  &   63 \\
CP-Odd oscillated          & -43   &  22\\
Beam background            &  0  &  0 \\
Detector bkg.              &  60  &  10\\
\hline
\end{tabular}
\caption{\it Event rates for a 4400~kt-y exposure. The signals are
computed for $\thetaot=3^\circ$, $\dmot=0.6\cdot10^{-4}eV^2$, 
$\delta=90^\circ$.}
\label{tab:beta:rates}
\end{center}
\end{table}

\subsubsection{Sensitivity to CP violation}

A search for leptonic CP violation can be performed running the beta beam
with $^{18}$Ne and $^6$He, and fitting the number of muon-like events to
the \pnuenumu probability. The fit can provide simultaneus determinations
of $\theta_{13}$ and the CP phase $\delta$.
Uncertainties in the mixing-matrix parameters can affect the
precise determination of $\theta_{13}$ and $\delta$. These effects
have been taken into account, considering as errors in these other 
parameters the values that can be expected by the near future experiments:
\begin{itemize}
        \item
        A 10\% error on the solar $\delta m^2$ and $\sin^2{2\theta}$,
as expected from the KamLAND experiment, after 3 years of data 
taking~\cite{KamLAND}.
        \item
  A 2\% error on the atmospheric $\delta m^2$ and $\sin^2{2\theta}$,
as expected from the JHF neutrino experiment~\cite{jhf}.
\end{itemize}
Only the diagonal contributions of these errors are considered in the
following. They contribute less than 5\% of the total error in
the final sensitivity.
Given the relative interaction rates for quasi-elastic events, a sharing
of 3 years of $^6$He and 7 years of $^{18}$Ne has been considered.

\begin{figure}[ht]
\centerline{\epsfig{file=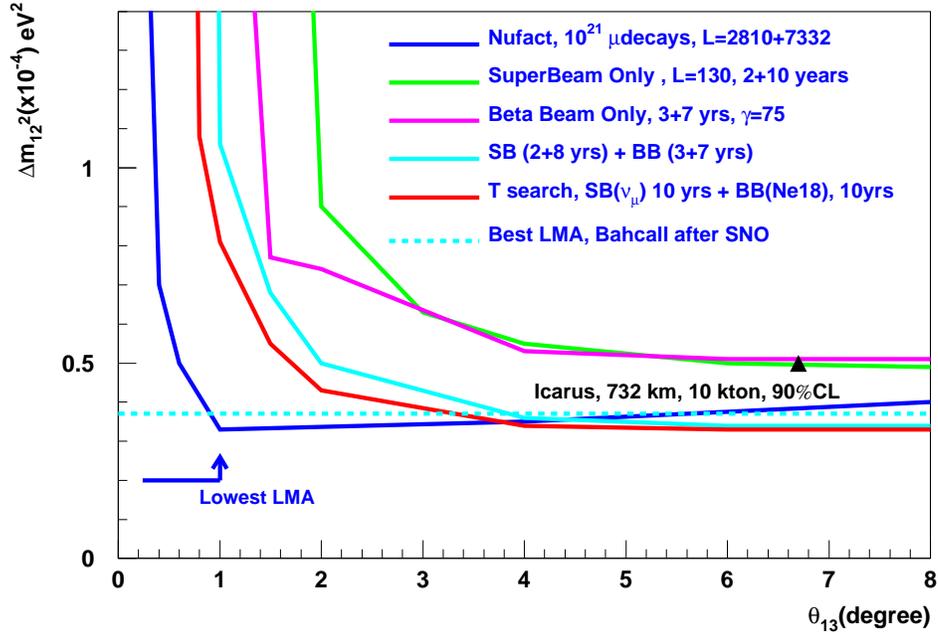,width=0.80\textwidth} }
\caption{\it The CP sensitivity of the beta beam, of the SPL super-beam, 
and of their combination. They are compared with a
Neutrino Factory experiment with two detectors~\cite{golden}.}
\label{fig:beta:sens}
\end{figure}

The results of this analysis are summarized in Table~\ref{tab:beta:rates},
for an arbitrary choice of the mixing matrix parameters.
Since the sensitivity to CP violation is heavily dependent on the true 
value of \dmot and \thetaot, we prefer to express the CP sensitivity for a
fixed value of $\delta$ in the whole \dmot, \thetaot parameter space.
The CP sensitivity to separate
$\delta=90^\circ$ from $\delta=0^\circ$ at the 99\%CL as a function
of \dmot and \thetaot, following the convention of~\cite{golden},
is plotted in Fig.~\ref{fig:beta:sens}.

\subsubsection{Synergy between the SPL super-beam and the beta beam}

The beta beam needs the SPL as injector, but consumes at most $\sim 3\%$
of the SPL total number of produced protons. The fact that the energies of
both the super-beam and the beta beam are below 0.5 GeV, with the beta
beam being tunable, offers the fascinating possibility of exposing the
same detector to two neutrino beams at the same time.

We recall that the SPL super-beam is a \numu (\nubarmu) beam, while 
the beta beam is a
\nue (\nubare) beam, so the combination of the two beams offers
the possibility of CP, T and CPT searches at the same time:
  \begin{itemize}
        \item
    Searches for CP violation, running the super-beam with \numu and 
\nubarmu, and the beta beam with $^6$He (\nubare) and $^{18}$Ne (\nue).
        \item
    Searches for T violation, combining neutrinos from the super-beam 
(\numunue) and from the beta beam using $^{18}$Ne ($\nue \rightarrow 
\numu$), or antineutrinos from the super-beam ($\nubarmu \rightarrow 
\nubare$) and from the beta beam using $^6$He ($\nubare \rightarrow 
\nubarmu$).
        \item
    Searches for CPT violation, comparing $P(\numu \rightarrow \nue)$ to
$P(\nubare \rightarrow \nubarmu)$ and $P(\nubarmu \rightarrow \nubare)$
to $P(\nue \rightarrow \numu)$.
\end{itemize}
It is evident that the combination of the two beams would not result
merely in an increase in the statistics of the experiment, but would offer
clear advantages in the reduction of systematic errors, and would offer 
the redundancy needed to establish firmly any CP-violating effect within
reach of the experiment.


We should also stress that this combination of beams offers the
only known possibility of measuring CPT violation combining signals 
from the same detector, a crucial issue for the control of systematic
errors.

Fig.~\ref{fig:beta:sens} summarizes the CP sensitivity of different
combinations of the super-beam and beta beam compared with the Neutino
Factory sensitivity. The combined sensitivity of the super-beam and beta
beam complex shown in Fig.~\ref{fig:beta:eff} is very competitive with
that of a Neutrino Factory, offering a truly complementary approach to the
search of of leptonic CP violation.

\newpage
\section{THE NEUTRINO FACTORY}\label{sec:nufact}

\subsection{\bf Overview}

In a Neutrino Factory, neutrinos are produced by the decays of muons
circulating in a storage ring. Most of what is known of muon storage rings
is due to the pioneering work of the Muon Collider
Collaboration~\cite{muoncollcoll}. They were able to formulate and to a
large extent simulate the basic concepts of a Muon Collider. The concept
of a Neutrino Factory was born from the observation that the beams of
neutrinos emitted by the decaying muons along the accelerator chain or in
the storage rings could be valuable physics tools~\cite{geer}, the
potential of which was emphasised in the ECFA prospective
study~\cite{ecfapersp}. Neutrino Factory design is presently being pursued
in the United States~\cite{nfusa}, at CERN~\cite{nfcern} and in
Japan~\cite{nfjap}. It owes much to the earlier Muon Collider studies, and
its very similar components are briefly recalled here.

\subsubsection{General principles}

The design includes a very high-power proton driver, delivering on target
typically 4 MW of beam power of protons with energy in excess of a few
GeV. A super-conducting linac at 2.2 GeV has been studied at
CERN\cite{spl}, while the US design calls for a rapid cycling proton
synchrotron at 16-24 GeV, and an upgrade of JHF is considered in Japan.  
Designing a target that can withstand the thermal shock and heat load
naturally leads to a liquid jet target design, although rotating high
temperature solids are also being considered. Pions produced are collected
as efficiently as possible by a magnetic channel, which involves a 20~T
solenoid or powerful magnetic horns.  Pions quickly decay into muons with
a similar energy spectrum.  At this point the beam is 0.6~m in diameter
and has an energy spread of more than 100\%.

A momentum interval near the largest particle density, typically
$250\pm100$ MeV/c, is monochromatized to within a few MeV by means of
phase rotation, using a strong variable electric field to slow down the
fastest particles and accelerate the slower ones. This requires
low-frequency ($\sim 50-100$~MHz) RF cavities or an induction linac. To
reduce the transverse emittance, cooling is necessary, and is provided by
ionization cooling~\cite{ioncool}. This involves energy loss of muons
through a low-Z material, e.g., liquid hydrogen, in a strongly focusing
magnetic field (solenoids of 5-10 Tesla), which reduces momentum in all
three dimensions, followed by accelerating RF cavities, which restore the
longitudinal momentum. The net effect is a reduction of emittance, leading
to a transverse beam size of a few centimetres.
 
\begin{figure}[tbhp]
  \begin{center}
    \epsfig{file=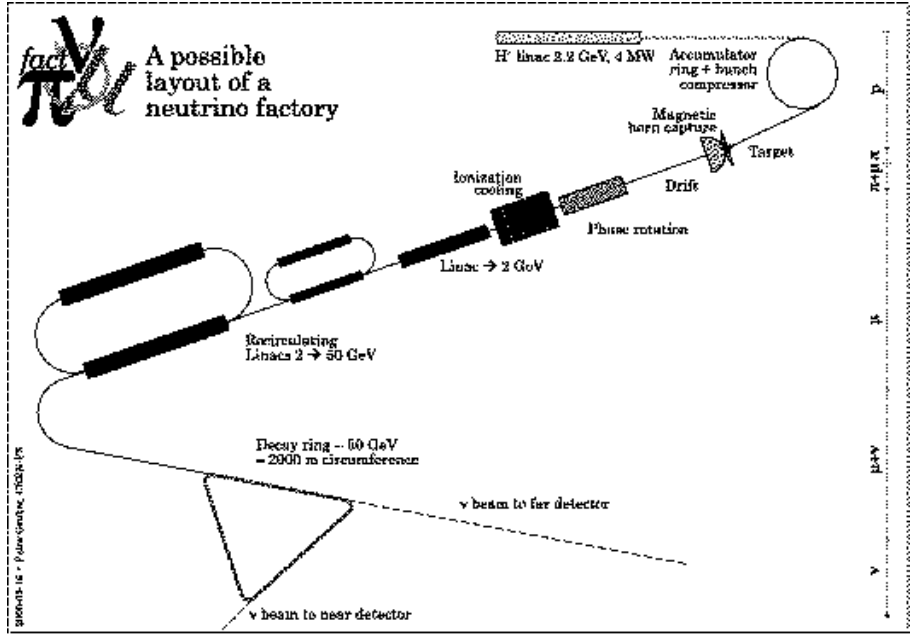,width=12cm}
    \caption{\it Schematic layout of the CERN scenario for a Neutrino 
Factory.}
    \label{fig:nflayout}
  \end{center}
\end{figure}

This leads to a linear configuration, as shown in Fig.~\ref{fig:nflayout},
for the initial muon beam preparation section, or {\sl {muon front-end}}.  
In this concept, each beam element is used only once. It could be
interesting, to save hardware, to be able to perform phase rotation and/or
transverse cooling in a recirculating configuration.  Indeed, a system of
large aperture FFAG accelerators with low frequency RF (around 1.5 MHz) is
the key to the Japanese Neutrino Factory design. Also, much progress has
recently been made on `ring coolers', which allow both transverse and
longitudinal cooling in a circular configuration~\cite{cooling-ring}.

Assuming that the delicate questions of optics can be solved, these `ring'
options share the difficulty of injecting or extracting from a ring the
very large emittance beam of muons available at the end of the decay
channel. The possibility of very large aperture and very fast kickers is
the major unknown and will be a key issue for these potentially
cost-saving developments.

\begin{figure}[tbhp]

\centerline{\includegraphics[width=0.9\textwidth]{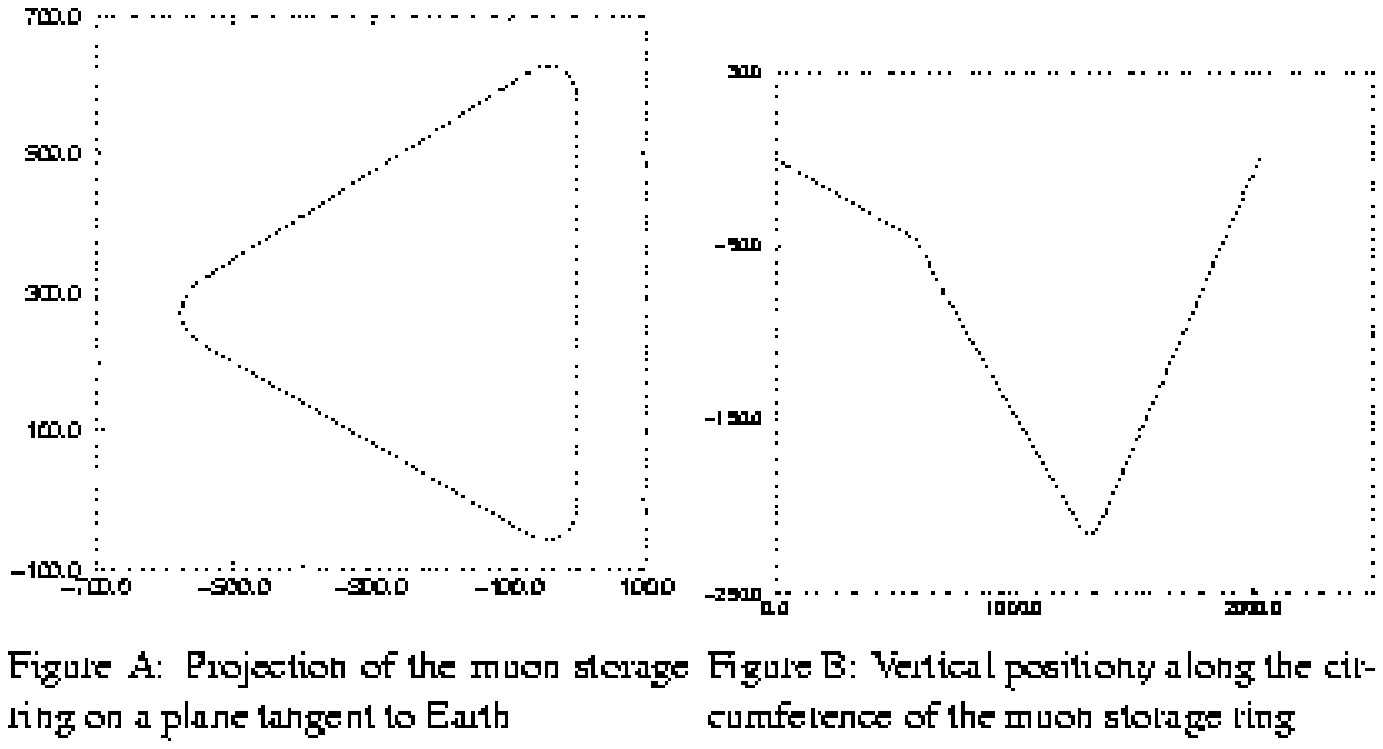}}
\centerline{\includegraphics[width=0.9\textwidth]{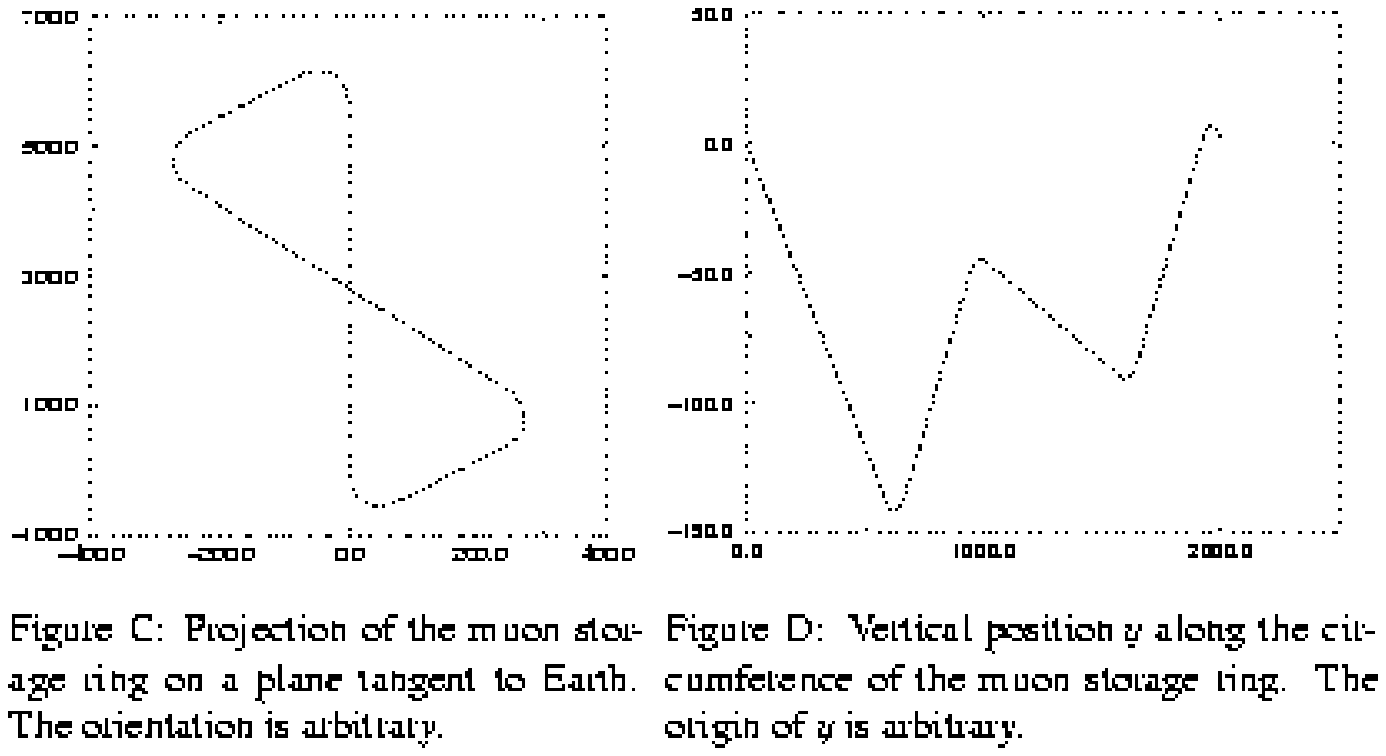}}

\caption{{\it
Description of the decay rings designed in~\cite{keil}. Top panel: the
triangular geometry, bottom panel: the bow-tie geometry. 
}}
\label{decayrings}
\end{figure}

Last but not least, a linac followed by recirculating linacs - or FFAG
accelerators - provide the fast acceleration of muons to an energy of 20
to 50 GeV, the optimal operating energy being presently under discussion.
Around $10^{21}$ muons per year of $10^7$ seconds could then be stored in
a ring, where they would circulate for a few hundred times during their
lifetime. The storage ring can take the shape of a racetrack, triangle or
bow-tie. These latter two configurations allow several beams of decay
neutrinos to be produced in the direction of short- and long-baseline
experiments. Optics have been designed~\cite{keil} for muon storage rings
of either triangular or bow-tie geometry, pointing for instance at
distances of 730 km (which would correspond to the CERN-Gran Sasso beam
line), and 2800 km (which would correspond to a more distant site in the
Canary Islands or the Nordic countries). The geometries of these rings 
are shown in Fig.~\ref{decayrings}.
 
As can be inferred from this brief description, Neutrino Factory design
involves many new components and extrapolations beyond state-of-the-art
technology. The first design studies~\cite{nfusa,nfcern,nfjap} have come
to the conclusions that, with the present designs and technology, such a
machine could indeed be built and reach the desired performance, but that
serious work was needed to bring the cost down. Assuming adequate funding,
it is considered that about five years of research and development will be
necessary to reach a point where a specific, cost-evaluated machine can be
proposed.

\subsubsection{Rates and backgrounds}

The large beam intensities envisaged in the framework of muon collider
studies would provide an unprecedented neutrino flux, that would allow one
to perform experiments with an unprecedented precision on the oscillation
measurements. We now take a closer look at the spectra and event rates
produced.

\begin{figure}[tbhp]
\begin{minipage}[h]{0.45\textwidth}
\begin{center}
\epsfig{file=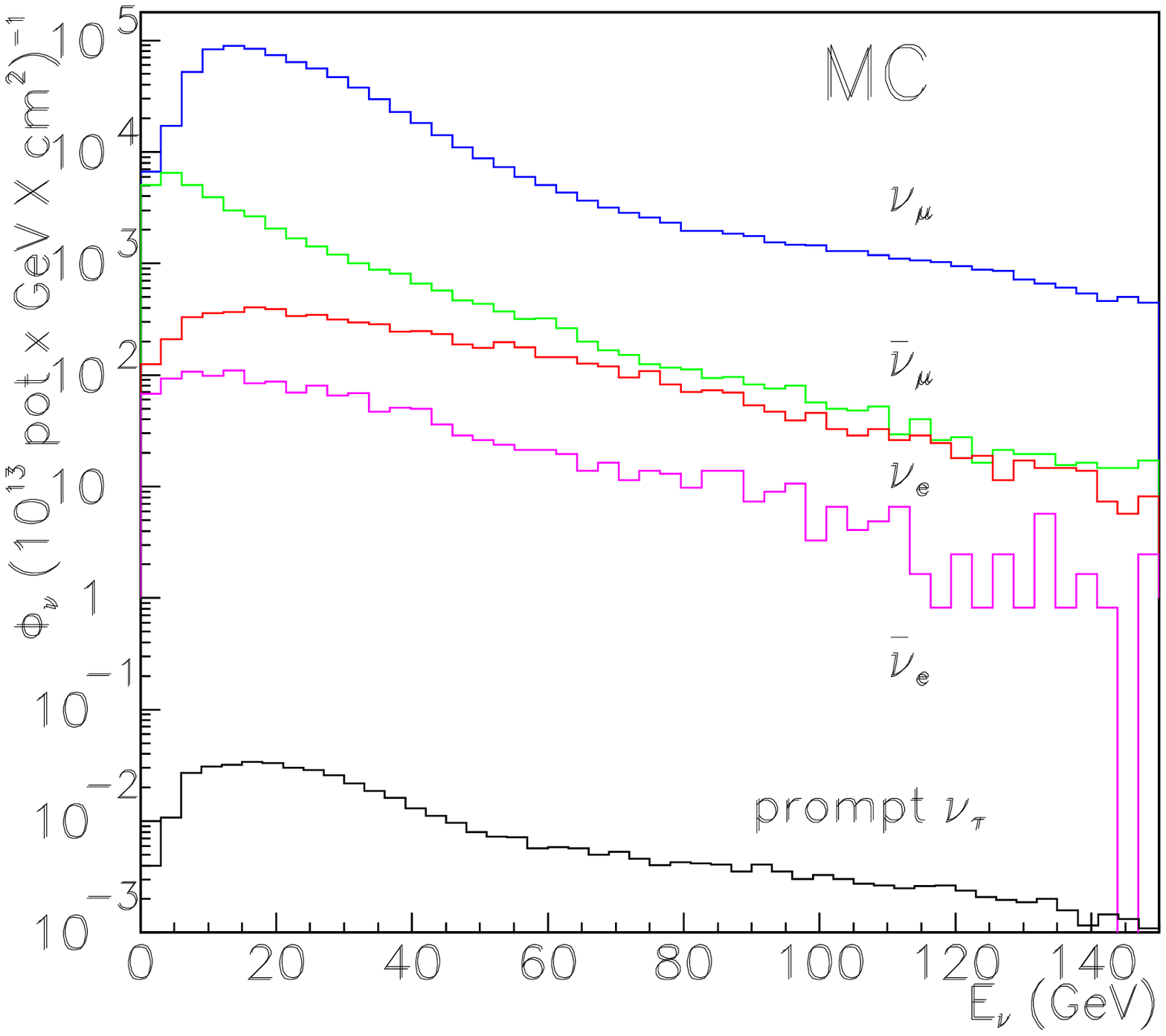,bb= 70 50 567 567,width=8cm}
\end{center}
\end{minipage}
\begin{minipage}[h]{0.45\textwidth}
\begin{center}
\epsfig{file=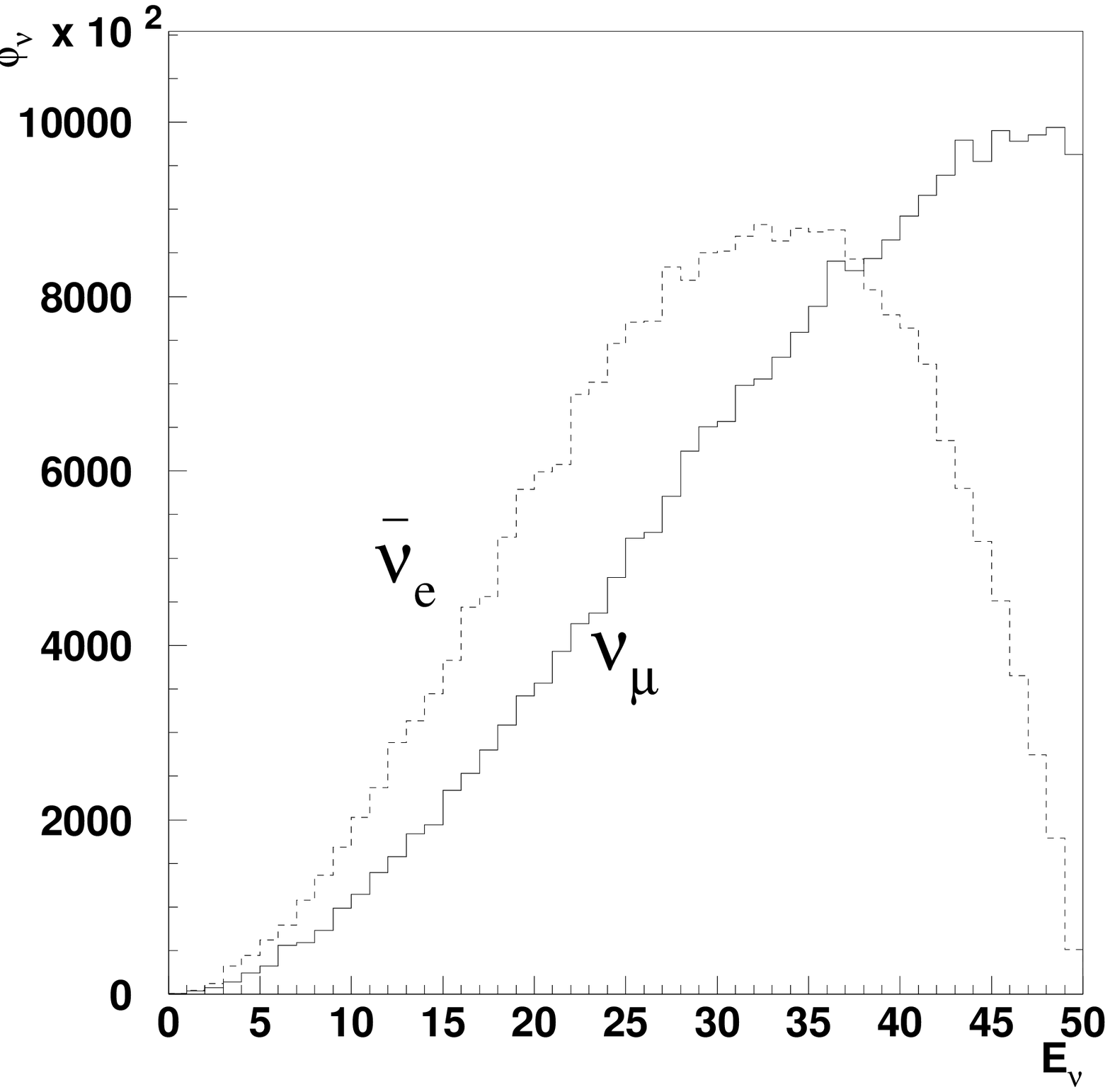,width=8cm}
\end{center}
\end{minipage}

\caption{\it Comparison of the neutrino spectra for a traditional
neutrino beam (WANF at CERN), on the left, and a neutrino factory,
on the right. The vertical axes are not to scale.} 
\label{fig:spectra}
\end{figure}

The neutrino energy spectrum in muon decay at rest follows the following
distribution:

\begin{eqnarray}
\frac{d^2 N_{\nu_\mu}}{dx d\Omega} & \propto &
\frac{2 x^2}{4\pi}[(3-2x)+(1-2x)P_\mu \cos\theta]\\
\frac{d^2 N_{\bar{\nu}_e}}{dx d\Omega} & \propto & \frac{12
  x^2}{4\pi}[(1-x)+(1-x)P_\mu \cos\theta]
\label{eq:fluxesfrommuon}
\end{eqnarray}
where $x \equiv 2E_\nu/m_\mu$, $P_\mu$ is the muon polarisation, and
$\theta$ is the angle between the muon polarisation vector and the
neutrino direction. In the laboratory frame, when considering a detector 
located
on the same axis as the Lorentz boost, as is the case for long-baseline
experiments where the detector size can be neglected with respect to the
baseline, the shape of the energy distribution is preserved. So, if muons
are accelerated to an energy $E_\mu$, the spectral shape will be described
by the same formulae as above, this time with $x=E_\nu/E_\mu$.
The spectral shape of a Neutrino Factory flux, compared to that of a
typical neutrino beam from pion decays, namely the WANF beam at CERN, is
shown in Fig.~\ref{fig:spectra}.

For different values of the muon energy, the total neutrino flux at the
far location will increase as $E^2$, because of the shrinking of the
angular opening of the neutrino beam due to the Lorentz boost
$\gamma^{-2}=(m_\mu/E_\mu)^2$. Moreover, since the deep-inelastic
scattering cross section rises approximately linearly with neutrino
energy, and the spectral shape only depends on $x=E_\nu/E_\mu$, the total
number of events observed in a far detector will grow as $E_\mu^3$.  
Geometrical solid-angle considerations suggest that, always assuming
negligible detector size with respect to the baseline, the flux goes like
$1/L^2$. Neglecting matter effects, the oscillation probabilities will
depend on $L/E_\nu$, so keeping the same oscillation probability and
maximising the number of events would ideally require very long baselines
and large muon energies. The limitation to this logic, apart from the
physical size of the Earth's diameter, comes from the matter effect, that
depresses oscillation probabilities for baselines above 4000 km. This
topic will be discussed later in more quantitative terms.

\begin{table}[tbhp]
  \begin{center}
    \begin{tabular}[h]{|l|l|l|}\hline
      $\mu$ charge&Ev.Class&Events\\ \hline
      &$\nu_\mu$ CC&134362\\
      $\mu^-$&$\nu_\mu$ NC&39952\\
      &$\bar{\nu}_e$ CC&52000\\
      &$\bar{\nu}_e$ NC&18054\\ \hline
      &$\bar{\nu}_\mu$ CC&60010\\
      $\mu^+$&$\bar{\nu}_\mu$ NC&21067\\
      &$\nu_e$ CC&117369\\
      &$\nu_e$ NC&34558\\ \hline
    \end{tabular}
    \caption{\it The number of events detected by a 10~kt detector for 
$10^{21}$ muons in the ring, assuming a baseline of 3000~km and a muon 
energy of 30~GeV.}
    \label{tab:rates}
  \end{center}
\end{table}

The total number of events for the non-oscillation case, observed per
kiloton of detector mass at a distance of 3000~km with a 30~GeV muon beam,
is shown in Table~\ref{tab:rates}. Rates for different values of the muon
energy, baseline, number of muons or detector mass can be obtained from
the relation

\[N_{Events}(E_\mu,L,N^\mu,M)=N_{Events}^{table}(\frac{E_\mu}
{30 GeV})^3(\frac{3000 km}{L})^2(\frac{N^\mu}{10^{21}})
(\frac{M}{10 kton}).\]
One of the main characteristics of the neutrino factory is that it
delivers a well-defined beam free of intrinsic background. For instance,
negative muons circulating in the ring will produce $\nu_\mu$, that in
turn will again produce negative muons in the interaction with the
detector. Positive muons are in principle only produced from the
oscillation of the $\bar{\nu}_e$ component of the beam. The reversed
argument applies for positive muons in the ring. In general, we talk about
Right-Sign Muons (RSM) coming from the beam, and Wrong-Sign Muons (WSM)
coming from the oscillation. In practice, the picture is not so simple,
since other processes can make contributions to the WSM sample. They are
quite rare, but they can become important when we want to be sensitive to
small oscillation probabilities, for instance if the value of
$\theta_{13}$ is very small.

The main background contributions to wrong-sign muons, in the
case of a beam produced by $\mu^-$ decays, are:
\begin{itemize}
\item $\bar{\nu}_\mu$ CC events where the right-sign muon
is lost, and a wrong-sign muon is produced by the decay of a
$\pi$, K or D. The most energetic muons are produced
by D decays.
\item $\nu_e$ CC events where the primary electron is
not identified. In this case, D decays are not a major problem since,
due to the neutrino helicity, they would produce right-sign
muons. However, wrong-sing muons can come from pion and
kaon decay.
\item $\bar{\nu}_\mu$ and $\nu_e$ NC events where charm
production is suppressed with respect to charged currents,
and therefore also the main contributions are given by
pion and kaon decays.
\end{itemize}
These backgrounds can be rejected using the facts that muons
coming directly from neutrino interactions are higher in
energy and more separated from the hadronic jets than
those produced in secondary decays. A cut on momentum and
on the transverse momentum $Q_t$ of the muon with respect to the jet can 
reduce the background to wrong-sign muons by several orders
of magnitude.

\subsection{\bf Flux Control and Resulting Constraints on the Decay Ring
Design}

One of the most significant qualities of the Neutrino Factory, and more
generally of a system where one stores a beam of decaying particles (such
as the beta beam) is the potential for excellent neutrino flux control.  
The main parameters that govern the systematic uncertainties on the
neutrino fluxes are as follows.

\begin{itemize}
\item 
The monitoring of the total number of muons circulating in the ring,
\item
Theoretical knowledge of the neutrino fluxes from muon decay,
including higher-order radiative effects,
\item
Knowledge of the muon beam polarisation,
\item 
Knowledge of the muon beam energy and energy spread,
\item 
The muon beam angle and angular divergence.
\end{itemize}

Beam shape parameters are crucial for the measurement of oscillation
length, while the absolute normalisation is essential for the measurement
of the mixing angle. The relative normalisation of the two muon charges
plays a crucial role in the measurement of CP asymmetries.

\subsubsection{Absolute flux monitoring}

Monitoring the total number of muons in the ring can be done in a
number of ways. The total beam current can be estimated using a Beam
Current Transformer (BCT), the total number of decay electrons can be
estimated using an electron spectrometer, the product of the flux and
cross section can be inferred from a near-by detector and, finally, the
absolute normalisation can be obtained from semi-leptonic neutrino
interactions in a nearby detector.

The operation of a BCT in the decay ring could provide fast-response
monitoring of the muons in the ring. There are, however, a few potential
difficulties that could limit the precision of such a device, which could
normally reach the $10^{-3}$ level.  The first one is the
presence of decay electrons in the ring, along with the muons. Since all
muons decay, the number of accompanying electrons could potentially be
much larger than the remaining muons after a few muon lifetimes.  A study
of such decay electrons has been made~\cite{keillosses}, with the
conclusion that for 50~GeV muon momentum, the decay electrons are lost in
the beam elements (or the collimators placed to protect them)  after less
than half a turn, either because they are momentum-mismatched or because
they lose energy in the arcs by synchrotron radiation.  Consequently their
number should be always less than about $2 \times 10^{-3}$ of the 
remaining
muons.  In addition, most of the losses arise in the straight sections or
in the early part of the arcs, so that a BCT situated just at the
beginning of a straight section would see an even smaller fraction of
them.  Another worry could be the existence of a moving electron cloud
created by beam-induced multipacting, or by ionization of the residual
gas or of the chamber walls.  This has been studied by
in~\cite{zimmermannlosses}, with the conclusion that the electron cloud
will be several orders of magnitude less than the muon flux itself. In the
absence of a significant parasitic current, it can be concluded that the
BCT readings should be precise to the level of a few $10^{-3}$, or better.
This seems the most practical way to compare the flux induced from $\mu^+$
and $\mu^-$ decays.

\begin{figure}[tbhp]

\centerline{\includegraphics[width=0.9\textwidth]{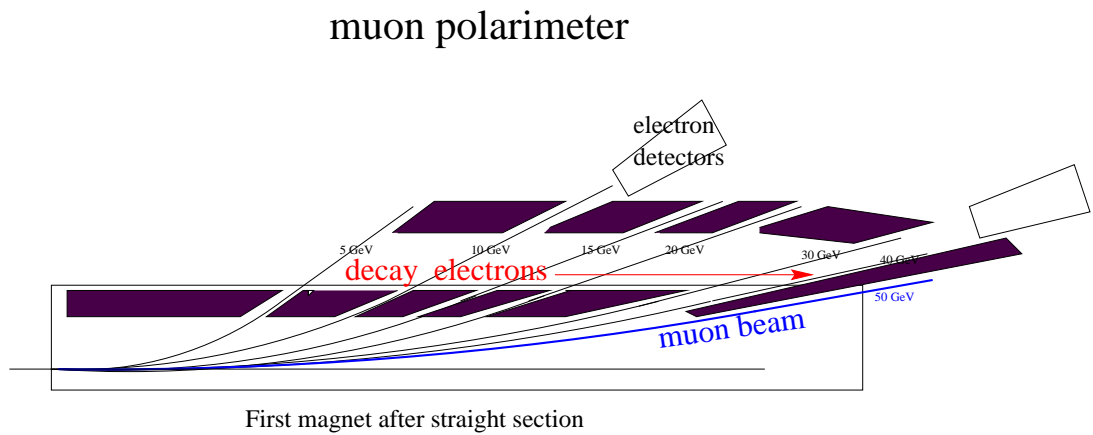}}

\caption{{\sl
A possible muon polarimeter design. The momenta of the decay electrons 
accumulated in a short straight section are analysed in a bending magnet 
in the muon decay ring. Slits in the shielding define the
acceptance of a number of momentum bins.
}}
\label{polarimeter}
\end{figure}

The decay electrons will be used to measure the polarisation of the beam
with a spectrometer as described below, and in Fig.~\ref{polarimeter}. The
same device could in principle be used to monitor the number of electron
decays in an absolute way, especially if one selects the momentum bite
where the electron spectrum is insensitive to the muon polarisation.  
Certainly this will be a useful tool, as a cross-check or for monitoring,
but a very detailed study of the dependence of the acceptance of this
device on the beam parameters must be performed before a conclusion can be
reached.

Knowledge of the flux does not provide knowledge of the cross sections 
folded
with the detector acceptance. This task is traditionally delegated to a
near detector.  The high flux should make things very easy. The event
numbers appearing in Table~\ref{tab:rates} would be obtained with a
detector of 10~kg situated on the neutrino beam axis at a distance of 3~km
from the middle of the decay tunnel. Given the high importance of
precision measurements in the Neutrino Factory, it is likely that a near
detector will be an important tool for beam normalisation. Unlike the
situation with conventional pion decay beams, the near detector will in
fact be able to measure absolute cross sections for a large number of
exclusive and inclusive processes.

It is worthwhile mentioning, finally, the possibility offered by the
measurement of purely leptonic interaction processes, which have been
discussed in~\cite{MMetal}. Of practical interest for normalisation is
the measurement of $\nu_{\mu} + e^- \rightarrow \mu^- \nu_e$, which
appears as a low-angle forward-going muon with no recoil.  Using the
standard electroweak theory, this purely leptonic charged-current process
can be calculated with high precision, and could be measured with a
dedicated detector aimed at measuring also the $\nu_{e} + e^- \rightarrow
e^- \nu_e$ and $\bar{ \nu_{e}} + e^- \rightarrow e^- \bar{\nu_e}$
processes.  The weakness of this method is that it only applies to the
$\mu^-$ decay beam, but it could be seen as an overall absolute
normalisation process for the muon flux.

To conclude, there are many tools to monitor and control the absolute flux
normalisation in a neutrino factory, so that the near detectors should be
able to provide very accurate measurements of inclusive and exclusive
cross sections, within the detector acceptance. A flux normalisation at
the level of a few $10^{-3}$ seems an achievable goal. The relative
normalisation of the $\mu^-$ and $\mu^+$ decay beams should be known with
similar precision.

\subsubsection{Theoretical knowledge of the neutrino fluxes from muon
decay}

The expressions given above for the neutrino flux in muon decay,
(\ref{eq:fluxesfrommuon}), do not include QED radiative corrections, which
have been calculated in~\cite{Broncano-Mena}. The dominant source of
corrections is, as can be expected, related to photon emission from the
decay electron. 
For the electron energy distribution, the corrections 
are of the order of 1\% due to terms proportionanl to 
${\frac{\alpha}{\pi}}\ln {({\frac{m_\mu}{m_e}})}$. It turns out that
the neutrino spectrum is insensitive to the electron mass, i.e., the
integration over the system of electron and photons cancels mass
singularities.  It can be seen that, in the forward direction, an overall
decrease of the neutrino flux of about $4 \times 10^{-3}$ is observed, 
with a
larger decrease near the end point. The global decrease can be understood
by the overall softening and angular widening of the neutrino decay
spectrum due to photon emission.
 
Since the overall size of the corrections is small, one can certainly
trust the calculated spectrum to a precision much better than $10^{-3}$.
 
\begin{figure}[tbhp]

\centerline{\includegraphics[width=0.9\textwidth,height=10cm]{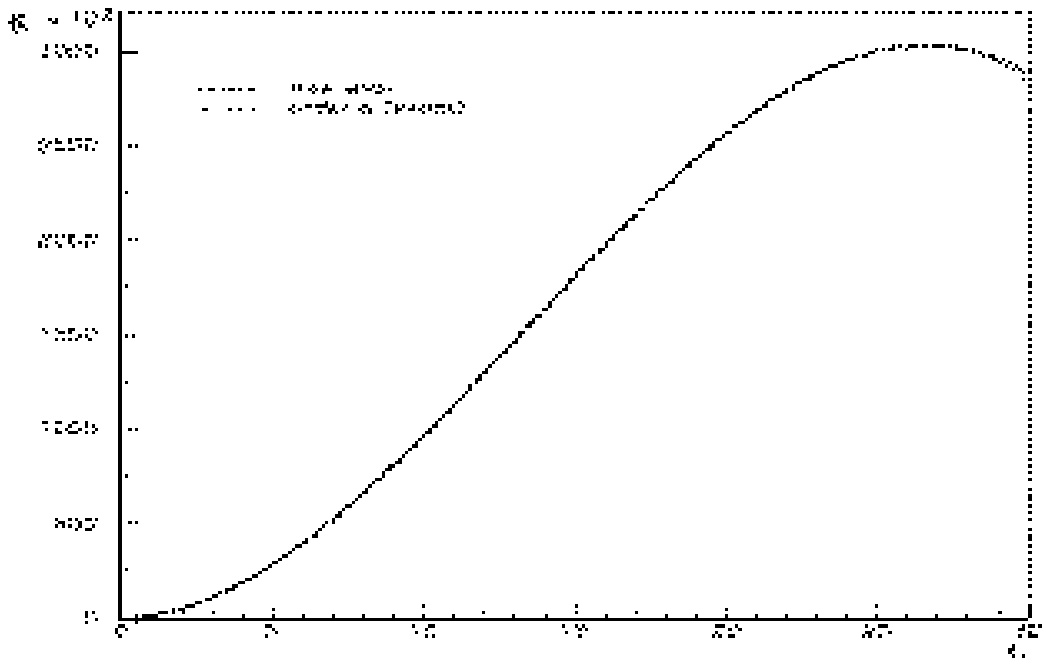}}
\centerline{\includegraphics[width=0.9\textwidth,height=10cm]{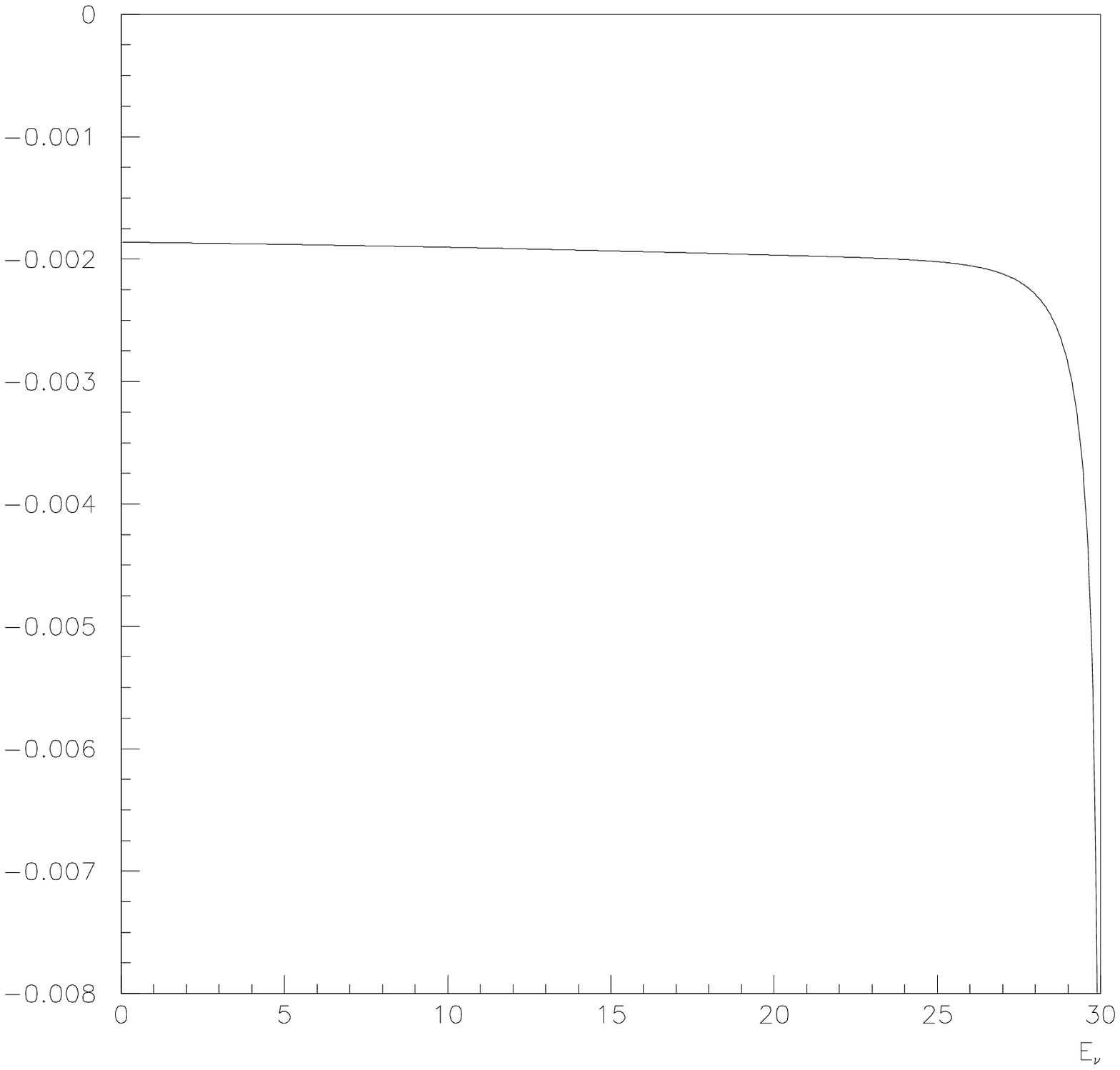}}

\caption{{\sl
Radiative corrections to the muon neutrino flux in $\mu^-$ decay. 
Top panel: the resulting energy distribution at zero angle, bottom 
panel: the
relative change due to the ${\cal O}(\alpha)$ correction. The 
overall reduction of flux is due to the additional energy taken away by 
photons, which slightly widens the angular distribution of the neutrinos. 
In order to avoid infinities at the end point, the quantity plotted is 
$\frac{\Phi({\cal O}(\alpha)) -\Phi_0}{\Phi({\cal O}(\alpha)) +\Phi_0}$. }}
\label{numufluxcorrection}
\end{figure}

\begin{figure}[tbhp]

\centerline{\includegraphics[width=0.9\textwidth,height=10cm]{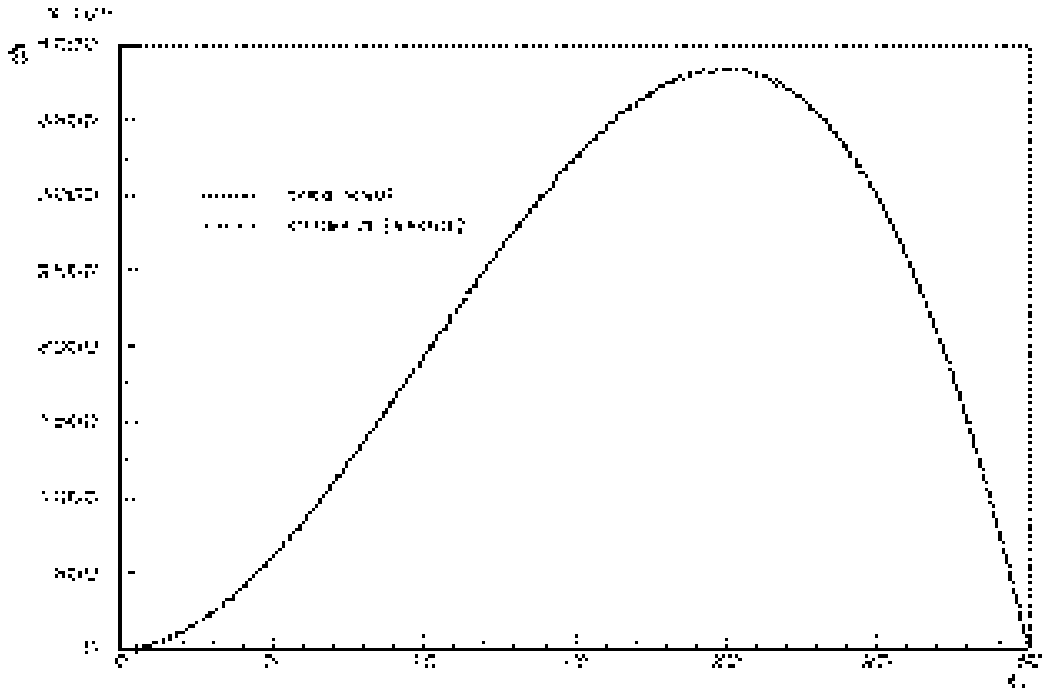}}
\centerline{\includegraphics[width=0.9\textwidth,height=10cm]{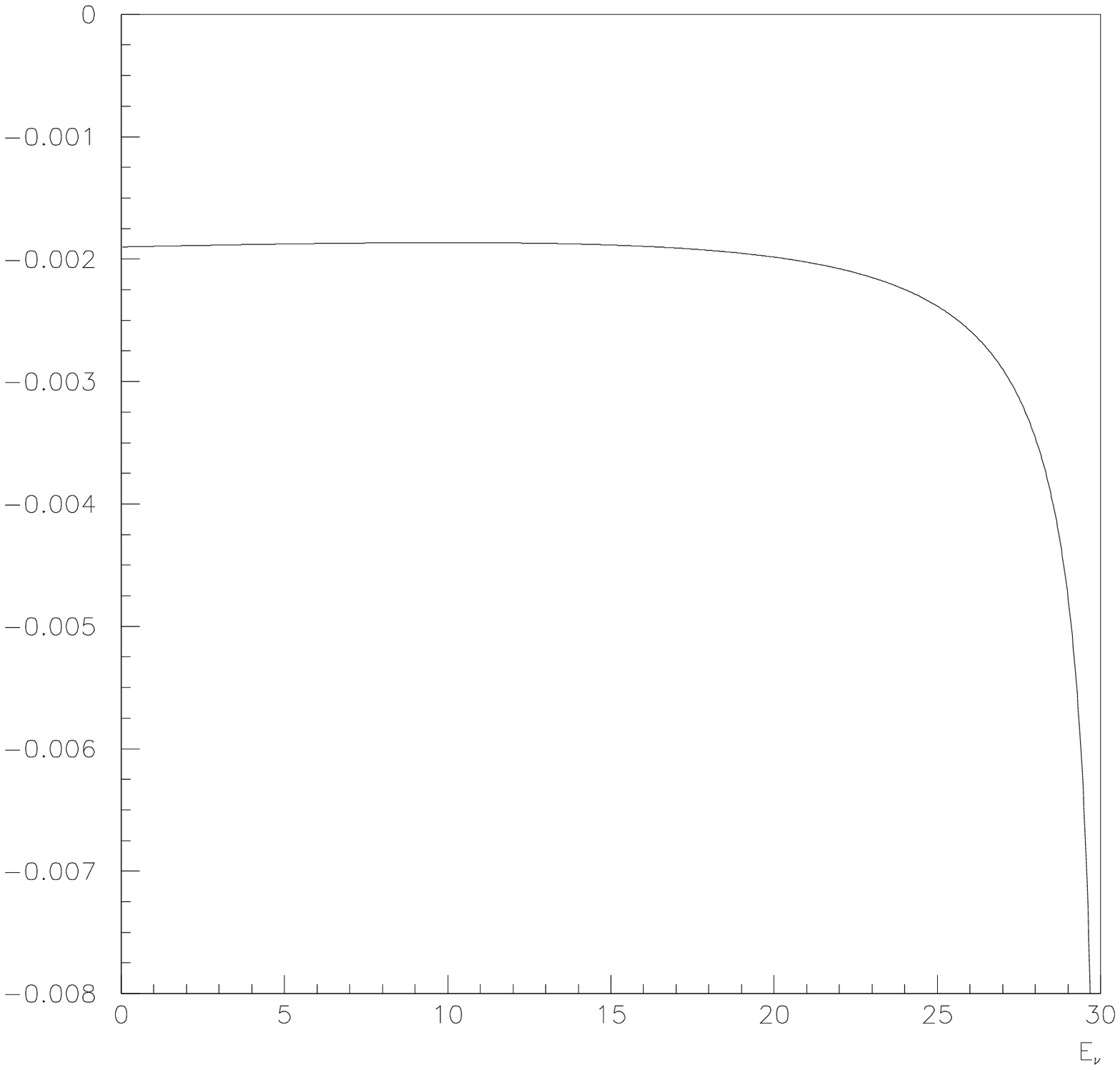}}

\caption
{{\it Radiative corrections to the
electron anti-neutrino flux in $\mu^-$ decay. 
Top panel: the resulting energy distribution at zero angle, bottom 
panel: the
relative change due to the ${\cal O}(\alpha)$ correction. The 
overall reduction of flux is due to the additional energy taken away by 
photons, which slightly widens the angular distribution of the neutrinos.In order to avoid infinities at the end point, the quantity plotted is 
$\frac{\Phi({\cal O}(\alpha)) -\Phi_0}{\Phi({\cal O}(\alpha)) +\Phi_0}$. 
}}
\label{nuefluxcorrection}
\end{figure}

\subsubsection{Muon polarisation}

Muons are naturally polarised in pion decay. In the
$\pi^+\rightarrow\mu^+\nu_{\mu}$ rest frame, both the $\nu_{\mu}$ and
$\mu^+$ have negative helicity. In the laboratory frame, the resulting
average helicity of the muon, or longitudinal polarisation, is reduced
from -100\% for a pion at rest to $<h>= -18\%$ for pions above 200-300
MeV momentum~\cite{FGF}. For a pion of given momentum, muon polarisation
is correlated with muon momentum.  It has been argued in~\cite{Palmer}
that monochromatisation of the pions followed by i)~a drift space to 
separate
muons of different momenta, and ii)~collection in successive RF buckets,
should allow separation in different bunches of muons of different
polarisations. This does not change the {\sl average} polarisation, but
creates bunches of different polarisation (up to 50\%), that can be of use
for physics, as long as the times of neutrino interactions are recorded
with a precision of a few nanoseconds.

The muon spin precesses in electric and magnetic fields that
are present during cooling and acceleration, but the muon
spin tune $\nu$ 
-- the number of additional spin precessions happening when the muon
makes a complete turn -- is very low:
$$
\nu = a_\mu \gamma = \frac{g_\mu-2}{2} \frac{\rm E_{ beam}}{m_\mu}
= \frac{\rm E_{beam} (GeV)}{90.6223(6)}\,\,.
$$
It has been evaluated~\cite{FGF} that 80 to 90\% of the original 
polarisation 
will survive all muon handling up to the injection into the storage ring. 
Its orientation will depend on the number of turns that the muons encounter
along the accelerator chain, and can be arranged to be longitudinal 
by an appropriate choice of geometry and of 
the energies in the recirculating linacs~\cite{lion}. 
As we will see, this is not necessarily important. 

What will happen to the muon polarisation in the decay ring depends in the
first instance on whether its geometry is a ring (race track or triangle)
in which the muons undergo one rotation per turn, or a bow-tie, in which
the muon undergoes zero net rotation at each turn.

{\it In the case of a ring}, the polarisation will precess. The
orientation of the polarisation vector will be rotated with respect to the
muon direction by an angle which increases each turn by $2 \pi \nu$.
Unless the energy is chosen very carefully, it will not be aligned, and
reduced on average by a factor 2. At a muon energy of precisely $E =
45.311$~GeV, the spin tune is 0.5 and the polarisation flips during 
each turn.
This would allow the most powerful use of the polarisation for physics
purposes, but absolutely requires that the orientation is correctly chosen
at injection, a condition which is otherwise unnecessary in a ring
geometry.  If no special measure is taken, however, depolarisation will
occur, since particles of different energies will have their spins precess
with different spin-tunes.

The muon polarisation can be monitored by momentum analysis of the decay
electrons, as discussed in~\cite{CERN9902:bdl}, in a polarimeter that
could look like that sketched in Fig.~~\ref{polarimeter}. One can expect
that this measurement will be difficult: the relative normalisation of
electron rates in the different energy bins will depend on various muon
beam parameters such as its exact angle and divergence, and on a precise
modelling of the beam-line geometry. In a ring geometry, the device will
be exposed to a succession of negative and positive helicity muon bunches,
so it will have to perform relative measurements.  These should be
sensitive to small effects, with a {\em relative} precision of a few
percent. In a bow-tie geometry, however, there will be no spin precession,
and one will be left to measure the polarisation based on the measured
electron spectrum.  A few \% absolute accuracy seems in this case already
very challenging.

The spin precession in a storage ring provides a means of high precision
($10^{-6}$ or better) for energy calibration~\cite{Raja}. As shown
in~\cite{CERN9902:bdl}, the measurement of the depolarisation can be used
to measure the energy spread with high precision. In this case, the 
combined effect of precession and depolarisation ensure
that the muon polarisation integrated over a fill averages out to 0 with
an excellent precision: simulations show that any residual polarisation is
less than $4 \times 10^{-4}$.

\begin{figure}[h]

\includegraphics[width=0.45\textwidth]
{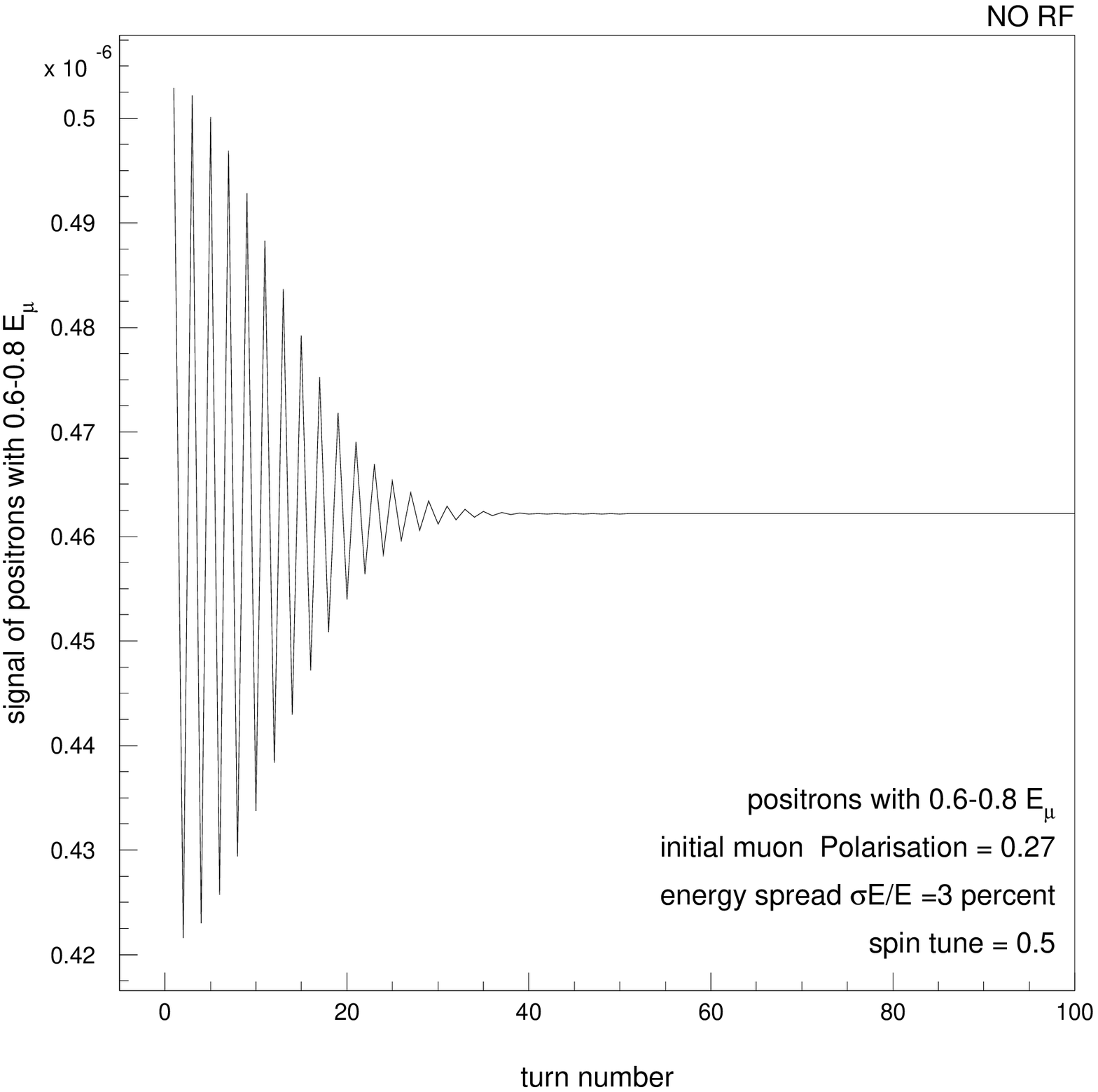}
\hfill
\includegraphics[width=0.45\textwidth]
{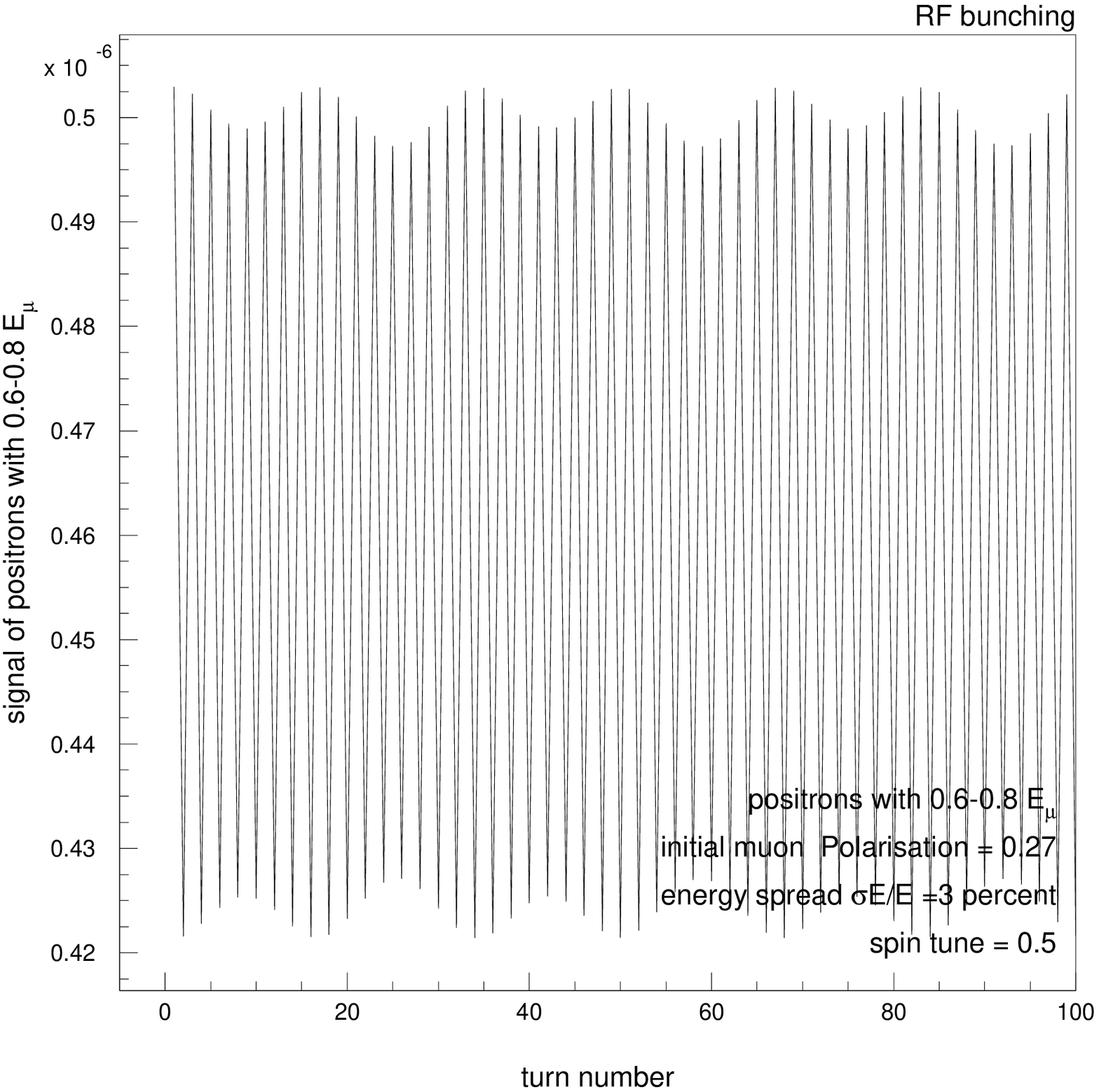}
\caption
{\sl Oscillation with turn number in a fill of the number of electrons
in the energy range 0.6-0.8 $E_{\mu}$, normalised to the 
total number of muon decays during the given turn.
The oscillation amplitude is a measure of the beam polarisation, 
its frequency a measure of the beam energy, 
and, if there is no RF bunching,  
its decrease with time is a measure of energy spread.
The muon lifetime corresponds here to 300 turns.
The beam energy is $E_{\mu}=45.311$~GeV and the energy spread is 
$3\times 
10^{-2}$.
On the left, there is no bunching  RF in the muon storage ring, 
on the right there is RF bunching with $Q_s=0.03$.}
\label{precession} 
\end{figure}

Depolarisation can be avoided, if the storage ring is equipped with an RF
system that ensures that the muons undergo synchrotron
oscillations~\cite{lion}. By doing this, one loses the possibility to
measure the energy spread from the depolarisation, but one can maintain
the muon polarisation.  The average is still essentially zero, but by
recording the exact time of neutrino events, one can infer their bunch
number and turn number, and deduce the polarisation of the decay muons. In
a ring geometry either mode of operation is left open, if one can run with
the required RF system on or off.

{\it In a case of a bow-tie}, the muons will not depolarise: 
spin precession is zero no matter what the muon energy is. This configuration
is not as convenient as the ring for several reasons.
\begin{itemize}
\item{It will be impossible to measure the spin precession, so that
 the energy and energy spread of the muon beam will not be calibrated.} 
\item{The polarisation will not average to zero. This, 
combined with the fact that 
the polarisation measurement will be more
challenging, means that the flux determination will be affected  by a sizeable 
uncertainty, due to the beam polarisation error.}
\item{It will be difficult to change the sign of the muon beam 
polarisation.}
\item{Unless the geometry is very carefully chosen, the beam 
polarisation will be different for the two long straight sections.}
\end{itemize}
For these reasons, and despite the fact that in principle
the useful beam polarisation is higher in the bow-tie geometry, 
{\it the ring geometry is preferred} from the point of view of beam 
control.  

\subsubsection{Neutrino fluxes and muon polarisation}

Neutrino spectra with different beam polarisations are given by the 
following
equations valid for $\mu^+$ decays in the 
muon centre-of-mass (reverse polarisations for $\mu^-$):
\\-- for the $\nu_{\mu}$:
$$\frac{d^2 N}{dx d\cos\theta}=Nx^2[(3-2x)-{{\mathcal P}}(1-2x)\cos\theta]$$ 
\\-- for the $\nu_{e}$:
$$\frac{d^2 N}{dx d\cos\theta}=6Nx^2[(1-x)-{{\mathcal P}}(1-x)\cos\theta]$$
\\where $\theta$ is the decay angle in the muon centre-of-mass frame, 
$\mathcal P$ is the muon longitudinal polarisation, and $x=2 E_\nu/m_{\mu}$.

In a long-baseline experiment, one is at extremely small angles, so that
$\cos\theta=1$. In this case, the $\nu_{e}$ component of the beam is
completely extinct for $ {\mathcal P}=+1$. This is due to spin
conservation in the decay: a right-handed muon cannot decay at zero angle
into a left-handed $\nu_{e}$.

\begin{figure}[tbhp]

\includegraphics[width=0.45\textwidth]
{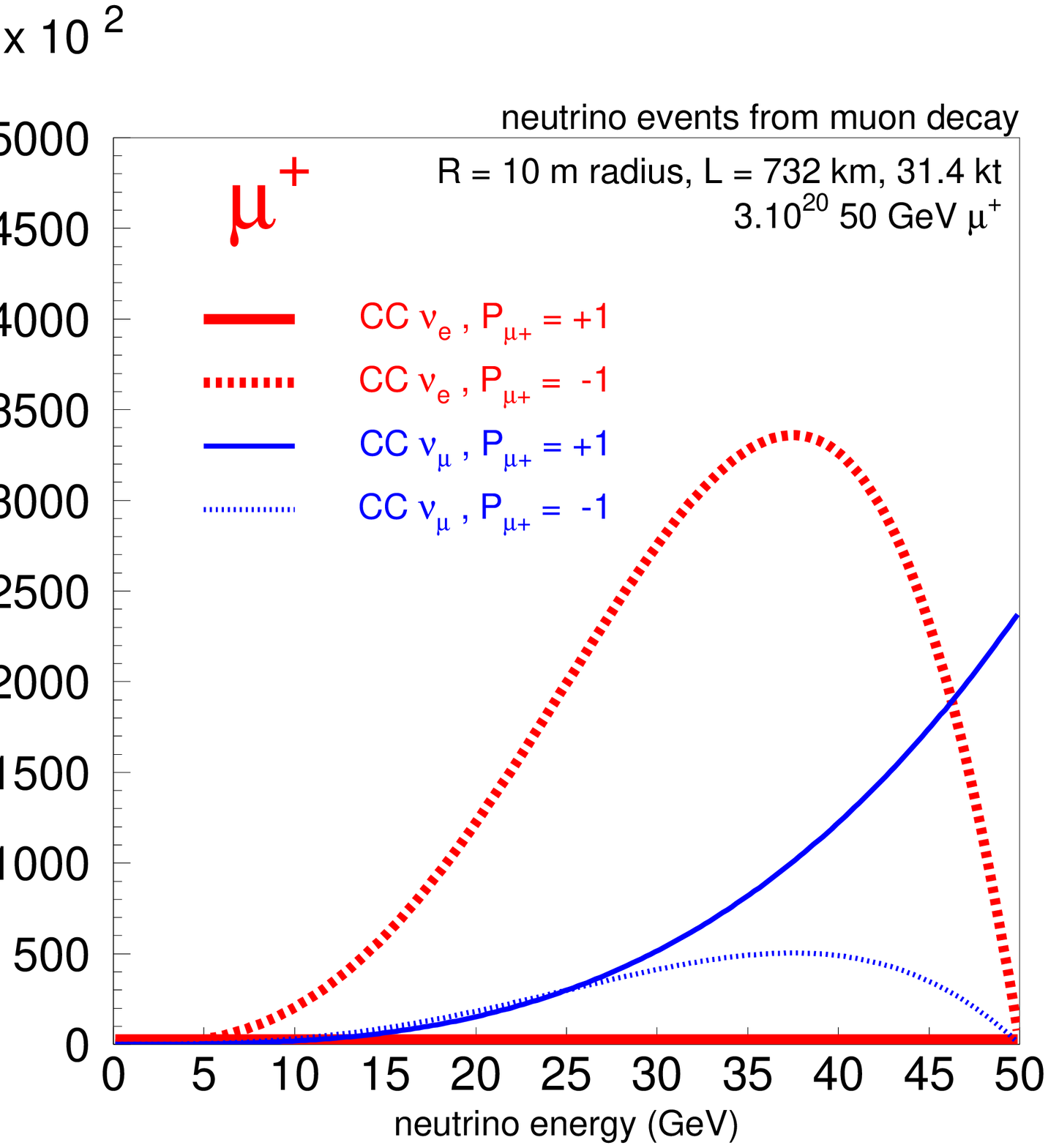}
\hfill
\includegraphics[width=0.45\textwidth]
{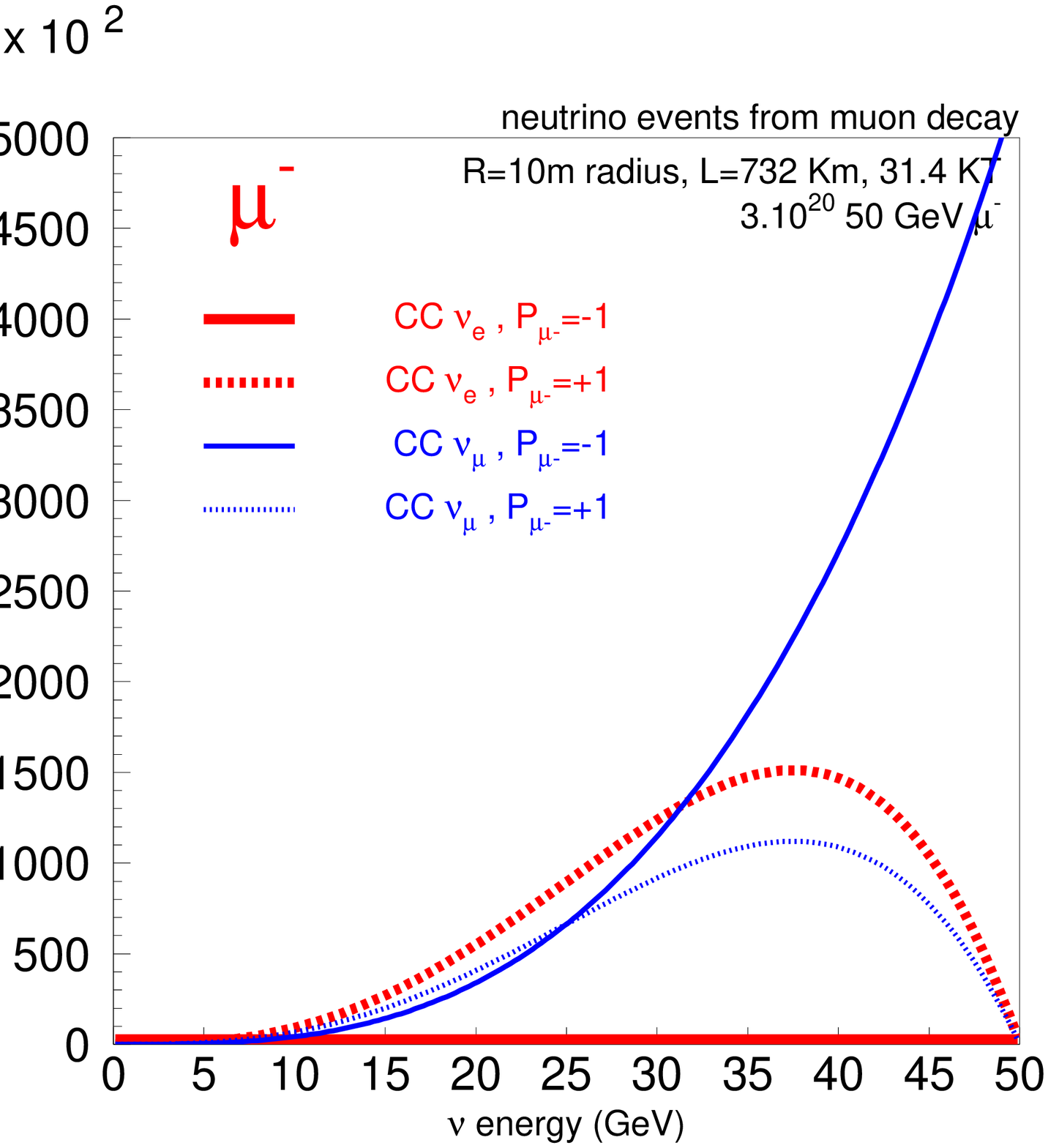}
\\

\caption
{\sl Event numbers for a detector of density 5 with 10~m radius that is
20~m long, situated 732~km away from the muon storage ring,
for $\mu^+  \rightarrow e^+ \nu_e \bar{\nu_{\mu}}$ (left) and  
$\mu^- \rightarrow e^- \bar{\nu_e} \nu_{\mu}$ (right) beams of 50 GeV.
Full lines show the spectra for the `natural' helicity
$\mathcal P =+1$ for $\mu^+$, and dashed ones for the opposite case. 
The CC $\nu_e$ for $\mu^+$ with  $\mathcal P =+1$ 
and  CC $\nu_e$ for $\mu^-$ with  $\mathcal P =-1$ are not visible, 
because the fluxes are almost exactly zero. The
vertical axis gives event numbers per bin of 250~MeV. 
This plot assumes no muon beam 
angular divergence and no beam energy spread. 
}
\label{events} 
\end{figure}

Event numbers can readily be obtained by multiplying by the cross section.
They are shown in Fig.~\ref{events} for a 10~m radius detector 20~m long
situated 730~km away.  Since the neutrino and anti-neutrino cross sections
are in the ratio 1/0.45, negative muons provide enrichment in $\nu_{\mu}$
and positive ones in $\nu_{e}$.

It is clear from Fig.~\ref{events} that the combination of muon sign and
polarisation allows large variations in the composition of the beam, in a
controlled way. Since detector studies show that the muon sign can easily
be determined in a charged-current (CC) (anti)neutrino event, but that the 
electron sign is
much more difficult, we have tried to use the variation of electron
neutrino flux with muon polarisation to infer a signal of $\nu_{\mu}
\rightarrow \nu_e$ oscillations to be compared (for a T-violation test)  
with the $\nu_{e} \rightarrow \nu_{\mu}$ oscillation measured with the
wrong-sign muons. Unfortunately, even for 40\% beam polarisation, the
improvement in the sensitivity to CP/T violation is no more than the
equivalent of a factor of 1.5 to 2 in statistics.  Certainly, it appears
that polarisation is more useful as a tool to measure the beam properties
than as a physics tool. Nevertheless, these statements might be 
parameter-dependent, and should be revisited once the oscillation 
parameters are
better known.

\subsubsection{Effect of beam divergence}

\begin{figure}[h]
\includegraphics[width=0.4\textwidth]
{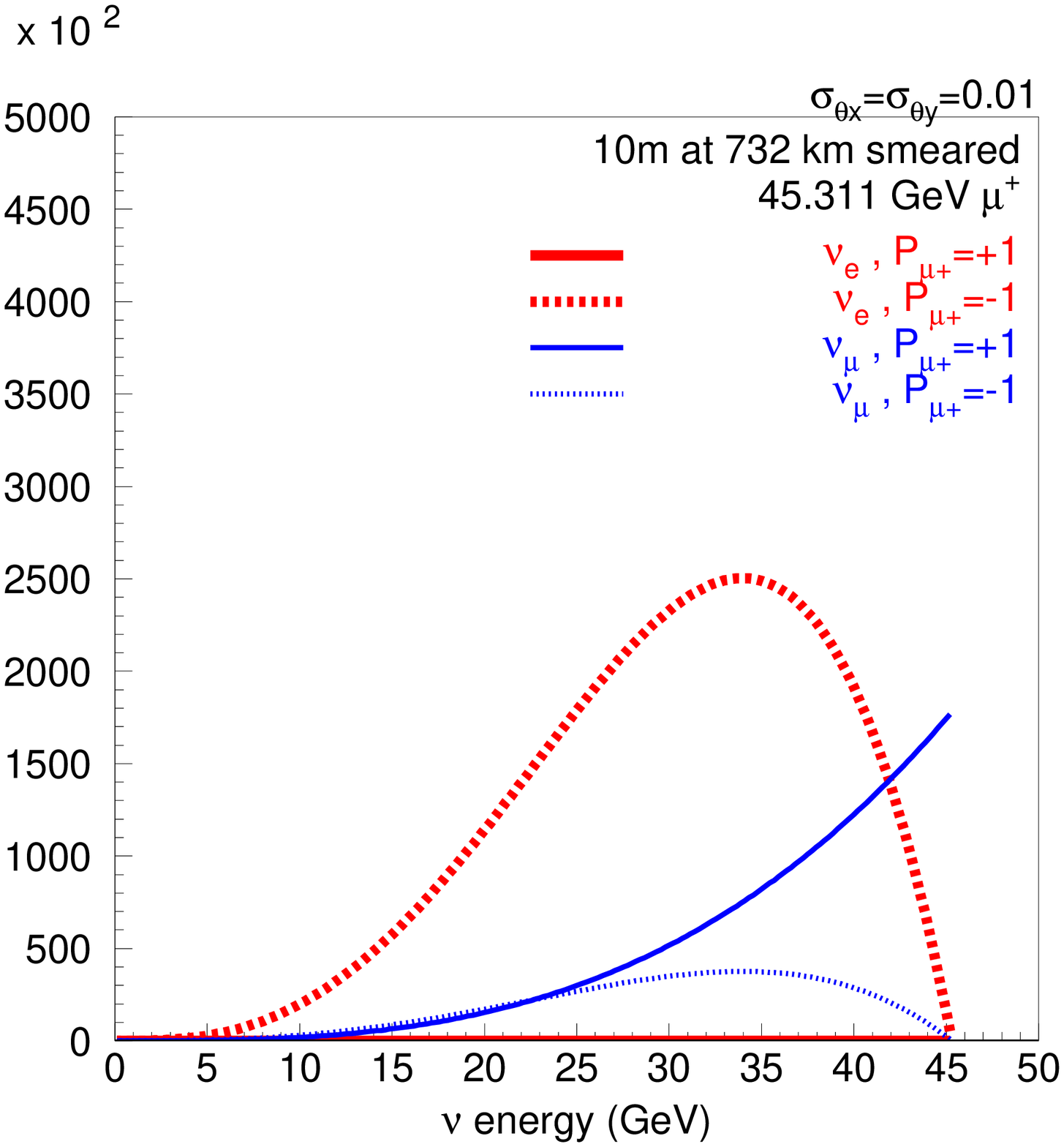}
\hfill
\includegraphics[width=0.4\textwidth]
{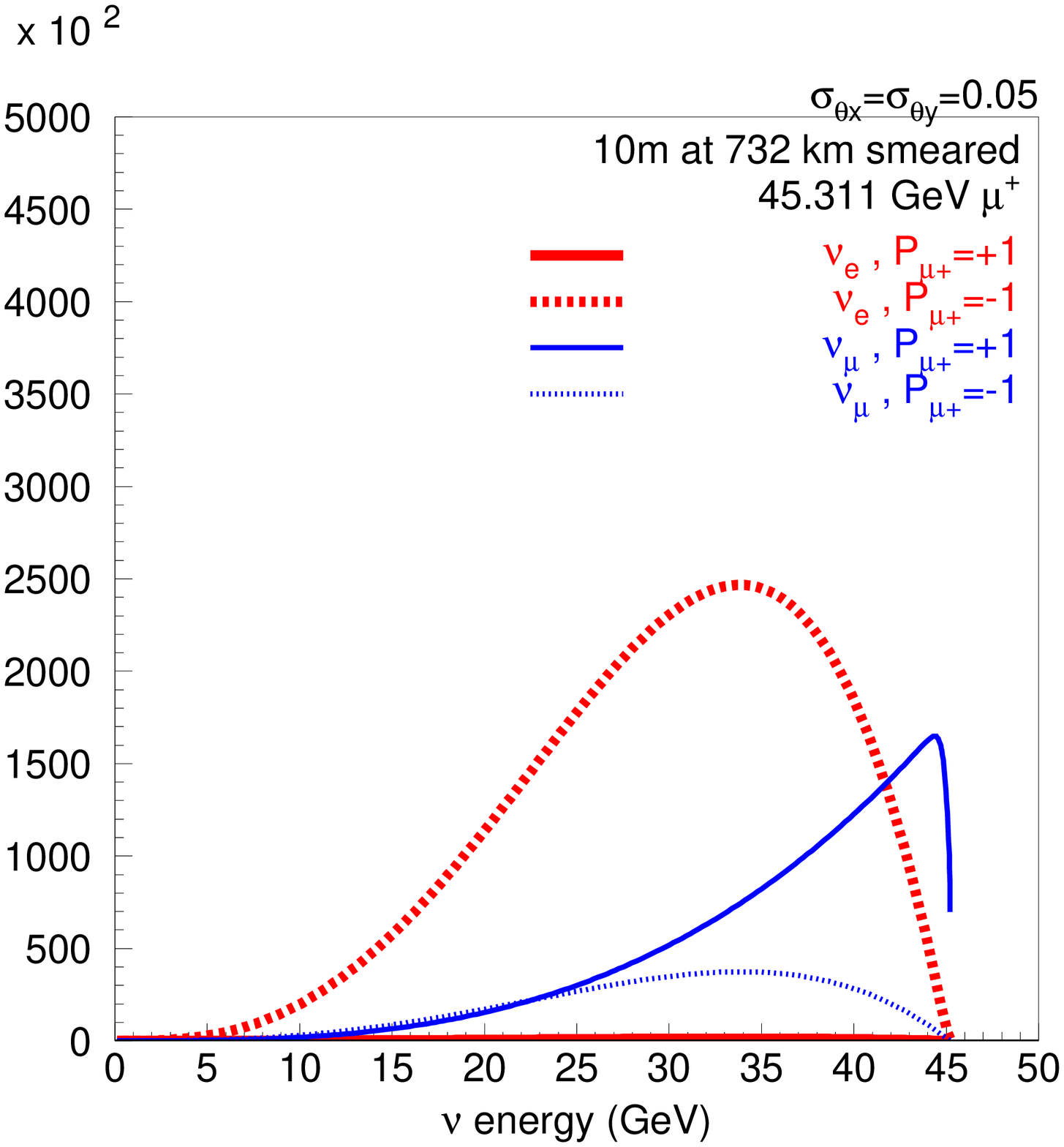}
\\
\includegraphics[width=0.4\textwidth]
{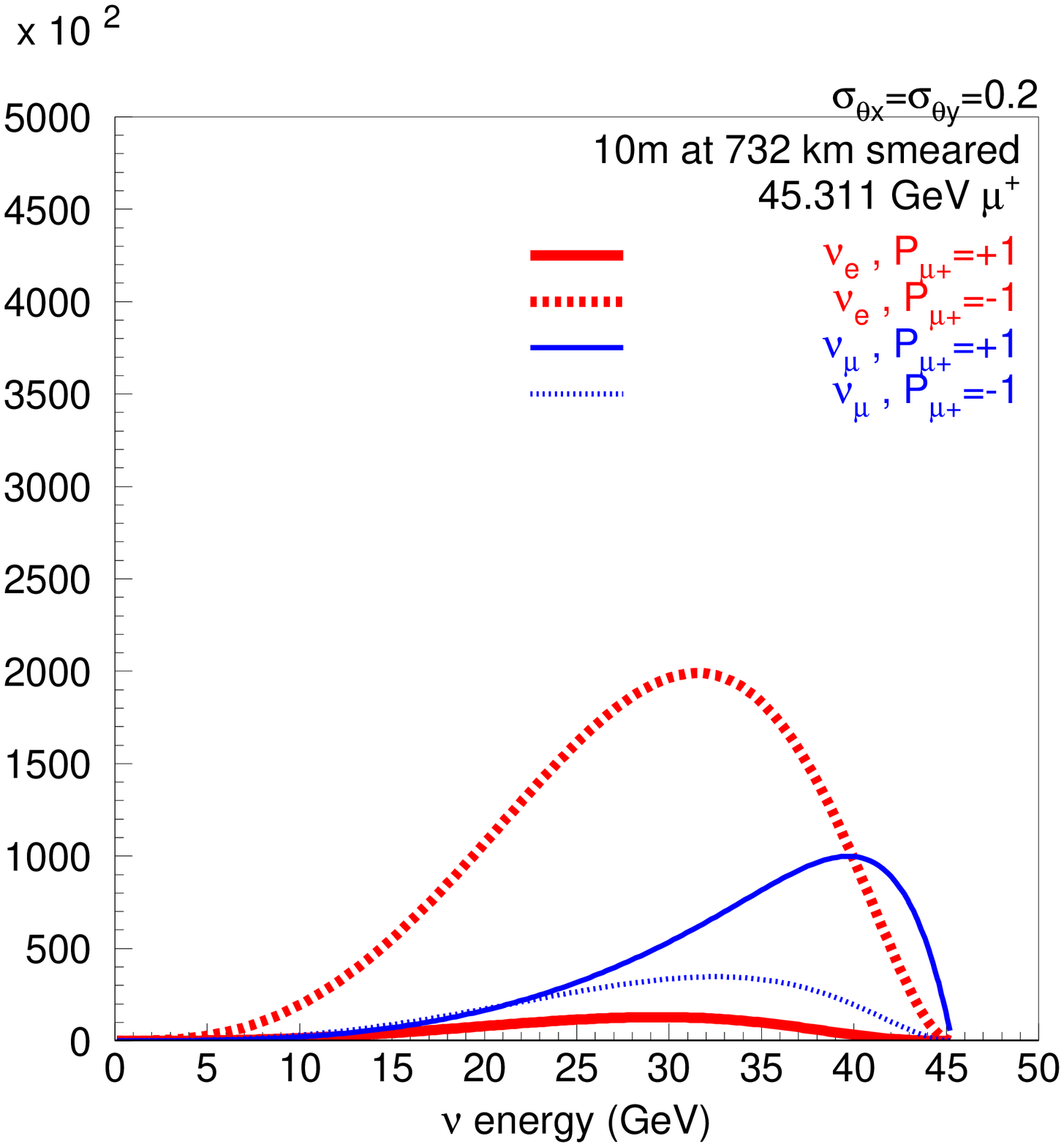}
\hfill
\includegraphics[width=0.4\textwidth]
{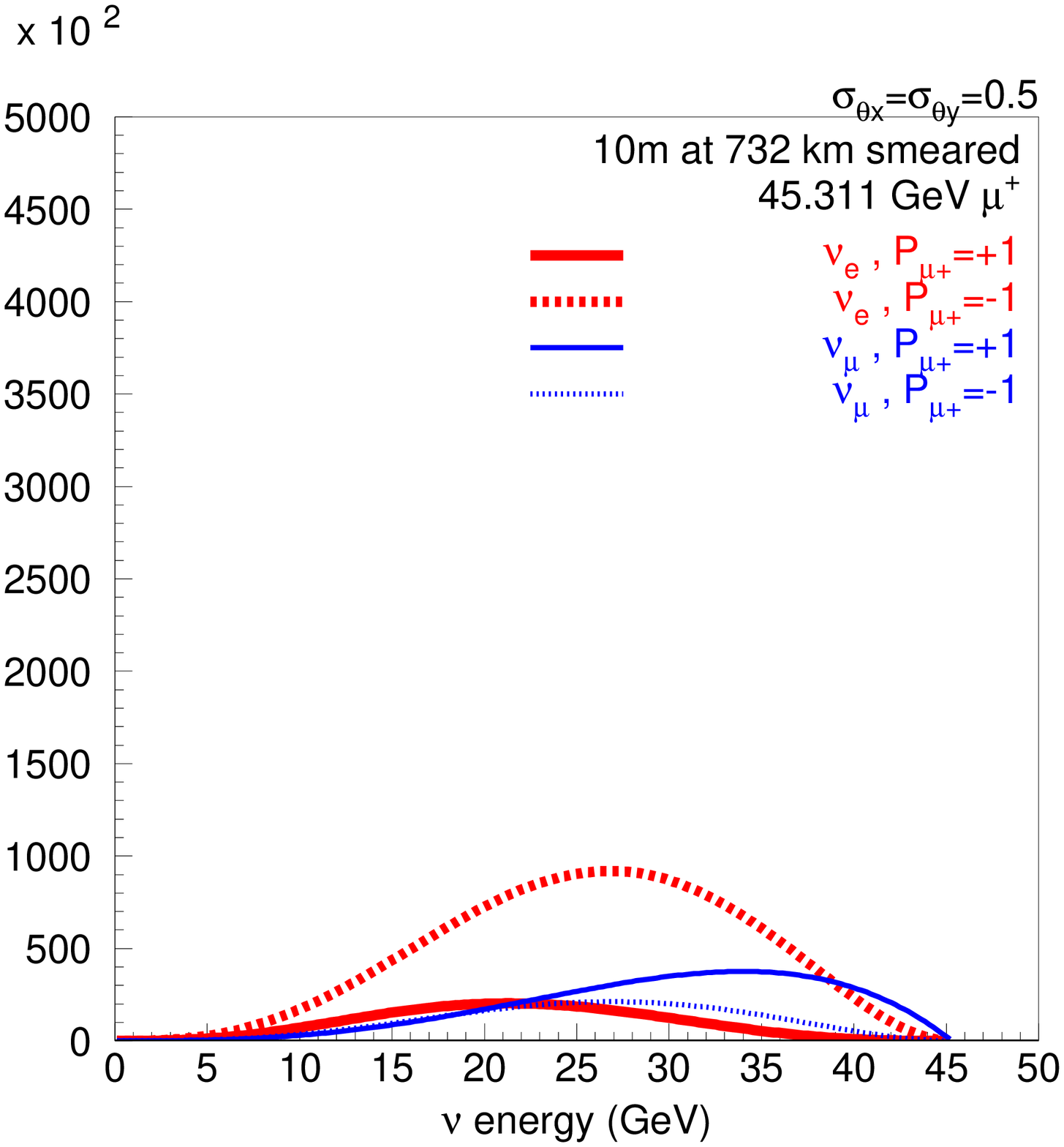}
\\
\caption
{\sl Neutrino event spectra for different beam divergences;
Upper left:  
$\sigma \theta_x = \sigma \theta_y = 0.01~{m_\mu}/{E_\mu}$;
upper right: 
$\sigma \theta_x = \sigma \theta_y = 0.05~{m_\mu}/{E_\mu}$;
lower left:  
$\sigma \theta_x = \sigma \theta_y = 0.2~{m_\mu}/{E_\mu}$ ;
lower right:  
$\sigma \theta_x = \sigma \theta_y = 0.5~{m_\mu}/{E_\mu}$.
It is clear that beam divergence results in a loss of events, and 
in a sizeable distortion of the spectra and of their 
muon polarisation dependence.}
\label{fig:smearedevents} 
\end{figure}

The opening angle of the neutrino beam is typically $1/\gamma$, where
$\gamma = E_{\mu}/m_{\mu}$. As soon as the beam divergence is comparable
with this natural opening angle, a large fraction of the flux will be
lost. This is shown for 45.311~GeV muons in Fig.~\ref{fig:smearedevents}.  It
is clear that beam divergence results in a loss of events, and in a
sizeable distortion of the spectra and of their muon polarisation
dependence. A beam divergence not larger than $\sigma \theta_x= \sigma
\theta_y=0.2{m_\mu}/{E_\mu}$ seems to be desirable, if one want to avoid a
large sensitivity of physics results upon the experimental determination
of the muon beam parameters.

\begin{figure}[ht]
\begin{center}
\includegraphics[width=17cm]
{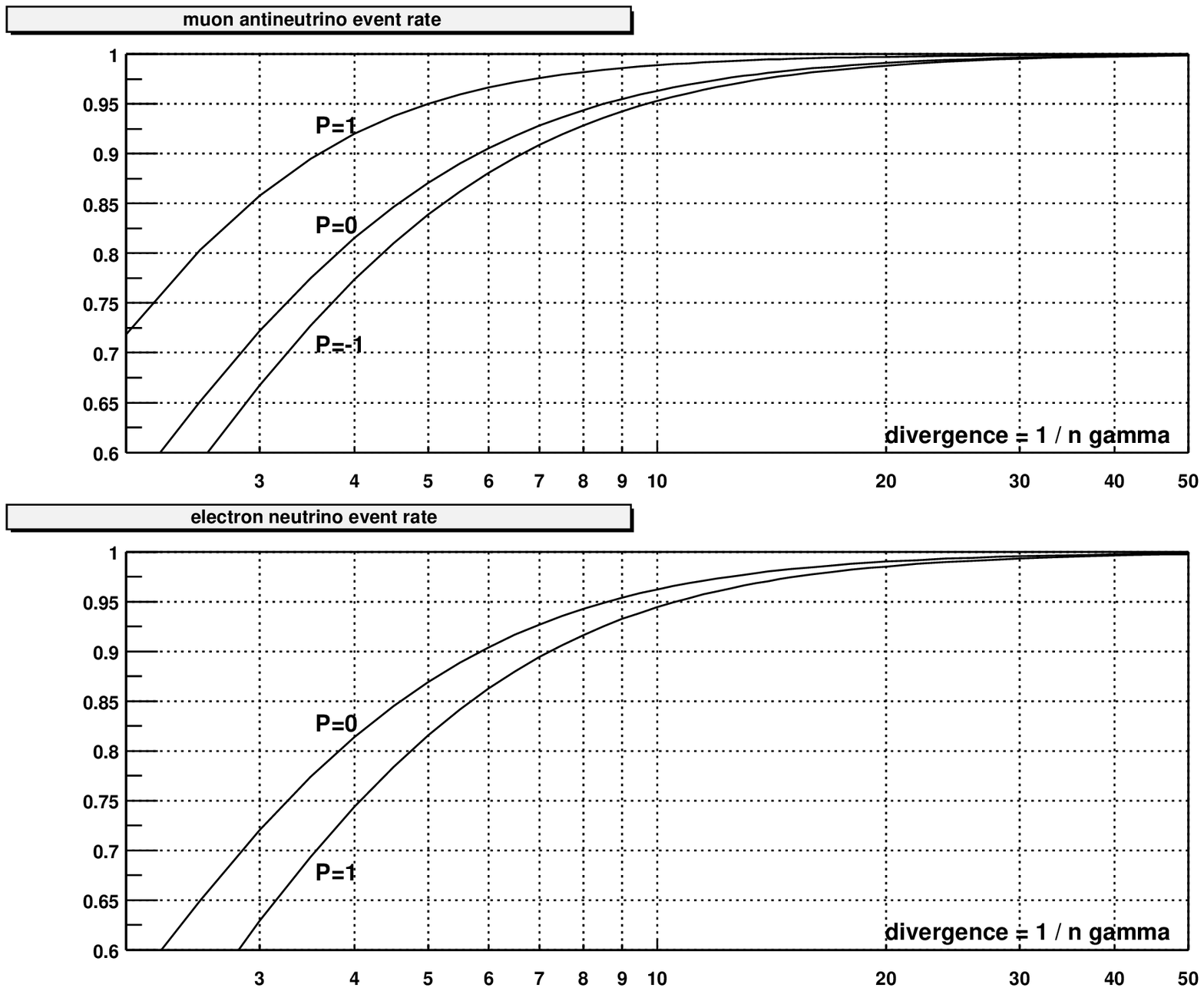}
\caption{ \sl The relative event rates for muon 
anti-neutrinos (top) and electron
neutrinos (bottom), for various polarisation values as a function of the 
beam divergence, parametrised as $1/n \gamma$.}
\label{fig:div}
\end{center}
\end{figure}

This effect has been studied more precisely in~\cite{Ioannis},
where event numbers are calculated for various polarisations and 
divergences. The impact of the muon beam divergence on the neutrino event rate 
can be seen in Fig.~\ref{fig:div}.
The first conclusion one can draw from these plots is that, for a given
number of muons, the highest flux is obtained for small muon beam
divergence. In order to keep the event rate loss due to the muon beam
divergence below 5\%, the divergence should be close to 0.1 /
$\gamma_{\mu}$.

\begin{figure}[ht]
\begin{center}
\includegraphics[width=17cm]
{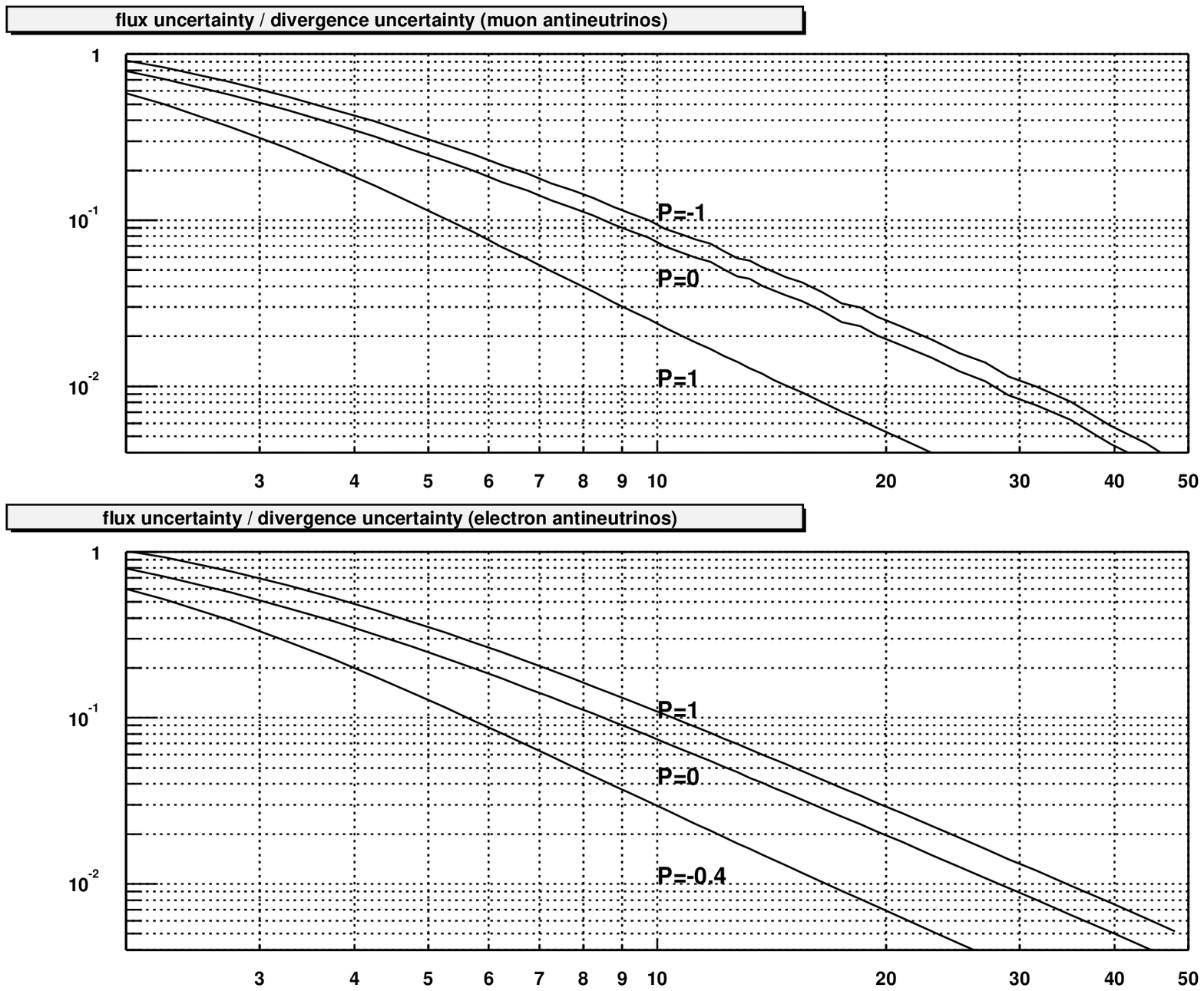}
\caption{\it The ratio of the uncertainty in the event rate over the 
uncertainty in the muon beam divergence as a function of the beam 
divergence, parametrised as $1/n \gamma$. The top
(bottom) plot corresponds to muon anti-neutrinos (electron neutrinos).}
\label{fig:diverror}
\end{center}
\end{figure}

From the curves in Fig.~\ref{fig:div}, one can determine the relative
error of the predicted event rate, given the uncertainty in the knowledge
of the beam divergence itself. For example, if the beam divergence is 0.1
/ $\gamma$ and is known with a relative precision of 10\%, the
$\overline{\nu}_{\mu}$ and $\nu_{e}$ event rates can both be predicted
with an accuracy of about 0.75\%.  For a divergence of 0.2~/~$\gamma$, the
uncertainty on the flux would be 2.5 \%. As we will see, however, the
knowledge of the beam divergence is unlikely to be a constant relative
fraction.

\begin{figure}[ht]
\begin{center}
\includegraphics[width=17cm]
{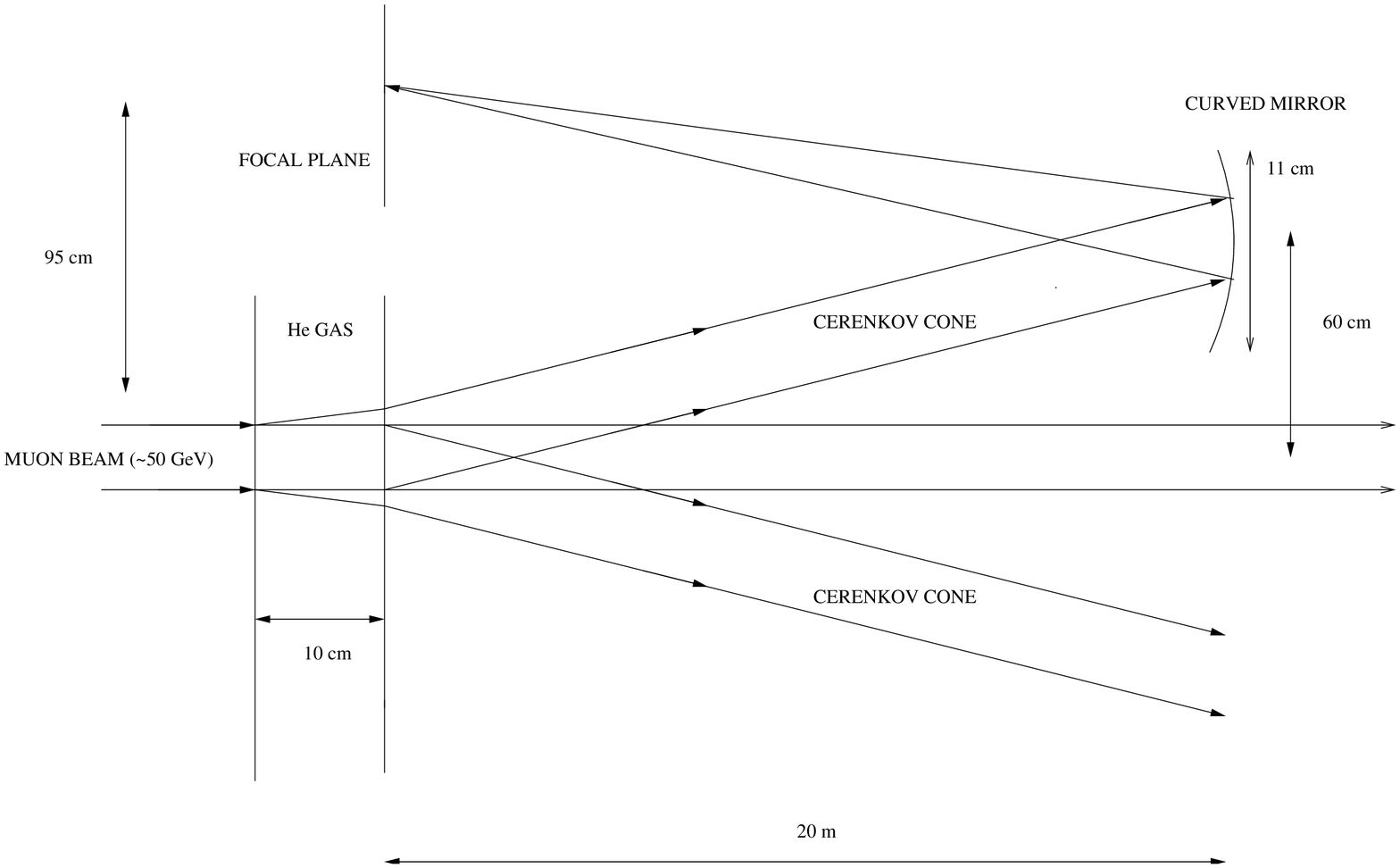}
\caption{\sl Schematic of a muon beam divergence measurement device. 
A low-pressure He gas volume is contained by windows 
(one of which must be transparent) within a straight section 
of the  the muon decay ring. The \v{C}erenkov light is collected by a 
parallel 
to point optics in the direction of interest, so as to provide an image of the
angular distribution of particles in the focal plane. }
\label{fig:beamcerenkov}
\end{center}
\end{figure}
 
One can turn the argument around, and request that the beam divergence be
0.1 / $\gamma$ and known to a relative precision of 1.5\%, so that the
corresponding uncertainty on flux is only $10^{-3}$.  It is clear that in
this case the muon beam divergence will need to be measured.  For a beam
of 50 GeV, the beam divergence is 200 micro-radians and the requirement is
that it should be known to 3 micro-radians.

As a measurement device, one could imagine a gas \v{C}erenkov detector
focusing the \v{C}erenkov radiation in such a way as to make an image of 
the
muon beam direction, as sketched in Fig.~\ref{fig:beamcerenkov}. This has
been studied in~\cite{Piteira}, with the conclusion that for 200
micro-radians divergence, a precision of a few \% can be achieved.  
The additional multiple scattering introduced by the device leads to a
growth of emittance during the muon fill, by a few tens of micro-radians,
which is small and will be measured.  Since the resolution is dominated by
optical imperfections, diffraction effects and heating effects in the gas
of the \v{C}erenkov detector, they act as an additional experimental
smearing $\sigma_{\mathrm{exp}}$ added in quadrature to the true beam
divergence $\sigma_{\mathrm{beam}}$.  In the scheme of
Fig.~\ref{fig:beamcerenkov}, the largest effect is optical diffraction,
which amounts to 30 micro-radians. It is easy to show that the correction
for experimental resolution is

\begin{equation}
\frac{\Delta \sigma_{\mathrm{beam}}}{\sigma_{\mathrm{beam}}}
=
\frac{\Delta \sigma_{\mathrm{exp}}}{\sigma_{\mathrm{exp}}}
\left( \frac{\sigma_{\mathrm{exp}}}{\sigma_{\mathrm{beam}}} \right)^2.
\label{differror}
\end{equation}
This makes the beam divergence progressively harder to measure as it
becomes smaller. Assuming that the experimental error is 30~micro-radians
and is known with a precision of 30\% of its value, the above gives a flux
uncertainty of $5 \times 10^{-4}$, more or less independent of the beam
divergence in the range of 0.05 to 0.2.

In conclusion, the requirement that the beam divergence be no greater than
$0.1/\gamma$ ensures that the corrections and uncertainties to the
neutrino fluxes remain small (below 1\%), even if one should rely on the
accelerator properties themselves.  In order to achieve a higher precision
a direct measurement of the beam divergence will be necessary -- and is
probably feasible. If relaxing this condition would allow a larger muon
flux, a divergence measurement device becomes mandatory, and would ensure
that the uncertainty on the neutrino flux remains well below $10^{-3}$.

\subsubsection{Summary of uncertainties in the neutrino flux}

A first look has been given to the sources of systematic uncertainties in
the neutrino fluxes and their possible cures.

\begin{itemize}
\item 
The monitoring of the total number of muons circulating 
in the ring can be inferred from a Beam Current Transformer 
with a precision of the order of $10^{-3}$ or better. 
The decay electrons vanish quickly and are not a problem. 
\item
The theoretical knowledge of the neutrino fluxes from muon decay
is not an issue. Radiative effects have been calculated: they
amount to around $4 \times 10^{-3}$, with an error that is much 
smaller.
\item
The muon beam polarisation determines the flux directly, 
both in shape and magnitude. Its seems delicate to determine its
value  with a precision much better than a few \%. In a ring geometry, 
however, polarisation precesses and averages out with high precision 
(a few $\times 10^{-4}$). This is  a strong argument in favour of a ring 
geometry against a bow-tie geometry. 
\item 
The event rate varies like the muon beam energy to the third power, but 
the muon
beam energy
can be inferred very precisely from the muon spin precession. A 
polarimeter idea has been outlined, and the measurement should cause
no difficulty. 
Beam polarisation can be preserved if an RF system is installed in the
decay ring. 
The energy spread can be derived from the depolarisation pattern, 
in special runs
with no RF if necessary. 
\item 
The muon beam angle and angular divergence have an important effect on the 
neutrino flux. For a given number of muons, the smaller the beam divergence, 
the higher the flux. 
Thus the beam divergence in the straight section of the muon decay ring 
should be made as small as possible, but should not constitute a limit 
on the number of stored muons. 
\item 
Measurement devices for the beam divergence
will be necessary, but they can probably be designed and built to ensure 
a flux uncertainty below $10^{-3}$.
 \end{itemize}

In addition, the near detector stations should allow measurements of 
cross sections with high precision. The inverse muon decay reaction 
$\nu_{\mu} + e^- \rightarrow \mu^- \nu_e$ offers the possibility of an 
absolute normalisation of the flux. 

We conclude that, provided the necessary instrumentation is foreseen, 
the Neutrino Factory flux should  be known with a precision 
of the order of $10^{-3}$.
 
\subsection{\bf Detector issues}
%
\newcommand{\bc}{\begin{center}}
\newcommand{\ec}{\end{center}}
\newcommand{\bi}{\begin{itemize}}
\newcommand{\ei}{\end{itemize}}
\newcommand{\hs}{\mbox{\hspace{10mm}}}
\newcommand{\lra}{\Leftrightarrow}
\def \ut{\underline}
\def \ot{\overline}
\def \n{$\nu_\mu$}
\def \l{\large}
\def \m{${\mbox}$}
\def \sth{{\mbox{sin}}^2 2\theta}
\def \ne{\nu_e}
\def \t{\tau}
\def \nm{\nu_\mu}
\def \nt{\nu_\tau}
\def \nb{\overline{\nu}}
\def \sn{\sigma^{\nu N}}
\def \sa{\sigma^{\nb N}}
\def \sr{\frac{\sa}{\sn}}
\def \mpt{mP_{T}}
\def \la{\Lambda_{\mbox{QCD}}}
\newcommand{\Uud}{\mbox{$U_{ud}$}}
\newcommand{\dbar}{\mbox{$\overline{d}$}}
\newcommand{\nbar}{\mbox{$\overline{n}$}}
\newcommand{\el}{\mbox{$e^{-}$}}
\newcommand{\tm}{\mbox{$\tau^{-}$}}
\newcommand{\numb}{\mbox{$\overline{\nu_{\mu}}$}}
\newcommand{\nueb}{\mbox{$\overline{\nu_{e}}$}}
\newcommand{\ds}{\mbox{$D_{s}^{\pm}$}}
\newcommand{\Pt}{\mbox{$P_T$}}
\newcommand{\Ptm}{\mbox{$P_T^m$}}
\newcommand{\dc}{\mbox{$^{\circ}$C}}
\newcommand{\dg}{\mbox{$^{\circ}$}}
\newcommand{\Wpm}{\mbox{$W^{\pm}$}}
\newcommand{\Zed}{\mbox{$Z^{o}$}}
\newcommand{\pbar}{\mbox{$\overline{p}$}}
\newcommand{\lmm}{\mbox{$l^{-}$}}
\newcommand{\lpl}{\mbox{$l^{+}$}}
\newcommand{\as}{\mbox{$\alpha_s$}}
\newcommand{\rs}{\mbox{$\sqrt{s}$}}
\newcommand{\ppm}{\mbox{$\pi^{\pm}$}}
\newcommand{\kmp}{\mbox{$K^{-}p$}}
\newcommand{\nanom}{\mbox{$\rm{nm}$}}                   
\newcommand{\microm}{\mbox{$\mu\rm{m}$}}                
\newcommand{\millim}{\mbox{$\rm{mm}$}}                  
\newcommand{\centim}{\mbox{$\rm{cm}$}}                  
\newcommand{\metre}{\mbox{$\rm{m}$}}                   
\newcommand{\kilom}{\mbox{$\rm{km}$}}                   

\newcommand{\eV}{\mbox{$\rm{eV}$}}                      
\newcommand{\keV}{\mbox{$\rm{keV}$}}                    
\newcommand{\MeV}{\mbox{$\rm{MeV}$}}                    
\newcommand{\GeV}{\mbox{$\rm{GeV}$}}                    
\newcommand{\TeV}{\mbox{$\rm{TeV}$}}                    
\newcommand{\Tesla}{\mbox{\rm T}}
\newcommand{\GeVc}{\mbox{{\rm GeV}/{\it c}}}

\newcommand{\nanos}{\mbox{$\rm{ns}$}}                   
\newcommand{\micros}{\mbox{$\mu\rm{s}$}}                
\newcommand{\millis}{\mbox{$\rm{ms}$}}                  
\newcommand{\seconds}{\mbox{$\rm{s}$}}                  
\newcommand{\dmtwothree}{\mbox{$\Delta{m^2_{23}}$}}
\newcommand{\thonetwo}{\mbox{$\theta_{12}$}}
\newcommand{\thonethree}{\mbox{$\theta_{13}$}}
\newcommand{\thtwothree}{\mbox{$\theta_{23}$}}

\newcommand{\grams}{\mbox{$\rm{g}$}}                    
\newcommand{\kilog}{\mbox{$\rm{kg}$}}                   
\newcommand{\pot}{\mbox{$\rm{p.o.t.}$}}                 
\newcommand{\mip}{\mbox{$\rm{m.i.p.}$}}                 
\newcommand{\refe}[1]{(\ref{#1})}
\newcommand{\micron}{$\mu$m}
\newcommand{\pion}{$\boldmath \pi$}
\newcommand{\kaon}{$\boldmath K$}
\newcommand{\proton}{$\boldmath p$}
\newcommand{\steracc}{$\mu$sr~\%$\Delta p/p$}
\newcommand{\yield}{(incident protons $\cdot$ sr $\cdot$ ($\Delta p/p$ \%))$^{-1}$} 
\newcommand{\ygif}{(incident protons $\cdot$ sr $\cdot$ ($\Delta p/p$ \%)) } 
\newcommand{\xsec}{$mb/GeV^2$}
\subsubsection{Magnetic calorimetric iron detectors}

The measurement of $ \nue \to \num$ oscillations through the
appearance of wrong-sign muons calls for massive detectors weighing ${\cal
O}(50)$~Kt, with the capability of $\mu$ identification and the
measurement of their charge. A large mass magnetized iron calorimeter with
active elements based on RPCs or scintillator bars can fulfill both tasks.  
The particle electric charge can be determined by an external magnetic
field $\overline{B}$. A charged particle traversing $N$ steel plates, each
of thickness $x$, will receive a $p_{t}$ kick of $0.03 N x (cm)  B (T) $
GeV. Multiple scattering will introduce a random $p_{t}$ of $0.014 N
\sqrt{x/X_{0}}$ GeV, with $X_{0}=1.76$ cm for the radiation length of
steel.  After $N$ steel absorbers, the significance of the charge
determination will be
$$ 2 \frac{0.003 B(T) x(cm)}{0.014 \sqrt{x/X_{0}}} \sqrt{N}. $$
For a typical field $B \sim 1 \ T$, a $3$ or $4-\sigma$ measurement can
easily be obtained for muons.

Based on the MINOS experience~\cite{MINOS}, proposals have been put 
forward to use the
design of the proposed atmospheric neutrino detector 
MONOLITH~\cite{MONOLITH}, or a novel design of
iron-scintillator detector (LMD)~\cite{lmd} as a far detector in 
a neutrino-factory beam.
In the following, we summarize the  known experimental
results about active read-out elements, and then outline the main 
aspects of the designs of MONOLITH and LMD.

\subsubsection{Summary of the properties of active elements in massive 
iron calorimetric detectors}

Candidate active elements for massive iron calorimetric detectors for
$\nu$ studies are usually Iarocci proportional tubes, RPCs, or
scintillator bars, and each solution has its avantages and disadvantages.  
Proportional tubes are well studied, and with strip read-out will have
good spatial resolution, but do not perform well for calorimetry. RPCs are
inexpensive and easy to manufacture, and have good timing properties, but
their usual gas is flammable and their operation for long periods of time
is still debatable. Scintillators have both good timing properties and
very good calorimetric properties.  On economic grounds, the use of
good-quality scintillator with long attenuation length can be precluded. A
relatively cheap solution, pionereed by the MINOS Collaboration, is to use
a polystyrene-based scintillator doped with $1 \%$ PPO and $0.03 \%$
POPOP, extruded into bars of the required dimensions~\cite{extr-scint}.
The bar can be coated with a layer of polystyrene mixed with titanium
oxide, which is co-extruded at the same time as the scintillator and
provides light-tightness and diffusion of the light back into the
scintillator.  Because of the poor attenuation length of the scintillator,
the light produced in the bar is channelled to the photo-detector via
green wavelength-shifting (WLS) fibres of small diameter $\sim 1$~m. This
solution allows the cost of the read-out to be reduced by using multipixel
PMTs with pixel dimensions comparable to the fibre size.

Other solutions include the use of liquid scintillator in a vessel with
wavelength-shifting fibre read-out, where optical separators can provide a
granularity of a few centimeters over a large volume~\cite{doucet}.
Individual cells can be formed from extruded rigid PVC, which is made
reflective by a layer of titanium dioxide. A WLS fibre along the side of
each cell can convert any absorbed blue scintillator light to green, a
fraction of which is then trapped inside the fibre and transported to
external multipixel PMTs at the very end of the detector.  With the
extruded scintillator or liquid scintillator option, one can obtain good
energy resolutions (both electromagnetic and hadronic), good spatial
resolution determined by the cell width $d$: $\sigma_{x} \simeq
d/\sqrt{12}$, and the possibility of fast timing.  Reasonable light yields
have been measured. As an example, with extruded scintillator and BCF91A
WLS fibers, 8~m long, up to 2 p.e./MIP crossing at the far end have been
measured.

\subsubsection{The Monolith design}

\begin{figure}
 \begin{center}
 \mbox{\epsfig{file=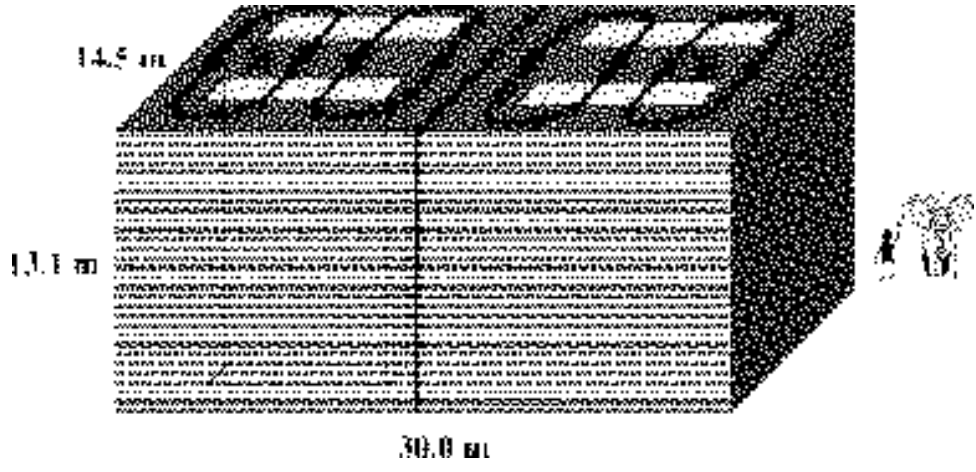,width=0.8\linewidth}} \\
 \end{center}
 \caption{\it Layout of the Monolith detector, showing the magnetic 
field.}
 \vspace{1cm}
 \label{fig:monolith}
 \end{figure}

The Monolith concept is based on a massive tracking calorimeter with a
modular structure, shown in Fig.~\ref{fig:monolith}, where the magnetic
field configuration is also shown: iron slabs are magnetized at $B \simeq
1.3$~T.  Each module would be a stack of 120 8~cm thick iron absorber
plates ($14.5 \times 15$ m$^2$), interleaved with sensitive elements in a
2~cm gap. The detector height is about 13~m, and a mass of 34~kt can be
obtained with two such modules.  External scintillator counters are
foreseen, as a veto, to reduce the background from cosmic-ray muons. As
active elements, glass spark counters (RPCs with glass electrodes) were
chosen. They provide $x - y$ coordinates with a 3~cm pitch and a time
resolution of about 1~ns. The usual bakelite electrodes are substituted by
commercially available high-resistivity float-glass electrodes, suitable
for operation in streamer mode in a low counting environment. They have
stable volume resistivity over long periods and avoid, due to better
planarity and homogeneity, surface treatment with linseed oil.  The
high-voltage supply is applied to the electrodes by means of water-based
graphite. Long-term stability of the chambers over several months show a
remarkable stability of the efficiency and the time resolution. Test-beam
results on a 8~t prototype have shown an hadronic energy resolution
$\sigma_{E}/E \sim 68 \%/\sqrt{E} + 2 \%$, sufficient for a full
reconstruction of energy and direction of interacting neutrinos, a time
resolution $\sim 1.7$~ns and a good tracking capability~\cite{mono-test}.

On average 95\% of the muons induced by atmospheric neutrino
interactions inside MONOLITH have their charge correctly assigned,
without specific selections on the track quality. This makes it
possible to test the neutrino mass hierarchy through the study of
earth-induced matter effects with atmospheric neutrinos
\cite{ttf01}.  In the context of a neutrino factory project, charge
discrimination could be significantly improved by selecting
only unambigous muon candidates \cite{PhD_MSelvi}.
The efficiency to discover wrong-sign muons was estimated to
be around 16\%, for a background rejection of order $10^{-6}$.
The hadron shower direction reconstruction has also been
studied in detail. \\
The  sensitivity region in the ($\Delta m_{23}^2$,$sin2(2\theta_{13})$)
parameter space is about two
order of magnitude better than the current Chooz limit in the whole
Super-Kamiokande $90\%~ C.L.$ allowed region of $\Delta m_{23}^2$, if
the baseline is $732~km$; better at $3500~km$.\\
This result is achieved also in the simplest version of the detector
configuration and of the analysis, i.e. horizontal-$8~cm$ planes and a
simple muon momentum cut. Among the possible improvements, the best
seems to be the vertical-$8~cm$ planes option, that provides a very
good hadronic background rejection. This choice of the detector has
also good performances with atmospheric neutrinos.\\

The results of the analysis of a test beam prototype confirm that the
requested hadronic angular resolution is achievable with the massive
magnetized calorimeter design.\\

\subsubsection{The Large Magnetic Detector (LMD)}
\label{subsec:LMD}

\begin{figure}[tbhp]
\begin{center}
\mbox{\epsfig{file=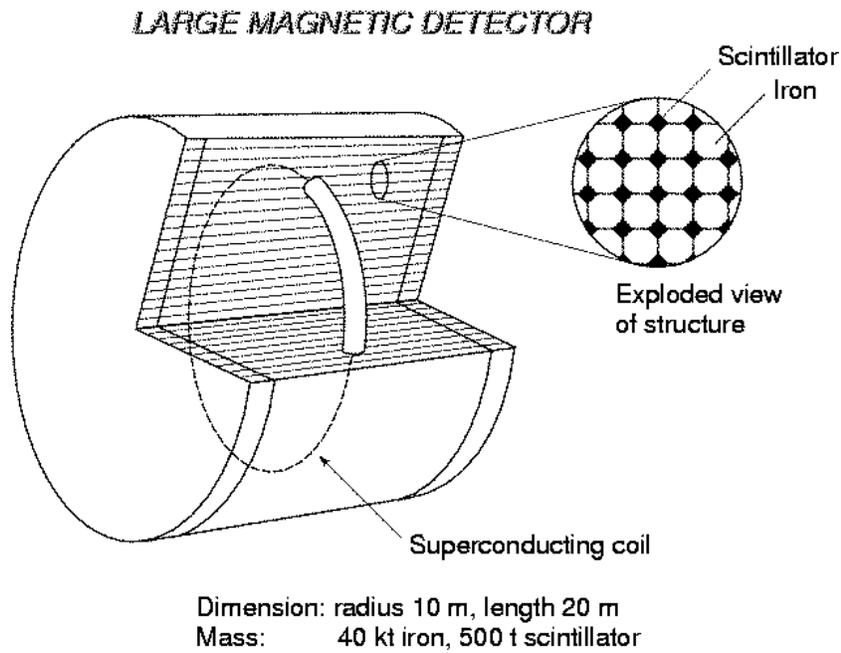,width=11cm}}
\end{center}
\caption{\it Sketch of the Large Magnetic Detector.}
\label{fig:detector}
\vspace{7mm}
\end{figure}

The proposed apparatus, shown in Fig.~\ref{fig:detector}, is a large
cylinder of 10~\metre\ radius and 20~\metre\ length, made of 6~\cm\ thick
iron rods interspersed with 2~\cm\ thick scintillator rods built of
2~\metre\ long segments. The light read-out on both ends allows the
determination of the spatial coordinate along the scintillator rod. The
detector mass is 40~kt, and a superconducting coil generates a
solenoidal magnetic field of 1~\Tesla\ inside the iron.  A neutrino
traveling through the detector sees a sandwich of iron and scintillator,
with the $x - y$ coordinates being measured from the location of the
scintillator rods, and the $z$ coordinate being measured from their
longitudinal segmentation.

The performance of the proposed detector would be similar to that of
MINOS~\cite{MINOS}. The main difference lies in the mass, which is an
order of magnitude larger, and in the smaller surface-to-volume ratio
which together, we believe, make the large magnetic detector superior for
the detection of \num\ and \numb\ events at the neutrino factory, and for
the simultaneous study of atmospheric neutrino events.

To study the performance of the Large Magnetic Detector, we have performed
a fast Monte Carlo simulation based on the GEANT 3 package~\cite{geant}.
The simulation assumes a sequence of 4~cm of iron followed by 1~cm of
scientillators.  The muon momentum and the hadronic shower are smeared
following the parameterizations of the MINOS proposal.
 
We consider the $\numubar + \nue$ neutrino beam originating from a
$\muplus$ beam of 50 \GeVc. The wrong-sign muon signal is generated
assuming the oscillation parameters $\dmtwothree = 2.8 \times 10^{-3}$,
$\thonetwo=45^\circ$, and $\thonethree=8^\circ$. We have simulated $25000$
\numu\ CC, $10^7$ \numubar\ CC, $10^7$ \numubar\ NC and $10^7$ \nue\ CC
events.

To study the charge misidentification background, we have considered the 
$10^7$
\numubar CC events, which have a realistic neutrino spectrum.  This
simulation includes energy loss and multiple scattering.  Finite
transverse resolution and loss of hits are simulated {\it a posteriori}.  
The momentum and the charge of the $\muplus$ were computed by means of a
Kalman Filter fit to the recorded hits. It was assumed that there would be
$2\%$ of lost hits, a transverse resolution of 0.5~cm and a distance
between measurement planes of 15~cm. These parameters were varied inside
logical limits, and it turned out that even for $20\%$, 2.0~cm and 25~cm,
this background was below $10^{-5}$ for a momentum cut of $5$~GeV.

The discrimination of physical backgrounds from the signal is based on the
fact that the $\muminus$ produced in a \numu CC signal event is harder
and more isolated from the hadron shower axis than in background events
(\numubar CC, \nue CC, \numubar NC and \nue NC).  Accordingly, we
performed an analysis based on the momentum of the muon, $P_\mu$, and a
variable measuring the isolation of the muon from the hadron shower axis,
$Q_{\rm t} = P_\mu \sin^2 \theta$. Obviously, optimal cuts depend on the
baseline.  Table~\ref{tab:LMD_optimal} gives the signal and background
events, after optimal cuts, at three different baselines. The
intermediate distance $\sim 3500$~\km turns out to be the best baseline
from the point of view of optimizing the $S/N$ ratio and thus the
sensitivity to the oscillation parameters.

\vspace{1cm}
\begin{table}[htbp]
\caption{\it Signal efficiency and fractional backgrounds for optimal
cuts in the LMD detector.}
\label{tab:LMD_optimal}
\vspace{4mm}
\begin{center}
\begin{tabular}{|c|c|c|c|c|c|c|}
\hline
Baseline(km) & $P_\mu$ cut &  $Q_t$ cut & \numu\ signal efficiency & \numubar\ CC &\nue\ CC & \numubar + \nue\ NC \\
\hline
\hline
732  & $5.0$  & $1.4$ & $0.24$ & $11 \times 10^{-7}$ & $1.0 \times 10^{-7}$ & $2.0 \times 10^{-7}$  \\
3500 & $5.0$  & $0.7$ & $0.45$ & $54 \times 10^{-7}$ & $1.0 \times 10^{-7}$ & $23 \times 10^{-7}$  \\
7332 & $5.0$  & $0.6$ & $0.52$ & $74 \times 10^{-7}$ & $1.0 \times 10^{-7}$ & $30 \times 10^{-7}$  \\
\hline
\end{tabular}
\vspace{7mm}
\end{center}
\end{table}

\subsubsection{A Liquid Argon detector}
\begin{figure}[tbhp]
\begin{center}
\begin{tabular}{cc}
\mbox{\epsfig{file=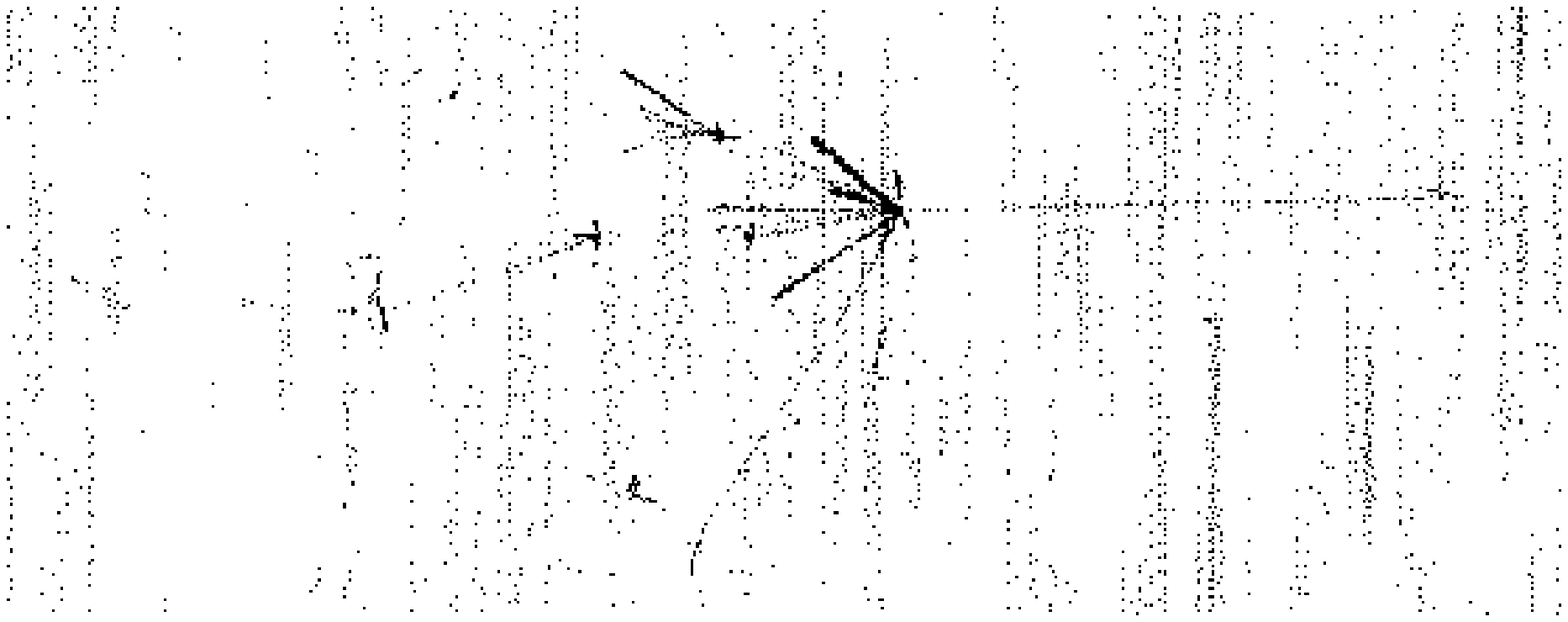,height=5cm,width=10cm}} &
\mbox{\epsfig{file=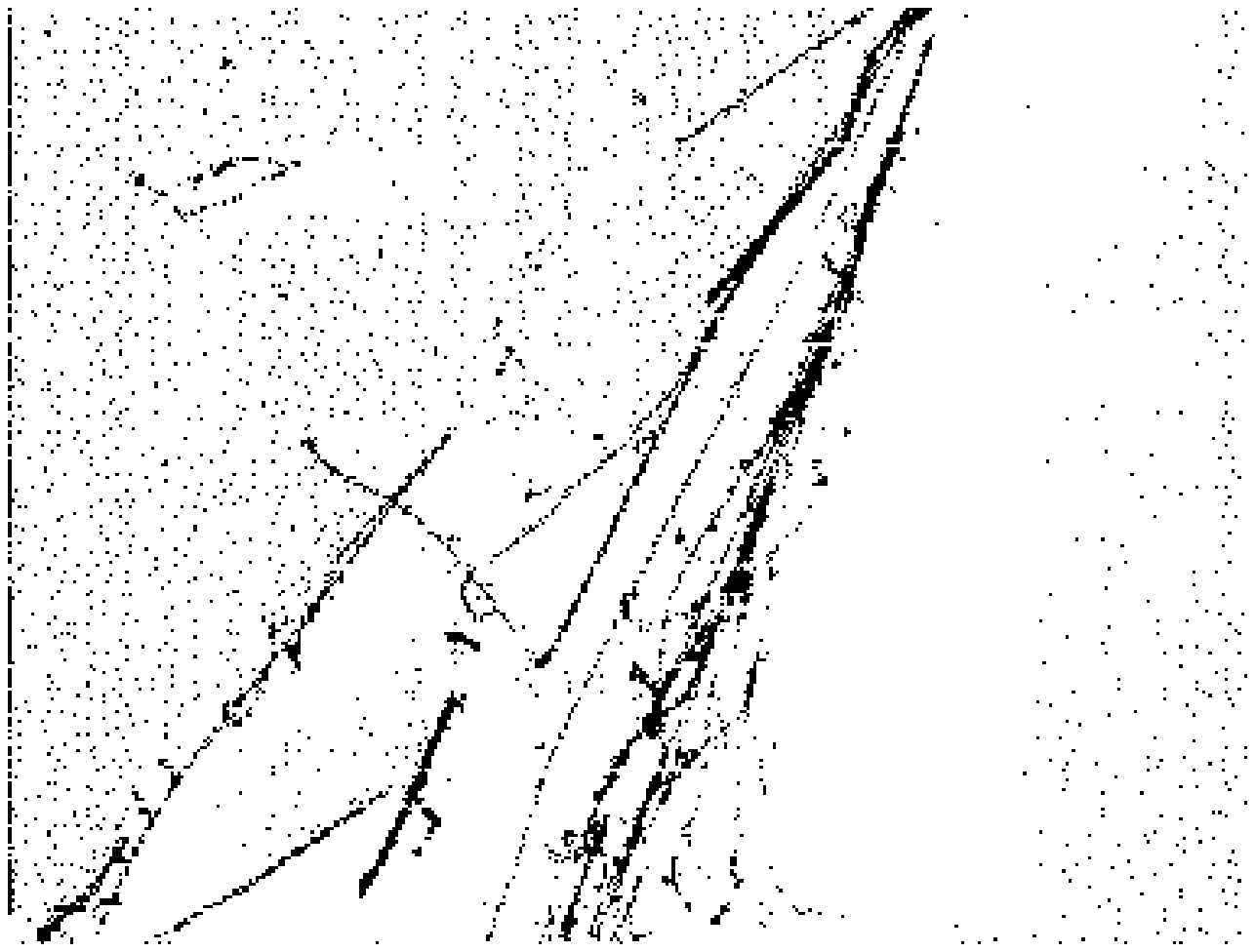,width=6.5cm}} \\
\end{tabular}
\vspace{0.5cm}
\caption{\it Left panel: electronic liquid argon imaging of a hadronic
interaction: the collection wire coordinate is represented on the horizontal 
axis
and the drift coordinate on the vertical axis (270$\times$80~cm$^2$ 
shown).
Right panel: a cosmic ray induced shower (40$\times$40~cm$^2$ shown).}
\label{fig:3ton} 
\end{center}
\end{figure}

A liquid argon time projection chamber (TPC), as developed by the ICARUS
Collaboration, is continuously sensitive and self-triggering, with the
ability to provide three-dimensional imaging of any ionizing event: see
Figure~\ref{fig:3ton}. This detector technology combines the
characteristics of a bubble chamber with the advantages of electronic
read-out. It offers the possibility of performing complementary and
simultaneous measurements of neutrinos, including those from cosmic ray
events, from the CERN to Gran Sasso beam (CNGS), and even those from the
Sun and from supernovae. The same class of liquid argon detector can also
be envisaged for high-precision measurements at a neutrino factory, and
can be used to perform background-free searches for nucleon decay. Hence,
an extremely rich and broad physics programme~\cite{ICARUS}, encompassing
both accelerator and non-accelerator physics, will be addressed. This will
be able to answer fundamental questions about neutrino properties and
about the possible physics of the nucleon decay.

A liquid argon detector is a precise tracking device with geometrical
resolution 3$\times$3$\times$0.2~mm$^3$, that also makes high-resolution
$dE/dx$ measurements and has full-sampling electromagnetic and hadronic
calorimetry. Imaging provides excellent electron identification and
electron/hadron separation. Energy resolution is excellent ($\sim
3\%/\sqrt{E}$~for electromagnetic showers) and the hadronic energy
resolution of contained events is also very good ($\sim 30\%/\sqrt{E_h}$).

The operating principle of the ICARUS liquid argon TPC is rather simple:
any ionizing event (from a particle decay or interaction) taking place in
the active liquid argon volume, which is maintained at a temperature $T
\sim 89$~K, produces ion-electron pairs. In the presence of a strong
electric field ($\sim 0.5$~kV/cm), the ions and electrons drift parallel
to the field in opposite directions. The motion of the faster electrons
induces a current on a number of parallel wire planes located at the end
of the sensitive volume. The knowledge of the wire position and the drift
time provides the three-dimensional image of the track, while the charge
collected on the wire provides information on the deposited energy.

Liquid argon imaging is now a mature technique, that has previously been
demonstrated up to the scale of a 15~ton prototype. The latest major
milestone has been the successful operation of the ICARUS 600~t prototype
({\tt T600}), constructed during 2000 with all its various components,
which operated during the summer of 2001. The detector is to be 
transported to the Gran Sasso underground laboratory to continue a long 
test period that will mainly feature the observation and study of 
atmospheric and solar neutrinos.

The overall performance of an underground detector is primarily determined
by two factors: its total mass and its geometrical granularity, which
determines the quality with which signal events can be reconstructed, and
hence separated from backgrounds.  The final phase of the ICARUS physics
program requires a sensitive mass of liquid argon of 5~kt or more. The
{\tt T600} detector stands today as the first proof that such large
detectors can be built, and that liquid argon imaging technology can be
implemented on such large scales. The superior bubble-chamber-like
features of the ICARUS detector will provide additional and fundamental
tools for elucidating in a comprehensive way the pattern of neutrino
masses and mixings.

\subsubsection{Magnetized liquid argon detectors}

\begin{figure}[H]
\begin{center}
\epsfig{file=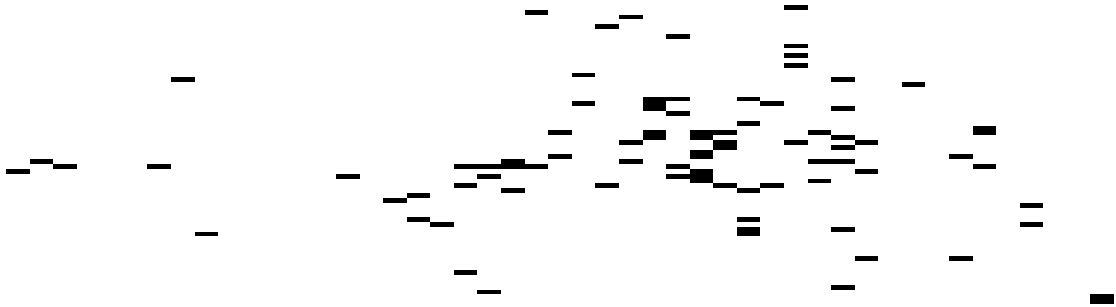,width=15cm,height=4cm}
\caption{\it Simulation of a 2.5~GeV electron shower in a magnetized 
liquid argon TPC. The magnetic field has strength $B = 1.5$~T, and is 
directed perpendicular to the plane of the figure.}
\label{eshower}
\end{center}
\end{figure}

The possibility of complementing the excellent $dE/dx$ measurements and
energy resolution with those provided by a magnetic field has been
studied. Embedding the volume of argon into a magnetic field would not
alter the imaging properties of the detector, and measurements of the
bending of charged hadrons or penetrating muons would allow a precise
determination of the momentum and a determination of the charge: see
Fig.~~\ref{eshower}.  Unlike muons or hadrons, the early showering of
electrons makes their charge identification difficult. It is found that
the determination of the charge of electrons of energy in the range
between 1 and 5~GeV is feasible with good purity, provided the field has a
strength in the range of 1~Tesla. Preliminary estimates show that these
electrons exhibit an average curvature sufficient to have electron charge
discrimination better than $1\%$ with an efficiency of 20\%.

\subsubsection{A hybrid emulsion detector}
\label{opera}
 
This technology is based on the Emulsion Cloud Chamber (ECC) concept (see
references quoted in \cite{operaproposal}), which combines the
high-precision tracking capabilities of nuclear emulsions and the large
mass achievable by employing metal plates as a target. It has already been
adopted by the OPERA Collaboration~\cite{operaproposal,operastatus} for a
long-baseline search for $\nu_{\mu}\leftrightarrow\nu_{\tau}$ oscillations
in the CNGS beam~\cite{cngsbeam}. The experiment uses nuclear emulsions as
high-resolution tracking devices for the direct detection of the $\tau$
produced in the $\nu_{\tau}$ CC interaction with the target.

The basic element of the ECC is the cell, which is made of a 1~mm thick
lead plate followed by a thin emulsion film. The film is made up of a pair
of emulsion layers $50~\mu$m thick on either side of a $200~\mu$m plastic
base. The number of grain hits in $50~\mu$m ($15$-$20$) ensures redundancy
in the measurement of particle trajectories and in the measurement of
their energy loss in the non-relativistic regime when they are about to
stop~\cite{operaproposal}. This is an important handle to reduce the
background coming from $\nu_\mu$CC charm production with a soft primary
muon.

Thanks to the dense ECC structure and the high granularity provided by the
nuclear emulsions, the OPERA detector is also suited for electron and
$\gamma$ detection~\cite{operaproposal}.  The resolution in measuring the
energy of an electromagnetic shower is about 20\%. Furthermore, the
nuclear emulsions are able to measure the number of grains associated to
each track. This allows an excellent two-track separation:
${\cal O}(1)~\mu$m or even better. Therefore, it is possible to
disentangle single-electron tracks from tracks produced by electron pairs
coming from $\gamma$ conversion in the lead.

The outstanding position resolution of nuclear emulsions can also be used
to measure the angle of each charged track with an accuracy of about
1~mrad. This allows momentum measurement by using multiple Coulomb
scattering with a resolution of about 20\%, and the reconstruction of
kinematical variables characterizing the event, i.e., the missing
transverse momentum at the interaction vertex, $p_T^{miss}$, and the
transverse momentum of a track with respect to hadronic shower direction,
$Q_T$. For details on the event reconstruction, both with the nuclear
emulsions and the electronic detector in an OPERA, we refer the interested 
reader to~\cite{operaproposal,operastatus}.

Concerning the capability of an emulsion-based detector to handle the high
interaction rate expected at a Neutrino Factory, we should stress that
only events with a wrong-sign muon (WSM) in the final state are analysed.  
Therefore, one expects an attainable number of events to be
scanned~\cite{donini}. Having in mind that in the past few years the
scanning power increased by a factor of 10 every two years, it could be
considered as realistic to assume that, by the time the Neutrino Factory
will become operational, it will be possible to scan a larger number of
events, say by a factor two. Therefore, a lead-emulsion mass a factor two
larger than the one proposed for OPERA could be envisaged for a detector
operating at a Neutrino Factory~\cite{donini}.

\def\ra{\rightarrow}
\def\numunue{\nu_\mu\rightarrow\nu_e}
\def\numunutau{\nu_\mu\rightarrow\nu_\tau}
\def\nuebar{\bar\nu_e}
\def\numubar{\bar\nu_\mu}
\def\nue{\nu_e}
\def\anue{\bar\nu_e}
\def\nutau{\nu_\tau}
\def\anutau{\bar\nu_\tau}
\def\anumu{\bar\nu_\mu}
\def\numu{\nu_\mu}
\def\numubarnuebar{\bar\nu_\mu\rightarrow\bar\nu_e}
\def\nuebarnumubar{\bar\nu_e\rightarrow\bar\nu_\mu}
\def\anul{\bar\nu_\ell}
\def\nul{\nu_\ell}

\subsection{\bf Oscillation Physics at the Neutrino Factory}

The possibility of having intense neutrino beams of well-known
composition opens the road to a large variety of physics
studies. Having a simultaneous beam of electron and muon
neutrinos, distinguished by helicity, allows the study of
several oscillation processes. If we consider negative muons
in the ring, the following transitions can occur:
\begin{itemize}
\item $\nu_\mu\to\nu_\mu$ disappearance
\item $\nu_\mu\to\nu_e$ appearance
\item $\nu_\mu\to\nu_\tau$ appearance
\item $\bar{\nu}_e\to\bar{\nu}_e$ disappearance
\item $\bar{\nu}_e\to\bar{\nu}_\mu$ appearance
\item $\bar{\nu}_e\to\bar{\nu}_\tau$ appearance
\end{itemize}
An important feature of the Neutrino Factory is the possibility of
having opposite muon charges circulating in the ring, therefore
allowing also the study of the charged-conjugated processes of
those above. 

The simultaneous presence of both neutrino
flavours in the beam poses the problem of separating neutrinos
due to oscillations from beam background. A simple 
identification of the lepton produced in charged-current 
interactions is not sufficient, since muons, for instance, could
come from the $\nu_\mu$ component of the beam, from the
oscillation $\bar{\nu}_e\to\bar{\nu}_\mu$, or even from the 
oscillation $\bar{\nu}_e\to\bar{\nu}_\tau$, followed by the
decay $\tau\to\mu$. The obvious way to distinguish neutrinos coming
from the beam from those coming from oscillations is to measure
the charge of the lepton produced in charged-current events.
The ideal case would be to be able to measure the charge for
both electrons and muons, and perhaps find a way also to identify taus.
Since the last two requirements are quite difficult to match,
we consider as a default case that the detector for the neutrino
factory will only be able to identify the charge of muons. 
If also electron identification can be performed, the detected
events can be classified in four classes:
\begin{itemize}
\item Charged-current electrons,
\item Right-sign muons,
\item Wrong-sign muons,
\item Events with no leptons.
\end{itemize}

\begin{figure}[p]
\begin{minipage}[h]{0.45\textwidth}
\begin{center}
\epsfig{file=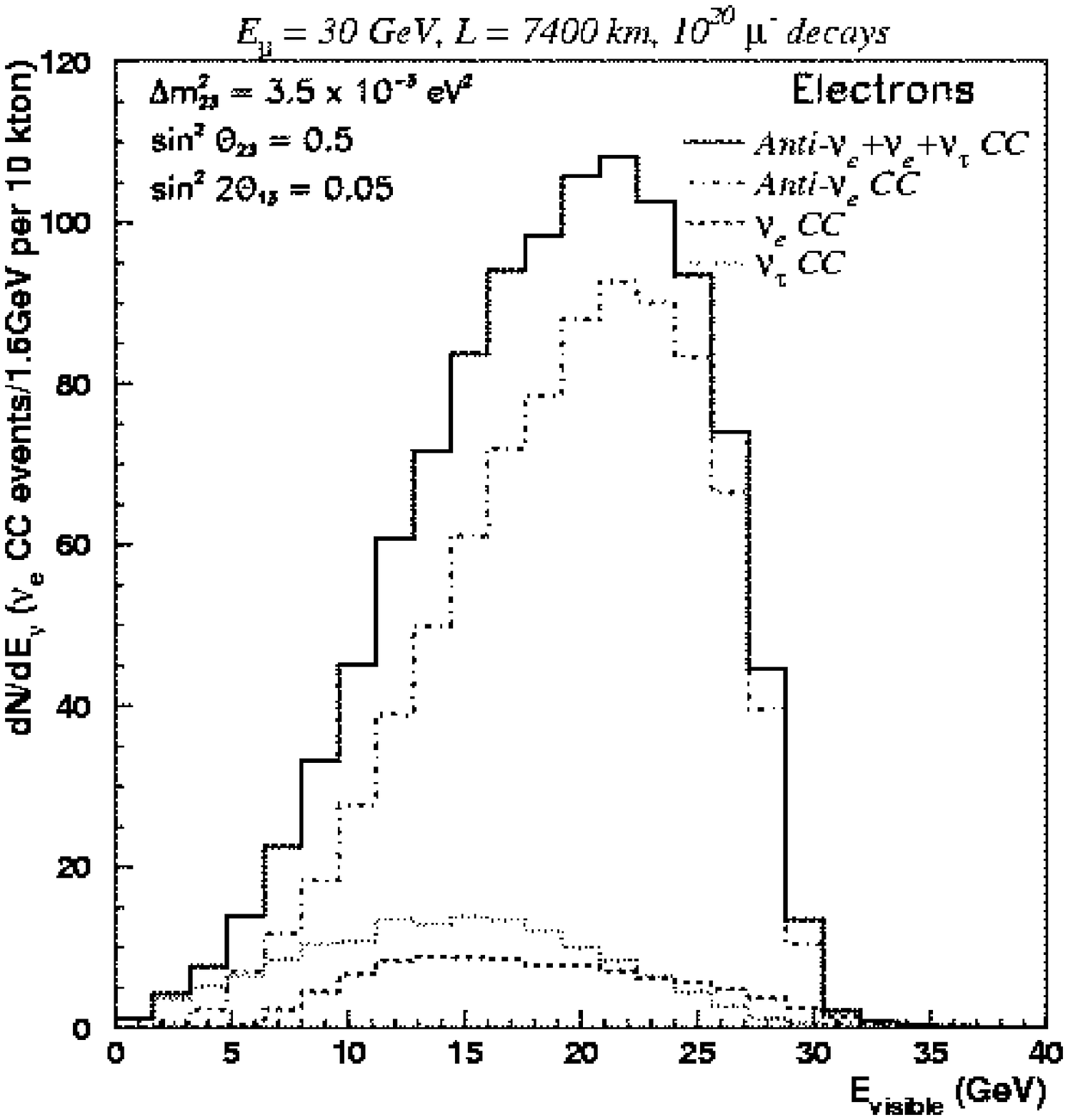,width=8cm}
\end{center}
\end{minipage}
\begin{minipage}[h]{0.45\textwidth}
\begin{center}
\epsfig{file=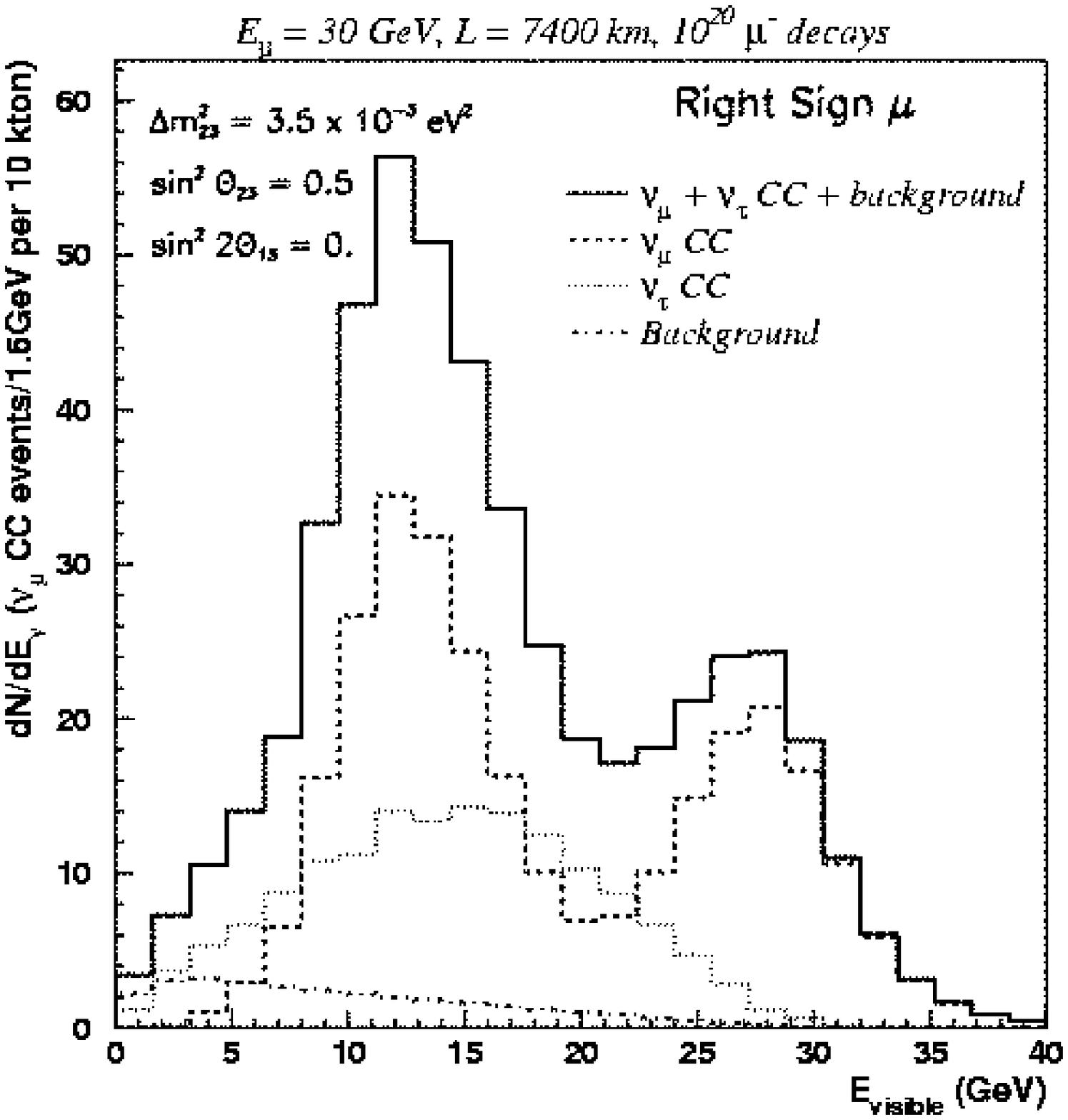,width=8cm}
\end{center}
\end{minipage}
\end{figure}

\begin{figure}[p]
\begin{minipage}[h]{0.45\textwidth}
\begin{center}
\epsfig{file=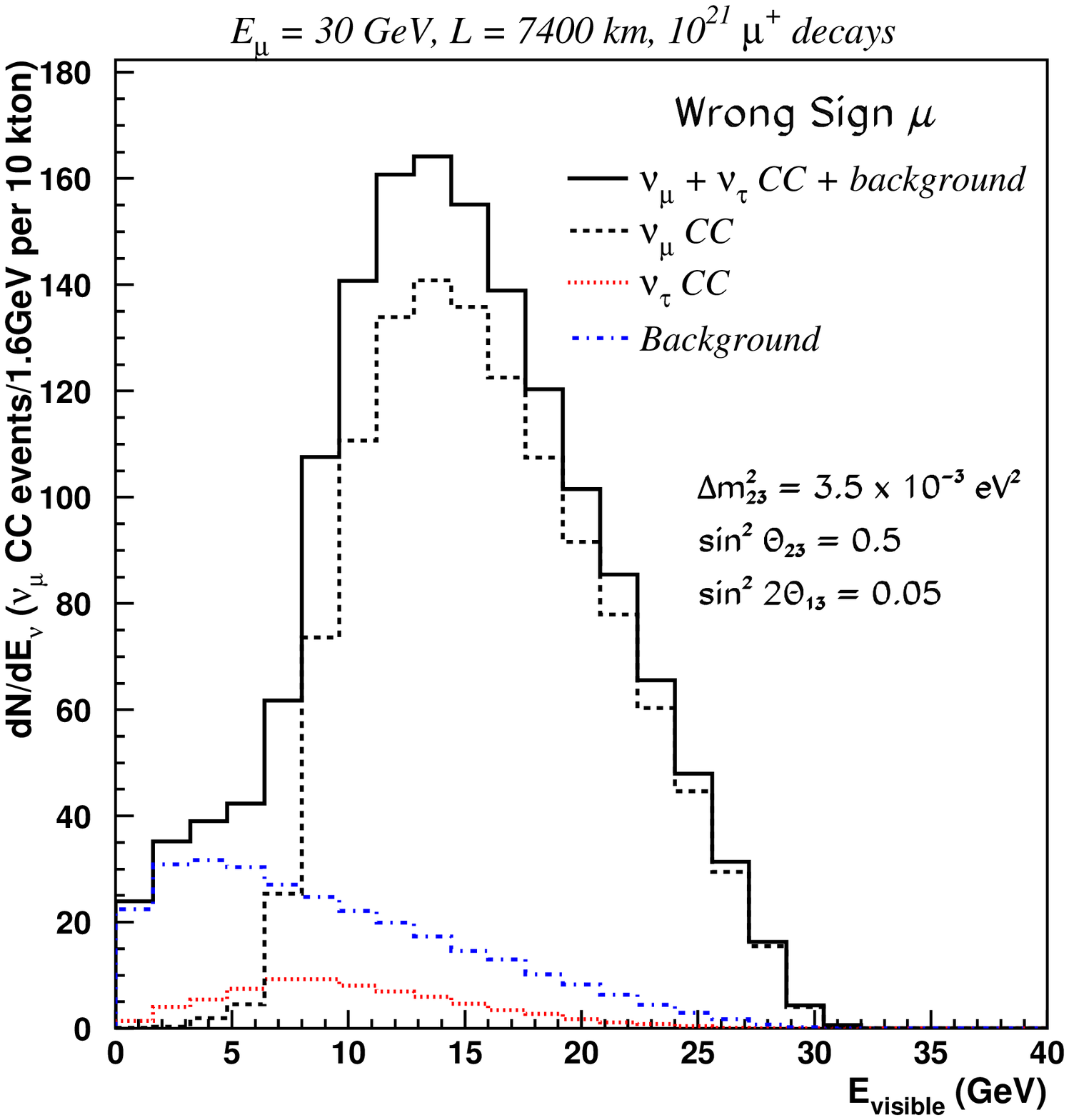,width=8cm}
\end{center}
\end{minipage}
\begin{minipage}[h]{0.45\textwidth}
\begin{center}
\epsfig{file=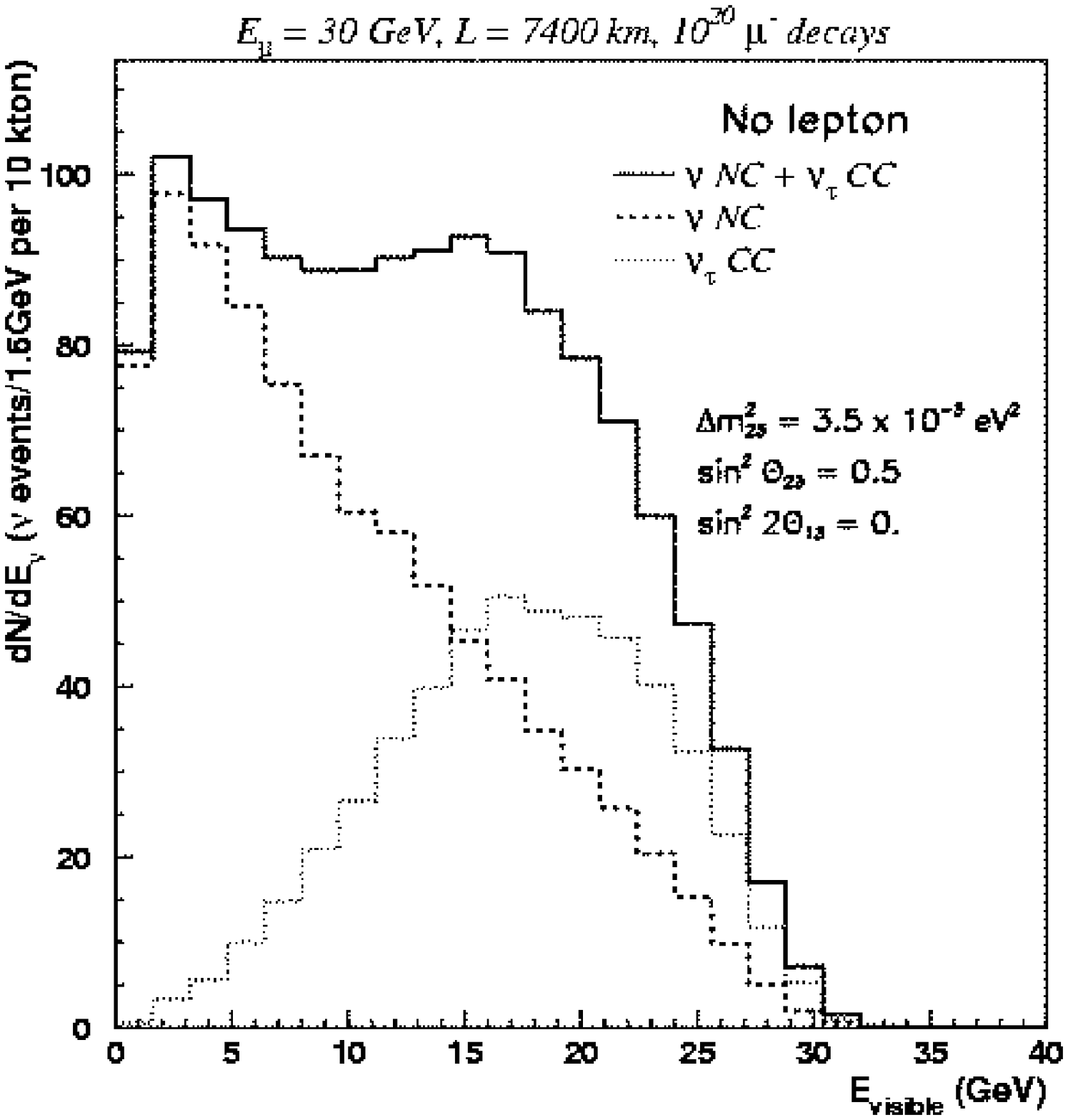,width=8cm}
\end{center}
\end{minipage}
\caption{\it The four classes of events that can be studied in a detector
with electron and muon charge identification. From the top, left to right:
events with high-energy electrons, right-sign muons, wrong-sign muons and
no charged leptons~\cite{physpot}.}
\label{fig:classes}
\end{figure}

An example of the set of energy spectra for these classes, for positive
and negative muons circulating in the ring, is given in
Fig.~\ref{fig:classes}. At a later stage, we discuss the physics
opportunities that will be opened by a detector able to also identify the
charge of the electron, and taus using kinematic criteria.

\subsubsection{Precision measurements of oscillations}

The first parameters to be measured are those governing
the leading atmospheric oscillation $\nu_\mu\to\nu_\tau$: $\theta_{23}$ and
$\Delta m^2_{23}$. These parameters are mainly determined from
the disappearance of muon neutrinos in the beam, observed using
right-sign muon events. The maximum of the oscillation probability
will produce a dip in the visible spectrum; the energy position 
of this dip will be correlated to the value of $\Delta m^2_{23}$,
and the depth to $\theta_{23}$. It is therefore favourable to choose
an energy and baseline such that the maximum of the oscillation
probability lies comfortably inside the detectable spectrum.\par
\begin{figure}[tbhp]
\begin{minipage}[h]{0.45\textwidth}
\begin{center}
\epsfig{file=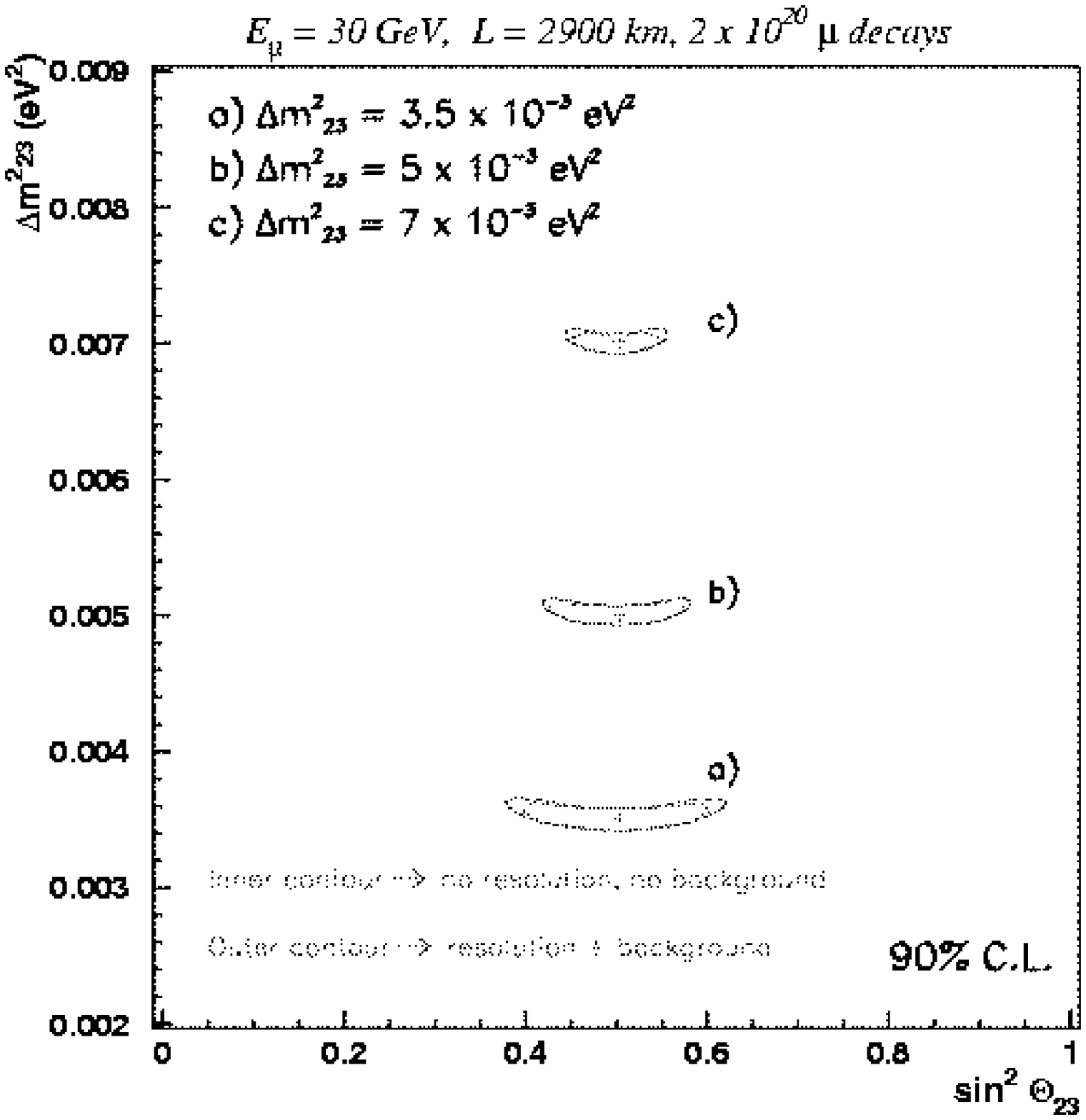,width=8cm}
\end{center}
\end{minipage}
\begin{minipage}[h]{0.45\textwidth}
\begin{center}
\epsfig{file=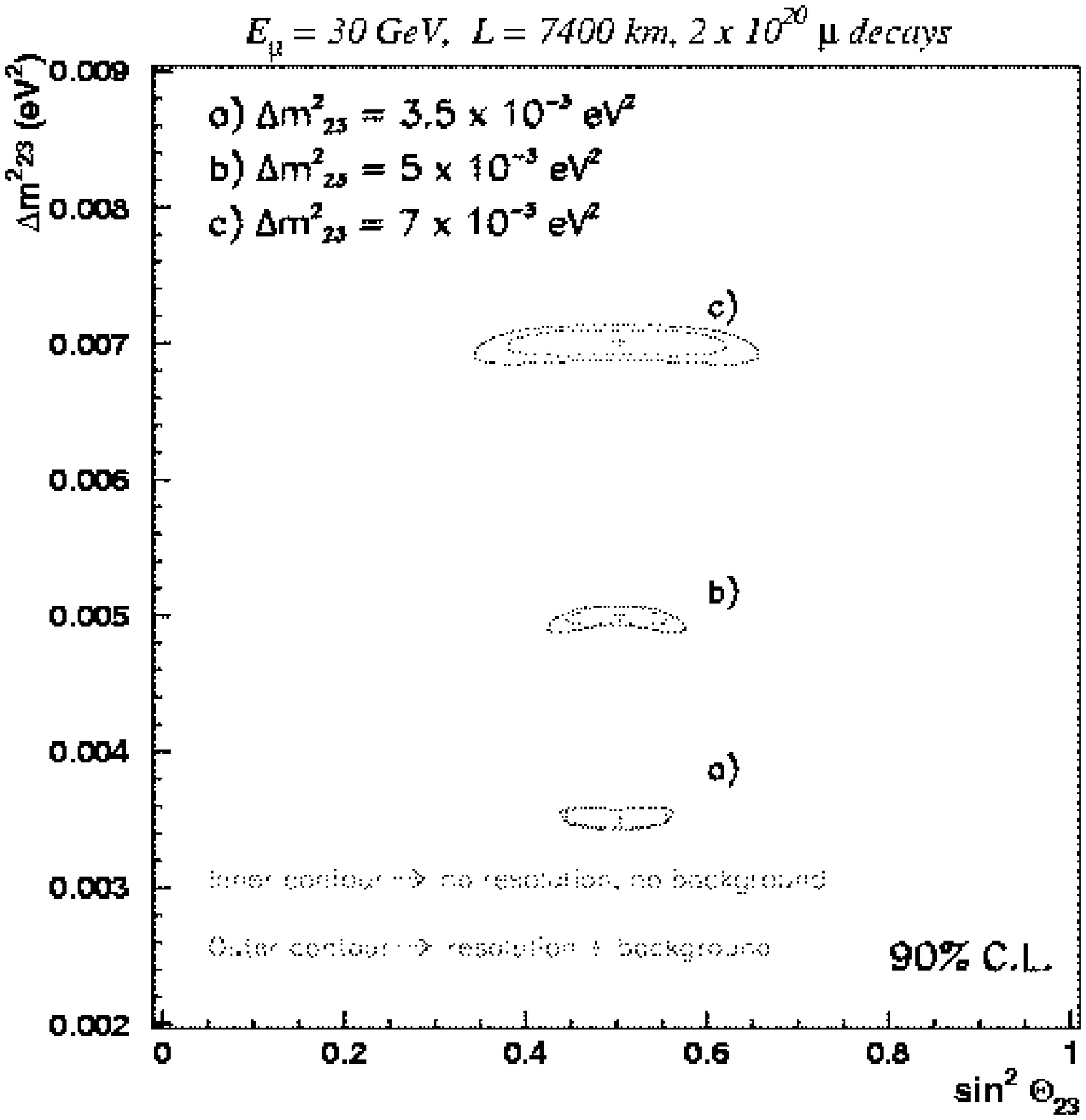,width=8cm}
\end{center}
\end{minipage}

\caption{\it Measurement of the parameters of the leading 
oscillation~\cite{physpot}.}
\label{fig:leading}
\end{figure}

The precision on the measurement of the oscillation parameters 
has been addressed by several groups~\cite{bargerpar,physpot,freund}, 
and  is normally performed by a fit of the energy spectra of the 
event classes (as previously said, this fit is largely dominated
by right-sign muon events). As shown in Fig.~\ref{fig:leading},
precisions of the order of 1\% on $\Delta m^2_{23}$ and of 10\%
on $\sin^2 \theta_{23}$ are foreseen.
\par

\subsubsection{Sensitivity to $\theta_{13}$}

So far, the most accurate information we have on the angle
$\theta_{13}$, i.e., the mixing of electron neutrinos and
the other two families (which are maximally mixed among themselves), is
that $\sin^2 2\theta_{13}<0.1$~\cite{osc_data_reactor}.
In a favourable case, a non-zero value of this parameter 
could be discovered before
the Neutrino Factory by experiments running in first-generation 
neutrino beams, such as ICARUS, OPERA and MINOS. Much larger sensitivity
will however be achieved by super-beams, for instance
the JHF-Super-Kamiokande
project. Experiments performed in `conventional' beams from
pion decays will, however, always be limited by the presence of
a $\nu_e$ component in the beam itself, representing an 
irreducible background to the search for $\nu_\mu\to\nu_e$
oscillations. It was first realized in \cite{CPviol} that the Neutrino Factory 
would have a significantly improved sensitivity to $\theta_{13}$ from the measurement of the wrong-sign muon signal.
 This measures the oscillation $\nu_e\to\nu_\mu$, where the oscillated
muon neutrinos are easily separated from the beam component
of opposite helicity by measuring the charge of the produced muon.
We have already seen that the decays of charmed particles, kaons and
pions can produce a background to the wrong-sign muon sample.
Applying strong cuts on muon momentum and isolation, it was
reported~\cite{golden} that the background could be reduced by as 
much as a factor $10^6$, keeping an efficiency of about 40\%.
An excellent background rejection with good efficiency could also be 
obtained using a hybrid emulsion detector~\cite{donini}.\par
\begin{figure}[p]
\begin{minipage}[h]{0.45\textwidth}
\begin{center}
\epsfig{file=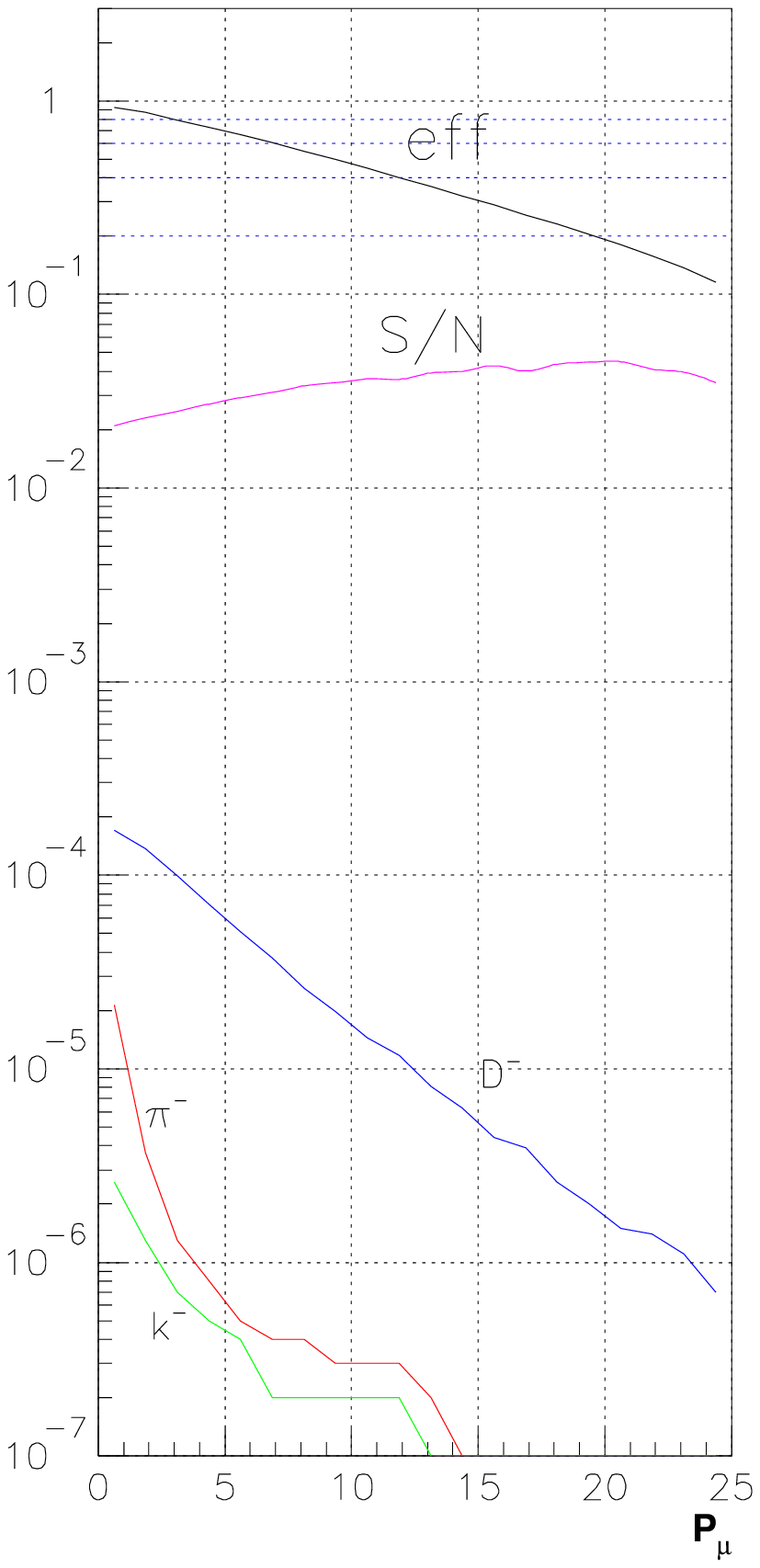,width=8cm,height=8cm}
\end{center}
\end{minipage}
\begin{minipage}[h]{0.45\textwidth}
\begin{center}
\epsfig{file=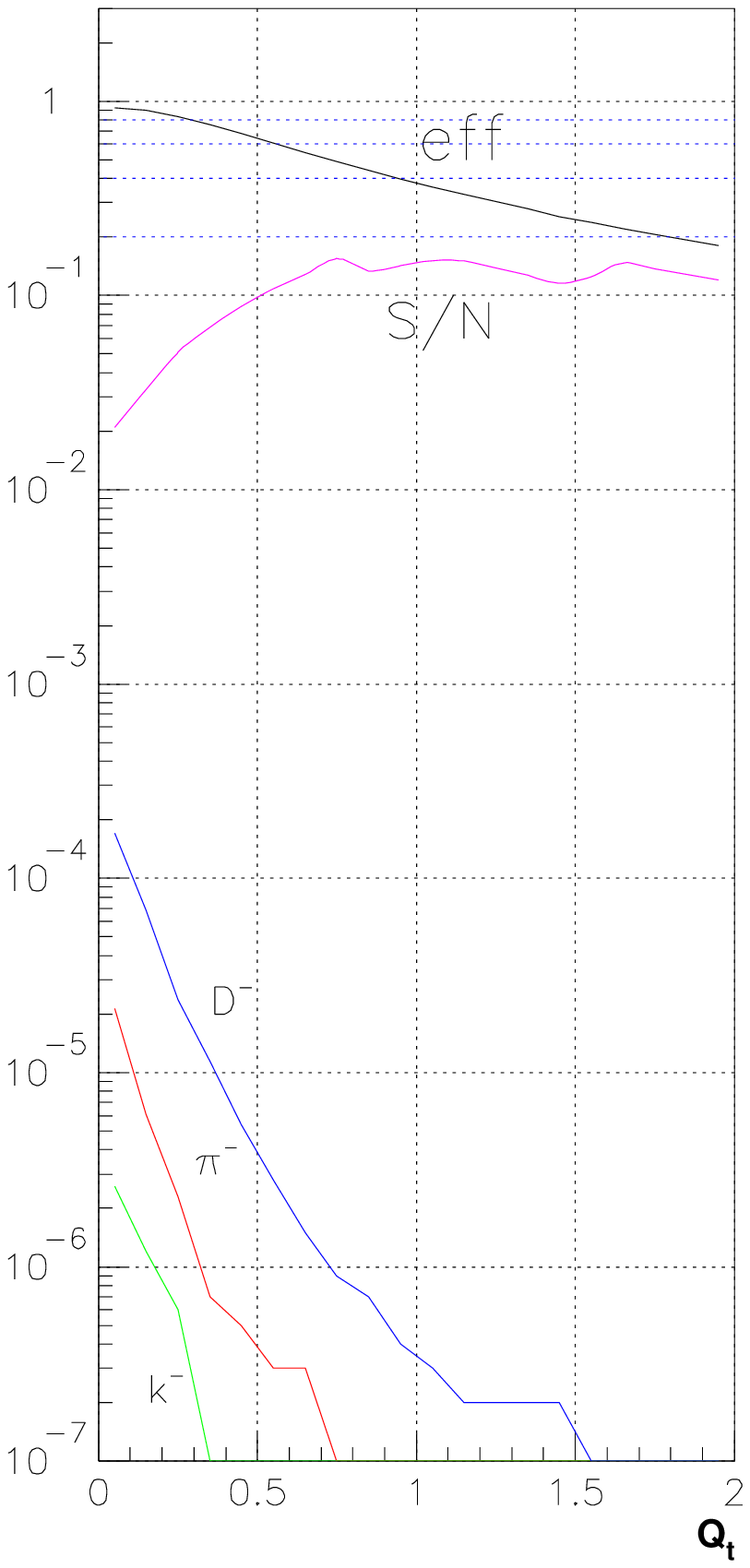,width=8cm,height=8cm}
\end{center}
\end{minipage}
\begin{minipage}[h]{0.45\textwidth}
\begin{center}
\epsfig{file=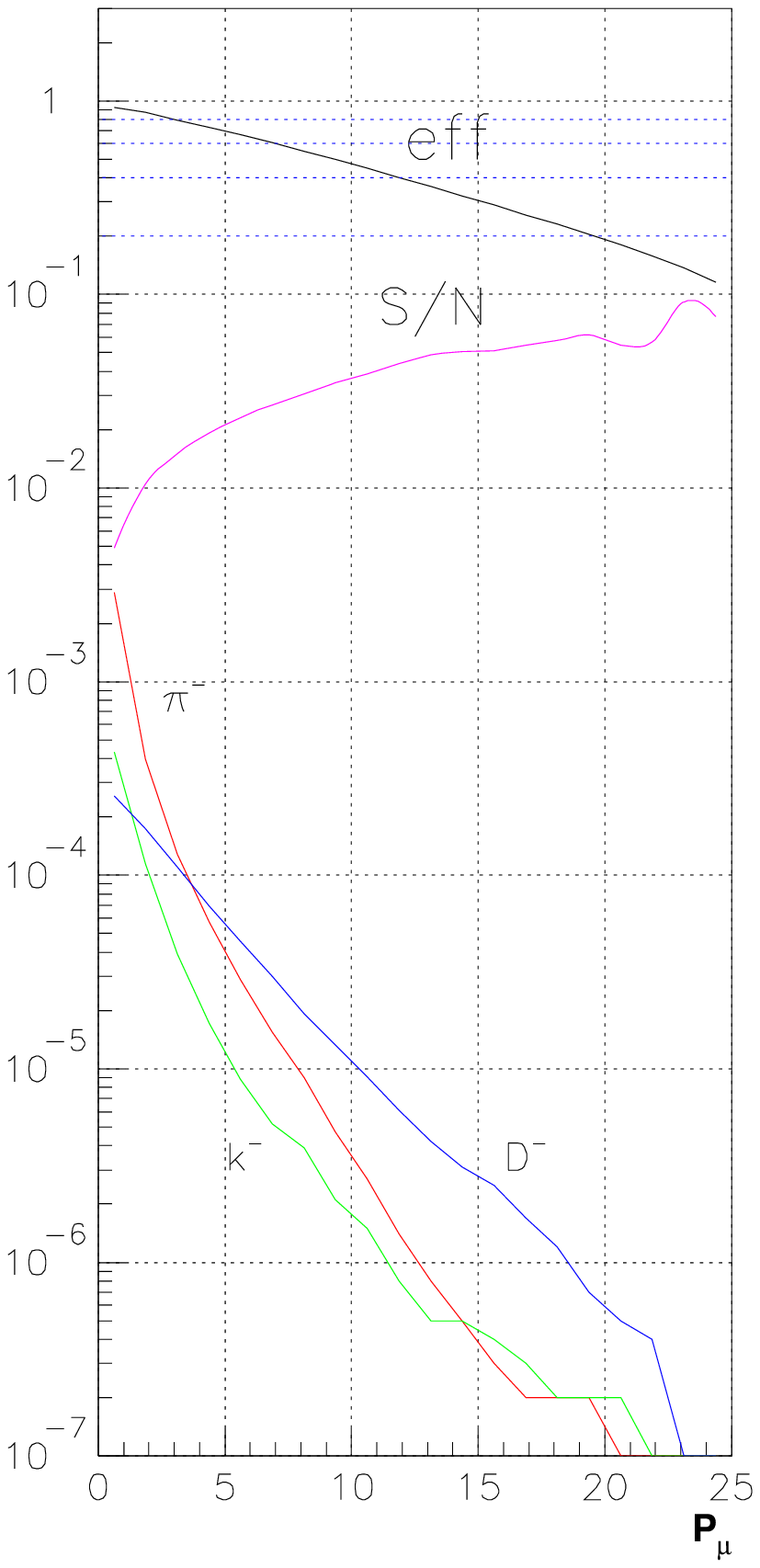,width=8cm,height=8cm}
\end{center}
\end{minipage}
\begin{minipage}[h]{0.45\textwidth}
\begin{center}
\epsfig{file=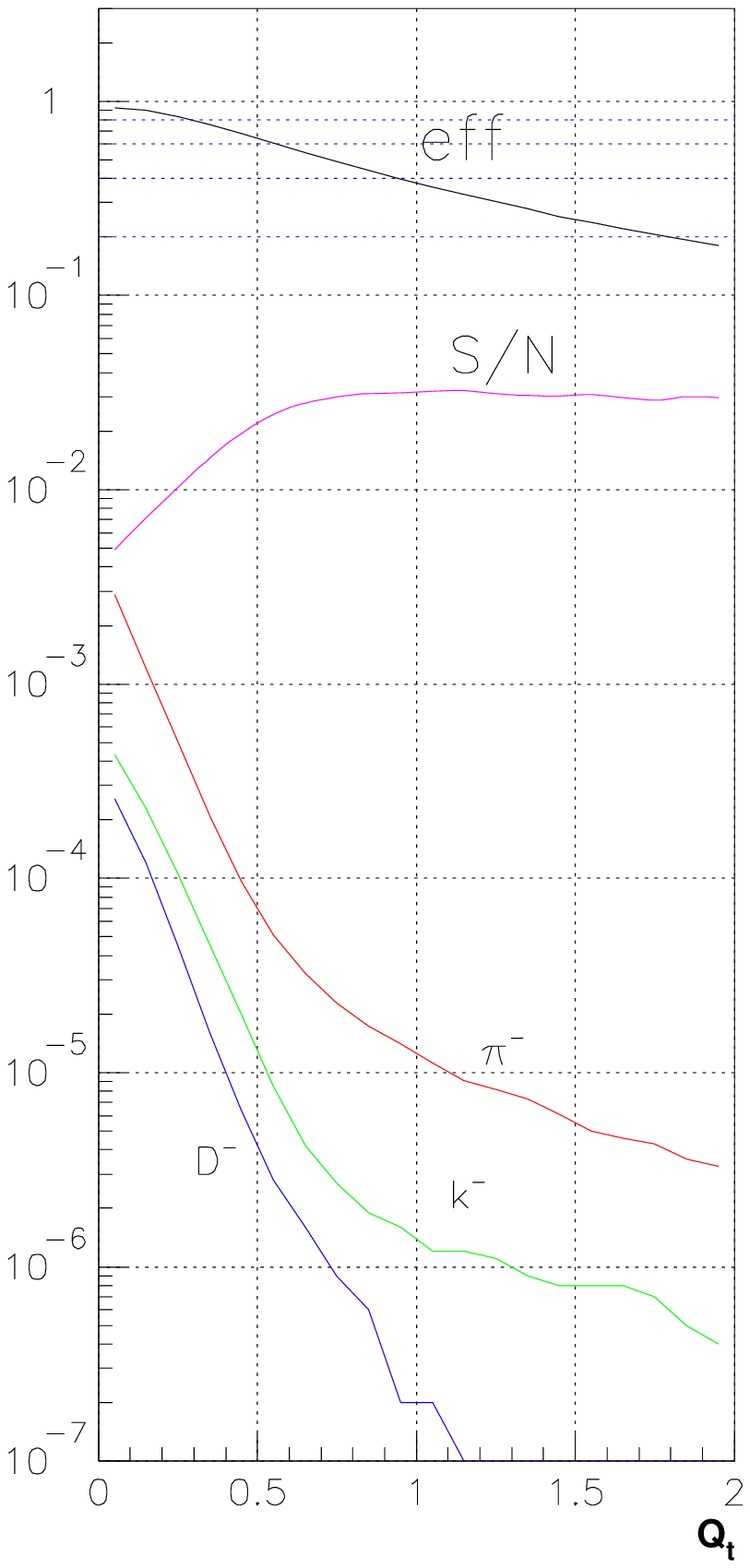,width=8cm,height=8cm}
\end{center}
\end{minipage}
\label{fig:bg}
\caption{\it Backgrounds and efficiencies for wrong-sign muons, 
for charged-current (upper plots) and neutral-current (lower
plots) events. We display the dependence on the cut on muon momemtum
(left plots) and on the $Q_t$ of the muon with respect to the hadronic
jet (right plots)~\cite{golden}.}
\end{figure}

As in the previous case, the parameter $\theta_{13}$ is
extracted from a fit to the energy distribution of the various
classes, and in this case the fit is largely dominated by the
wrong-sign muons. Moreover, from the formula of the oscillation 
probability we see that the value of $\theta_{13}$ has a limited 
influence on the spectral shape, and even factorizes out from the
energy dependence in the approximation $\Delta m^2_{12}=0$, see (\ref{eq:nuenumusimp}),
so most of the information actually comes from just counting
wrong-sign muon events.\par
\begin{figure}[tbhp]
\begin{minipage}[h]{0.45\textwidth}
\begin{center}
\epsfig{file=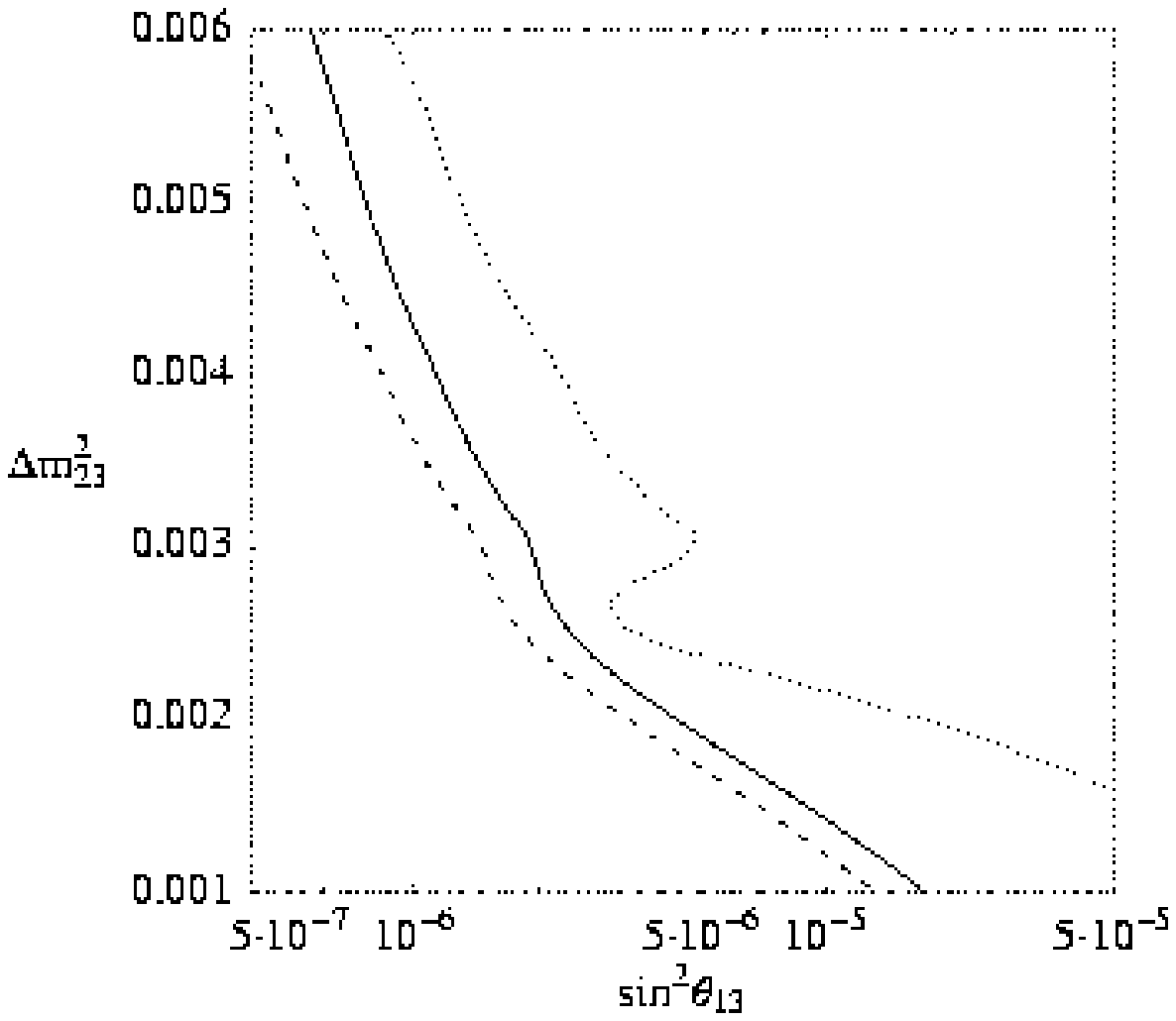,width=8cm}
\end{center}
\end{minipage}
\begin{minipage}[h]{0.45\textwidth}
\begin{center}
\epsfig{file=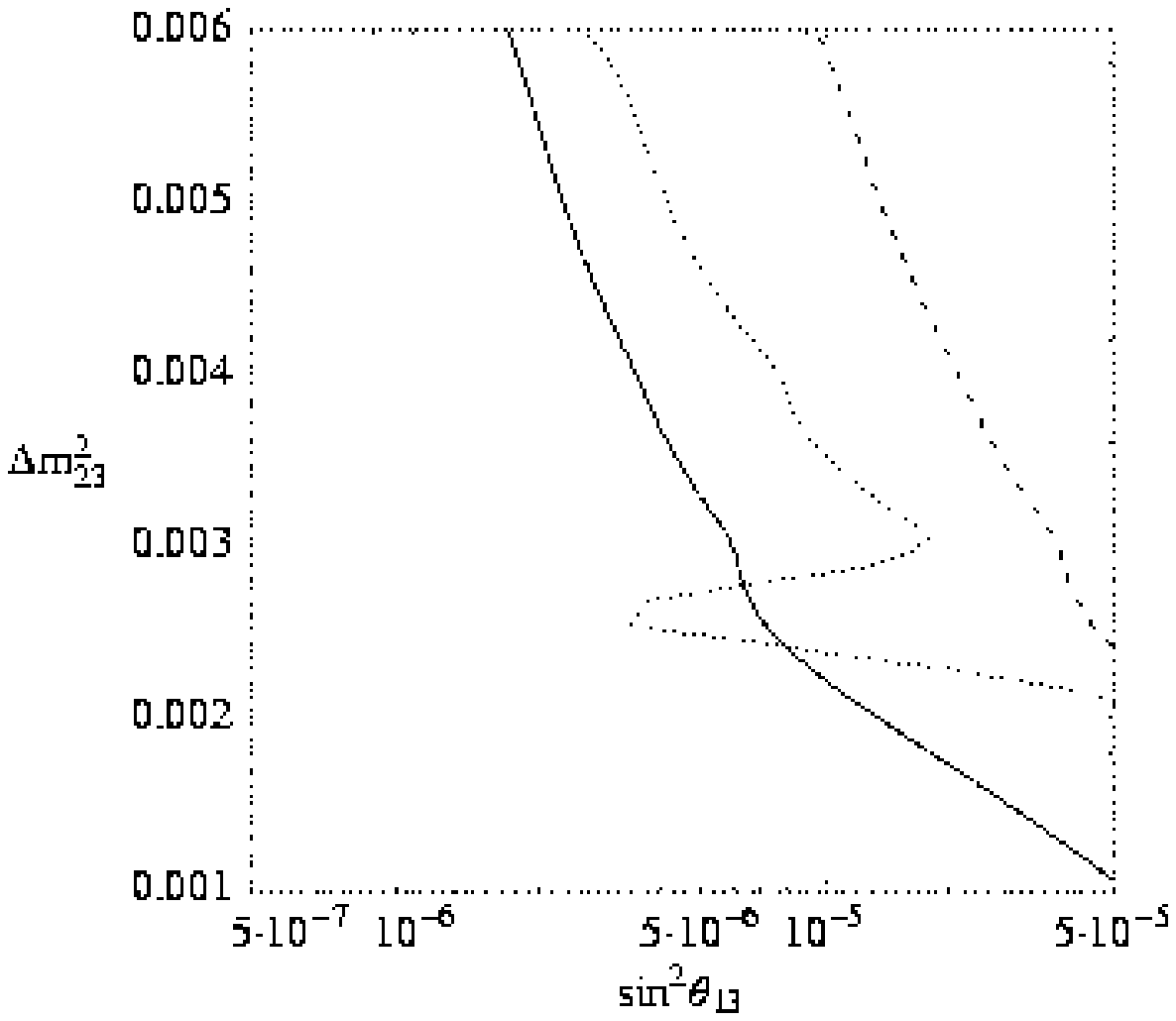,width=8cm}
\end{center}
\end{minipage}

\caption{\it Sensitivity to $\sin^2 \theta_{13}$ for a magnetized iron 
detector,
in the case of no backround (left) and when a realistic background is
considered (right). The three lines correspond to baselines of 730 (dashed), 3500 (solid) and
7300~km (dotted). Both plots were made assuming negligible effects of the solar parameters (i.e. for the small angle or small $\Delta m^2_{12}$ solutions of the solar neutrino deficit)~\cite{golden}.}
\label{fig:t13}
\end{figure}

The background level is the ultimate limiting factor for
this measurement, and pushing background rejection to the high
values quoted above allows one to reach sensitivities of the order of 
$\sin^2 \theta_{13}<10^{-5}$ for the optimal intermediate baseline (see 
the right plot of Fig.~\ref{fig:t13}). 
In the ideal case of no background and 100\% efficiency for the signal, 
almost another order of magnitude on the sensitivity could be
gained (left plot of Fig.~\ref{fig:t13}).

\subsubsection{Matter effects}

For practical reasons, long-baseline experiments can only be
performed with neutrinos passing through the Earth. We know that
electron neutrinos traversing matter interact with the electrons
of the material, leading to a modification of the effective mixing matrix,
and therefore of the oscillation probability, in a way which is asymmetric
 between neutrinos and antineutrinos. We have seen in section 
\ref{sect:oscmatter} that the oscillation probabilities including matter 
effects in the context of three-family oscillations are rather simple 
when the solar mass difference, $\Delta m^2_{12}$, can be neglected. 

According to (\ref{eq:nuenumumatter}), the oscillation probability $\nu_e
\rightarrow \nu_\mu$ has the same form as in vacuum (\ref{eq:nuenumusimp})
if $\theta_{13}$ is modified to:
%
\begin{equation}
\sin^2 2 \theta^m_{13}(D)=\frac{\sin^2 2 \theta_{13}}{\sin^2 2 \theta_{13}+(\frac{D}{\Delta m_{23}^2}-\cos 2\theta_{13})^2}
\label{eq:t13matter}
\end{equation}
and the effective atmospheric mass splitting is modified to
\begin{equation}
\Delta M^2_{23}\simeq\Delta m^2_{23}\sqrt{{\sin^2 2 \theta_{13}+
(\frac{D}{\Delta m^2_{23}}-\cos 2\theta_{13})^2}}.
\end{equation}
where
\[D=\pm2\sqrt{2} G_F n_e E=7.65\times 10^{-5} eV^2 (\frac{\rho}
{g cm^{-3}})(\frac{E}{GeV}),\]
with $n_e$ being the average electron density in the path crossed, 
$\rho$ the average electron density, and the sign $\pm$ standing
for neutrinos and antineutrinos, respectively. The quantity
$\Delta m_{23}^2$ can also be positive in the case of a hierarchical spectrum or negative in the case of the inverted hierarchy. In the first case,
 the denominator of (\ref{eq:t13matter}) is always larger than 1 for antineutrinos, and thus 
the effective angle is always smaller than the vacuum one.
For neutrinos, the resonance condition will be met for
\[D\simeq \Delta m_{23}^2 \cos 2\theta_{13},\]
for which value the effective angle becomes unity, no matter what
the vacuum value was. It is therefore possible to have large
enhancements of the oscillation probability, around the resonance
energy
\[E^{res}\approx\frac{1.32\times 10^4 \cos 2\theta \Delta m^2(eV^2)}
{\rho(g/cm^3)}.\]
Note however that the effective mass splitting at resonance decreases
linearly  with the vacuum value of $\sin 2 \theta_{13}$, so, in order
to get large probabilities, one needs to go to larger baselines for 
smaller
$\sin 2 \theta_{13}$. 
If the sign of $\Delta m_{23}^2$ is negative, the same reasoning
applies, interchanging neutrinos and antineutrinos.

The asymmetry between neutrinos and antineutrinos due to the matter
effect is a very effective way to measure the sign of $\Delta m_{23}^2$, and
establish the degenerate or hierarchical structure of neutrino
masses. It has never been observed, and will be hard 
to detect with conventional 
super-beams, since it requires the comparison of runs with neutrinos
and antineutrinos, and baselines larger than 1000 km. 
On the other hand, the crossing of such an MSW 
resonance is a plausible explanation for reconciling the large
disappearance of solar neutrinos from the Sun with the lack of a
significant seasonal variation, as would be expected from the eccentricity 
of
the Earth orbit if the oscillation was taking place in the vacuum.
Unfortunately, solar neutrino experiments cannot tell us anything
about the sign of $\Delta m_{23}^2$.

\begin{figure}[tbhp]
\begin{center}
\epsfig{file=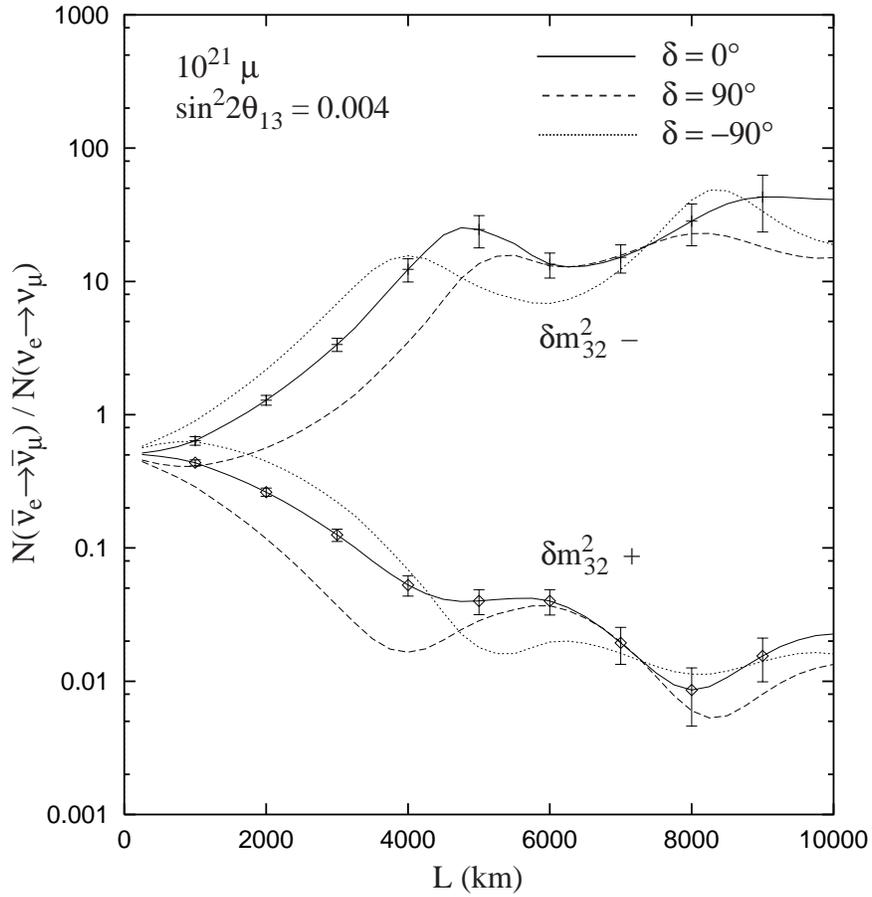,width=12cm}
\caption{\it Ratio of the oscillation probabilities of neutrinos and
antineutrinos as a function of the baseline. The different lines 
correspond to values of the CP-violating phase $\delta$ of $0^\circ$,
$90^\circ$ and $-90^\circ$. After about 1000~km, matter effects
dominate over CP violation, whose effects even go to zero
at a particular baseline around 7500 km~\cite{bargermsw}.}
\label{fig:mswcpvsl}
\end{center}
\end{figure}

Matter asymmetries on earth are relatively easy to detect at
a Neutrino Factory~\cite{mswfirst,bargermsw}.
Although this is good news for the measurement of the sign of $\Delta m^2_{23}$, these asymmetries actually interfere with the 
measurement of leptonic CP violation, which is the most ambitious goal 
of this machine. The relative amplitudes of the two effects as
a function of the baseline can be seen in Fig.~\ref{fig:mswcpvsl}.
It shows, on a logarithmic scale, the ratio of events that have 
oscillated 
into muon antineutrinos to events that have oscillated into muon 
neutrinos, 
as a function of the baseline. The measurement of such a ratio requires
running the machine alternatively with positive and negative
muons in the ring. The upper lines correspond to negative values
of $\Delta m_{23}^2$, and the lower ones to positive values.
The three lines are obtained for different values of the 
CP-violating parameter $\delta$, that will be discussed in more
detail in the next section, and the error bars correspond to
statistical errors for a 40~kt detector after $10^{21}$ muon
decays in the ring for each muon charge. We see that the
asymmetry due to matter effects is quite large, and can easily be
detected for baselines larger than about 1000~km. 
Matter effects and CP violation interfere, and one effect can mask the 
other, until baselines of about 2000 km. Above that distance matter
effects are always larger than CP violation, which is even completely
cancelled for one particular distance around 7500~km.\par

\subsubsection{CP violation}

\begin{figure}[tb]
\begin{center}
\epsfig{file=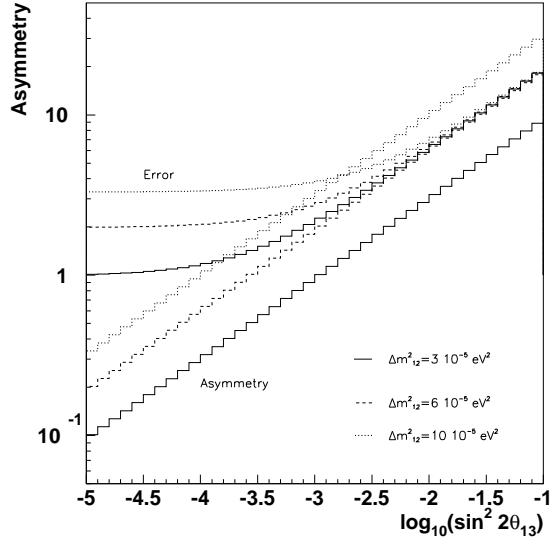,width=8cm}
\caption{\it The CP-violating asymmetry and its error as a 
function of $\theta_{13}$, for different values of 
$\Delta m^2_{12}$.}
\label{fig:errorcp}
\end{center}
\end{figure}

Detecting the presence of a complex phase in the leptonic
mixing matrix is one of the most ambitious goals of neutrino
physics, and would justify the effort of building a neutrino
factory~\cite{CPviol,dickfreund,dongav,Romanino:1999zq,golden,physpot}.
In vacuum, we saw in (\ref{eq:nuenumuvacuum}) that the difference 
between the oscillation
probabilities for the transitions $\nu_e\to\nu_\mu$
and $\bar{\nu}_e\to\bar{\nu}_\mu$ can be written as
\begin{eqnarray}
\Delta_{CP} \sim \frac{1}{2} c_{13}\sin 2\theta_{12}\sin 2\theta_{13}\sin 2\theta_{23}
\sin\delta[\sin \Delta_{12}\sin \Delta_{13}\sin \Delta_{23}], 
\end{eqnarray}
which shows that, for the large mixing angle solution of the solar 
deficit, 
it is suppressed only in the parameters $\Delta_{12}$ and $\theta_{13}$.
Since the CP-even parts of the probabilities are always larger than the CP-odd
parts, they dominate the number of events and thus the error on the measured asymmetry. For small enough $\Delta_{12}$, these terms are approximately proportional 
to $\sin^2 2\theta_{13}$ and independent of $\Delta_{12}$ (they are driven by the atmospheric oscillation, i.e. the first term in (\ref{vacexpand})). 
Assuming Gaussian errors, the
significance of CP violation will then be proportional to
\begin{eqnarray}
  \frac{\Delta_{CP}}{\sqrt{N_{ev}}}\propto \frac{\sin 2\theta_{13}} 
{\sqrt{\sin^2 2\theta_{13}}}  \;\Delta_{12},
\label{eq:errorcp}
\end{eqnarray}
and will not depend on $\theta_{13}$~\cite{Romanino:1999zq}, while it is 
linearly proportional to the small mass splitting. 
The fact that $\Delta_{12}$
is actually different from zero means that the subleading oscillation
will become important and eventually dominate for small $\theta_{13}$.
 In this regime the role of $\sin 2 \theta_{13}$ and $\Delta_{12}$ are 
interchanged and the significance of the CP asymmetry is suppressed 
linearly in $\sin 2 \theta_{13}$ and approximately independent of the solar mass splitting
\begin{eqnarray}
  \frac{\Delta_{CP}}{\sqrt{N_{ev}}}\propto \frac{\Delta_{12}} 
{\sqrt{(\Delta_{12})^2}}  \;\sin 2 \theta_{13}. 
\end{eqnarray}
  This is shown in 
Fig.~\ref{fig:errorcp}, where the CP asymmetry at the maximum of the 
oscillation probability is compared to its error (computed for a 
relatively modest neutrino flux) as a function of $\sin^2 2\theta_{13}$.
We see that the effect and the error decrease together with 
$\sin^2 2\theta_{13}$ until a value (that depends on $\Delta_{12}$)
where the subleading oscillation becomes dominant, and the number of
oscillated events goes to a constant even for very small $\theta_{13}$.
In this regime the significance of the asymmetry decreases linearly 
with $\sin 2 \theta_{13}$. 

A possible measurement of this asymmetry is given by the integrated 
asymmetry~\cite{CPviol}:
\begin{eqnarray}
A^{CP}_{e\mu}=\frac
{\{N[\mu^-]/N_0[e^-]\}_+-\{N[\mu^+]/N_0[e^+]\}_-}
{\{N[\mu^-]/N_0[e^-]\}_++\{N[\mu^+]/N_0[e^+]\}_-},  
\end{eqnarray}
where the suffix defines the sign of the decaying muon, 
$N[\mu^\pm]$ is the measured number of wrong-sign muons
and $N_0[e^\mp]$ the expected number of $\bar{\nu}_e$ and
$\nu_e$ charged-current interactions in the absence of
oscillations. The statistical significance of the
asymmetry, i.e., the asymmetry divided by its error, assumed
to be Gaussian, in vacuum is proportional to \cite{dongav}
\begin{eqnarray}
  \frac{A^{CP}}{\Delta A^{CP}}\propto \sqrt{E}\left|\sin\left(
\frac{\Delta m^2_{23} L}{4 E_\nu}\right)\right|.
\end{eqnarray}
This is maximal at the peak of the atmospheric oscillation.

\begin{figure}[tb]
\begin{minipage}[h]{8cm}
\begin{center}
\epsfig{file=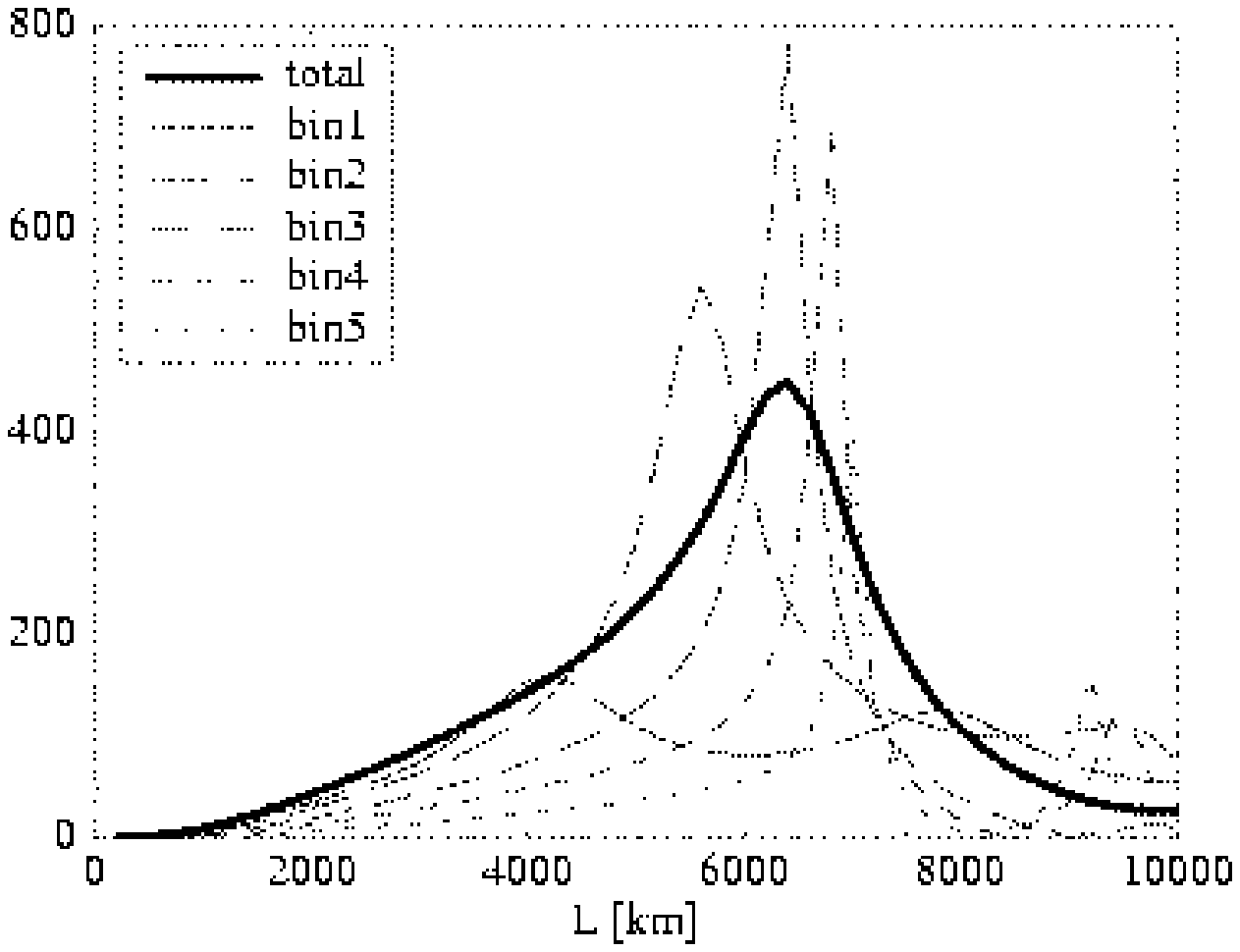,width=8cm}
\end{center}
\end{minipage}
\begin{minipage}[h]{8cm}
\begin{center}
\epsfig{file=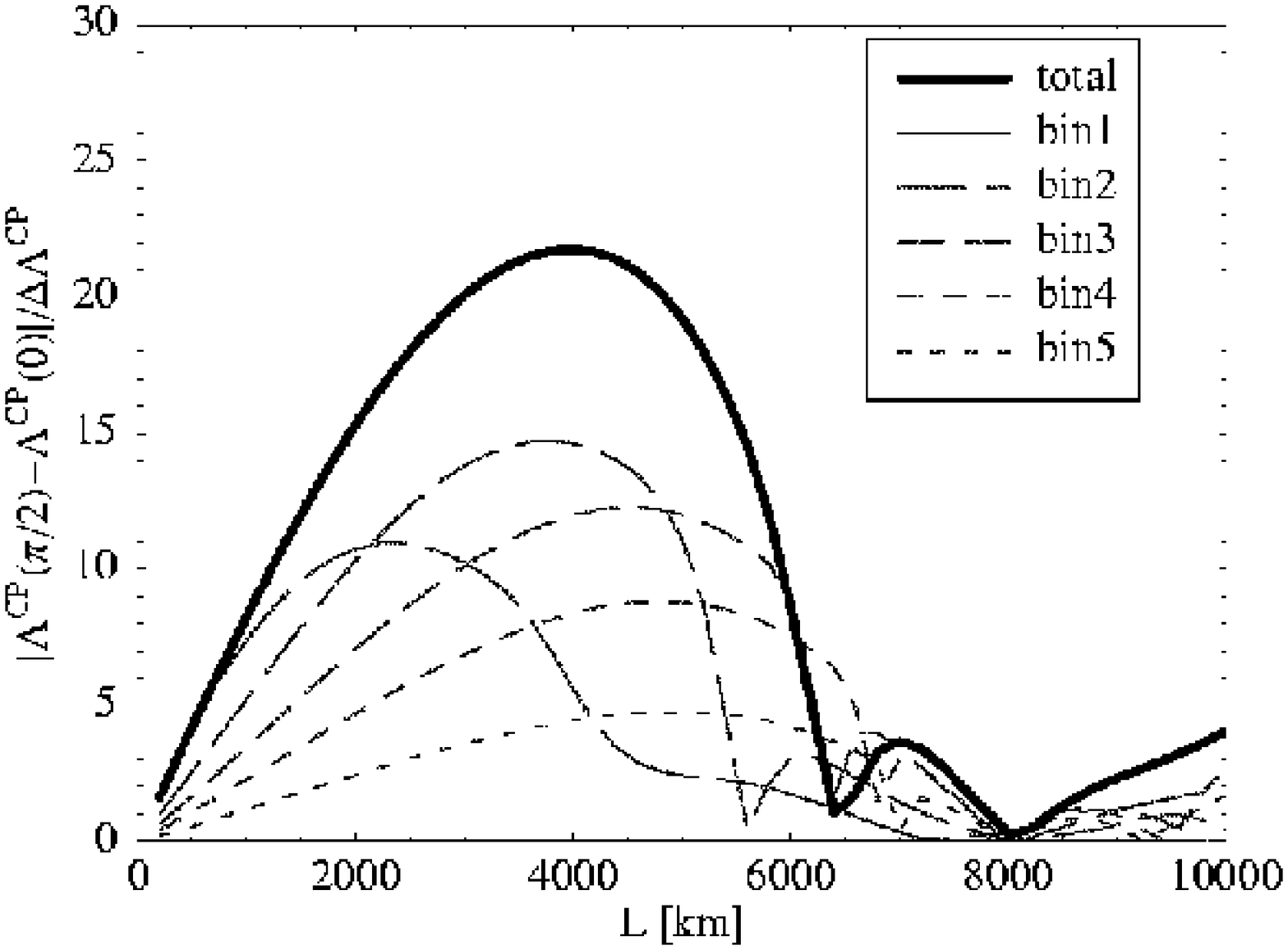,width=8cm}
\end{center}
\end{minipage}
\caption{\it The significance of the CP asymmetry 
as a function of the baseline~\cite{golden}. On the left, the asymmetry 
is just the matter-induced asymmetry (SMA solution parameters). On the right
we show, for LMA parameters, a subtracted asymmetry directly measuring 
$\delta$. }
\label{fig:asymplot}
\end{figure}

These considerations, valid in vacuum, get however modified by matter
effects, as already seen in Fig.~\ref{fig:mswcpvsl}.
In matter, very large asymmetries are created by the MSW
effect. The left plot of Fig.~\ref{fig:asymplot} shows the total 
asymmetry in matter as a function of the baseline, while the one on the right shows the asymmetry obtained after subtracting the matter-induced asymmetry for $\delta=0$. Clearly 
the significance of the matter-induced asymmetry and the intrinsic one peak 
at different baselines. 

Although the right plot shows that for the parameters in the LMA range, 
intrinsic CP violation should be 
measurable at distances ${\cal O}(3000)$ km, this analysis assumes only 
experimental errors and not `theoretical' ones. Given that the 
matter-induced asymmetry depends
very sensitively on the oscillation parameters, mainly $\theta_{13}, \theta_{23}$ and $\Delta m^2_{23}$ as well as on the Earth matter density, it is 
very important to clarify whether the expected uncertainties 
in those `theoretical' parameters will be small enough not to wash out 
the effect. 

Concerning the atmospheric parameters $|\Delta m^2_{23}|$ and $\sin^2 
2\theta_{23}$, we have seen that they can be measured
with very good precision using disappearance measurements at the neutrino 
factory. 
It was shown \cite{ms} that the knowledge of the precise profile of the
matter density along the neutrino path is not very relevant, but for a neutrino
experiment only the knowledge of the mean value is needed. This average 
density can probably be known with $5\%$ accuracy  
\cite{geller}, or even directly measured with a $10\%$ precision 
\cite{physpot}.\par
\begin{figure}[t]
\begin{center}
\includegraphics*[width=10cm,clip]{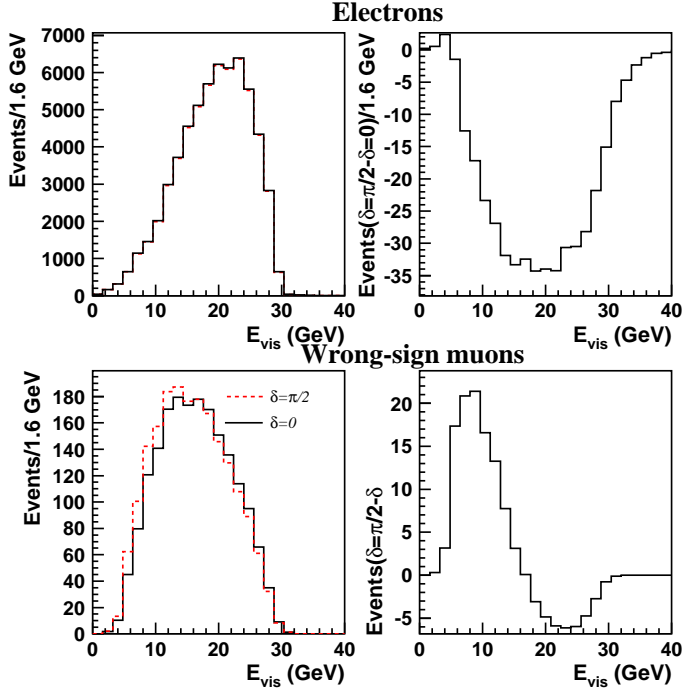}
\caption{\it Left plots: number of oscillated events for $\delta=0$
(black histogram) and $\delta=\pi/2$ (red histogram).
The right plots represent the difference of the two, for 
$\nu\mu\to\nu_e$ and $\nu_e\to\nu_\mu$ transitions, respectively.
If the difference is similar, the signal in the electron channel
is invisible due to the vey large background. In the muon channel,
we notice the the shift of the oscillation maximum generated by CP
violation~\cite{physpot}.}
\label{fig:cpdiff}
\end{center}
\end{figure}

On the other hand the angle $\theta_{13}$ as well as the discrete choices of the sign of 
$\Delta m^2_{23}$ and  $\cos \theta_{23}$ if $\theta_{23} \neq \pi/4$, will have to be 
determined simultaneously using the wrong-sign muon signals.  The analysis 
has to be refined to include these unknowns and, as we will see in the next
sections, this will reveal the existence of
\begin{itemize}
\item Strong parameter correlations: mainly between $\theta_{13}$ and 
$\delta$ \cite{golden},
\item Degenerate solutions~\cite{jordi,bargerdeg}.
\end{itemize}
In order to minimize and resolve these, it is important to use all  
the spectral information available and choose appropiately the baseline. 
Fig.~\ref{fig:cpdiff} shows the energy dependence of the intrinsic effect 
of $\delta \neq 0$. In the global fit of course we consider the wrong-sign 
muon signals for both $\mu^+$ and $\mu^-$ and not only the CP asymmetries, 
since also the
CP-even parts contain very valuable information on the unknowns. 

\subsubsection{Correlations and choice of baseline}
\begin{figure}[tbhp]
\begin{center}
\includegraphics*[width=12cm]{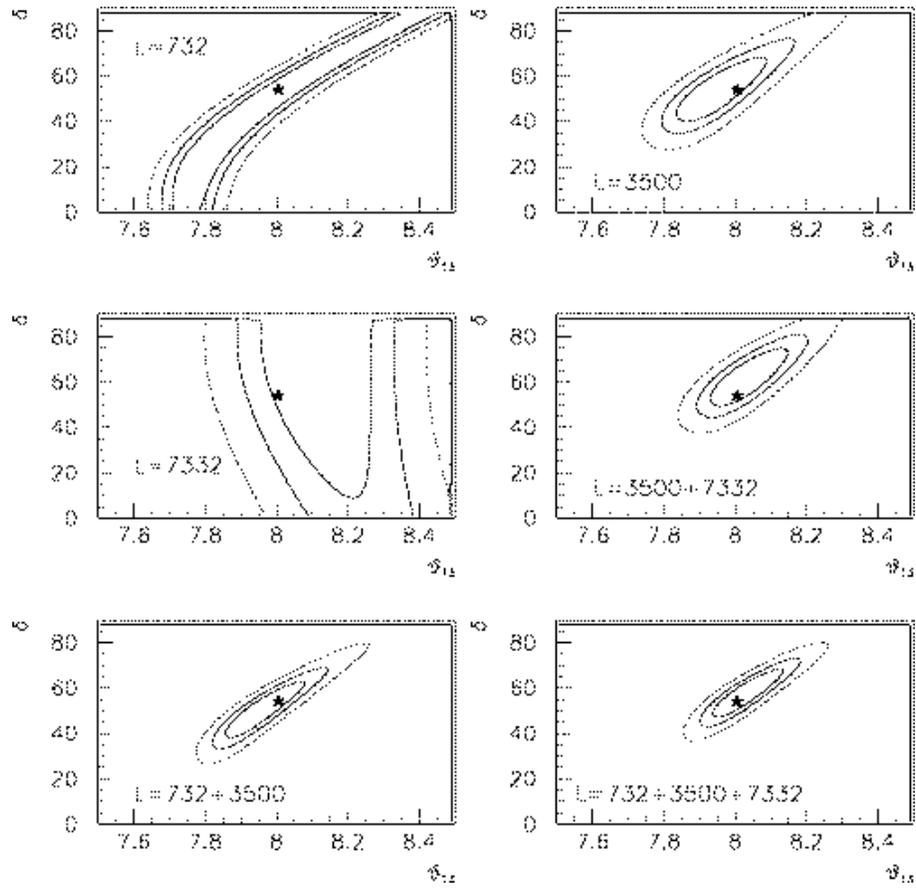}
\caption{\it Correlations in a simultaneous fit of $\theta_{13}$ and 
$\delta$,
using a neutrino energy threshold of about 10 GeV.
Using a single baseline correlations are very strong, but can be 
largely reduced by combining information from different baselines and detector 
techniques~\cite{golden}.}
\label{fig:correlations}
\end{center}
\end{figure}

\begin{figure}[tbhp]
\begin{center}
\includegraphics*[width=10cm,clip]{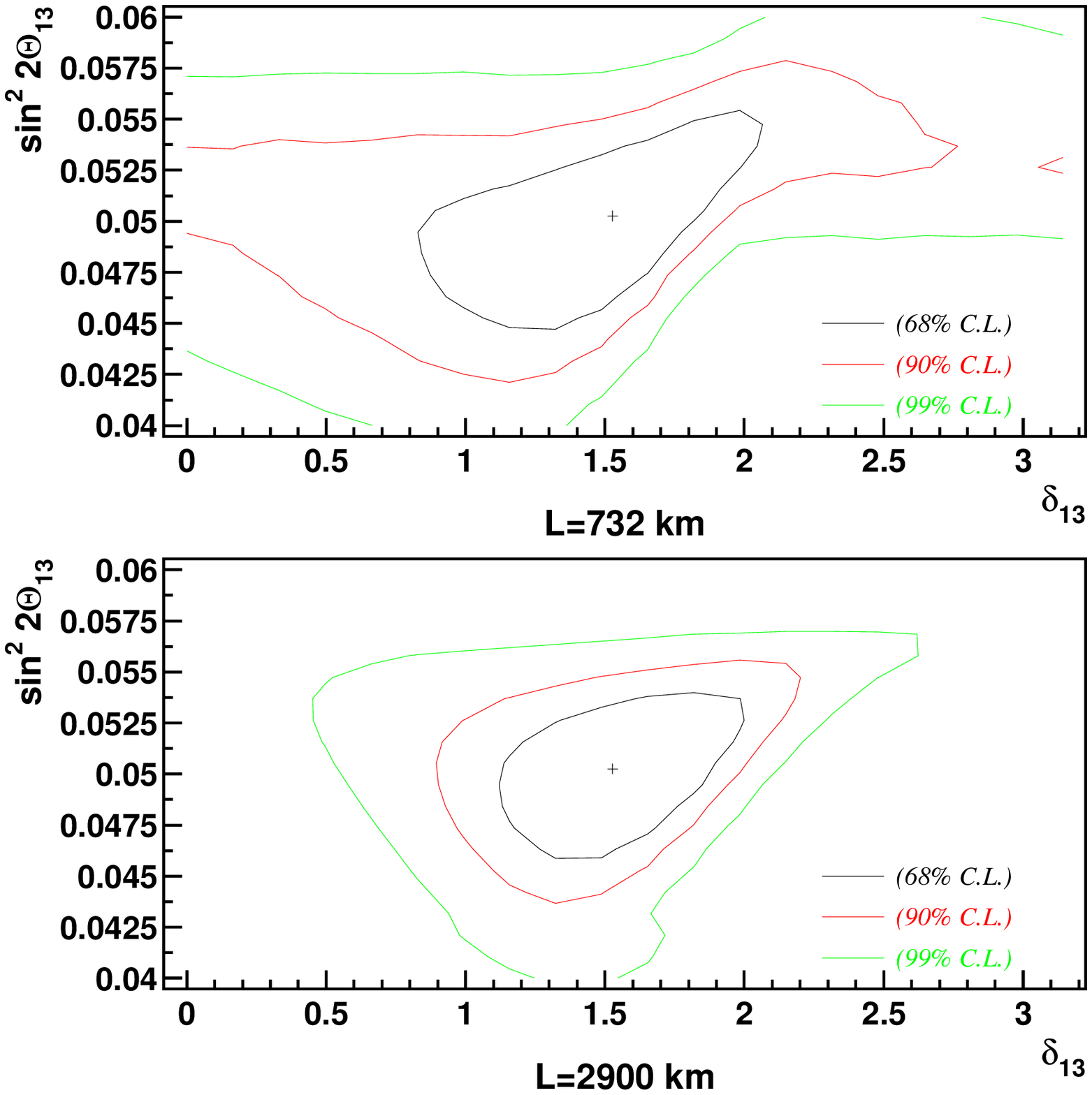}
\caption{\it Contours in the $\sin^2 2\theta_{13}-\delta$ plane assuming
very low energy threshold (2~GeV muon energy) at distances of
732 or 2900~km. Even if not as pronounced as in the case of the previous
plot, done with
much higher neutrino energy thresholds, the short baseline starts to 
show correlations between the two parameters~\cite{physpot}.}
\label{fig:deltavst13}
\end{center}
\end{figure}

Due to the small energy dependence induced by CP violation,
the use of spectral information to have a simultaneous measurement of
$\delta$ and $\theta_{13}$ is not very effective, and only helps under
some conditions.
In fact, the first simultaneous fits of $\theta_{13}$ and $\delta$ perfomed in 
\cite{golden,physpot} revealed for most of the cases a strong correlation 
between the two parameters, which was more important at shorter 
baselines, see Figs.~\ref{fig:correlations} and ~\ref{fig:deltavst13}. 
 The reason for this is easy to understand. The neutrino and antineutrino 
oscillation probabilities in vacuum 
(\ref{eq:nuenumuvacuum}), have the following dependences on
$\theta_{13}$ and $\delta$:

\begin{eqnarray}
P(\nu_e\to\nu_\mu) P(\bar{\nu}_\mu\to\bar{\nu}_e) &=& A \sin^2 2 \theta_{13} + B \sin 2 \theta_{13} \cos \delta \pm C \sin 2 \theta_{13} \sin \delta + D
\label{eq:cpprob2} 
\end{eqnarray}
where the coefficients $A,B,C,D$ depend on the energy, baseline and the 
remaining oscillation parameters. In matter a similar structure is found,
though the
coefficients are different for neutrinos and antineutrinos and depend 
also on the earth matter density. In the limit of large $E_\nu$ or small 
$L$ ($\Delta_{12} < \Delta_{23} \ll 1$), the coefficients $A, B$ and $D$ have 
the same energy dependence, while $C$ is 
suppressed by an extra factor $\Delta_{23}$ compared to the others. 
This is so both in vacuum and in matter, because matter effects are 
irrelevant in this limit. In this situation it is clear that the measurements 
of the 
oscillation probabilities for neutrinos and antineutrinos give information on 
the same combination of the two unknowns: $\theta_{13}$ and $\delta$
\begin{eqnarray}
A \sin^2 2 \theta_{13} + B \sin 2 \theta_{13} \cos \delta + D,
\label{eq:corr} 
\end{eqnarray}
 resulting in the strong correlation~\cite{golden} 
that can be nicely seen in the left plot of Fig.~\ref{fig:nufactsuperb}, at
a baseline of $L=732$ km. Even though the spectral information has been 
used for this baseline,  it is only relevant in the low-energy region. 
The same fit performed with a lower energy threshold in
Fig.\ref{fig:deltavst13}
shows that correlations are smaller, but still present.
As the baseline is increased  the CP-odd term, i.e., the coefficient $C$, 
grows and the neutrino and antineutrino probabilities measure
two independent combinations of the two parameters. Also the energy 
dependence in the accessible energy range becomes more important. Indeed,
the 
fits at $L=3500$~km in Fig.~\ref{fig:correlations} and at $L=2900$~km in
Fig.~\ref{fig:deltavst13} show that 
$\delta$ and $\theta_{13}$ can be disentangled. At even larger baselines,  
sensitivity to $\delta$ is lost.\par 
How much can the neutrino energy threshold actually be lowered in order to
reduce correlations, using spectral information in the important 
low-energy
region? The answer to this question depends of course on the detector,
but since we want to discriminate the oscillation signal from a constant
background, also on the amplitude of the oscillation signal, i.e., on 
the
oscillation parameters. For small values of $\theta_{13}$,
the good purity needed to discriminate the wrong-sign muon signal from
background requires very hard cuts on the muon momentum, and therefore the
loss of the most interesting part of the neutrino spectrum. 
For instance, a magnetized iron
detector can in principle reach background rejection factors of the order
of $10^6$, but at the price of an effective cutoff on neutrino energies of
about 10~GeV.

The conclusion of this study was that in the case of one baseline, the optimal choice to study 
CP violation was in the intermediate range $L= {\cal O}(3000)$ km 
for a machine of $E_\mu = 30-50$~GeV~\cite{golden}.\par

\par

All the above considerations pointing to the optimal distance of 
${\cal O}(3000)$km
have assumed a constant beam flux. It is possible to argue that a
lower-energy machine could profit from savings on the accelerator
and provide a larger flux than a high-energy one. If we assume that, for
various possible Neutrino Factory designs, the flux can be inversely
proportional to the beam energy (in other words, the total beam power is
constant), this fact compensates the linear sensitivity growth, and it can
be shown \cite{cptoptim} that all energy/baseline combinations that 
keep constant the ratio of these two quantities are equivalent, with a
slight preference for the lower energy-baseline combinations, due to the
smaller influence of matter effects. This study assumes a
neutrino energy threshold of 2 GeV, for which a baseline of 732 km and a 
maximal neutrino energy of 7.5 GeV are slightly better than a baseline of
2900 km and an energy of 30 GeV, if the flux of the first case is 4 times 
larger than the second one. However, a coarse detector with larger energy
thresholds will always benefit from a longer baseline and higher neutrino
energy~\cite{freund}.\par
Correlations of the phase $\delta$ with other oscillation parameters have
also been studied in several works \cite{sato,jordi,freund,yasuda,physpot}.
The main conclusions are:
\begin{itemize}
\item The parameters of the leading oscillation will be measured 
independently
by the Neutrino Factory itself using the same-sign muon sample, and the
precision reached will be much superior than that producing any sizeable 
systematic effect on the measurement of $\delta$. An exception are sign and
octant ambiguities, that can produce degenerate solutions, and will be 
discussed in the next section.
\item Knowledge of the parameters governing solar neutrino oscillations 
($\theta_{12}$ and $\Delta m_{12}$) is essential, but the indications are
that KamLand results should be precise enough to reduce to negligible levels
the impact of the uncertainties on these measurements to the determination of
$\delta$.
\item The average earth density can probably be known up to 5\% using 
geophysical techniques \cite{geller}; however, a precision at least a
factor 2 better would be desirable.
\end{itemize}


\subsubsection{Degeneracies}

We have seen in  the previous section that it is very hard to disentangle
$\theta_{13}$ and $\delta$ at short baselines of ${\cal O}(1000)$ km, 
because 
the neutrino and antineutrino probabilities are sensitive to approximately 
the same combination of the two parameters. At larger baselines the correlation is resolved thanks mainly to the larger $\sin \delta$ term in (\ref{eq:corr}). However, even in this case, 
the impossibility of measuring precisely the energy dependence of the 
wrong-sign muon signals can result in the existence of degenerate 
solutions for the $\theta_{13}$ and $\delta$, i.e., disconnected areas of 
the parameter space that lead to experimentally 
indistinguishable signatures~\cite{jordi}. The origin of this is that, at 
fixed
neutrino energy and baseline, the equations 
\begin{eqnarray}
  P_{\nu_e\nu_\mu}(\theta'_{13},\delta',\Theta')\simeq P_{\nu_e\nu_\mu}(\theta_{13},\delta,\Theta)\nonumber\\
  P_{\bar{\nu}_e\bar{\nu}_\mu}(\theta'_{13},\delta',\Theta')\simeq P_{\bar{\nu}_e\bar{\nu}_\mu}(\theta_{13},\delta,\Theta)
\label{eq:probeq}
\end{eqnarray}
have more solutions than the true one: $\theta'_{13}= \theta_{13}$ and $\delta'=\delta$. We use the symbols $\Theta$ and $\Theta'$
to denote the other sets of oscillation parameters. The reason why in 
general 
$\Theta'$ might be different from $\Theta$ is due to the discrete ambiguities
of the type sign$(\Delta m^2_{23})$ and sign($\cos \theta_{23}$). The three main types of 
degeneracies correspond to different choices of $\Theta'$:

\begin{itemize}
\item Intrinsic, when $\Theta'=\Theta$, so all other parameters are the 
same, but the conditions (\ref{eq:probeq}) are satisfied for sets of 
$\theta_{13}, \delta$ different from the true one~\cite{jordi}.
\item The sign of $\Delta m^2_{13}$ is different in the two sides of the equations (\ref{eq:probeq}) \cite{bargerdeg}. In vacuum, a change of sign of 
$\Delta m^2_{13}$ is approximately equivalent to the substitution
$\delta\rightarrow \pi-\delta$ \cite{minakata}. 
\item The value of $\theta_{23}$ on the left side of (\ref{eq:probeq}) is $\pi/2-\theta_{23}$ of that on the right side~\cite{bargerdeg}. The fake 
solutions obviously collapse to the intrinsic ones for maximal mixing:  
$\theta_{23}=\pi/2$.
\end{itemize}

\begin{figure}[tbph]
\begin{center}
\includegraphics*[width=15cm]{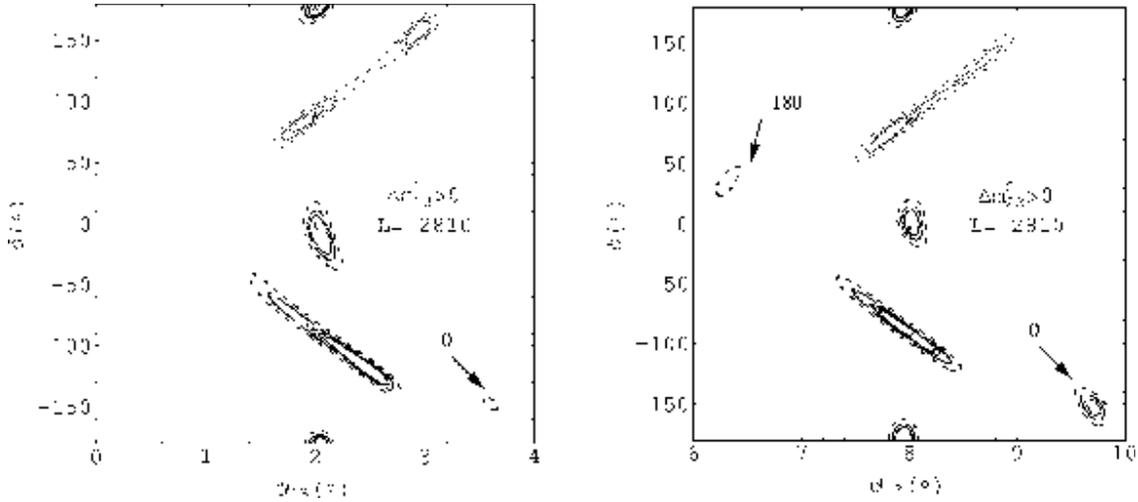}
\caption{\it Simultaneous fits of $\delta$ and $\theta_{13}$ at $L = 
2800$~km, 
for $\theta_{13}=2^\circ$ (left) and $8^\circ$ (right). True values are
indicated by stars, and many degenerate solutions are 
present~\cite{jordi}.}
\label{fig:intrinsic}
\end{center}
\end{figure}

An example of intrinsic degeneracies is shown in Fig.~\ref{fig:intrinsic},
obtained for a simultaneous fit of $\theta_{13}, \delta$ in a large
magnetised iron detector. The true values of the parameters are indicated
by the stars, while a fit to the observed data would find several
disconnected regions. These plots were made for a baseline of 2800~km.

\begin{figure}[tbhp]
\begin{center}
\includegraphics*[width=10cm,bb=180 290 432 478]{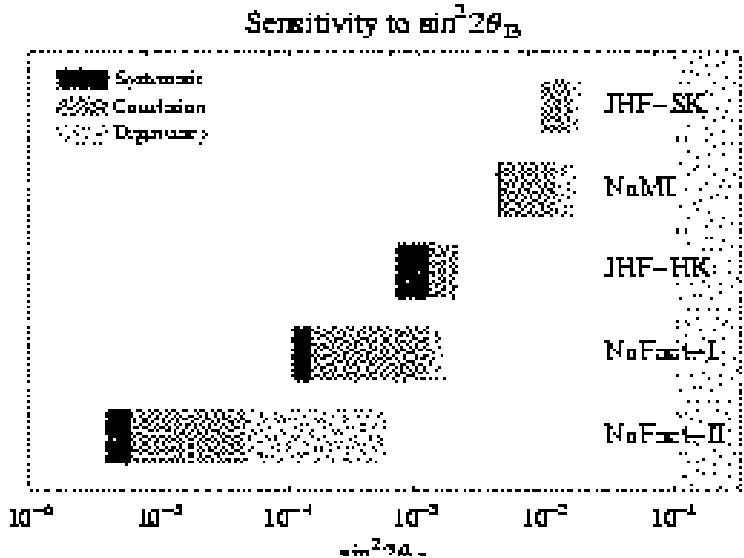}
\caption{\it Limitations on the achievable sensitivity to $\sin^2 2 \theta_{13}$ for
various assumptions on future experiments, which are described in~\cite{freund2}. The sensitivity to $\sin^2 2 \theta_{13}$ is
defined as the largest value of $\sin^2 2 \theta_{13}$ which can be fit to
$\sin^2 2 \theta_{13}=0$, which means that the degenerate solutions enter
as an additional source of uncertainties. In this interpretation, the
importance of resolving the degeneracies from the combination of various
experiments is very clear.}
\label{fig:huberdeg}
\end{center}
\end{figure}

Degeneracies can be a very serious source of uncertainties in the
determination of the parameters. 
If, for instance, we define (conservatively) the sensitivity to $\sin^2 2
\theta_{13}$ as the largest value of $\sin^2 2 \theta_{13}$ which can be
fit to $\sin^2 2 \theta_{13}=0$, we will find that the degenerate solution
can be interpreted as an additional uncertainty which makes it almost
impossible to push the sensitivity to $\sin^2 2 \theta_{13}$ much below
$10^{-3}$~\cite{freund2} (see Fig.~\ref{fig:huberdeg}). 

\begin{figure}[tbhp]
\begin{minipage}[h]{8cm}
\begin{center}
\epsfig{file=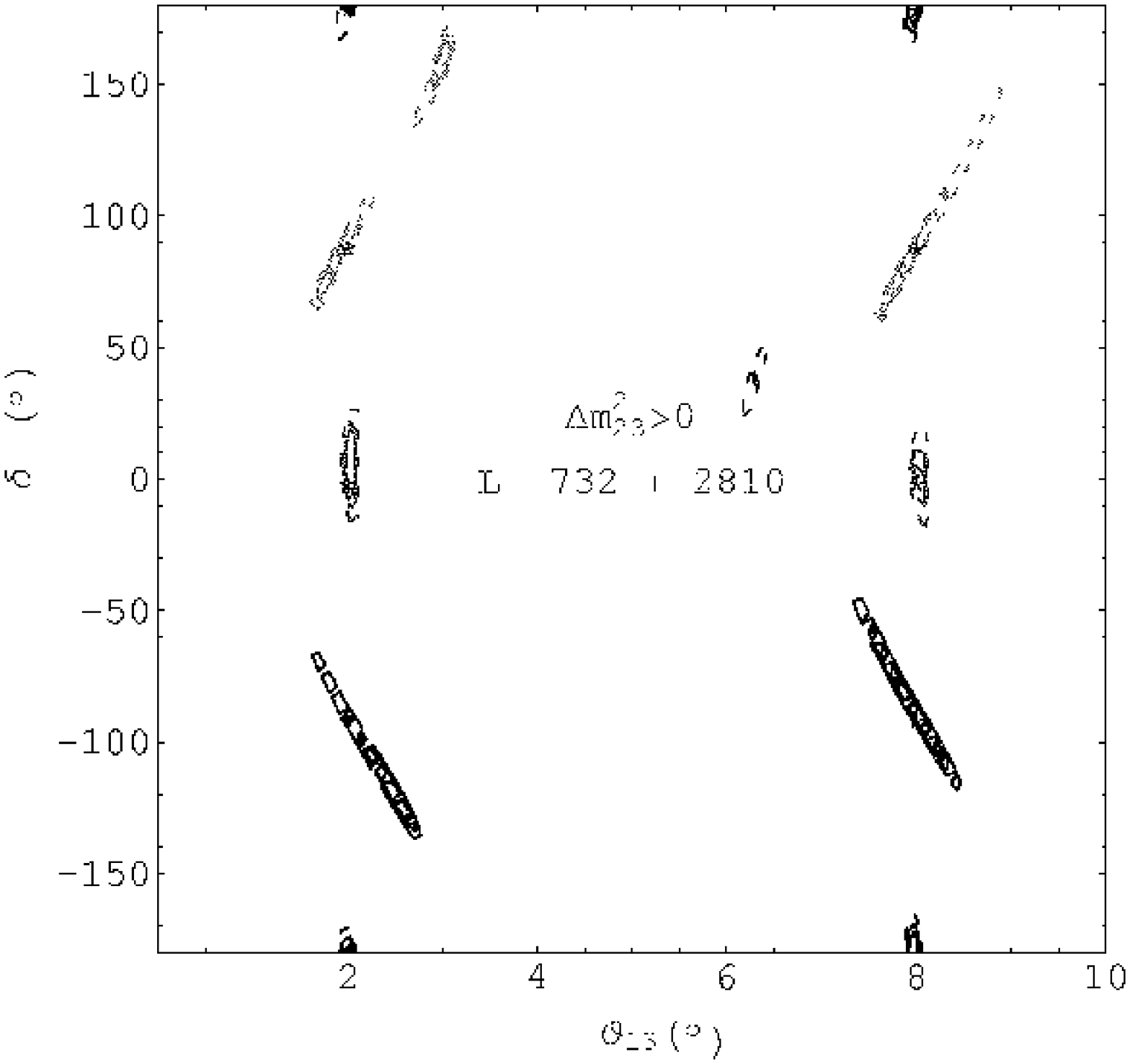,width=8cm}
\end{center}
\end{minipage}
\begin{minipage}[h]{8cm}
\begin{center}
\epsfig{file=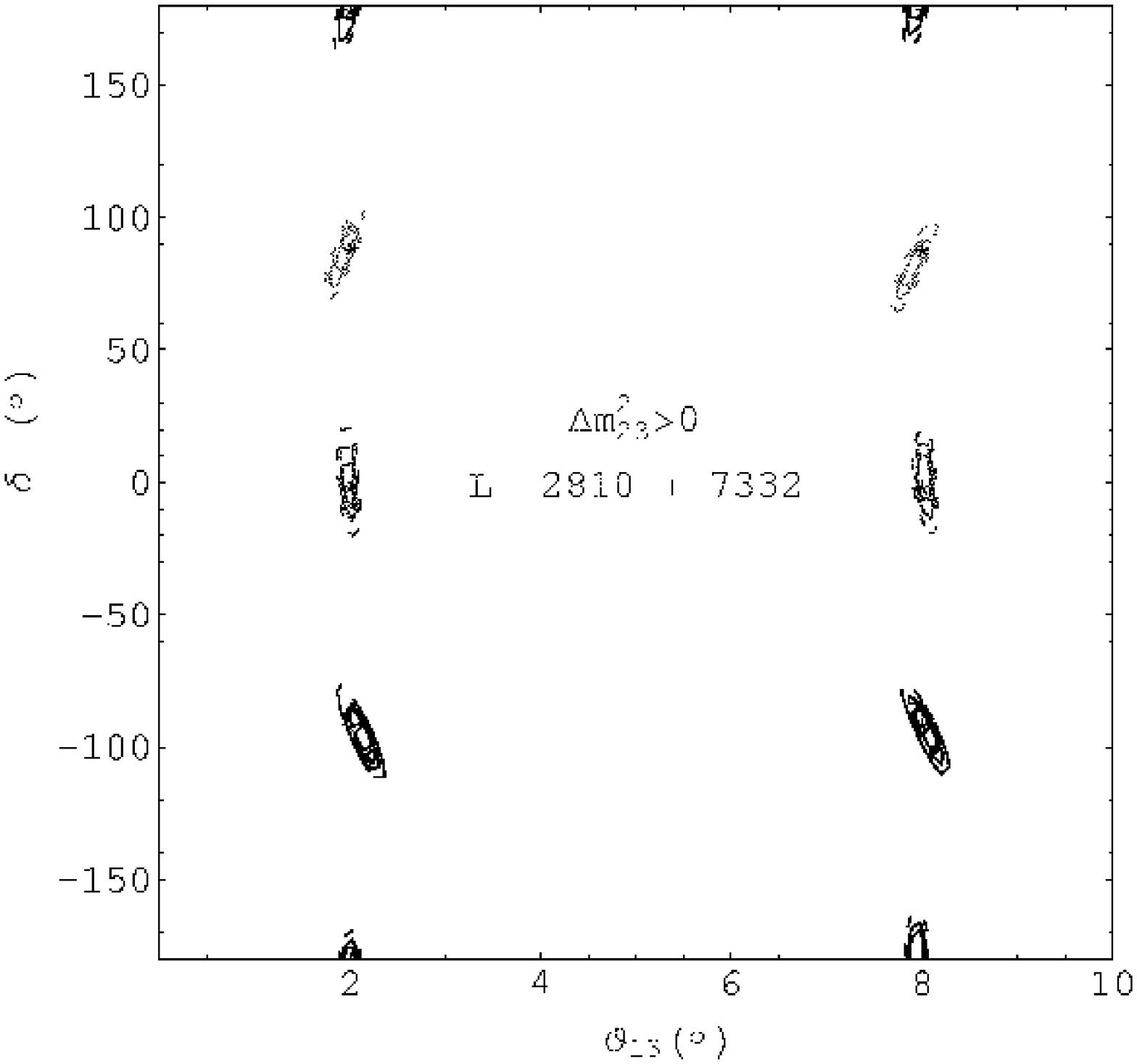,width=8cm}
\end{center}
\end{minipage}

\caption{\it For a wide choice of the parameters, in order to reduce 
correlations in a simultaneous fit to $\theta_{13}$ and 
$\delta$, the combination of the intermediate and the very long baseline
is preferred over that of the short and intermediate 
baselines~\cite{jordi}.}
\label{fig:correl2}
\end{figure}

There are fortunately several handles to deal with these degeneracies. The
position of the fake solutions is very sensitive to the value $L/E$ and
the presence or not of matter effects.  The first possibility is then to
have two baselines at the Neutrino Factory, since combining the
information of two experiments allows the exploration of a different $L/E$
regime, and matter effects behave differently at two baselines for the
disconnected sets of parameters. We see from Fig.~\ref{fig:correl2} that
the combination of the intermediate and the very long baseline eliminates
all degeneracies. The combination of the intermediate and a short
baseline, e.g., 732~km is not so good, because of the strong correlations
present in the shorter baseline.

\begin{figure}[tbh]
\begin{center}
\includegraphics*[width=10cm]{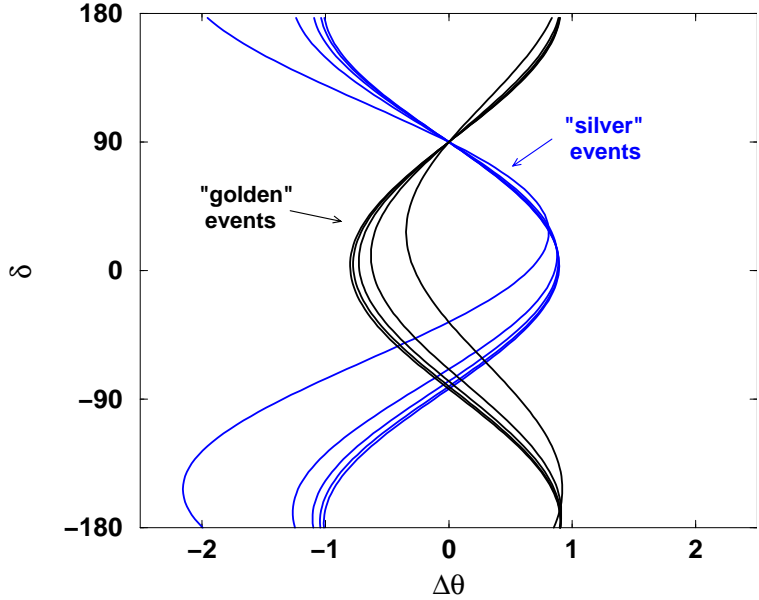}
\caption{\it Curves of equal oscillation probability in the 
$\delta, \Delta\theta$
plane (where $\Delta \theta$ is the difference between $\theta_{13}$ and
its nominal value of $5^\circ$), for $\nu_e\to\nu_\mu$ and
$\nu_e\to\nu_\tau$. We see that these curves, generated for
different energy bins, all meet at the nominal value of the 
parameters~\cite{donini}.}
\label{fig:silver}
\end{center}
\end{figure}

If a baseline of the order of 732 km is used in addition to one around
3000 km, it is more interesting to consider at the shorter baseline a
detector with capabilities to detect $\tau$'s from $\nu_e\to\nu_\tau$
oscillations. The measurement of this transition would be a great help in
resolving degeneracies, since the two processes $\nu_e \to \nu_\mu$ and
$\nu_e \to \nu_\tau$ give complementary information\cite{donini}. In fact,
the formulae for oscillation probabilities in vacuum of these processes
are very similar: the $\nu_e \rightarrow \nu_\tau$ probability can be
obtained from (\ref{eq:nuenumusimp})  by swapping $s^2_{23}
\leftrightarrow c^2_{23}$ and changing the sign of the $\delta-$dependent
terms. In Fig.~\ref{fig:silver}, we plot the curves of equal $\nu_e
\rightarrow \nu_\mu$ and $\nu_e \rightarrow \nu_\tau$ probabilities in the
($\Delta \theta_{13}\equiv \theta'_{13}-\theta_{13}$, $\delta'$) plane for
some fixed values of the true parameters:  $\theta_{13}=5^\circ, \delta =
90^\circ$. The different curves correspond to different energy bins. The
curves all meet only at the true values of the parameters:
$\delta'=90^\circ$ and $\Delta \theta=0$, while all the curves for the
wrong-sign muons are approximately degenerate.  This illustrates the
potential of the combination of golden and silver events to exclude the
fake solutions.

In \cite{donini} a detailed study of 
a realistic experimental setup has been presented. The detector considered
is an OPERA-like hybrid with lead and emulsion~\cite{operaproposal}. It has been shown that the $\tau$ vertex
recognition in the emulsion detector allows one to separate 
$\nu_e \to \nu_\tau$ from
$\nu_e \to \nu_\mu$ muon events, and to reduce strongly the effect of 
the charm decay
background.  The outcome of this study is that a 4~kt detector at $L =
732$~km in combination with the 40~kt magnetised iron detector at $L =
3000$~km would solve the intrinsic ambiguity for $\theta_{13} \geq
1^\circ$ and any value of $\delta$.

Another interesting approach to resolve degeneracies is the combination of 
information from the Neutrino Factory and a super-beam~\cite{olga}. The
synergy of the super-beam project with the Neutrino Factory has already
been discussed from the point of view of the accelerator infrastructure, 
since we saw
that several components needed for building a super-beam are the same as
those for the Neutrino Factory. Most likely, at the time of the operation
of this facility, results from the super-beam will already be available,
and the synergy between the two projects will emerge in the physics.  Due
to the very different $L/E$ regime - super-beams are designed to operate
around the oscillation maximum - and matter effects - which are much
smaller for super-beams due to the shorter baseline, the knowledge of the
super-beam results will help resolve the degeneracies. 

\begin{figure}[tbhp]
\begin{minipage}[h]{0.45\textwidth}
\begin{center}
\epsfig{file=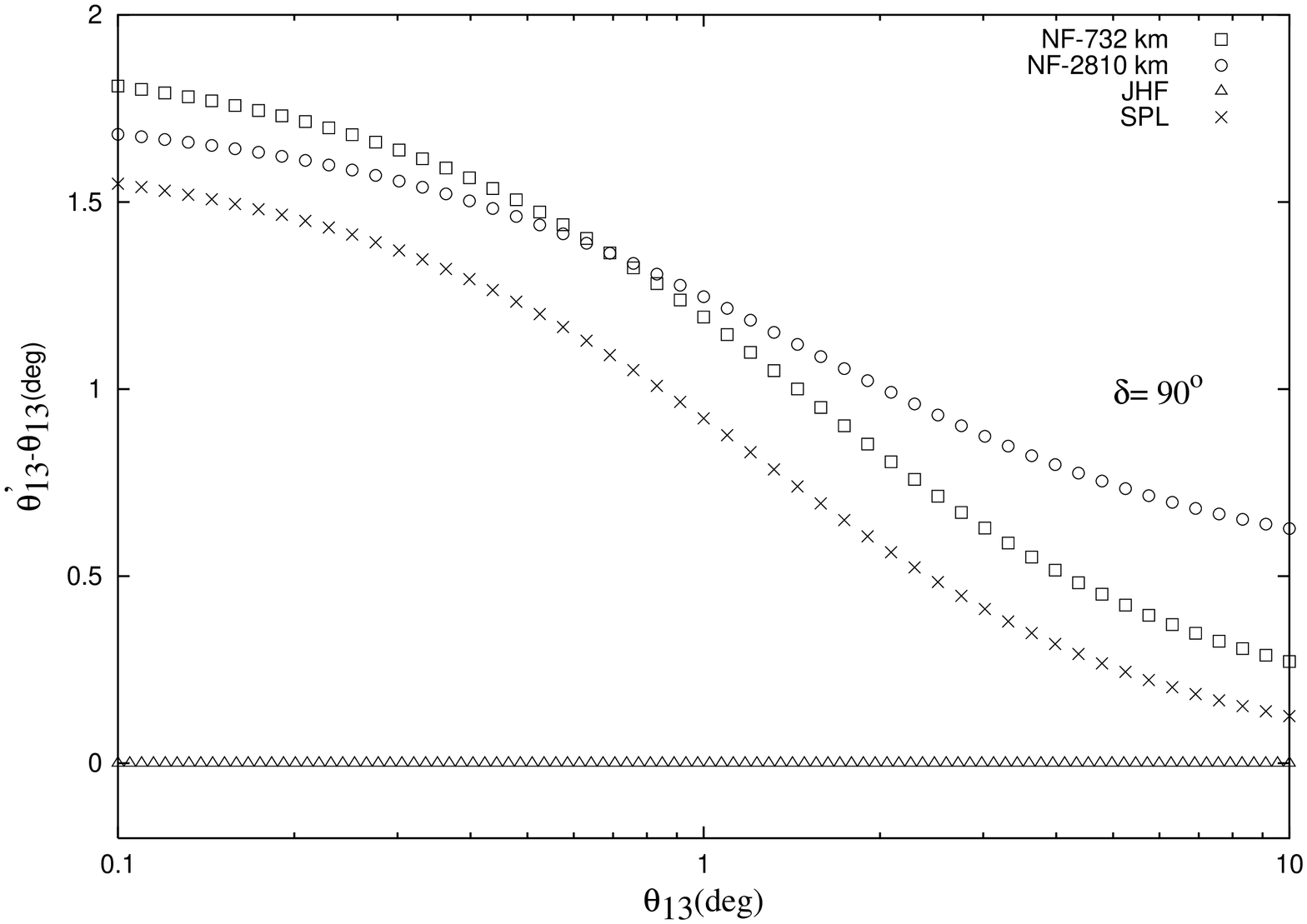,width=5cm,angle=270}
\end{center}
\end{minipage}
\vspace{.5cm}
\begin{minipage}[h]{0.45\textwidth}
\begin{center}
\epsfig{file=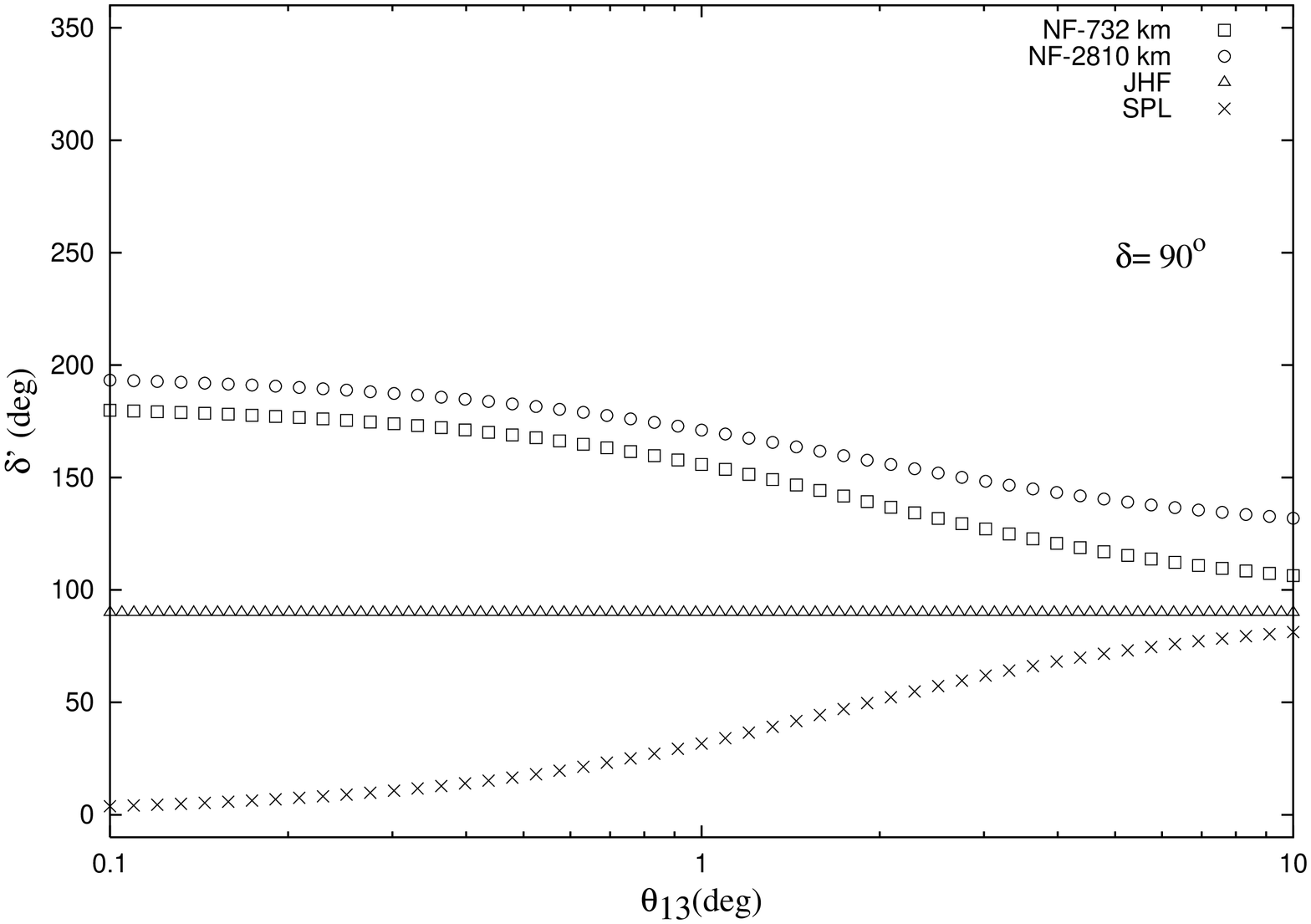,width=5cm,angle=270}
\end{center}
\end{minipage}
\caption{\it Values of $\theta_{13}'-\theta_{13}$ (left) and $\delta'$ 
(right) for 
the intrinsic fake solution for $\delta=90^\circ$. We see that fake solutions,
albeit present for all kinds of experiments, appear for different values 
of the parameters, and therefore a combination of experiments allows one 
to exclude them~\cite{olga}.}
\label{fig:degsb}
\end{figure}

Sufficiently powerful super-beams can search for CP violation in a
favourable corner of the parameter space. It was shown~\cite{olga} that a
combination of the results from the proposed CERN-Fr\'ejus super-beam and
a Neutrino Factory with a single baseline could eliminate almost
completely all degeneracies. The reason is easy to understand from
Fig.~\ref{fig:degsb}: the fake solutions, present in both experiments,
appear in different areas of the parameter space, and therefore are
excluded once the experiments are combined.

\begin{figure}[tbhp]
\begin{minipage}[h]{0.45\textwidth}
\begin{center}
\epsfig{file=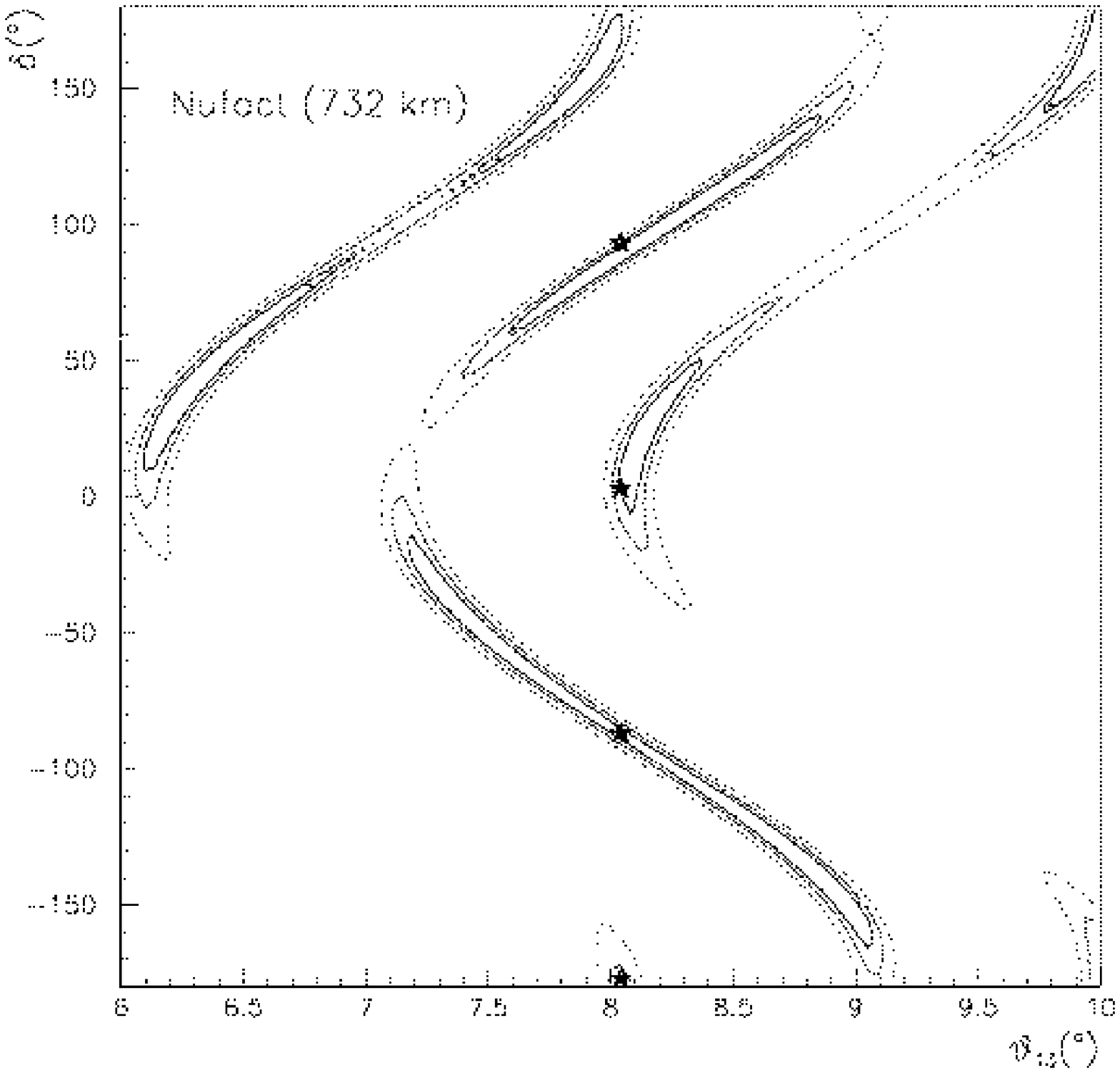,width=8cm}
\end{center}
\end{minipage}
\begin{minipage}[h]{0.45\textwidth}
\begin{center}
\epsfig{file=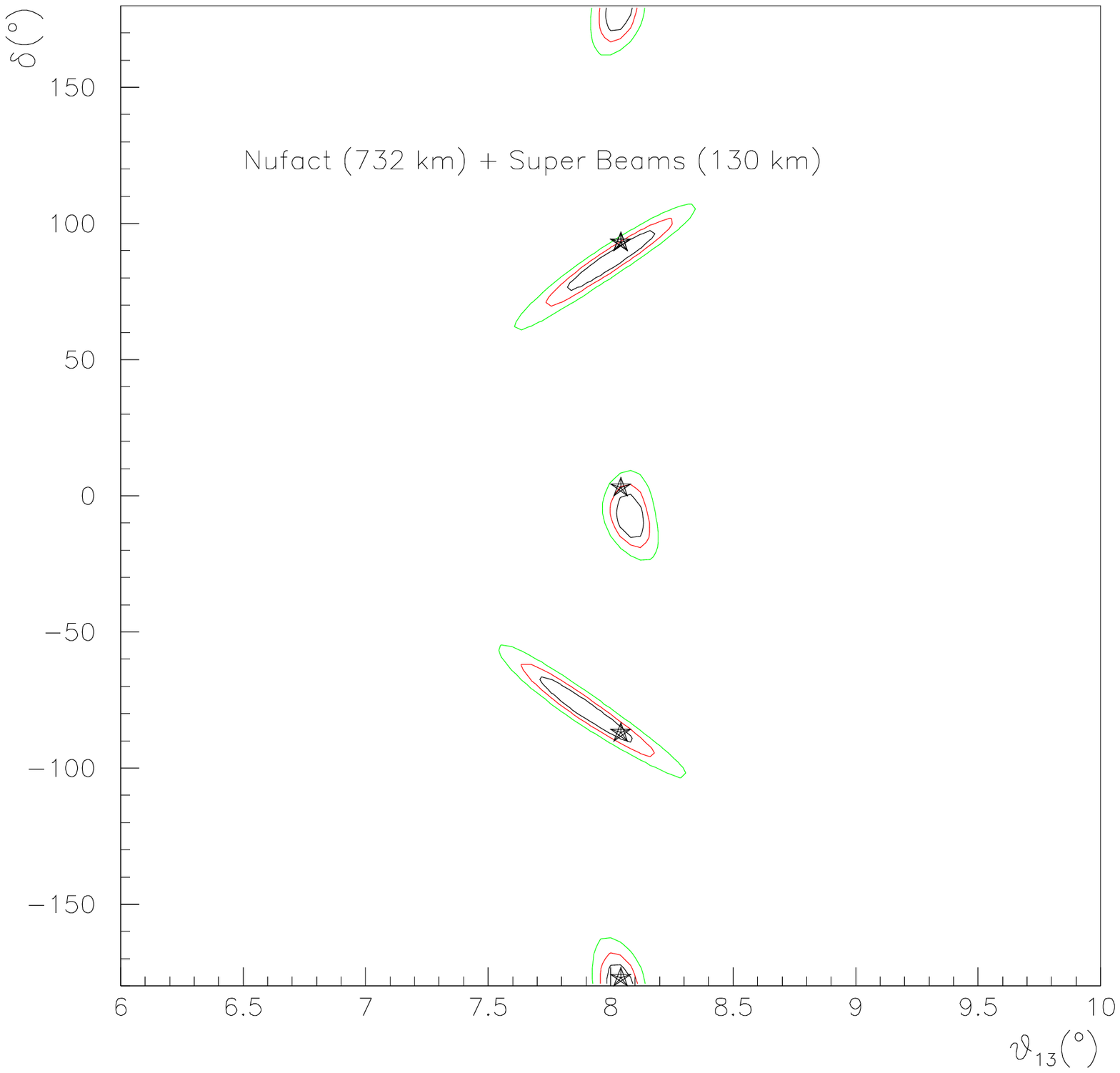,width=8cm}
\end{center}
\end{minipage}
\caption{\it Simultaneous measurements of $\theta_{13}$ and $\delta$ and of the
CP-violating phase $\delta$. The left plot is produced using only the 
information from a Neutrino Factory with an iron magnetised detector placed
at 732 km: there are clear correlations between the parameters and 
degenerate solutions. In the right plot, the Neutrino Factory is combined 
with the results from the super-beam, and both correlations and 
degeneracies disappear~\cite{olga}.} 
\label{fig:nufactsuperb}
\end{figure}

Combined with the results of the super-beam, even the baseline of 732~km,
normally too short to give definitive results on its own, can provide good
measurements of the oscillation parameters. This is shown by the contours
in the plane $(\theta_{13},\delta)$ in Fig.~\ref{fig:nufactsuperb}. We
already saw that a high-energy Neutrino Factory coupled with a detector
with high threshold for neutrino detection reaches its maximal sensitivity
for a baseline of about 3000~km. For a shorter baseline, the experiment
measures only the combination (\ref{eq:corr}) as we already discussed
(left plot in Fig.~\ref{fig:nufactsuperb}). The addition of the super-beam
information, which also is not able to resolve the degeneracies on its
own, leads to a very good determination of the two parameters as shown in
the right plot of Fig.~\ref{fig:nufactsuperb}.

This result is extremely important for the definition of the Neutrino
Factory design itself. It is common belief that an intermediate baseline
of about 3000 km is the best compromise between being long enough to have
the maximum of the oscillation at sufficiently high energy, and not too
far not to have the CP violation suppressed by matter effects. Since a
low-energy super-beam will probably be built before the Neutrino Factory,
the energy/baseline optimization of the latter will have to take into
account the combination with the super-beam results.

\subsubsection{T violation}

\begin{figure}[tbhp]
\begin{center}
\epsfig{file=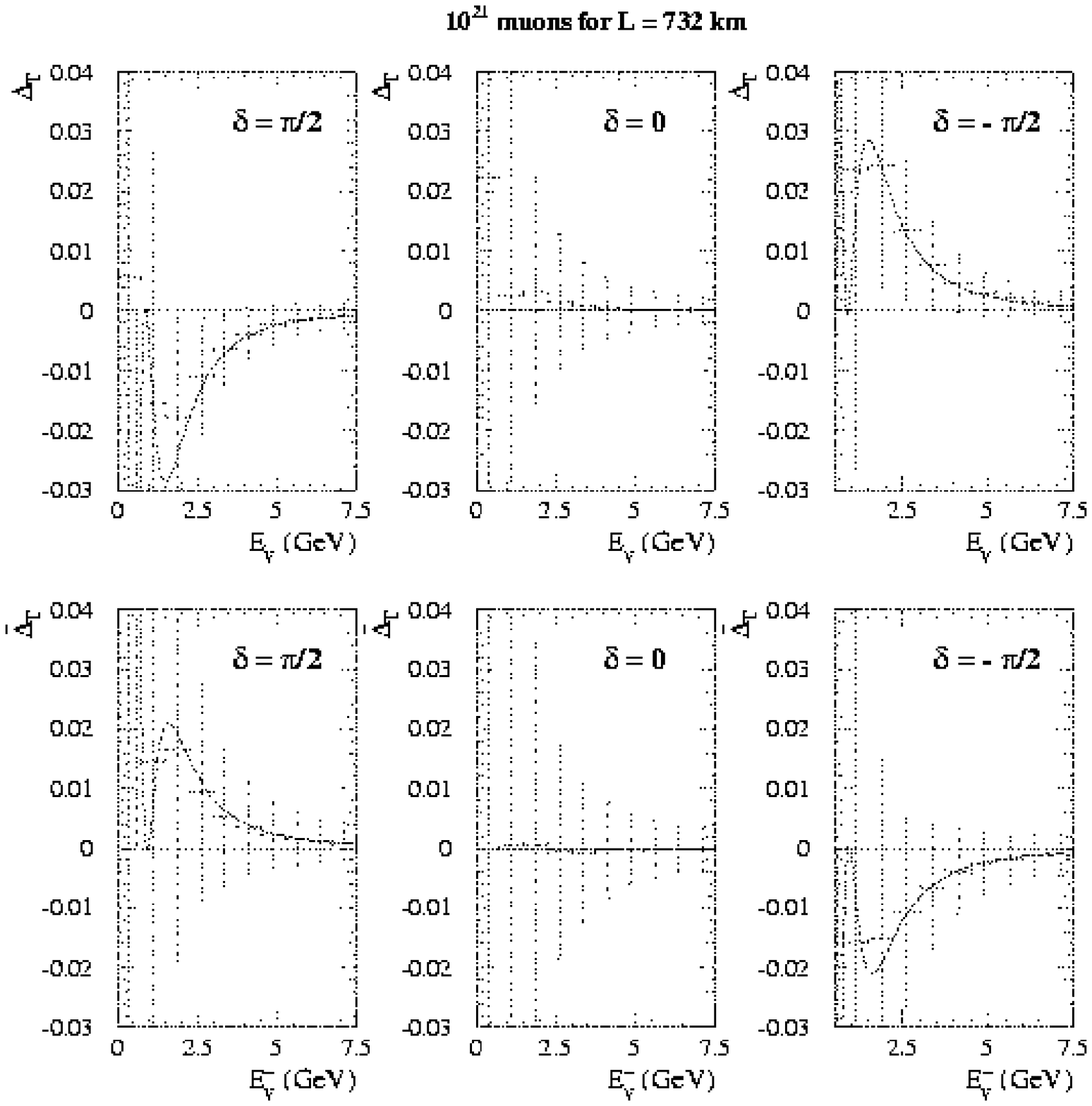,width=16cm,height=9cm,bb=50 20 560 750,clip}
\end{center}\vspace{-1cm}
\caption{\it Oscillation probabilities for T-conjugate processes measured 
with a 10~kt detector with electron charge identification capabilities. 
The error bars correspond to $10^{21}$ muons decays, for an electron 
charge efficiency of 20\% and misidentification of 0.1\%~\cite{cptoptim}.}
\label{fig:tviol}
\end{figure}

The problem of the neutrino-antineutrino asymmetry generated by
matter effects acting as a background to CP violation could be 
solved if, instead of the CP asymmetry, we
could measure the T asymmetry, i.e., the difference in oscillation
probability between the processes $\nu_e\to\nu_\mu$ and $\nu_\mu\to
\nu_e$. This is in principle possible at the Neutrino Factory, since if
the  machine is operated with two different beam polarities, both
processes (and even their complex conjugates) occur. The problem is
that, in order to discriminate oscillated events from the beam component 
of
the same flavour, the charge has to be measured for both leptons. While this is
possible in a relatively straightforward way for muon final states, the
measurement of the electron charge is much more complicated, due to the 
early
formation of electromagnetic showers. The possibility of measuring
the electron charge has been studied~\cite{cptoptim} for a magnetised 
liquid argon detector. A charge confusion at the per mille
level with an efficiency of about 10\% for electrons of less than 5~GeV 
energy should be reached with values of the magnetic field
exceeding 1~Tesla. In that case, it would be possible to measure directly
the two T-conjugated oscillation probabilities, and detect the difference
between the two (see Fig.~\ref{fig:tviol}) with a statistical
significance similar to that of the CP violation. Due to the impossibility
of detecting the charge of high-energy electrons with realistic values of
the magnetic field, an experiment looking for T violation would require a
low-energy, high-intensity muon beam.

\subsubsection{Search for $\nu_e\to\nu_\tau$}

We saw in the previous section that the detection of $\tau$ leptons could
be useful in resolving the degeneracies in the search for CP violation,
due to the different behaviour of the $\delta-$dependent term in the
oscillation probability. The detection of taus, both from
$\nu_\mu\to\nu_\tau$ and $\nu_e\to\nu_\tau$ oscillations, is also
extremely important as a test of the unitarity of the mixing matrix.
Comparing the yields from these channels to those expected from the
disappearance, it is possible to put strong limits on the conversion of
neutrinos into another species, for instance sterile neutrinos. If we call
$\alpha$ and $\beta$ the ratios of the oscillation probabilities
$\nu_\mu\to\nu_\tau$ and $\nu_e\to\nu_\tau$ with respect to their expected
value, the precisions on these parameters, with the assumptions made
in~\cite{physpot}, are shown in Table~\ref{tab:constraining}.

\begin{table}[tbhp]
  \begin{center}
    \begin{tabular}{|c|c|c|}\hline
      Baseline&$\Delta\alpha/\alpha$&$\Delta\beta/\beta$\\ \hline
      2900&2\%&25\%\\
      7400&2\%&20\%\\ \hline
    \end{tabular}
    \caption{\it Precision in the determination of the parameters $\alpha$ 
and 
$\beta$, that quantify the amount of $\nu_\mu\to\nu_\tau$ and 
$\nu_e\to\nu_\tau$ present in the data~\cite{physpot}.}
    \label{tab:constraining}
  \end{center}
\end{table}

\subsection{\it Search for New Physics at the Neutrino Factory}

As discussed earlier, it is not possible to accommodate all the present
experimental results within the standard three-family oscillation
framework. Up to now, we have assumed that the unconfirmed result from
LSND could be explained by some misunderstood experimental effect, rather
than by neutrino oscillations. We have already commented that Neutrino
Factory physics would be even richer if the LSND result is indeed due to
neutrino oscillations.

If the effect is due to some other interactions violating lepton flavour,
distinct from neutrino oscillations, again the Neutrino Factory would be
the ideal machine to study them. Because of the shape of the produced
neutrino spectrum, this machine would also be an ideal place to look for a
modification of the MSW potential for matter oscillations, since these
effects would mainly be visible in the high-energy part of the spectrum. 

Discussions of these two possibilities are presented in the following two
subsections. We note also that a deviation from the standard Lorentz
metric due to extra dimensions could be explored by deviations of the
neutrino velocity from the speed of light, which could in principle be
measurable if one equipped a far detector with RPCs having a time
resolution around 200~ps~\cite{AmVo}.

\subsubsection{New physics in short-baseline experiments}

The search for lepton-number-violating (LNV) muon decays with $\Delta 
L=2$, of the type:
\begin{eqnarray}
& \mu^-\ra e^- + \nue + \nul \label{eq:dk2}
\end{eqnarray}
where $ (\ell=e,\mu,\tau) $, would be possible by
looking for wrong-sign electrons, which would produce the 
following standard charged-current weak interactions in the 
detector:
\begin{eqnarray}
\nue + N \ra e^- + X \\
\nul + N \ra l^- + X.
\end{eqnarray}
We assume that $\mu^-$ are circulating in the ring, since the signature 
sought has two
neutrinos with negative helicity in the final state, 
so that one profit from the enhanced
cross sections, compared with the positive-helicity case. 

The signal for such anomalous muon decays does not depend on the distance
between the neutrino source and the detector, 
unlike the oscillation case, and
therefore is better studied with relatively small detectors
at short distances, where one benefit from the higher neutrino flux. 
A 10~ton fiducial mass Liquid Argon detector located
at a distance of 100~m from the muon storage ring was considered 
in~\cite{Bueno:2000jy} as a realistic case.

Because the signal sought consists in the appearance of final-state 
electrons, while standard events have positrons in the final state,
an experiment with electron charge discrimination 
is mandatory.
This is an experimental challenge, given 
the short radiation length in liquid argon.
Realistic efficiencies and experimental backgrounds
in a detector with the characteristics
of the ICARUS Liquid Argon imaging TPC\cite{Vignoli:2000yn} have been 
studied. The detector 
simulated was a medium-sized Liquid Argon vessel surrounded by a dipole 
magnet.

Table~\ref{tab:elesearch} shows the effects of the cuts applied for a
normalization of $10^{19}$ muon decays. To compute the expected number of
signal events, a branching probability of $2.5\times 10^{-3}$ was
considered, compatible with the LSND excess.

Given the low muon energies considered, most of the events will be
quasi-elastic. Therefore, a final state configuration containing an
electron and a reconstructed proton and no additional hadronic particles
is required. These criteria reduce the quas-ielastic background (where we
expect a neutron rather than a proton in the final state) by almost three
orders of magnitude, while keeping more than $50\%$ of the signal. The
neutral-current background, where electron candidates come from $\pi^0$
conversion, is in general soft. After a cut on the electron candidate
momentum, this kind of background becomes negligible. Preliminary studies
indicate that, for a Liquid Argon detector inside a 1~T magnetic field, a
fit to the direction of the electromagnetic shower could provide a good
determination of the charge of leading electrons.
Table~\ref{tab:elesearch} shows that applying loose criteria in the
determination of the lepton charge is enough to eliminate $\anue$
charged-current background.

\begin{table}[ht]
\begin{center}
\begin{tabular}{|c|c|c|}
\hline
Cuts & $\nue$ CC & $\anue$ CC \\
\hline\hline
 \multicolumn{3}{|c|}{$E_{\mu^-}=2\rm\ GeV$}\\
\hline
Initial & 540 & 62500 \\
One proton & 367 & 11000 \\
No pions & 323 & 100 \\
$E_e >1$ GeV & 103 & 17 \\
Candidate charge & 21 & 0.4 \\
\hline\hline
 \multicolumn{3}{|c|}{$E_{\mu^-}=5\rm\ GeV$}\\
\hline
Initial & 5290 & 802000 \\
One proton & 3390 & 212160 \\
No pions & 2112 & 495 \\
$E_e >3$ GeV & 351 & 163 \\
Candidate charge & 71 & 4 \\
\hline\hline
 \multicolumn{3}{|c|}{$E_{\mu^-}=1\rm\ GeV$}\\
\hline
Initial & 76 & 6300 \\
One proton & 53 & 529 \\
No pions & 48 & 8 \\
$E_e >0.2$ GeV & 43 & 4 \\
Candidate charge & 10 & 0.1 \\
\hline\hline
\end{tabular}
\caption{\it The effects of cuts on background and signal. We assumed
a negative muon ring energy $E_{\mu^-}$ of
1, 2 and 5~GeV and a total of $10^{19}$ standard
decays. The detector was assumed to be located 100~m from the center of 
the straight section. The lepton-number-violating decay was assumed to 
have the branching probability $2.5\times 10^{-3}$. We denote by
$E_e$ the energy of the identified electron in the event.}
\label{tab:elesearch}
\end{center}
\end{table}

\subsubsection{New physics in long-baseline experiments}

Exotic physics effects may appear in neutrino production, neutrino
interactions or neutrino propagation in matter. Whilst the first two cases
are better studied in a near detector, benefiting from the higher flux,
the effects of new physics on matter propagation are better observed in a
long-baseline experiment, and might even interfere in the interpretation
of long-baseline oscillation experiments.

We recall that the
standard MSW effect gives rise to diagonal contributions to the
neutrino squared-mass matrix proportional to the neutrino energy.
A flavor-changing neutrino
interaction would give rise to a non-diagonal term which, although
smaller than the diagonal ones, would also grow with the neutrino
energy. The effective neutrino squared-mass matrix $M_{eff}^2$ would 
become
\begin{equation}
  \label{eq:MM}
  M_{eff}^2 = U 
  \begin{pmatrix}
    0 & 0 & 0 \\
    0 & \Delta m^2_{21} & 0 \\
    0 & 0 & \Delta m^2_{31}
  \end{pmatrix} U^\dagger
  + 2 E V
  \begin{pmatrix}
    1+\epsilon_{ee} &
    \epsilon_{e\mu} &
    \epsilon_{e\tau}  \\
    \epsilon_{\mu e} &
    \epsilon_{\mu\mu} &
    \epsilon_{\mu\tau}  \\
    \epsilon_{\tau e} &
    \epsilon_{\tau\mu} &
    \epsilon_{\tau\tau} 
  \end{pmatrix} \; ,
\end{equation}
where $E$ is the neutrino energy, $U$ is the neutrino mixing matrix in the
usual parameterization, and $\Delta m^2_{21}>0$ is the smaller 
squared-mass difference. In the standard case, all the $\epsilon$ 
parameters are equal to zero.

A model-independent limit on $\epsilon_{\tau\mu}$ can be inferred from
atmospheric neutrino data~\cite{osc_anal_atm}: $\epsilon_{\tau
\mu}\lesssim 0.05$. Significant limits on the single
$\epsilon^{e,u,d}$ quantities can be obtained if one assumes that the new 
physics operators originate from SU(2)$_W$-invariant operators. Then
experimental bounds on charged-lepton processes
imply~\cite{Bergmann:98a,Bergmann:99a,Bergmann:00a}
\begin{align}
  \epsilon^e_{\mu e} & \lesssim 10^{-6} &
  \epsilon^e_{\tau \mu} & \lesssim 3\cdot 10^{-3} &
  \epsilon^e_{\tau e} & \lesssim 4\cdot 10^{-3} \\
  \epsilon^{u,d}_{\mu e} & \lesssim 10^{-5} &
  \epsilon^{u,d}_{\tau \mu} & \lesssim 10^{-2} &
  \epsilon^{u,d}_{\tau e} & \lesssim 10^{-2} \; .  
\end{align}
Since SU(2)$_W$ is broken, the limits above can be evaded. In the case of 
the exchange
of a SU(2)$_W$ multiplet of bosons with SU(2)$_W$ breaking 
masses~\cite{Bergmann:99a,Bergmann:00a}, the 
the SU(2)$_W$ breaking masses can relaxed the limits by a factor of about 
7, without a conflict with the electroweak precision data.

\begin{figure}[tbhp]
\begin{minipage}[h]{0.47\textwidth}
\begin{center}
\epsfig{file=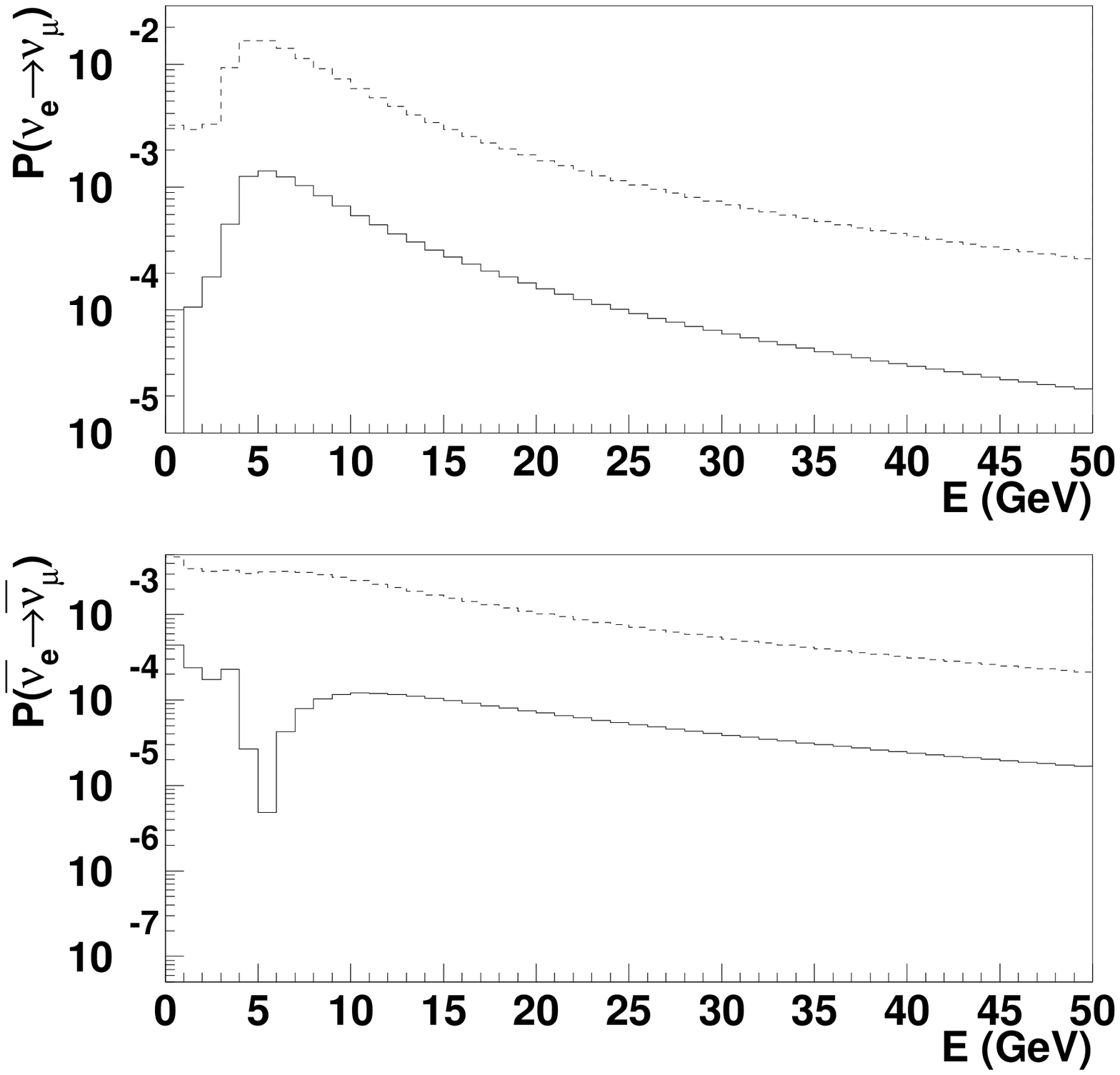,width=8cm}
\end{center}
\caption{\it The $\nu_e\rightarrow\nu_\mu$ oscillation probability in the 
standard
MSW model (full line) and in the presence of flavour-changing interactions
(dashed line), for $\sin^2 2\theta_{13}=0.001$ and $\epsilon_{\tau 
e}=0.07$~\cite{CampRomanino}.}
\label{fig:probem}
\end{minipage}
\hfill
\begin{minipage}[h]{0.47\textwidth}
\begin{center}
\epsfig{file=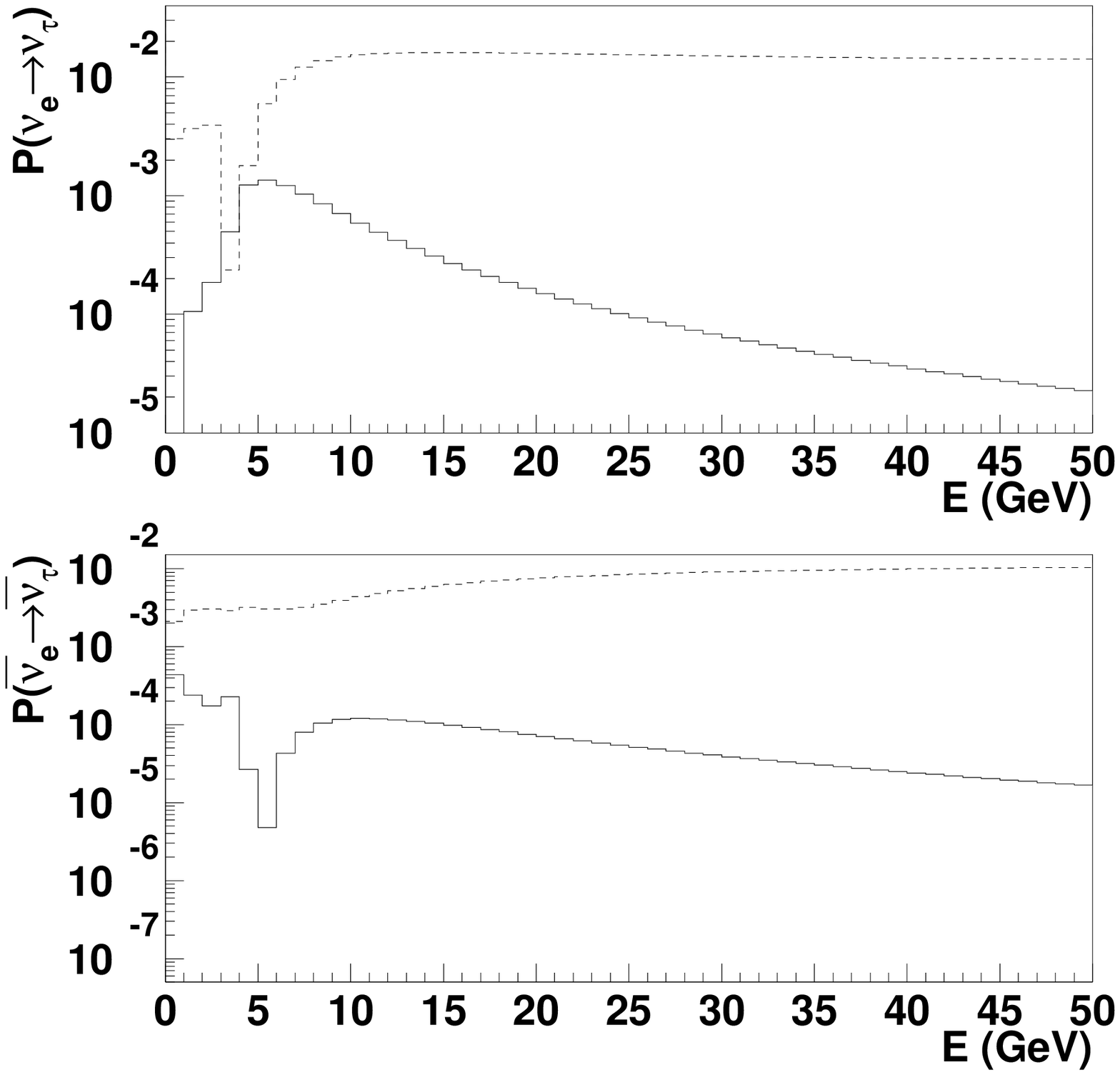,width=8cm}
\end{center}
\caption{\it The $\nu_e\rightarrow\nu_\tau$ oscillation probability in the 
standard
MSW model (full line) and in the presence of flavour-changing interactions
(dashed line), for $\sin^2 2\theta_{13}=0.001$ and $\epsilon_{\tau e}=0.07$.}
\label{fig:probet}
\end{minipage}
\end{figure}

In general, the interactions involving the third family are far less
constrained than those involving the first two. The constraint on
$\epsilon_{\mu e}$ is so strong that no neutrino experiment would be able
to improve it, since even at the limit the contribution it gives to
neutrino oscillation is negligible. On the other hand, a value of
$\epsilon_{\tau e}$ close to the experimental bound affects in a
significant way both $\nu_e\to\nu_\mu$ and $\nu_e\to\nu_\tau$
oscillations, as seen in Fig.~\ref{fig:probem}.

The probability for oscillation into muons gets multiplied by an almost
energy-independent factor. The presence of new physics leads to confusion
with the case of normal oscillations, and an experiment could be led to
infer a value of $\sin^2 2\theta_{13}$ higher than the true
quantity~\cite{Huber:01b}. However, the strong rise of the
$\nu_e\to\nu_\tau$ probability, a clear indication of the presence of
nonstandard interactions, can be seen from the decay $\tau \to \mu$, even
with a detector only able to measure the muon momentum. The spectral shape
of the muons will be so different that it is possible to distinguish the
case of normal oscillation from those of new physics in large region of
the parameter space~\cite{CampRomanino}, inferring the value of
$\theta_{13}$ from event counting, and then comparing the observed
spectrum with the theoretical one for the estimated $\theta_{13}$. It has
been shown that, for $\sin^2 2\theta_{13}\gsim 10^{-3}$, or
$\epsilon_{13}\gsim 10^{-2}$, the spectral differences are such that the
two cases can be clearly separated.

\newpage
\def\beq{ \begin{equation}}
\def\eeq{\end{equation} }
\def\bea{\begin{eqnarray}}
\def\eea{\end{eqnarray}}
\def\r{\gamma}
\def\eps{\epsilon}
\def\be{\bar{\epsilon}}
\def\d{\delta}
\def\a{\alpha}
\def\b{\beta}
\def\n{\nu}
\def\la{\lambda}
\def\ra{\rightarrow}
\def\k{m_{eff}}
\def\e{\epsilon}
\def\p{\pi}
\def\th{\theta}
\hyphenation{author another created financial paper re-commend-ed Post-Script}

\section{SUMMARY AND OUTLOOK}

Neutrino physics has been making great strides during the past few years.  
It has provided us with a unique window on physics beyond the Standard
Model, and significant progress has recently been made in measuring the
parameters of neutrino oscillations. The atmospheric mixing angle
$\theta_{23}$ seems to be near maximal, and a large value of the solar
mixing angle $\theta_{12}$ is now also favoured. These notable differences
from the small mixings observed previously among quarks whet our appetites
for further information on neutrino oscillations. These may cast light on
fundamental aspects of physics at the grand unification scale where all
the interactions may be unified. Just as the past generation has witnessed
the establishment of the Standard Model and quark physics, neutrinos and
physics beyond the Standard Model seem destined to be at the forefront of
experimental and theoretical physics for the coming generation.

Experiments now underway or being prepared will advance our knowledge of
neutrino masses and oscillations in many ways. They will have improved
sensitivity to neutrino masses and neutrinoless double-$\beta$ decay. They
will establish whether solar neutrino oscillations occur in the LMA, LOW,
SMA or VAC region of parameters. They will clarify whether the LSND
anomaly is due to neutrino oscillations. They will presumably confirm the
existence of oscillation patterns, and that atmospheric $\nu_\mu$
oscillate mainly into $\nu_\tau$. They will provide more precise
measurements of the solar and atmospheric neutrino oscillation parameters,
including the mass-squared differences, $\theta_{12}$ and $\theta_{23}$.
They will have improved sensitivity to $\theta_{13}$.

The neutrino physics communities in Europe, Japan and the United States
are all considering actively their options for the following step in
oscillation physics. These should be addressed towards the many questions
that are unlikely to be resolved in the approved round of experiments.  
These include pinning down $\theta_{13}$, seeing intrinsically
three-generation effects in mixing, searching for CP violation, and fixing 
the sign of
$\Delta m_{23}^2$ via matter effects. Short-term options include a JHF
neutrino beam in Japan and off-axis long-baseline experiments in the
United States and Europe. Specific experiments can be devised which
address many of the individual open questions, but there is no guarantee
that any or all of them will be answered by this next round.

We have presented in this report a set of options for Europe that would
provide powerful tools for addressing all the open questions, by providing
forefront experimental possibilities. 

\begin{itemize}

\item{A first step could be an intense proton source at CERN, the SPL,
that would provide a low-energy neutrino beam of unprecedented power. This
study also includes a summary report on the SPL design. The Fr\'ejus
tunnel is at just the right distance to observe oscillation effects with
an SPL beam, and placing there a megaton-class water \v{C}erenkov detector
such as UNO would be an attractive option. The SPL-UNO combination would
have a chance of observing CP violation in neutrino oscillations.}

\item{A second possibility that has emerged from this study group is the
concept of $\beta$ beams, coming from the decays of radioactive isotopes
produced by the SPL. Many accelerator questions vital for the viability of
this concept remain to be investigated, but it may provide an exciting
alternative option for studying CP violation in neutrino oscillations, 
and has interesting synergies with other options considered here.}

\item{The best long-term option for neutrino physics is the neutrino
factory based on muon decays in a storage ring. This offers unequalled
reach for all the basic open questions: the magnitude of $\theta_{13}$, CP
violation and matter effects. Several important questions about the design
of a neutrino factory remain to be answered, notably muon cooling. Even
without cooling, there is an exciting programme of experiments with slow
or stopped muons that complements the neutrino oscillation programme
described here. If muon cooling can be perfected beyond the requirements 
of the neutrino factory, a muon collider may become feasible, offering 
exciting options in Higgs physics and/or high-energy lepton collisions, as 
discussed elsewhere.}

\end{itemize}

This report is not the place for detailed studies of the accelerator
questions that must be answered before these exciting physics options
could become realities. However, we hope that this report communicates
sufficient enthusiasm for neutrino physics to provide the motivation
needed for research to make these options feasible.

\end{document}